\documentclass[12pt,hyperref]{book}
\usepackage[letterpaper, top=2.5cm, bottom=2.5cm]{geometry}
\usepackage{amsmath}
\usepackage{amssymb}
\usepackage{bm}
\usepackage{CJK}

\usepackage{graphicx}
\usepackage{cancel}
\usepackage{xcolor}
\usepackage{listings}
\usepackage{boxedminipage}
\usepackage{url}
\usepackage{indentfirst}
\usepackage{titletoc}
\usepackage{titlesec}
\usepackage[rotateright]{rotating}
\usepackage{lscape}
\usepackage{multirow}
\usepackage{makecell}  
\titleformat{\chapter}[display]{\Huge\bfseries}{Chapter\,\thechapter\,}{1em}{}

\usepackage[CJKbookmarks=true,
pdftex, pdfstartview=FitH, bookmarksnumbered=true, hyperindex=true, 
colorlinks=true, pdfborder=001, citecolor=blue, linkcolor=red,
anchorcolor=green,bookmarks={false}
]{hyperref}

\usepackage{fancyhdr}
\usepackage{makeidx}
\makeindex

\begin{document}

\lstset {
    basicstyle=\footnotesize, 
    numbers=left,
    numberstyle=\tiny,
    keywordstyle=\color[RGB]{0, 0, 255},
    commentstyle=\color[RGB]{0, 128, 0},
    frame=shadowbox,
    rulesepcolor=\color{red!20!green!20!blue!20},
    showspaces=false,
    showstringspaces=false,
    extendedchars=false,
    showtabs=false,
    tabsize=4, breaklines
}

\begin{CJK*}{UTF8}{gbsn}
\pagestyle{empty}

\title{\textbf{\Huge{Introduction to Fusion Ignition Principles: Zeroth Order
Factors of Fusion Energy Research\footnote{Note: This English version is primarily a translation generated using ChatGPT based on the original Chinese book, ``谢华生，聚变点火原理概述，中国科学技术大学出版社，合肥，2023" and can be cited as "Huasheng Xie, Introduction to Fusion Ignition Principles, USTC press, Hefei, 2023." For any inaccuracies or ambiguities, please refer to the original Chinese version for clarification.}}\\\large{ 聚变点火原理概述：聚变能源研究的零级量 }}}
\author{Huasheng XIE (谢华生)\footnote{Contact: huashengxie@gmail.com, https://github.com/hsxie/fusionbook, http://hsxie.me/fusionbook}\\ENN (新奥能源研究院)}

\date{\today}
\maketitle

\renewcommand{\chaptername}{}
\renewcommand{\chaptermark}[1]{\markboth{Chapter\,\thechapter\,\ #1}{}}
\renewcommand{\sectionmark}[1]{\markright{\textsection~\thesection\ #1}{}}

\makeatletter
\renewcommand*{\@makechapterhead}[1]{
  \vspace*{50pt}
  {\Huge\bfseries Chapter\,\thechapter\,\ #1}
  \vspace{40pt}
}
\makeatother

\renewcommand{\figurename}{Figure}
\renewcommand{\tablename}{Table}
\renewcommand{\appendixname}{Appendix}
\renewcommand{\bibname}{Reference}
\renewcommand{\indexname}{Index}

\renewcommand*{\contentsname}{Content}
\renewcommand*{\listfigurename}{List of Figures}
\renewcommand*{\listtablename}{List of Tables}

\chapter*{Book summary}
The physical goal of fusion energy research is to confine fusion fuel in a certain way so that the energy released from fusion exceeds the energy consumed to sustain the fusion process, thereby achieving economically viable energy production. Based on fundamental physics, this work focuses on the physics of the core region in fusion energy reactors, discussing the relevant limiting factors and parameter ranges. By examining the zeroth-order quantity system of the fundamental principles of fusion ignition, we review the current progress and challenges in fusion energy research. These challenges encompass various aspects such as physics, engineering, materials, and economics, providing better insights for the future development of fusion as an energy source.

\vspace*{50mm}

{\bf About the author:} Huasheng Xie is a researcher born in Hengyang, Hunan Province, China in 1987. He obtained a Bachelor's degree in Physics from Zhejiang University in 2010 and a Ph.D. in Plasma Physics in 2015. He worked as a postdoctoral researcher at Peking University from October 2015 to February 2018. Currently, he serves as the Chief Scientist of Fusion Theory and Simulation at the ENN Energy Research Institute. His research primarily focuses on theoretical and simulation studies of basic plasma and fusion plasma physics. He has conducted comprehensive research on the history and schemes of fusion energy. He has published more than 50 SCI papers, including over 30 papers as the first author or corresponding author in journals such as Physical Review Letters, Computer Physics Communications, Nuclear Fusion, and Physics of Plasmas. He has also authored the book "Introduction to Computational Plasma Physics" (Science Press, 2018).

 \thispagestyle{empty}

\newpage

\setcounter{page}{0} \pagenumbering{roman}

\pagestyle{fancy}

\chapter*{Preface}
\addcontentsline{toc}{chapter}{Preface}

The realization of fusion energy has always been "thirty years away." So, what are the missing pieces? This requires a systematic assessment to identify the problems and help find the direction for tackling them. This book focuses on the zeroth-order quantity of the basic principles of fusion ignition, conducting quantitative calculations and comparisons. All the data and codes of the established models are open and can be accessed at (http://hsxie.me/fusionbook, https://github.com/hsxie/fusionbook), allowing anyone to verify and reproduce the results.

"Zeroth-order quantity" comes from mathematical perturbation expansion or Taylor expansion, which is commonly used in physics research and familiar to fusion physicists. It represents the primary influencing factor, or the main contradiction, while first-order and second-order quantities become significant after addressing the zeroth-order quantity and can be considered as secondary factors. First-order and second-order quantities in fusion research, such as instability and turbulent transport, are often too complex and require excessive efforts from fusion researchers, which in turn leads to a neglect of a macroscopic global perspective in addressing the issues in fusion energy research. The purpose of this book is to identify the key factors in fusion energy research, which can usually be understood without complex mathematical physics foundations. When assessing the feasibility of a fusion technology pathway, the priority should be given to determining whether these key factors can be addressed.

This book originated from the question that the author wanted to clarify several years ago: "Is fusion energy really feasible and how is it feasible?" After studying from various perspectives, the conclusion was eventually reached: "It is challenging, but it is feasible." This book is a brief summary of the conclusion and the process. Special thanks are due to the research environment provided by ENN Energy Research Institute, as well as colleagues who participated in discussions and research on fusion pathways, and colleagues from domestic and international visits and exchanges. I would like to express my sincere gratitude to colleagues such as Li Yang, Michel Tuszewski, Bai Yukun, Chen Bin, Luo Di, Zhao Hanyue, Cai Jianqing, Deng Bihe, Guo Houyang, Liu Minsheng, Chen Peipei, and others for numerous discussions and joint analyses, as well as comments from senior experts such as Wang Xiaogang, Xu Guosheng, Feng Kaiming, Wu Songtao, Dong Jiaqi, and Hu Xiwei. It should be noted that the content provided in this book is a basic outline that analyzes the difficulty and feasibility but does not draw absolute conclusions. Based on different considerations and inclinations, different individuals may come to their own conclusions. For example, the viewpoint in this book tends to prioritize the commercialization of deuterium-deuterium-helium-3 fusion, while the author's organization, ENN, has chosen proton-boron fusion as its research and development goal based on environmental considerations. The industry mainstream, on the other hand, prioritizes deuterium-tritium fusion based on the ease of physics. These different choices and inclinations are not contradictory.

\vspace*{5mm}\vspace*{5mm}
\begin{flushright}
Huasheng Xie\\
ENN Energy Research Institute\\
July 2022
\end{flushright}

\chapter*{Recommendation}
\addcontentsline{toc}{chapter}{Recommendation}

Dr. Huasheng Xie's book, "Introduction to Fusion Ignition Principles" (also known as "Zeroth-Order Factors of Fusion Energy Research"), is a new work that bridges the gap between academic monographs and popular science literature. It provides a comprehensive understanding of the fundamental principles and necessary conditions for ignition and power generation in nuclear fusion energy research. It is suitable for educated general readers seeking knowledge of the basic principles of nuclear fusion energy ignition and generation, as well as a professional reference for scientists engaged in nuclear and fusion physics research.

Nuclear fusion energy has long been regarded as the primary avenue for fundamentally addressing humanity's energy needs. Especially in the face of significant energy and environmental challenges today, the importance of nuclear fusion energy development is increasingly evident. It has attracted widespread attention and is showing promising commercial prospects for fusion power generation. However, the attention received by fusion research has also led to many unscientific claims and misconceptions. The public, corporations, and researchers all need a deeper understanding of the fundamental principles and necessary conditions for fusion ignition. The publication of an academic work that explores fusion ignition principles and conditions with popular science elements is not only necessary but also timely.

Dr. Huasheng Xie's manuscript provides a comprehensive exposition of the fundamental principles of fusion ignition, particularly by summarizing the essential conditions that must be achieved for fusion ignition (the key issues of zeroth-order factors). It offers unique analysis and summarization of the scientific validity and feasibility of fusion power generation. This manuscript stands out among various fusion research books both domestically and internationally and is expected to receive great attention and appreciation from readers.

The manuscript is well-structured, clear in its explanations, and strikes a balance between academic rigor and accessibility. It is an outstanding work that is rarely seen. I highly recommend that this manuscript be published as soon as possible.

\begin{flushright}
Member of the National Magnetic Confinement Fusion Expert Committee

Professor at Harbin Institute of Technology

Xiaogang Wang
\end{flushright}

\vspace*{10mm}

Nuclear fusion energy is the future's sustainable energy source, and scientists worldwide have dedicated over half a century to arduously exploring this field. The implementation of the International Thermonuclear Experimental Reactor (ITER) has provided a glimpse of the dawn of commercial fusion energy applications. Controlled nuclear fusion energy technology presents the most significant technological and scientific challenges in human history. Even if the ITER project achieves its anticipated success, there remain numerous critical scientific and technological issues that require continued in-depth research, such as materials resilient to neutron irradiation, tritium self-sufficiency, and plasma control during burning.

The utilization of nuclear fusion energy requires the design and construction of fusion reactors. In this process, the physics design of fusion reactors is the first and crucial step. This book provides a detailed exposition and analysis of zeroth-order factors in the physics design of fusion reactors. It presents the fundamentals and principles of physics design, including the basics of fusion reaction physics and the Lawson criterion, parameter design selection for magnetically confined fusion reactor cores, parameter ranges, and design limitations for different types of fusion reactors. The book summarizes the latest achievements in this field from relevant literature, offering a systematic and comprehensive coverage with clear and accessible language.

This book serves as a valuable reference for senior undergraduate and graduate students, teachers, and researchers in relevant university programs, as well as for scientific and technological professionals engaged in controlled nuclear fusion research.

Dr. Huasheng Xie, the author, is currently the Chief Scientist of Fusion Theory and Simulation at ENN Energy Research Institute. He has long been involved in theoretical and numerical simulation work on fusion plasma physics, including fusion reactor physics, and has published a monograph on computational plasma physics. He is an outstanding expert in the field of fusion reactor physics design.

I strongly recommend the publication of this book.

\begin{flushright}
Former Deputy Chief Engineer, Institute of Fusion Science

Researcher, Southwest Institute of Physics, CNNC

Kaiming Feng
\end{flushright}

\vspace*{10mm}

In the past 20 years, particularly in the last decade, the development of fusion energy has gained significant momentum worldwide. The launch of the ITER project and the continuous progress in key fusion technologies have fueled the global enthusiasm for fusion energy development. Private capital has also flocked to this field, leading to the emergence of numerous technology startups focusing on fusion energy development. The scale of investment in fusion energy has been continually surpassed. However, most of these fusion energy technology startups have intentionally or unintentionally avoided the widely recognized most promising Tokamak concept for achieving fusion energy utilization in the earlier stages. Instead, they have focused more on alternative approaches with overall engineering simplicity compared to the complexities of the Tokamak system. Examples of such alternative approaches include field-reversed configuration, magnetized target fusion, spherical tokamaks, and target plasma compression. The consideration of these alternative systems appears to be driven by the complexity associated with the Tokamak system.

Over the past 20 years, the Advanced Research Projects Agency-Energy (ARPA-E) of the U.S. Department of Energy has provided funding to multiple private startup companies to advance the development of novel fusion technologies with high potential and impact. While the exploration of new fusion systems has been advocated by governments and private capital worldwide, the scientific feasibility of these new fusion systems remains to be proven.

Currently, there is a lack of reports that systematically analyze and compare the key parameters of fusion plasma for different fusion pathways, starting from the most fundamental elements of fusion plasma. This book fills this gap by starting from the fundamental elements of fusion plasma physics and presenting the most crucial arguments for different fusion systems under development. It helps provide a basic assessment of the scientific feasibility or difficulty of different fusion pathways.

The book does not delve into highly complex plasma physics theories, but it offers a comprehensive analysis and discussion of different fusion technology pathways from the perspective of fundamental principles and logical reasoning. It can serve as a valuable resource for researchers involved in fusion energy development to gain a overview understanding of the basic principles of fusion. It is also an excellent popular science book for the general public interested in fusion energy.

\begin{flushright}
Researcher and Former Deputy Director of the Institute of Plasma Physics, Chinese Academy of Sciences

Former Deputy Director-General and Chief Engineer of the ITER Organization Tokamak Engineering Department

Songtao Wu
\end{flushright}

\vspace*{10mm}

Nuclear fusion energy is considered one of humanity's most promising ultimate energy sources. Over the past half-century, major countries worldwide have invested significant human and material resources in controlled nuclear fusion research, accumulating abundant knowledge and experience, and achieving remarkable accomplishments. However, to realize the commercial application of nuclear fusion energy, humanity still faces several formidable challenges. This book provides a systematic introduction to the basic principles of nuclear fusion energy, delineates the conditions, advantages, and key scientific and technological challenges in implementing fusion energy through different nuclear fusion reactions. Furthermore, it applies the principles introduced to analyze the achievements, optimal plasma parameter ranges, and challenges encountered in various fusion research approaches, including magnetic confinement, inertial confinement, and magneto-inertial confinement.

This book holds significant reference value for policymakers and research planning and management personnel in fusion energy, enlightens professionals (especially device designers), students, and general readers who are engaged or interested in fusion research and the development and application of fusion energy. It serves as an excellent introductory and broadening-of-horizons reading material.

I strongly recommend its publication.

\begin{flushright}
Researcher, Southwest Institute of Physics, China National Nuclear Corporation

Jiaqi Dong
\end{flushright}

\vspace*{10mm}

After a preliminary reading of the manuscript "Zeroth-Order Factors of Fusion Energy Research" and at the request of the author, Huasheng Xie, I would like to strongly recommend the publication of this book by your publishing house for the following reasons:

1. If the book can be published now, it would perfectly meet the growing curiosity about fusion energy generated by various "fusion" commercial enterprises due to the influence of risk investments in innovative technologies both domestically and internationally.

2. The book provides in-depth analysis and comparison, particularly from a numerical modeling perspective, on the scientific feasibility, challenges, and difficulties of various fusion approaches. What sets this book apart from others is its innovative approach of retrospectively deducing the technical challenges and feasibility based on the perspective of energy economics and future environmental permissibility.

3. Therefore, the book has a broad potential readership, including the general public (ranging from high school and college students to young adults) who want to gain knowledge about fusion, investors and decision-makers in fusion enterprises, and professionals currently engaged in various fusion research activities.

4. The author has been involved in frontline nuclear fusion research for many years, diligently collecting and comparing various fusion approaches, and conducting rigorous numerical studies. Therefore, the book's professional standard is guaranteed.

5. Finally, but not least, it is worth noting that there is currently a lack of scientific literature (including popular science and discussions based on the most fundamental scientific and technological requirements) that reflects the latest trends in fusion research, both domestically and internationally. This book is timely and fills this gap. I hope that your publishing house will seize this opportunity.

\begin{flushright}
Professor, Huazhong University of Science and Technology

Xiwei Hu
\end{flushright}

\addcontentsline{toc}{chapter}{Content}

\tableofcontents

\setcounter{page}{0} \pagenumbering{arabic}

\chapter{Why Fusion Energy is Worth Pursuing}\label{chap:intro}

The core of human technological history can be summarized as energy and information. From learning to use fire, to fossil fuels, to nuclear energy, each breakthrough in the total amount of energy that humans can control and utilize has brought about significant progress and leaps in technology. Similarly, each breakthrough in the total amount and speed of information dissemination, represented by the domestication of horses, beacon towers, the birth of writing, papermaking, printing, trains and automobiles, aviation and aerospace, radio, and the internet, has also represented a leap in human technology. Since the 1940s, when humans released nuclear fission energy, there has not been a revolutionary change in the energy sector in the past few decades; during this time, the main theme of technological progress for humans has been information, represented by mobile communications, personal computers, and the internet, which have undergone exponential changes and greatly expanded the boundaries of human capabilities.

Statistics show that energy consumption is positively correlated with economic development (GDP), and people are increasingly expecting the "Bell Laboratories" of the energy sector to bring about new revolutions in the same way as in the communications field. In fact, the annual energy consumption by humans has not increased by orders of magnitude in the past few decades. The next leap in energy breakthroughs will inevitably be fusion energy in nuclear energy, as only fusion energy can lead humans beyond the solar system.

The history of human technology teaches us two experiences or lessons: (1) as long as it does not violate the laws of physics, it is difficult, but it can be achieved; (2) as long as it violates the laws of physics, no matter how beautiful the vision is, it cannot be achieved. The former includes the detection of gravitational waves, and the latter includes perpetual motion machines.

No scientific law states that fusion energy cannot be achieved, and furthermore, since stars are large natural fusion energy devices and thermonuclear weapons based on fusion principles have already been realized, fusion energy will eventually be achieved. However, the path to controllable fusion energy for humans is still long and full of obstacles. We need to understand the extent of the challenges, which is the main purpose of this book.

\section{The Current Energy Situation}In addition to nuclear fission energy, the energy currently used by humans on Earth is essentially derived from the solar nuclear fusion energy. The density at the center of the sun is approximately $1.5\times10^5{\rm kg/m^3}$ (number density is about $10^{31}{\rm m^{-3}}$), and the temperature is about 1.3 keV (15 million degrees). Its enormous mass is constrained by gravity. Referring to the data of Dolan (2013), the power of energy directly radiated from the sun to the Earth is 178000 TW (approximately $1.4 kW/m^2$), of which 62000 TW is directly reflected, 76000 TW is re-radiated after being received by the ground, 40000 TW of the received energy forms water vapor, 3000 TW is converted into wind energy, less than 300 TW is converted into waves, and 80 TW is converted into biomass energy through photosynthesis. These energies are then converted into various forms of water energy, geothermal energy, fossil energy (coal, oil, natural gas, etc.), and other forms of energy. In addition to the fusion energy directly radiated, the sun also provides about 3 TW of tidal energy to the Earth, which is essentially gravitational potential energy.
\begin{table}[htp]
\caption{World Energy Approximate Estimates [ref. Dolan (2013)], 1 ZJ=$10^{21}$J=31.7 TW$\cdot$year}
\begin{center}
\scriptsize
\begin{tabular}{c|c|c||c|c}
\hline\hline
& \thead {\bf \scriptsize Economically \\ \bf \scriptsize Exploitable (ZJ)} & {\bf Reserves (ZJ)} &  & \thead {\bf \scriptsize Technological \\ \bf \scriptsize Potential (ZJ/year)} \\\hline
{\bf Fossil Fuels}  &  & & {\bf Renewable Energy} &\\
Coal & 20 & 290-440 & Biomass Energy & 0.16-0.27\\
Oil & 9  & 17-23 & Geothermal Energy &  0.8-1.5\\
Natural Gas & 8 & 50-130 & Hydropower & 0.06\\
{\bf Nuclear Fission}  &  & & Solar Energy & 62-280\\
${\rm ^{238}U+^{235}U}$& 260  & 1300 & Wind Energy & 1.3-2.3\\
${\rm ^{232}Th}$ & 420 & 4000 & Ocean Energy & 3.2-11\\
{\bf Nuclear Fusion}  &  & &  & \\
Lithium in Seawater &  & $1.40\times10^{10}$ &  & \\
Lithium on Land &  & 1700 &  & \\
Deuterium &  & $1.60\times10^{10}$ &  & \\
\hline\hline
\end{tabular}
\end{center}
\label{tab:worldenergy}
\end{table}

The current global estimates of energy reserves are shown in Table 1. Different data sources may yield varying results, but the order of magnitude remains unchanged. The current global energy consumption by human activities is about 20 terawatts (TW), where 1 TW is equivalent to $10^{12}$ watts. It can be seen that fossil energy is still sufficient for human use for several decades to several centuries. Nuclear fission energy is enough for human use for 1,000 years. Renewable energy is also sufficient for human use at the current stage.

However, if humans want to have control over energy on a magnitude that goes beyond Earth, they can only rely on nuclear energy, especially nuclear fusion energy, which has an energy density millions of times higher than other energy sources. Moreover, as energy consumption increases significantly, new technologies are needed to economically obtain energy. Clean, carbon-free, and safe energy are also goals pursued by humans. Fusion energy has many advantages, such as high energy density, cleanliness, carbon-free nature, theoretical safety, and virtually unlimited raw materials. Therefore, it is considered the "ultimate energy source" for humans, which has been a long-term motivation for researching fusion energy. Here, "ultimate energy source" does not mean that no superior energy source can be found, but rather that it is already sufficient to be considered an ideal energy source for humans, capable of meeting the range of human imagination.

\section{Current Stage of Fusion Energy Research}

\begin{figure}[htbp]
\begin{center}
\includegraphics[width=15cm]{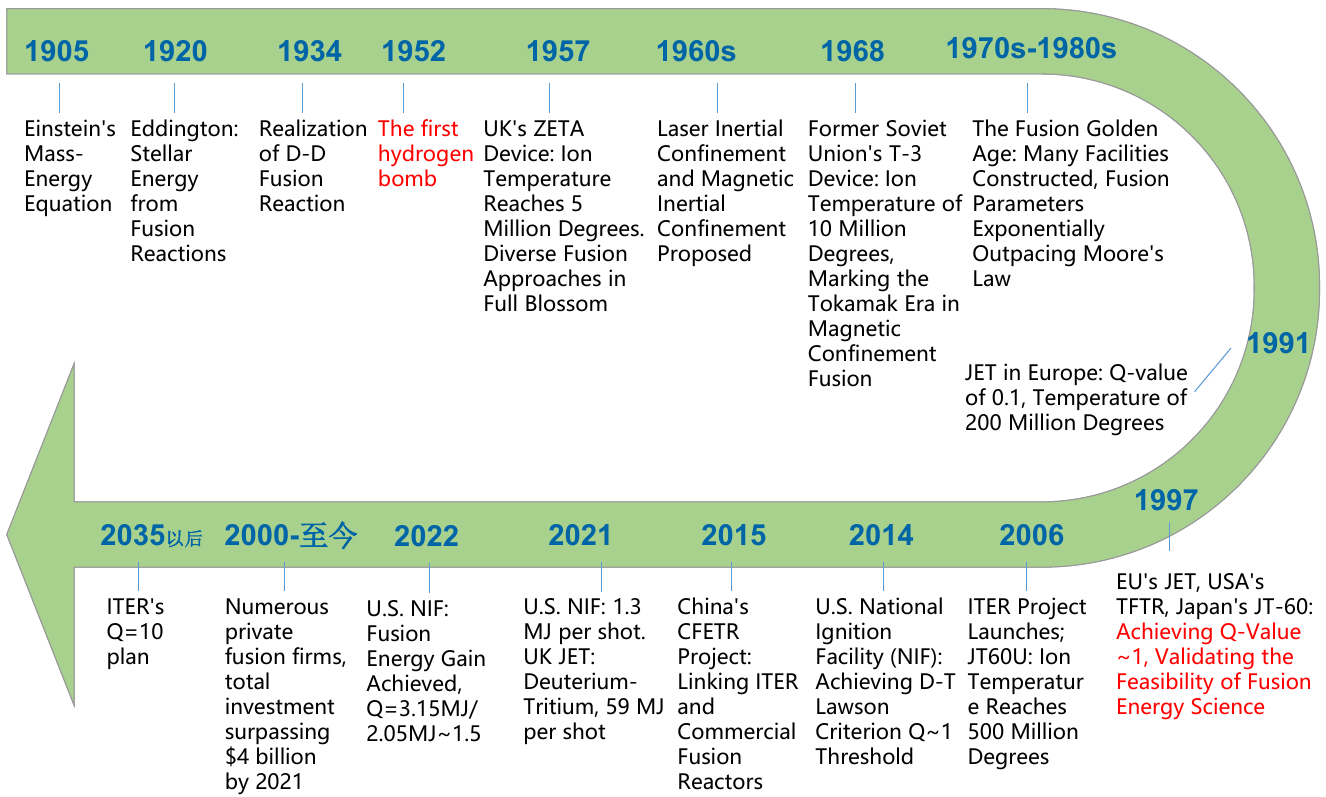}\\
\caption{Partial important milestones in the history of fusion energy research.}\label{fig:fusionmilestone}
\end{center}
\end{figure}\begin{figure}[htbp]
\begin{center}
\includegraphics[width=15cm]{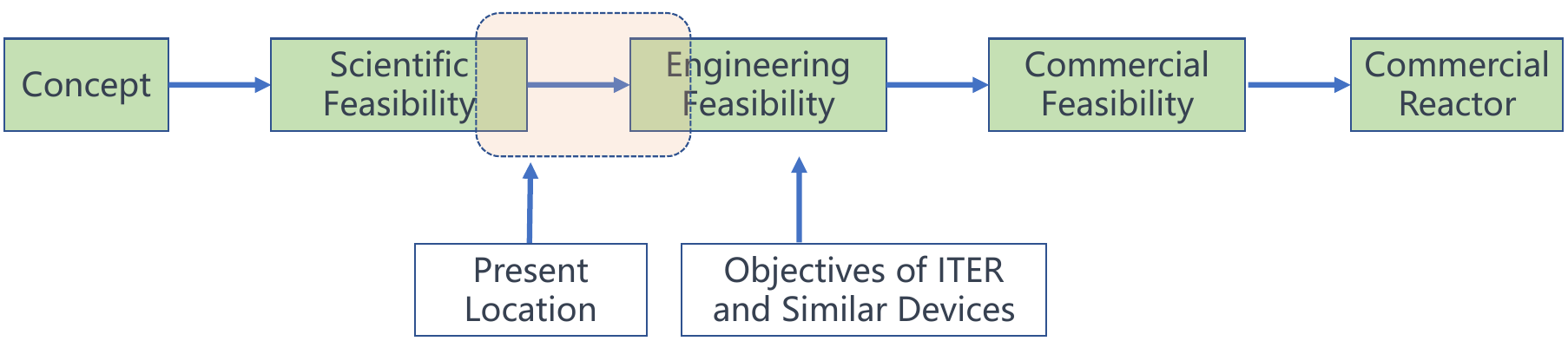}\\
\caption{Stages of fusion energy research at present.}\label{fig:fusionstage}
\end{center}
\end{figure}

Any form of energy or technology that can enter the market goes through four stages: concept feasibility, scientific feasibility, engineering feasibility, and commercial feasibility. The concept feasibility of fusion energy refers to the concept that fusion can become a form of energy, a point that people have long had no doubt about. Scientific feasibility is mainly measured by the Lawson criterion, which measures the gain or loss of energy. It is characterized by three parameters: temperature, density, and energy confinement time. The details of this will be discussed later in the text. Only when these three parameters reach certain values can the energy output of fusion be greater than the energy input required to sustain fusion, that is, the gain factor Q = fusion output energy/energy input to sustain fusion $>$1
. It is generally believed that representative magnetic confinement fusion, such as tokamak, achieved the conditions for energy gain close to deuterium-tritium fusion through D-T and D-D equivalence experiments conducted in the British JET, American TFTR, and Japanese JT60U devices around 1998, thus verifying its scientific feasibility. Representative inertial confinement fusion, such as NIF in the United States, also achieved basic energy gain in experiments conducted in 2014, 2021, and 2022, especially in 2022, when energy gain in fusion was achieved for the first time. Fusion energy equivalent to 3.15 MJ was produced from a laser energy of 2.05 MJ, resulting in Q$\simeq$1.5. Devices like the International Thermonuclear Experimental Reactor (ITER), representing international fusion energy research, will further verify scientific feasibility and partially verify engineering feasibility to confirm that certain engineering technologies can meet the requirements under fusion reactor conditions. It is expected that the first plasma discharge experiment at ITER will take place after 2030. As for the verification of the commercial feasibility of fusion, that is, whether the energy or power generation cost can compete with existing energy sources, it is currently difficult to determine a reliable timetable. Historically, all claims that ``fusion energy can be achieved in 30-50 years" have not fully assessed the difficulty of achieving fusion energy. Important milestones in the history of fusion energy research are shown in Figure \ref{fig:fusionmilestone}, and the stages of fusion energy research at present are shown in Figure \ref{fig:fusionstage}.

Various fusion schemes have been proposed, which can be mainly categorized according to the confinement method as gravity confinement, magnetic confinement, and inertial confinement. Magnetized target fusion can be classified as a combination of magnetic and inertial confinement and is also known as magneto-inertial confinement fusion. Electrostatic confinement is classified as inertial confinement. Some fusion schemes can be categorized as wall confinement, such as cold fusion. These schemes will be discussed and evaluated in detail later in the text.

\section{Confidence in Fusion Energy}We need to clarify that fusion itself is relatively easy, requiring only a commercially available kilovolt power supply and some deuterium to achieve what is often reported in the news as "fusion achieved by a high school student in their basement"; while fusion energy is very difficult, controlled fusion energy is even more difficult, and mankind has been striving for nearly seventy years to bridge the gap between fusion and fusion energy. The enormous energy of stars and the sun comes from fusion energy, and the realization of hydrogen bombs has convinced people that controlled fusion energy will eventually be achieved. There is no strict definition of what "controlled" means, but an energy release device that does not consider the hydrogen bomb, which releases energy equivalent to thousands of tons of TNT (trinitrotoluene) in an instant (on a microsecond scale), is not considered controlled because of its massive destructive power; on the other hand, the release of a few megajoules of energy in 1 nanosecond through inertial confinement is considered controlled because it does not cause destructive damage to the equipment.

The reason why fusion energy has not succeeded so far, besides its inherent difficulty, is that other energy sources are sufficient for human use in the current and future decades, and their prices are not yet at an unacceptable level, that is, it is not yet the time when fusion energy must be used. Therefore, the annual research and development funding for fusion is relatively small, about $1-3$ billion US dollars, with less than 30 thousand personnel involved. Survival crises usually bring a strong sense of urgency, such as the sense of urgency for many human technologies brought about by warfare, which leads to cost-unconstrained investment and hastens breakthroughs. The sense of urgency for fusion energy has not yet reached an indispensable level for mankind, and it is at most a "bonus" rather than a "life-saving straw".

\section{Positioning of this Book}

This book establishes a rigorous logical process to ensure that the logic itself has no issues, while also ensuring the integrity of models and data, so that everyone can reach the same conclusions according to the same logic. If breakthroughs are to be made, it is necessary to break through the assumptions in the models. This book does not involve overly complex theories, such as anomalous turbulence transport mechanisms, plasma instability, and so on. Instead, it aims to sort out key points through simple logic and calculations, namely the zeroth order factors, to help readers quickly assess the feasibility of various confusing plans, without being overly disturbed by less feasible ones. The readers here are not limited to non-fusion personnel who are interested, beginners, and energy policy makers, but also include fusion professionals. This is because, most fusion professionals only study a certain detailed branch of fusion, and may not have fully considered the complete aspects involved in fusion energy.

\vspace{30pt}

Key points of this chapter:
\begin{itemize}
\item Fusion energy is considered to be the ultimate energy source for humanity.
\item Fusion energy is theoretically feasible and will eventually be realized.
\item Fusion energy is extremely challenging, and other energy sources are still sufficient for human use for thousands of years, thus fusion energy is not yet indispensable.
\end{itemize} 
\chapter{Fusion Nuclear Reaction Basics}\label{chap:fuels}

In the ideal state, elements with atomic numbers below iron can undergo fusion reactions and release energy. However, when considering fusion energy, the ease of fusion reactions (described by reaction cross-sections and corresponding energies) and the amount of energy released per reaction are the first factors to be considered. This limits our choices. This is the most crucial reason why the development of fusion energy is so difficult and challenging.

\section{Fusion Nuclear Reactions of Interest}

Nuclear reactions refer to the processes in which incident particles collide with atomic nuclei (referred to as target nuclei), causing changes in the nuclear states or the formation of new nuclei. These incident particles include high-energy protons, neutrons, $\gamma$-rays, high-energy electrons, or other nuclear particles. Nuclear reactions obey the laws of conservation of nucleon number, charge, momentum, and energy. Common nuclear reactions include decay reactions (such as $\alpha$ decay, $\beta$ decay, $\gamma$ decay), heavy nuclear fission, light nuclear fusion, and so on.

\begin{figure}[htbp]
\begin{center}
\includegraphics[width=15cm]{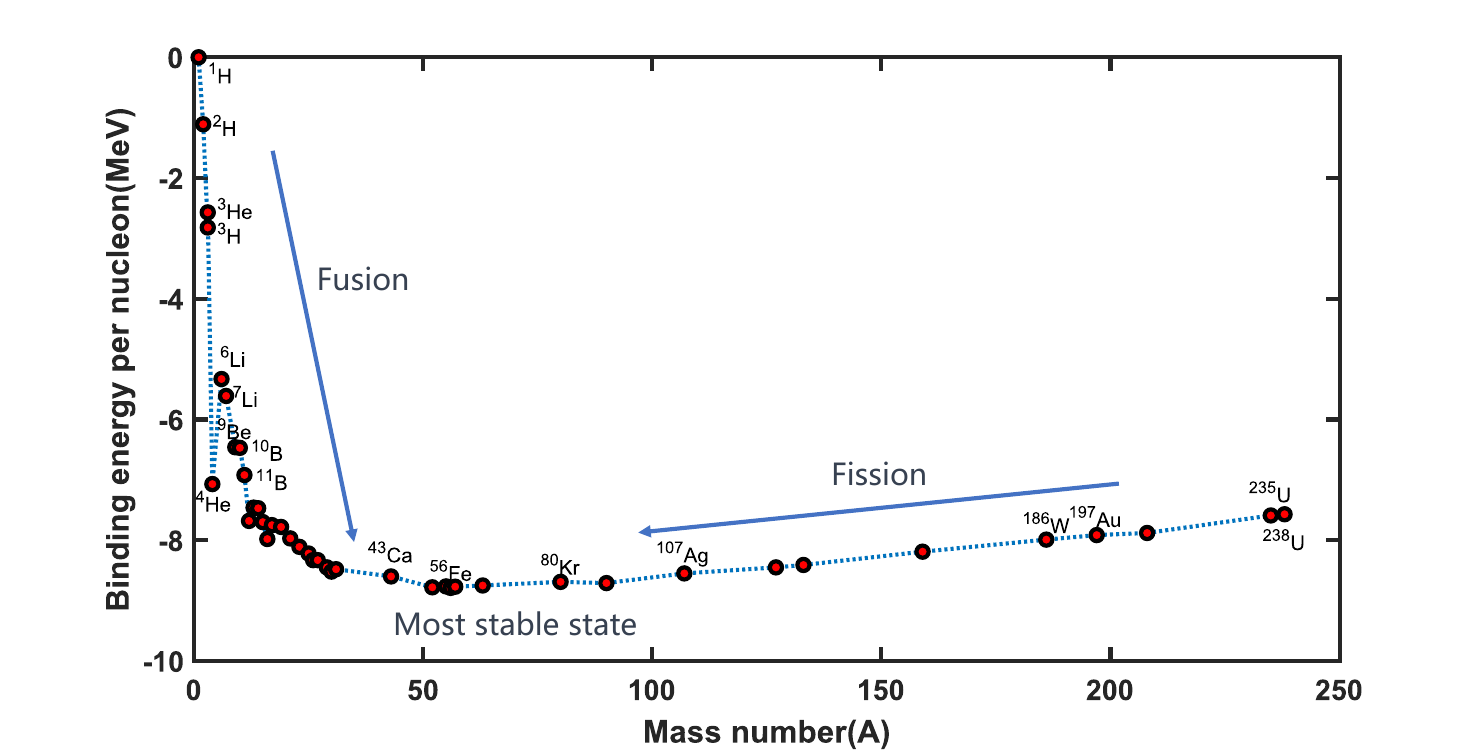}\\
\caption{Binding energy per nucleon versus mass number for some stable isotopes. The data primarily come from Ghahramany (2012).}\label{fig:bindingenergy}
\end{center}
\end{figure}

The binding energy of a nucleus represents the difference between the total energy of the nucleus $m_x$ corresponding to its mass number ${\rm ^{A}_{Z}X}$ and the total energy of independent protons $m_p$ and neutrons $m_n$, based on Einstein's mass-energy equation, which is
\begin{equation}
\Delta E=m_xc^2-[Zm_p+(A-Z)m_n]c^2=\Delta m c^2,
\end{equation}
where $c$ is the speed of light. The curve of the binding energy per nucleon $\Delta E/A$ versus the mass number $A$ is shown in Figure \ref{fig:bindingenergy}. In the figure, we can see that iron element ${\rm ^{56}_{26}{Fe}}$ is in the state of the lowest binding energy, and thus it is the most stable. Light nuclei can approach the lowest energy state through fusion reactions, while heavy nuclei can approach the lowest energy state through fission reactions. The typical energy release of a single nuclear reaction is in the order of MeV, which is approximately $1.6\times10^{-13}$J. In comparison, the energy release of a single chemical reaction involving the change of an outer electron's state, such as the combustion of hydrogen ${\rm H_2+\frac{1}{2}O_2 \to H_2O+2.96eV}$, is only in the order of eV, differing by about six orders of magnitude. In other words, the energy density of nuclear energy is approximately a million times that of fossil energy of the same mass.Typical fission reactions are described by the equation:
\begin{eqnarray}
{\rm ^{235}_{92}U+^1_0n} &\to& {\rm ^{141}_{56}Ba+^{92}_{36}Kr+3^1_0n+202.5MeV+8.8MeV},
\end{eqnarray}
where the 8.8MeV energy is carried by the antineutrino. The aforementioned and other types of fission reactions form the basis of current nuclear power plants. The main disadvantages are: (1) strong radioactivity, requiring specialized nuclear waste treatment; (2) limited availability of extractable materials on Earth, which is insufficient to support the long-term (over 10,000 years) rapidly growing energy demands of humanity. Therefore, nuclear fission is not considered as the ultimate energy source for humans.

In this study, we adopt a broad definition of "fusion" in which two atomic nuclei collide and undergo a nuclear reaction to produce new particles, without requiring the mass number of the resulting nucleus to be greater than that of the reactant nuclei. In contrast, "fission" refers to the process in which an atomic nucleus interacts with a neutron to produce new particles. Neutrons do not need to overcome the strong Coulomb potential in the atomic nucleus like charged particles, so fission reactions are usually easier to occur and control.

\begin{figure}[htbp]
\begin{center}
\includegraphics[width=15cm]{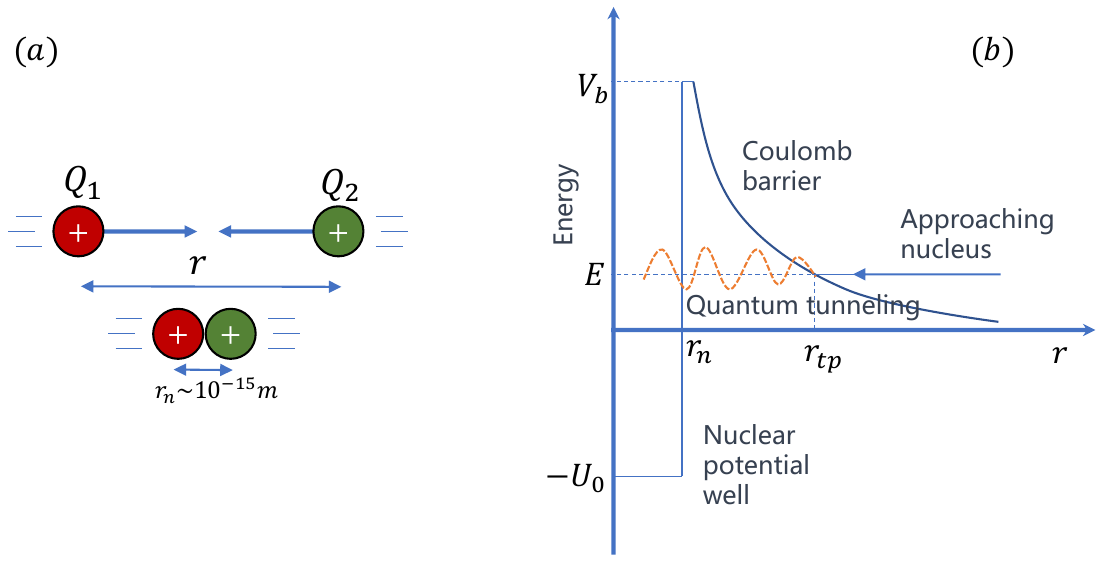}\\
\caption{Coulomb potential barrier in atomic nucleus. Due to quantum tunneling effect, fusion can occur with a certain probability when colliding with energy lower than the Coulomb potential barrier.}\label{fig:coulomb}
\end{center}
\end{figure}

\subsection{Coulomb Scattering Cross Section}

For fusion reactions to occur, the atomic nucleus itself needs to have a certain kinetic energy to overcome the Coulomb potential barrier. In classical physics, the electrostatic barrier is given by
\begin{equation*}
U(r)=\frac{1}{4\pi\epsilon_0}\frac{Z_1Z_2e^2}{r}=1.44\frac{Z_1Z_2}{r[{\rm fm}]}[{\rm MeV}],~~r>r_n,
\end{equation*}
where ${\rm 1fm=10^{-15}m}$, $r_n\simeq1.44\times10^{-15}(A_1^{1/3}+A_2^{1/3}){\rm m}$ represents the distance between the radii of the two nuclei. Taking $Z_1,Z_2$ as the charge numbers and $A_1,A_2$ as the mass numbers, when they are all equal to 1, the barrier potential $V_b=U(r_n)\simeq1{\rm MeV}$. From the above expression, it can also be seen that the larger the charge numbers $Z_1,Z_2$ of the atomic nuclei, the higher the Coulomb potential barrier, and the more difficult the fusion reaction is. Fortunately, due to quantum physics effects, there is quantum tunneling, which allows energies lower than the barrier to still tunnel through and undergo nuclear reactions near the nucleus. The potential well after passing through the Coulomb barrier attracts energy $U_0\simeq30-40{\rm MeV}$. Figure \ref{fig:coulomb} shows a schematic diagram of the potential barrier outside the atomic nucleus.\begin{table}[htp]
\caption{Coulomb screening parameter $\Lambda$ under typical fusion conditions.}
\begin{center}
\begin{tabular}{c|c|c|c}
\hline\hline
Temperature $T_e(eV)$ & Density $n(m^{-3})$  & $\ln\Lambda$ & Typical Device  \\\hline
 0.2 & $10^{15}$ & 9.1 & Q-machine  \\
2 & $10^{17}$ & 10.2 & \thead {\normalsize Experimental \\ \normalsize low-temperature plasma}  \\
 100 & $10^{19}$ & 13.7 & Typical tokamak  \\
 $10^4$ & $10^{21}$ & 16.0 & Fusion reactor  \\
 $10^3$ & $10^{27}$ & 6.8 & Laser plasma  \\
\hline\hline
\end{tabular}
\end{center}
\label{tab:logLambda}
\end{table}

By integrating the Coulomb differential scattering cross section, the effective total Coulomb scattering cross section is obtained [Chen (2015), Spitzer (1956)]
\begin{equation*}
\sigma=\pi b_{90}^2\ln\Lambda,~~~b_{90}=\frac{1}{4\pi\epsilon_0}\frac{Z_1Z_2e^2}{E},
\end{equation*}
where $\Lambda=12\pi n\lambda_D^3$ represents the shielding effect of the plasma on long-range electrostatic forces, $n$ is the particle number density, $\lambda_D$ is the Debye length, and $b_{90}$ represents the incident radius for a 90-degree scattering deflection. For fusion-related parameters, the variation range of $\ln\Lambda$ is not large, as shown in Table \ref{tab:logLambda}. For fusion reactor parameters, we can take a typical approximate value of $\ln\Lambda\simeq16$.

\subsection{Optional Fusion Nuclear Reactions}

\begin{figure}[htbp]
\begin{center}
\includegraphics[width=15cm]{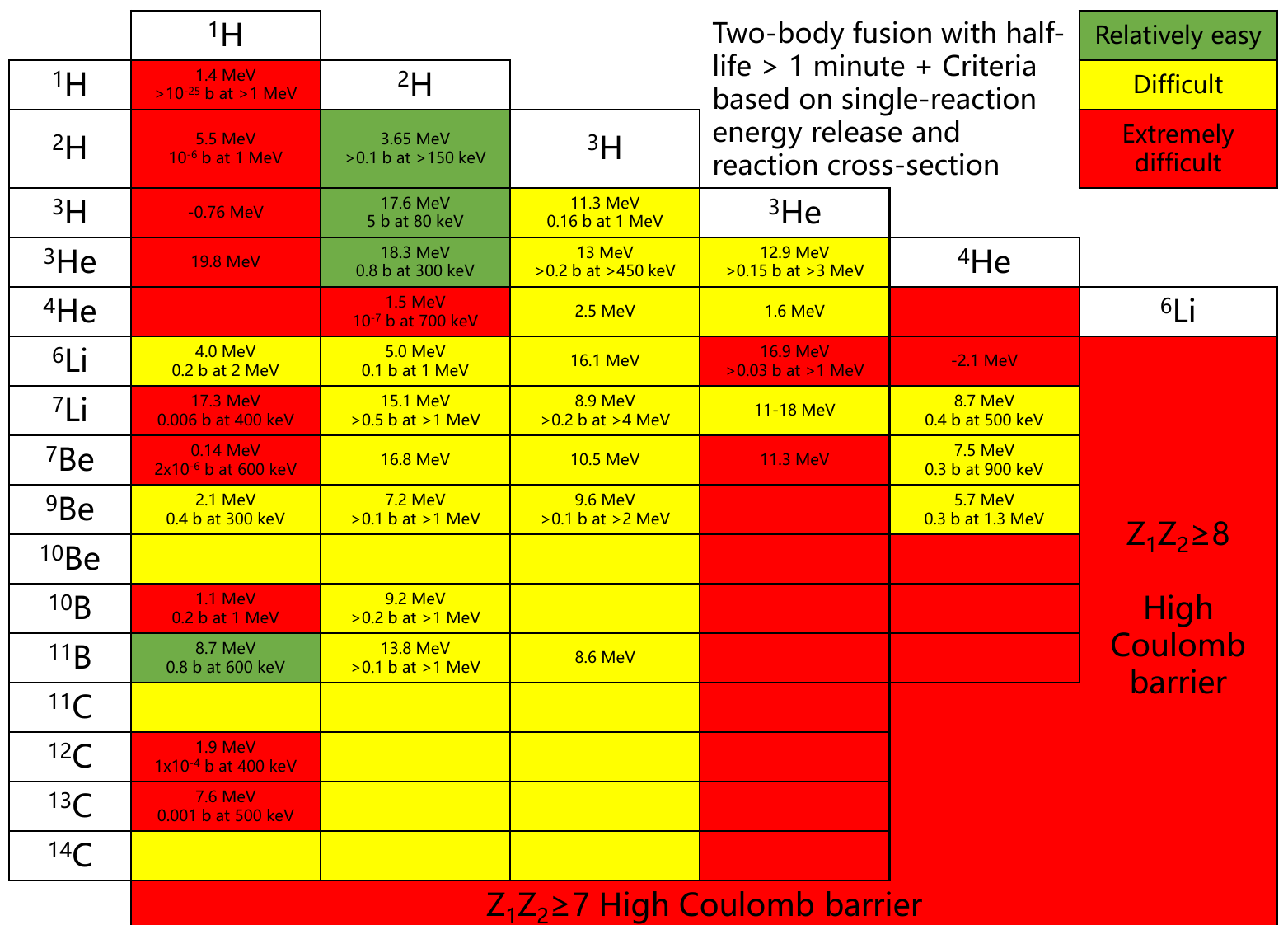}\\
\caption{There are few fusion reactions to choose from for fusion energy research.}\label{fig:fusionfuels}
\end{center}
\end{figure}

We base our selection of suitable fusion fuels on several criteria: value of the reaction cross section, reaction temperature, and value of the single energy release. The options for nuclear reactions are limited. In this section, we assume that the existing reaction cross section measurement data or theoretical data do not have orders of magnitude deviation. We can easily inventory the results in Figure \ref{fig:fusionfuels}, i.e., there are only a few relatively easy nuclear reactions, while others have reaction cross sections that are too low (less than 0.5b) or require too high reaction energy (greater than 500keV), where $1{\rm b}=10^{-28}{\rm m}^2$ and $1{\rm keV}\simeq1.16\times10^{7}{\rm K}$. From the figure, we can also see that nuclear reactions between nuclei with high atomic numbers $Z$ are indeed more difficult to occur. As we will see later, nuclei with high $Z$ also have the disadvantage of high radiation loss. Therefore, multiple factors restrict the prioritization of fusion nuclear reactions to only a few related to hydrogen and its isotopes.

\begin{figure}[htbp]
\begin{center}
\includegraphics[width=15cm]{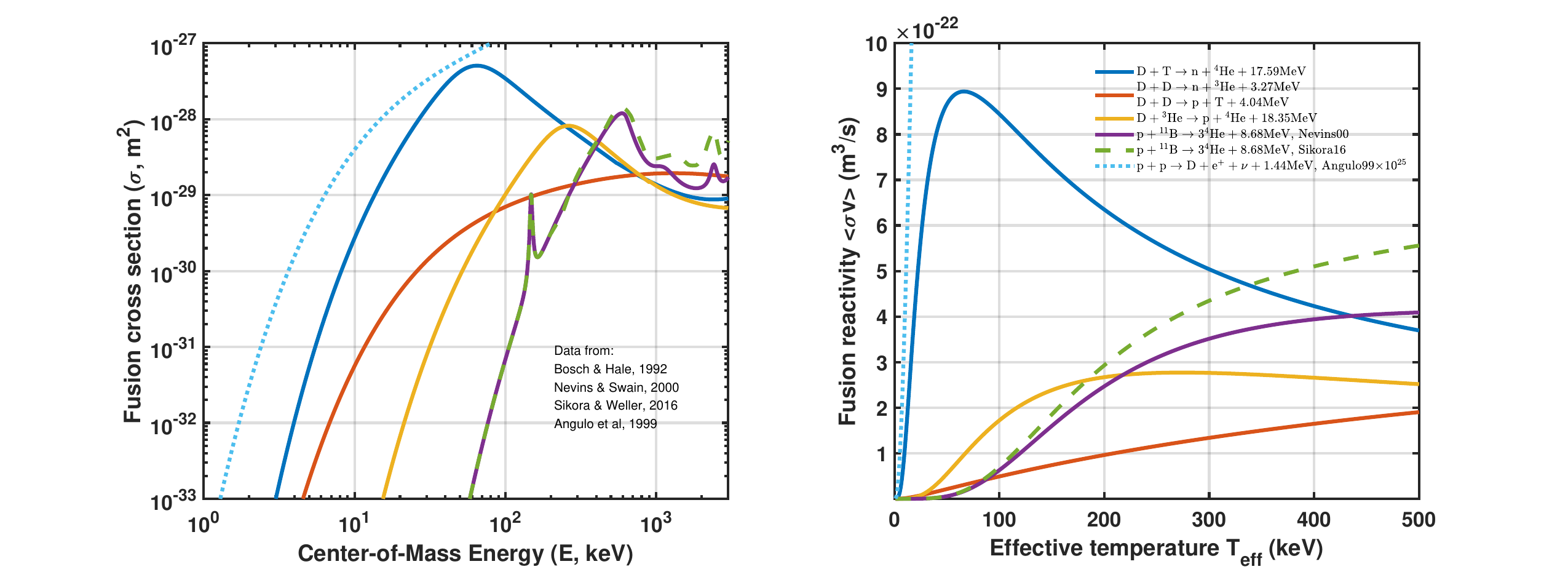}\\
\caption{Comparison of reaction cross-sections and reaction rates for several common nuclear reactions suitable for controlled fusion energy research, as well as the pp reaction in the sun, where the data for the pp reaction has been multiplied by $10^{25}$.}\label{fig:fusion_cross_section}
\end{center}
\end{figure}

The nuclear reactions on which the sun is based are believed to be the p-p (proton-proton) and CNO (carbon-nitrogen-oxygen) cycle. The temperature is about 15 million degrees (1.3keV) and the reaction cross-section is extremely low, less than $10^{-20}$b. It is precisely because of the extremely low reaction cross-section that the sun can burn for billions of years. Despite the low reaction rate, the energy release is enormous due to the large volume and mass. "Artificial sun" refers to the release of energy based on fusion reactions, rather than relying on fusion fuel from the sun. According to the aforementioned criteria, the fusion reactions worth considering are as follows:

\begin{eqnarray}
{\rm D+T} &\to& {\rm n (14.07MeV) +{}^4He (3.52MeV)},\\
{\rm D+D} &\to& {\rm n (2.45MeV)+{}^3He (0.82MeV)}  (50\%),\\\nonumber
{\rm D+D }&\to& {\rm p (3.03MeV) +T (1.01MeV)} (50\%),\\
{\rm D+{}^3He} &\to& {\rm p (14.68MeV) +{}^4He (3.67MeV)},\\
{\rm p+{}^{11}B} &\to& {\rm 3{}^4He +8.68MeV},
\end{eqnarray}

\begin{figure}[htbp]
\begin{center}
\includegraphics[width=15cm]{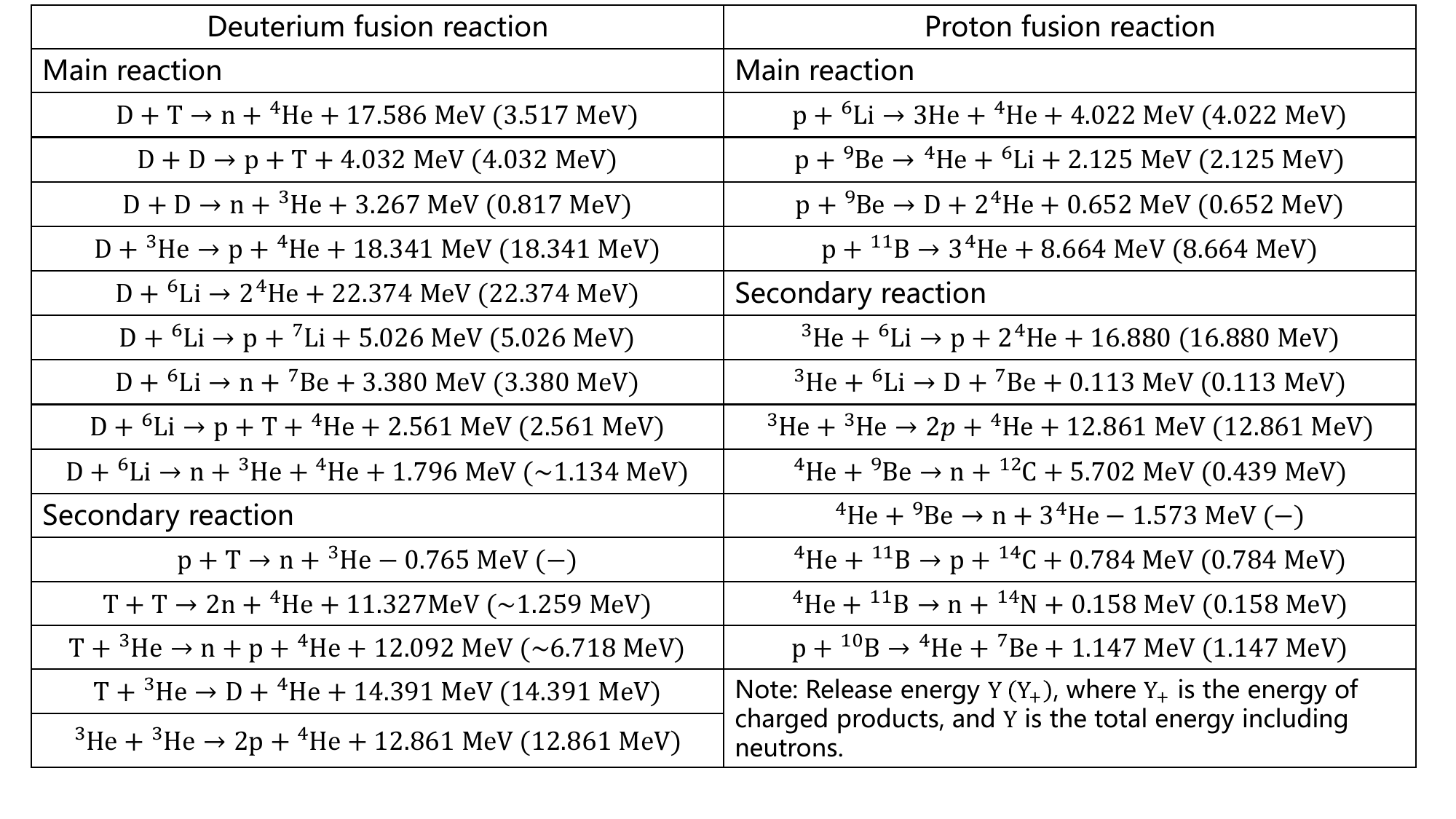}\\
\caption{Some main fusion reactions and minor reactions related to fusion energy [McNally82].}\label{fig:majorfusionreact}
\end{center}
\end{figure}

The 50\% in the D-D reaction represents that the cross sections of the two channels are approximately equal at the peak, and the energy allocation on the right side of the reaction equation does not include the initial kinetic energy of the left side nucleus. As for why we do not need to consider other nuclear reactions, on the one hand, if the fusion gain of the listed nuclear reactions can be successfully achieved, it is already sufficient for use; on the other hand, in the discussion of the Lawson criterion, we can see that the difficulty of the nuclear reactions listed here increases by orders of magnitude compared to those outside the listed nuclear reactions. Figure \ref{fig:fusion_cross_section} shows the cross section $\sigma$ and reaction rate $\langle\sigma v\rangle$ under the Maxwell distribution function for several major fusion reactions mentioned above. For the specific physical meanings of cross section and reaction rate, please refer to the appendix and Chapter \ref{chap:lawson}. In addition to the above-mentioned nuclear reactions, other major and minor fusion reactions of interest are shown in Figure \ref{fig:majorfusionreact}.

Note that the energy distribution of the two-body products (deuterium-tritium, deuterium-deuterium, deuterium-helium) can be uniquely determined by momentum conservation and energy conservation. Under the condition of neglecting the center of mass energy, the energy distribution of the products is inversely proportional to their masses. However, the product of hydrogen-boron (${p-{}^{11}B}$) fusion is three $\alpha$ particles (${^4He}$), and their energies are not equal and even have a certain energy spectrum distribution, which is different from the two-body nuclear reaction situation. This also leads to the possibility that the measured hydrogen-boron reaction cross section may not be accurate at present. For example, the data from Sikora (2016) show an increase of more than 50\% compared to the data from Nevins (2000) in the 0.2-2MeV range. Unlike deuterium-tritium (D-T), deuterium-deuterium (D-D), and deuterium-helium (D-${}^3He$) reactions, the accurate cross section data for hydrogen-boron reactions still need to be further determined by nuclear physicists.\begin{figure}[htbp]
\begin{center}
\includegraphics[width=15cm]{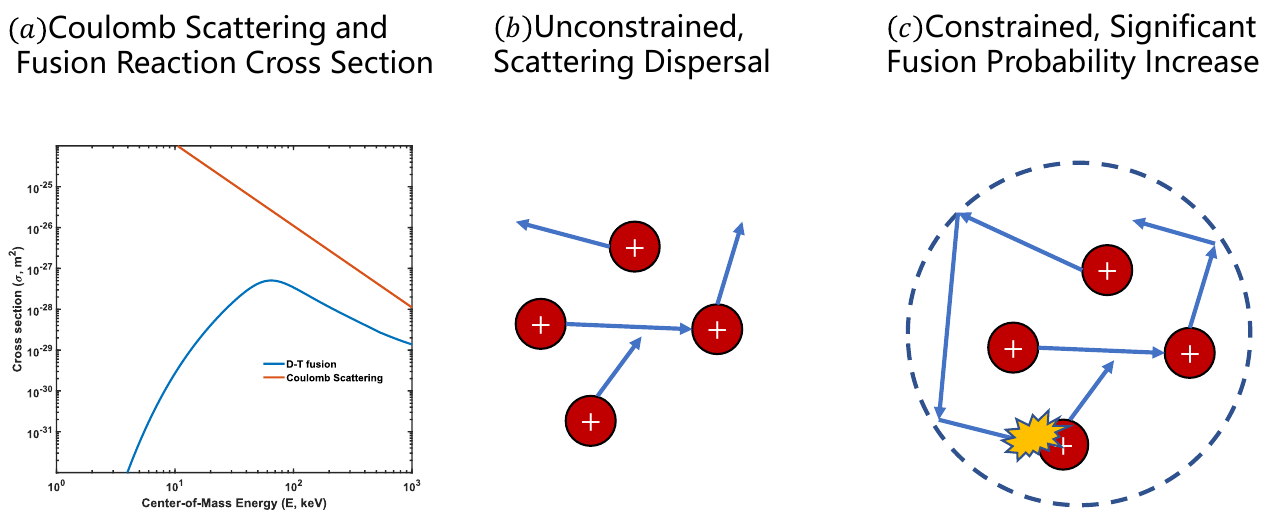}\\
\caption{The Coulomb scattering cross section is much larger than the fusion reaction cross section. Only when the fusion fuel is confined to effectively increase the collision probability, can fusion energy be viable.}\label{fig:coulombcross}
\end{center}
\end{figure}

Anyway, as shown in Fig. \ref{fig:coulombcross}, due to the fact that in the energy range we are concerned with $E<1{\rm MeV}$, the Coulomb scattering cross section is much larger than the fusion reaction cross section. In order to achieve fusion energy, it is necessary to confine the fusion fuels to allow them to react fully. This rules out the possibility of achieving fusion energy through target confinement. For the high-energy range $E\geq1{\rm MeV}$, it is difficult to use the kinetic energy of the reacting particles to achieve cost-effective fusion energy, because the kinetic energy itself is already close to the energy of the fusion products. In the energy range above 0.511MeV, which is higher than the electron relativistic energy, various other complex quantum and relativistic physics, such as the production and annihilation of positron-electron pairs, occur. Although the assessment of the system is currently lacking, there is no obvious reason to indicate that they have enough positive effects to offset the problem of excessive energy consumption due to their high energy.

By comparison, the reaction cross section of a thermal neutron (0.025eV) colliding with ${\rm ^{235}_{92}U}$ is 600b, the required fuel energy is small, and the reaction cross section is large, making it much easier for fission energy to be successful than fusion energy. This is the main reason why fission energy has succeeded quickly while fusion energy is still far from successful.

\section{The main advantages and disadvantages of several fusion fuels}

There are obvious disadvantages to various fusion fuels, so it is uncertain which fuel will be used for the first commercial fusion reactor in the future. However, it can be almost certain that it will be one of the following four fuels, or a combination of them. This is because other fuels are not superior to hydrogen boron here in terms of fuel abundance, no-neutron products, value of the reaction cross section, and required reaction temperature, so there is no need to focus on other more advanced fusion fuels before achieving hydrogen boron fusion energy.

\subsection{Deuterium-Tritium}
Deuterium-Tritium (${D-T}$) fusion is the easiest fuel to sustain a thermonuclear reaction. It has obvious advantages: (1) At low energies, its fusion reaction cross section is the largest; (2) It releases a large amount of energy in a single nuclear reaction (17.6MeV, of which 3.52MeV is carried by charged $\alpha$ particles); (3) It contains only singly charged nuclei, resulting in low bremsstrahlung radiation. Based on these advantages, its ignition temperature is the lowest among all fuels, about 5-10keV. This makes the required magnetic field for magnetic confinement fusion relatively low, while the fusion power density is high. For these reasons, Deuterium-Tritium fusion is given priority in research. However, it also has serious drawbacks. First, the half-life of tritium is only 12.3 years, 
\begin{eqnarray*}
{\rm T\to {}^3He+e^{-}+\nu_e},
\end{eqnarray*}
where $\nu_e$ represents an electron neutrino, 
which means that tritium (${}^3\text{H}$ or T) is practically non-existent on Earth and must be produced through other means, such as the following reactions:
\begin{eqnarray*}
\text{${}^6\text{Li}+n$} &\to& \text{T+${}^4\text{He} +4.7\text{MeV}$},\\
\text{${}^7\text{Li}+n$} &\to& \text{T+${}^4\text{He} +n -2.6\text{MeV}$},
\end{eqnarray*}
or
\begin{eqnarray*}
\text{${}^{10}\text{B}+n$} &\to& \text{T+2${}^4\text{He} +0.367\text{MeV}$},\\
\text{${}^{10}\text{B}+n$} &\to& \text{$^7\text{Li}+{}^4\text{He} +2.9\text{MeV}$},
\end{eqnarray*}
and these reactions also require neutron multiplication, such as using beryllium:
\begin{eqnarray*}
\text{${}^9\text{Be}+n$} &\to& \text{2${}^4\text{He} +2n -1.9\text{MeV}$}.
\end{eqnarray*}
Original tritium is mainly produced from fission reactors by using the neutron reaction with heavy water, and the nuclear reaction is:
\begin{eqnarray*}
\text{n+D} &\to& \text{T+$\gamma$}.
\end{eqnarray*}
Currently, the global annual production of tritium is approximately a few kilograms to several tens of kilograms. The feasibility of tritium breeding in fusion reactors is still inconclusive. If the aforementioned lithium is used as the breeding material, the limited lithium reserves on Earth would only be sufficient for deuterium-tritium fusion for about a thousand years\footnote{Seawater contains a certain proportion of lithium, and if it can be efficiently and cost-effectively extracted, it is estimated to supply use for hundreds of thousands of years.}. In this case, the advantage of deuterium-tritium fusion energy compared to fission energy is not significant.
Secondly, in addition to tritium itself being radioactive, another serious drawback of deuterium-tritium fusion is the damage caused by high-energy neutrons to the first wall structure. Currently, materials cannot guarantee to withstand neutron bombardment under fusion reactor conditions.
This makes deuterium-tritium fusion energy, although the easiest in terms of science, pose great challenges for materials, engineering, and commercialization. It is currently uncertain whether it is easier for deuterium-tritium fusion to overcome these engineering and commercialization challenges, or if fusion energy can be achieved more easily through fuels other than deuterium and tritium.

\subsection{Deuterium-Helium}

Deuterium-Helium (${\rm D-{}^3He}$) fusion is a fusion fuel with reaction conditions second only to deuterium-tritium. Its advantage lies in the production of no neutron byproducts, making it a candidate for advanced fuel. The main drawback is that the reserves of helium-3 on Earth are limited and expensive. Additionally, secondary reactions still produce neutrons. Therefore, even if the scientific feasibility issues are resolved, deuterium-helium fusion will still face constraints in terms of raw material costs and engineering. In addition to methods such as increasing the abundance of helium-3 through other nuclear reactions, based on the development of space technology, it is possible to mine helium-3 on the Moon, but reliable evaluations are currently lacking regarding its extractable quantity and cost.

Deuterium-Helium reactions have the potential to serve as catalytic reactions for other fusion reactions. For example, by adding a small amount of helium-3 to deuterium feedstock, a reaction can occur that increases the fusion temperature to the level required for deuterium-deuterium fusion, thus maintaining a deuterium-deuterium fusion reactor. This means that only a small amount of helium-3 feedstock is needed.

\subsection{Deuterium-Deuterium}

The main advantage of Deuterium-Deuterium (${\rm D-D}$) fusion is the abundance of raw materials and relatively large reaction cross-section. The abundance of deuterium (${\rm {}^2H}$ or ${\rm D}$) in natural hydrogen is 0.0139\%-0.0156\%, and 30 mg/L in seawater. The raw material is plentiful and extraction is inexpensive. The main drawbacks of deuterium-deuterium fuel are that the fusion reaction conditions are two orders of magnitude higher than deuterium-tritium (as discussed in the Lawson diagram), and neutron damage to the wall material still exists, although the neutron energy is much lower compared to deuterium-tritium fusion.

It should be noted that the products of deuterium-deuterium reactions include tritium and helium-3, with the reaction rate of secondary reactions higher than that of deuterium-deuterium reactions. This allows for rapid further reactions to generate more energy. We refer to this as catalytic deuterium-deuterium fusion. The total reaction rate is limited by the rate of deuterium-deuterium reactions, but the released energy, including primary and secondary reactions, ideally becomes:
\begin{eqnarray}
    {\rm 6D} &\to& {\rm 2n + 2p  +  2{}^4He+43.25MeV}.
\end{eqnarray}
Assuming that the intermediate processes involving ${\rm T}$ and ${\rm ^3He}$ occur slowly and transfer energy to the main plasma before undergoing fusion reactions to release energy, the energy allocated to the charged products is approximately $43.25-2.45-14.07=26.73$MeV, accounting for 62\% of the total energy. Based on the aforementioned data, on average, one deuterium nucleus releases a total energy of $(3.27+17.59+4.04+18.35)/6=7.21$MeV, which means that 1 gram of deuterium can release an energy of $3.45\times10^{11}$J, approximately equivalent to the energy produced by 83 tons of TNT explosion, or $9.6\times10^4$ kilowatt-hours. This is 3-5 times higher than the primary reaction of $3.27/2=1.64$MeV and $4.04/2=2.02$MeV. Through secondary reactions, the reaction conditions for deuterium-deuterium fusion can be significantly reduced. Furthermore, the neutrons produced in the products can react with lithium, releasing more energy.According to calculations, the deuterium contained in 1 liter of seawater can release energy equivalent to the energy released by burning 300 liters of gasoline (energy release of 47 MJ/kg, density of 0.71 kg/L) through deuterium-deuterium and secondary fusion reactions. If all the fusion energy released by deuterium can be fully utilized, it can supply human beings for nearly billions of years at the current rate of energy consumption. Calculated at a raw material cost of \$2 per gram, the cost of raw materials for each kilowatt-hour of electricity is only 0.002 cents or about 0.015 RMB, which is very low and can be neglected in the total generation cost.

\begin{table}[htp]
\scriptsize
\caption{Comparison of advantages and disadvantages of several major fusion reactions.} 
\begin{center}
\begin{tabular}{c|c|c|c|c}
\hline\hline
 Fusion Reaction & D-T & D-D & D-$^3$He & p-$^{11}$B \\\hline
Neutrons & Yes, 14 MeV & Yes, 2.45 MeV & Few & Very few \\
Tritium Breed & Required & Not required & Not required & Not required \\
\thead {\scriptsize Optimal Fusion \\ \scriptsize Temperature} & 10-30 keV & $\sim$50-100 keV & $\sim$50-100 keV & $\sim$100-300 keV \\
Reaction Rate ($m^3s^{-1}$) & 6e-22 @ 25 keV & 5e-23 @ 100 keV & 2e-22 @ 100 keV & 4e-22 @ 300 keV \\
Raw Materials & \thead {\scriptsize Scarce, \\ \scriptsize radioactive, controlled} & Abundant & \thead {\scriptsize Scarce, \\ \scriptsize partially controlled} & \thead {\scriptsize Abundant, \\ \scriptsize inexpensive} \\
\thead {\scriptsize Direct \\ \scriptsize Power Generation} & Not feasible & Not feasible & Feasible & Feasible \\
\thead {\scriptsize Single Reaction \\ \scriptsize Energy Release} & 17.59 MeV & 3.27-4.04 MeV & 18.35 MeV & 8.68 MeV \\
\hline\hline
\end{tabular}
\end{center}
\label{tab:fusioncompare}
\end{table}
\begin{table}[htp]
\caption{Fusion fuel prices and approximate reserves, note that the reserves and prices are only estimates and may have significant deviations.}
\scriptsize
\begin{center}
\begin{tabular}{c|c|c|c|c}
\hline\hline
Fuel & Earth's Reserves & \thead {\scriptsize Price \\ \scriptsize (USD/gram)} & \thead {\scriptsize 20TW$\cdot$years \\ \scriptsize Required Amount} & \thead {\scriptsize Years Available \\ \scriptsize for Human Use} \\\hline
Hydrogen &  Crust abundance: 1.4e-3 & $\sim$0.01 & 7.6e5 kg &  Infinite\\
Deuterium & Seawater 45 trillion tons & $\sim$2  & 7.2e6 kg  &  $\sim$billions of years\\
Tritium & $\sim$1-10kg & $\sim$1 million & 1.1e6 kg & $<$1 day \\
$^3$Helium & \thead {\scriptsize Easily extractable \\ \scriptsize quantity: approximately 500kg} &  $\sim$10,000 & 1.1e6 kg & $<$1 day \\
$^6$Lithium & \thead {\scriptsize Approximately \\ \scriptsize 1 million tons} &  $\sim$1 & 9.8e6 kg & $\sim$hundreds of years \\
$^{11}$Boron & Crust abundance: 9e-6 & $\sim$5 & 8.3e6 kg & $>$millions of years\\
\hline\hline
\end{tabular}
\end{center}
\label{tab:fuelcost}
\end{table}

\subsection{Hydrogen-boron}

On Earth, the natural abundance of boron consists of 80\% $^{11}$boron and 20\% $^{10}$boron. The crust abundance of boron is 9e-6, and it is found in seawater at a concentration of 4.8e-6. About 150 types of boron minerals have been discovered, and the world's borate reserves are estimated to be around 335-748Mt. Hydrogen-boron  (aka, proton-boron) fusion has the largest cross-section among fusion reactions that satisfy the condition of abundant fusion fuel and produce no neutrons. Therefore, achieving hydrogen-boron fusion energy is already sufficient to meet human needs, and there is no need for further research on fusion reactions with lower cross-sections before its realization. However, its drawbacks are also obvious. The fusion conditions for hydrogen-boron are much higher than deuterium-tritium fusion, even nearly 3 orders of magnitude higher without considering radiation. If radiation is considered, it is believed to be impossible to achieve controllable fusion energy. The hydrogen-boron (${p-{}^{11}B}$) reaction is not purely neutron-free, but has two main side reactions:
\begin{eqnarray*}
{\rm p+^{11}B} &\to& {\rm \gamma+^{12}C},\\
    {\rm p+^{11}B} &\to& {\rm n+^{11}C-2.765 MeV},
\end{eqnarray*}
The latter reaction involves neutrons, but they are soft neutrons with energy less than 3MeV, and in the temperature range of interest, this reaction accounts for only about $10^{-5}$. Overall, the neutron production from hydrogen-boron fusion is extremely low and can usually be neglected. The former reaction produces highly energetic gamma radiation in the product (97\% at 12MeV and 3\% at 16MeV), but this reaction branch also accounts for only about $10^{-4}-10^{-6}$, and can be effectively shielded with certain materials. Due to the fact that boron is typically in a solid state, it tends to accumulate on the walls of the fuel cycle system, which is a drawback of hydrogen-boron fusion compared to deuterium-tritium, deuterium-deuterium, and deuterium-helium fusion.

In addition, for the hydrogen-lithium reaction:
\begin{eqnarray*}
{\rm p+^6Li\to\alpha+^3He+4.02MeV},
\end{eqnarray*}
although the total reaction cross section is not as high as that of hydrogen-boron, there is a significant reaction cross section higher than hydrogen-boron in the low energy range around 100keV. Also, lithium has an atomic number of 3, which is lower than boron's atomic number of 5, resulting in weaker radiation. However, the abundance of $^6$ lithium is much less than that of $^{11}$ boron, and natural lithium on Earth is composed of 96.25\% $^7$ lithium and only 3.75\% $^6$ lithium. The lithium on land is only sufficient for human consumption for 1,000 years, and the demand for lithium in electronic products is high, which makes hydrogen-lithium based fusion reactors not significantly superior to fission reactors. Additionally, the side reaction product $\rm ^3He$ reacts with $\rm ^6Li$ to produce neutrons with a certain probability. Similarly, in deuterium-tritium fusion, the tritium breeding also requires $^6$ lithium, and the limitation of accessible $^6$ lithium resources on Earth means that deuterium-tritium fusion can at most be used as fuel for first-generation fusion power reactors. In view of the scarcity of tritium breeding, neutron handling, and helium-3 resources, we will prioritize considering the feasibility and difficulty of achieving hydrogen-boron fusion\footnote{In the actual power generation process as an energy source, there are other factors to consider, such as the condensability of the fuel. Higher atomic number elements such as boron and lithium are prone to condense in the pipeline, requiring additional means to address this. These factors are not considered as primary factors in this book.}. Then, we will consider deuterium-helium, deuterium-deuterium, and finally deuterium-tritium. This is because deuterium-tritium fusion has already been scientifically proven to be feasible, with the main challenges lying in engineering feasibility and commercialization costs, including the size of the fusion reactor, tritium breeding, and neutron shielding issues. A comparison of the advantages and disadvantages of several major fusion reactions is shown in Table \ref{tab:fusioncompare}. Reference reserves and prices of some major fusion materials are shown in Table \ref{tab:fuelcost}.

\section{Triple Product Requirements for Fusion}

In this section, we will see how fusion reaction cross-section data affects the required fusion energy gain conditions.

In the earliest Lawson's criterion [Lawson (1955)], based on energy balance, the required temperature, density, and confinement time for deuterium-tritium fusion were calculated considering fusion power, confinement, radiation, and power generation efficiency. Later, the triple product of temperature, density, and confinement time became the measure of scientific feasibility for fusion.

\begin{figure}[htbp]
\begin{center}
\includegraphics[width=15cm]{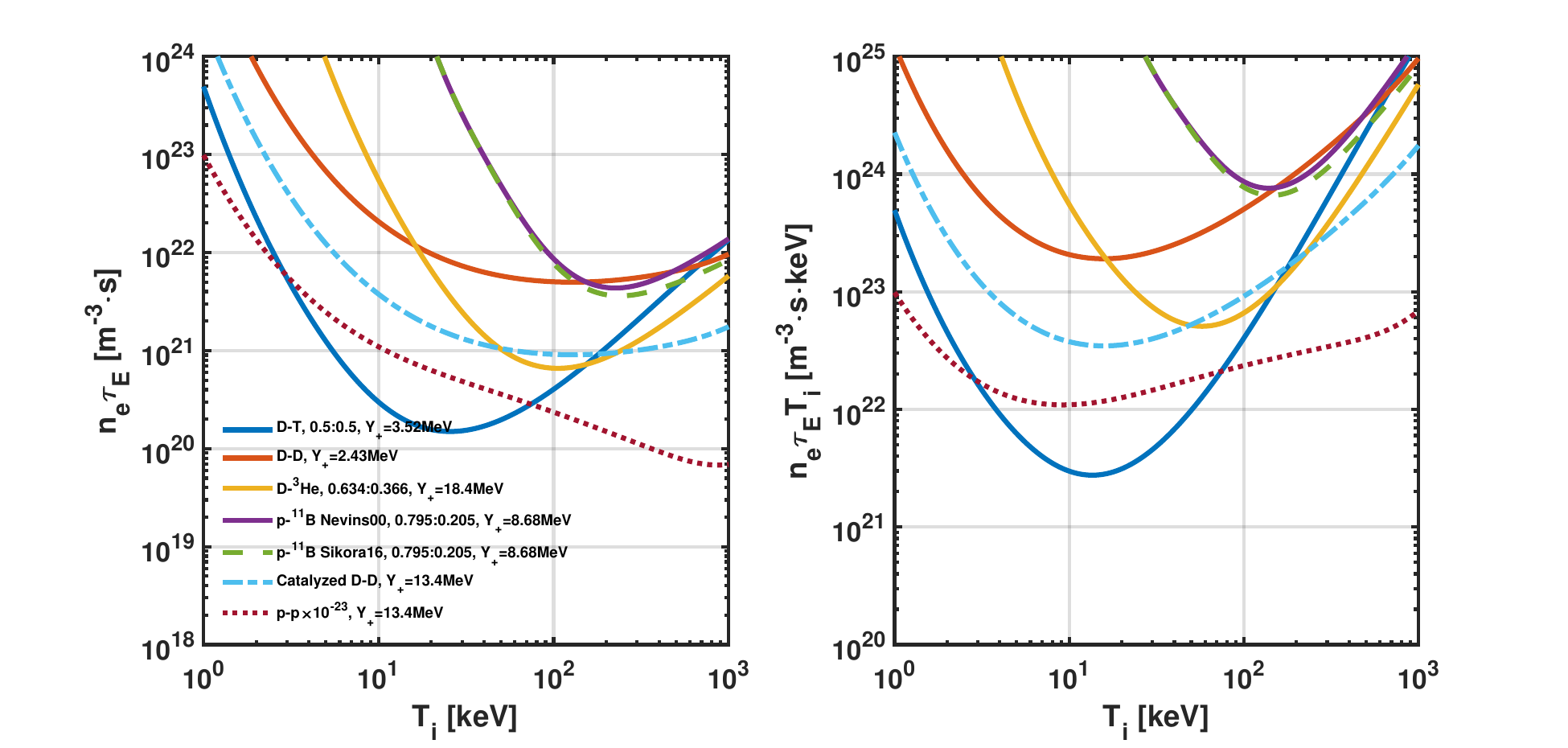}\\
\caption{Minimum requirements for Lawson's triple product for fusion, where $\tau_E$ here measures the energy confinement time that takes into account all losses such as radiation and transport and their combined effect on fuel reheating.}\label{fig:lawsonntauEnorad}
\end{center}
\end{figure}In this section, we present the simplest requirement for the triple product. In the subsequent chapters of this book, particularly Chapter \ref{chap:lawson}, the Lawson criterion and cases involving more factors will be discussed in more detail. In the case of steady-state self-sustaining, it is assumed that all the charged ions produced from fusion are used to heat the fuel, and all losses, such as radiation and transport losses, and the re-heating of the fuel due to these losses, are measured by the energy-confinement time $\tau_E$. Therefore, we have
\begin{eqnarray}
    \frac{E_{th}}{\tau_E}=f_{ion}P_{fus},
\end{eqnarray}
where the energy per unit volume $E_{th}$ and fusion power $P_{fus}$ are given by
\begin{eqnarray}
    E_{th}=\frac{3}{2}k_B\sum_jn_jT_j,\\
    P_{fus}=\frac{1}{1+\delta_{12}}n_1n_2\langle\sigma v\rangle Y.
\end{eqnarray}
Here, $Y$ represents the energy released in a single nuclear reaction, $f_{ion}$ is the proportion of energy released by the charged ions, i.e., $Y_+=f_{ion}Y$. $n_1$ and $n_2$ are the number densities of the two types of ions, and $T_j$ is the temperature of each component (including electrons and ions). When the two ions are different, $\delta_{12}=0$, and when they are the same, $\delta_{12}=1$. The above equations give the minimum parameter requirements for fusion power generation. The fusion reaction rate is assumed to follow the Maxwellian distribution in the appendix, yielding
\begin{eqnarray}
\langle\sigma v\rangle=\langle\sigma v\rangle_M.
\end{eqnarray}
Considering the case where the temperatures of the electron and ion are the same, $T_e=T_1=T_2=T$, we obtain
\begin{eqnarray}
 n_e\tau_E=\frac{3}{2}k_BZ_i(1+Z_i)({1+\delta_{12}})\frac{T}{\langle\sigma v\rangle Y_+}.
\end{eqnarray}
Here, it should be noted that $n_i=n_1+n_2$ and the quasi-neutral condition is expressed as $n_e=Z_1n_1+Z_2n_2=Z_in_i$. The minimum value of $n_e\tau_E$ obtained from this calculation, along with the corresponding $T_i$, is referred to as the "ignition" condition.
\begin{table}[htp]
\caption{Minimum $n_e\tau_ET_i$ requirements for various fusion reactions according to the Lawson ignition criterion.}
\begin{center}
\small
\begin{tabular}{c|c|c|c|c|c|c|c}
\hline\hline
Fusion Reaction & $n_1/n_i$ & $Y_+$ & $Y$ & $T_i$ & $n_e\tau_ET_i$ & $n_e\tau_E$ & \thead {\small Difficulty \\ \small Coefficient}\\
(Unit) &  &  MeV & MeV &  keV & ${\rm m^{-3}\cdot s\cdot keV}$ & ${\rm m^{-3}\cdot s}$ & \\\hline
D-T & 0.5 & 3.52 & 17.6 & 13.6 & 2.8e21 & 2.0e20 & 1 \\
D-D & 1.0 & 2.43 & 3.66 & 15.8 & 1.9e23 & 1.2e22 & 69 \\
D-$^3$He & 0.634 & 18.4 & 18.4 & 57.7 & 5.1e22 & 8.8e20 & 18 \\
p-$^{11}$B Nevins00 & 0.795 & 8.68 & 8.68 & 138 & 7.6e23 & 5.5e21 & 275 \\
p-$^{11}$B Sikora16 & 0.795 & 8.68 & 8.68 & 144 & 6.6e23 & 4.6e21 & 239 \\
Catalyzed D-D & 1.0 & 13.4 & 21.6 & 15.8 & 3.5e22 & 2.2e21 & 13 \\
p-p Cycle & 1.0 & 13.4 & 13.4 & 9.2 & 1.1e45 & 1.2e44 & 3.9e23 \\
\hline\hline
\end{tabular}
\end{center}
\label{tab:lawsonntauEnorad}
\end{table}

The choice of ion density ratio can have some influence on the results. Optimizing the density ratio minimizes $n_e\tau_E$, and the corresponding results can be seen in Figure \ref{fig:lawsonntauEnorad} and Table \ref{tab:lawsonntauEnorad}. We can see that the requirement for the Lawson triple product $n_e\tau_ET_i$ is lowest for the D-T fusion, at approximately $2.8\times10^{21}{\rm m^{-3}\cdot{}s\cdot{}keV}$; followed by D-$^3$He and catalyzed D-D, then pure D-D, and finally p-$^{11}$B. Here and subsequently in this paper, unless otherwise stated, the reaction cross section used for p-$^{11}$B is based on Sikora (2016). Based on these data, we can conclude that D-$^3$He fusion is 18 times harder than D-T fusion, D-D fusion is 69 times harder, catalyzed D-D is 13 times harder, and p-$^{11}$B fusion is 230-280 times harder than D-T fusion.However, in reality, it is not that simple. Bremsstrahlung and cyclotron radiation are not significant at temperatures of around 5-15 keV, which are required for deuterium-tritium fusion. However, they increase dramatically at temperatures of 40-300 keV, required for other fuels. In addition, with the low reaction cross-section of the fuel itself, the ratio of radiative energy loss to fusion energy release increases significantly. Therefore, the difficulty of maintaining the same $\tau_E$ for different fuels varies. Subsequent more complete calculations will further demonstrate the more accurate conditions required for non-deuterium-tritium fusion, where the energy-confinement time $\tau_E$ will change. For example, radiation losses will not be measured by $\tau_E$ and will be listed separately. 

Using the criteria mentioned above, one can better understand why stars can release energy in a steady state. For a solar core temperature of 1 keV, the pp reaction rate is about $10^{-48} \rm{m^3/s}$. Consequently, $n_e \tau_E \simeq 10^{46} \rm{m^{-3} \cdot s}$ is obtained. The density in the solar core $n_e \simeq 10^{32} \rm{m^{-3}}$, resulting in an energy-confinement time of $\tau_E \simeq 10^{14} \rm{s} \sim 3$ million years. In other words, to utilize the energy released from nuclear fusion with such a low cross-section reaction as pp, radiation and internal energy must be confined for millions of years or more. The probability of achieving this on Earth is evidently extremely low.

Because the requirements for deuterium-deuterium, deuterium-helium, and hydrogen-boron fusion are higher than those for deuterium-tritium fusion, more advanced techniques are needed. Therefore, they are also referred to as "advanced fuels".

\section{Possible Power Generation Methods}

\begin{figure}[htbp]
\begin{center}
\includegraphics[width=15cm]{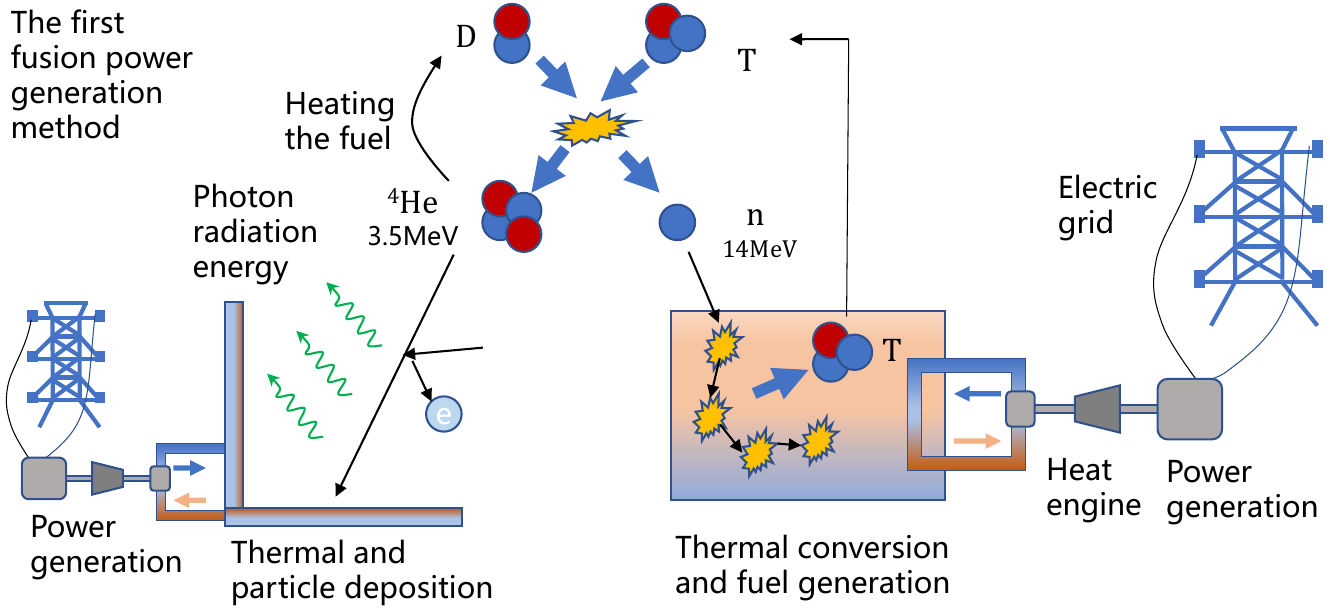}\\
\caption{First type of fusion power generation (deuterium-tritium fusion).}\label{fig:electricgrida}
\end{center}
\end{figure}

\begin{figure}[htbp]
\begin{center}
\includegraphics[width=12cm]{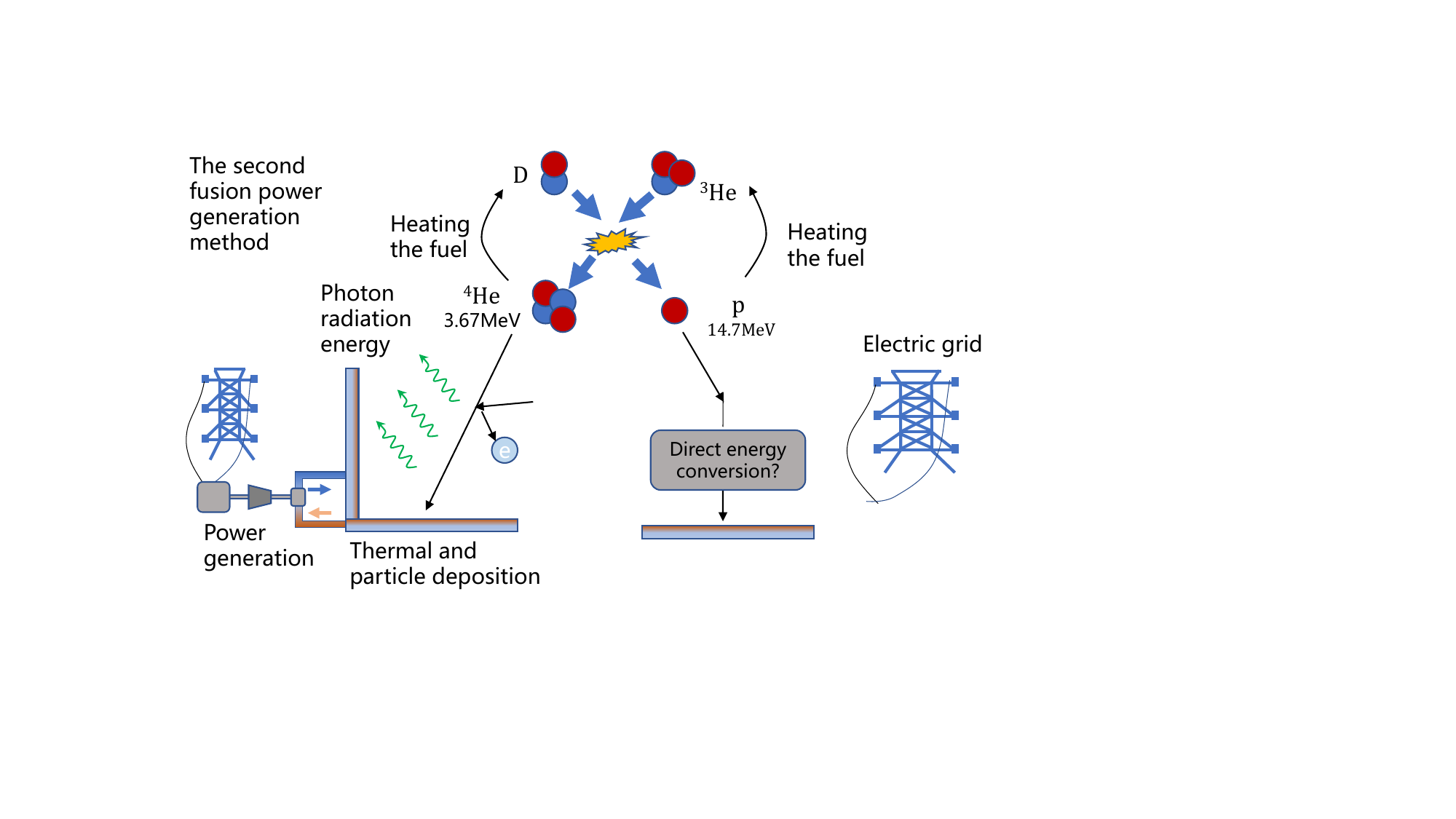}\\
\caption{Second type of fusion power generation (advanced fuels, deuterium-helium, hydrogen-boron).}\label{fig:electricgridb}
\end{center}
\end{figure}

The type of power generation is determined by the fusion products. If there are neutrons in the products (deuterium-tritium, deuterium-deuterium), a similar approach to that of fission reactors is needed, as shown in Figure \ref{fig:electricgrida}. If the products only consist of charged particles (deuterium-helium, hydrogen-boron), in addition to power generation through heat exchange (with a current conversion efficiency of about 30-40\%), direct power generation (with an expected conversion efficiency of over 80\%) may also be possible, as shown in Figure \ref{fig:electricgridb}. However, it should be noted that neutrons can penetrate the first wall and deposit in the blanket, thereby reducing the heat load on the first wall. On the other hand, all charged particles hit the first wall directly, resulting in a greater thermal load on it. In terms of heat load, non-neutron fusion may not necessarily be easier than neutron-producing fusion. 

At the same time, a large amount of energy is emitted in the form of radiation in the fusion reactor, and how to generate electricity or recycle this energy is a question worth considering. Due to the wide spectrum of Bremsstrahlung radiation, it can usually only be utilized through thermal conversion. On the other hand, synchrotron radiation has a relatively narrow frequency range and its wavelength is usually within the range that can be reflected by materials. Therefore, besides reducing losses by using high reflectivity (e.g., greater than 90\%), there may be other ways to utilize synchrotron radiation.

\section{Possible ways to improve fusion reaction rate}

From the analysis of the reaction cross section and fusion triple product discussed above, it can be seen that the key difficulty in achieving fusion energy lies in the low fusion reaction rate $\langle\sigma v\rangle$. If the fusion reaction rate could be increased by a factor of 100, or if there could be a higher reaction cross section $\sigma(E)$ in the low energy region, fusion energy might have been realized long ago, just like fission energy.

\subsection{Some basic ideas}

Since the fusion reaction rate is an integral over the reaction cross section, the most natural idea is to use non-thermalized distribution functions, such as beam distribution or high-energy ion tail distribution, to make the most of the resonance peak and improve the reaction rate as much as possible. Another feasible method is muon-catalyzed fusion and spin polarization, etc. These methods, when optimized, may achieve a tens of percent improvement in reaction rate or even a 1-2 times increase. However, no feasible method for a significant improvement in reaction rate has been found so far. For example, spin polarization is easily depolarized even in a thermalized plasma, so its improvement on fusion yield is minimal.

The negative muon has a mass about 212 times that of an electron, and its distance from the atomic nucleus when combined into a neutral state is also about the same multiple closer than an electron, which reduces the Coulomb barrier that the two atomic nuclei have to overcome, thereby increasing the fusion reaction cross section. However, the lifetime of a muon is short, about $2\times10^{-6}$ s, and it can only catalyze a few fusion reactions within its survival time. The total energy released is lower than the energy used to produce the muon. Therefore, as an energy source, it is currently not practical.

Some unconfirmed directions include anomalous gain (hydrogen-boron avalanche) and cold fusion (low-energy nuclear reactions). There may also be a significant improvement in reaction rate in high-density plasma states above $10^{28}$ m$^{-3}$.

Efforts to improve the fusion reaction rate are still an important direction being explored internationally. Finding a method that can significantly enhance the fusion reaction rate would greatly reduce the difficulty of achieving fusion energy.

\subsection{Reaction rate with non-Maxwellian distribution}

In this section, we will discuss the reaction rate with non-thermalized distribution based on the beam distribution.

\begin{figure}[htbp]
\begin{center}
\includegraphics[width=15cm]{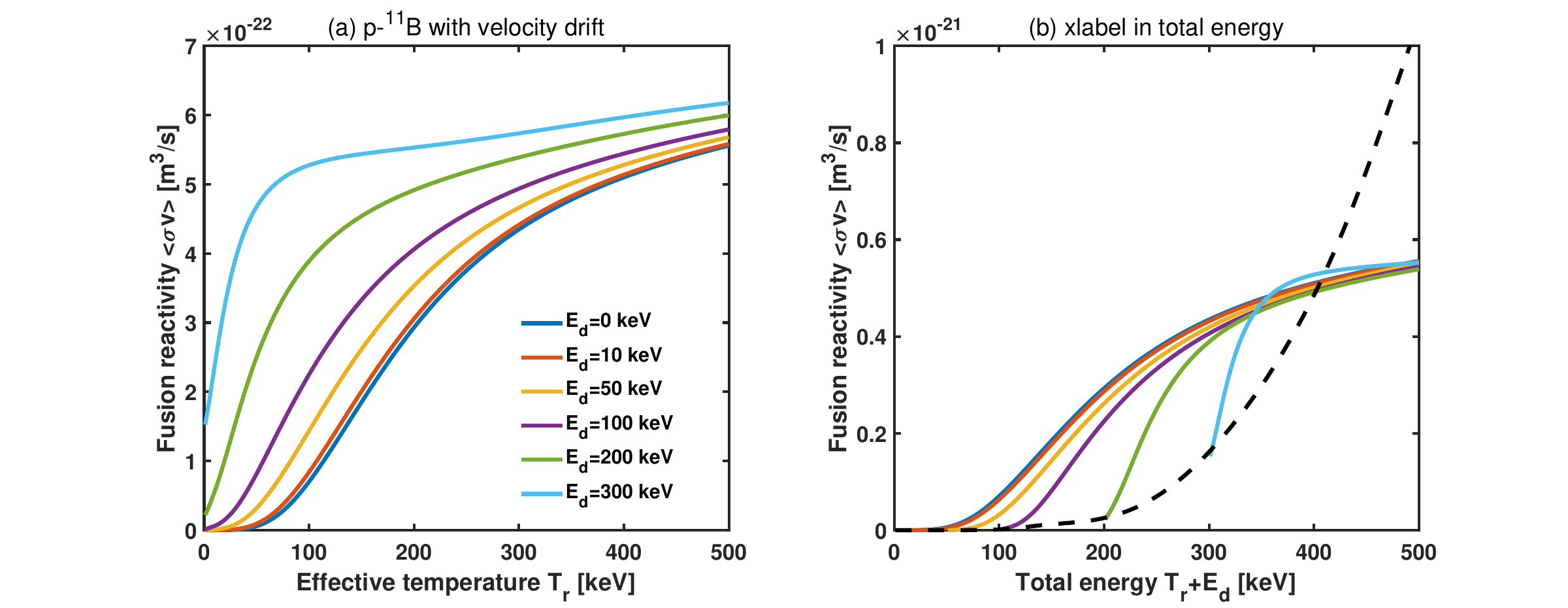}
\caption{The influence of the relative drift velocity difference of hydrogen and boron on the fusion reaction rate, where the dashed line represents the reaction rate when the beam enters a low temperature ($T_r=2$ keV) target plasma.}\label{fig:sgmvdmpb}
\end{center}
\end{figure}\begin{figure}[htbp]
\begin{center}
\includegraphics[width=14cm]{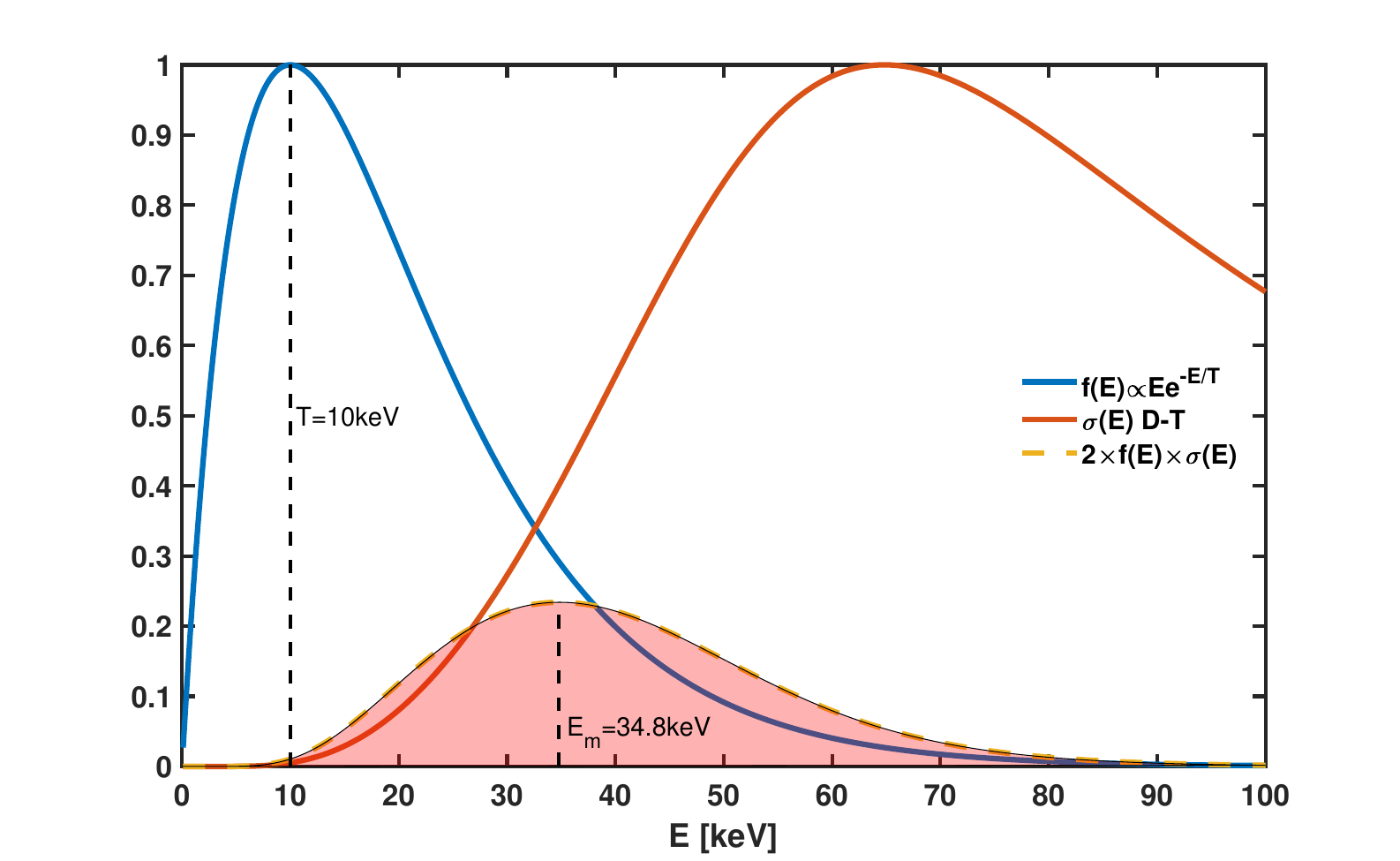}\\
\caption{When the Maxwellian distribution is used, the main contribution to the reaction rate comes from the high-energy tail of the distribution function.}\label{fig:fEsgm}
\end{center}
\end{figure}

\begin{figure}[htbp]
\begin{center}
\includegraphics[width=15cm]{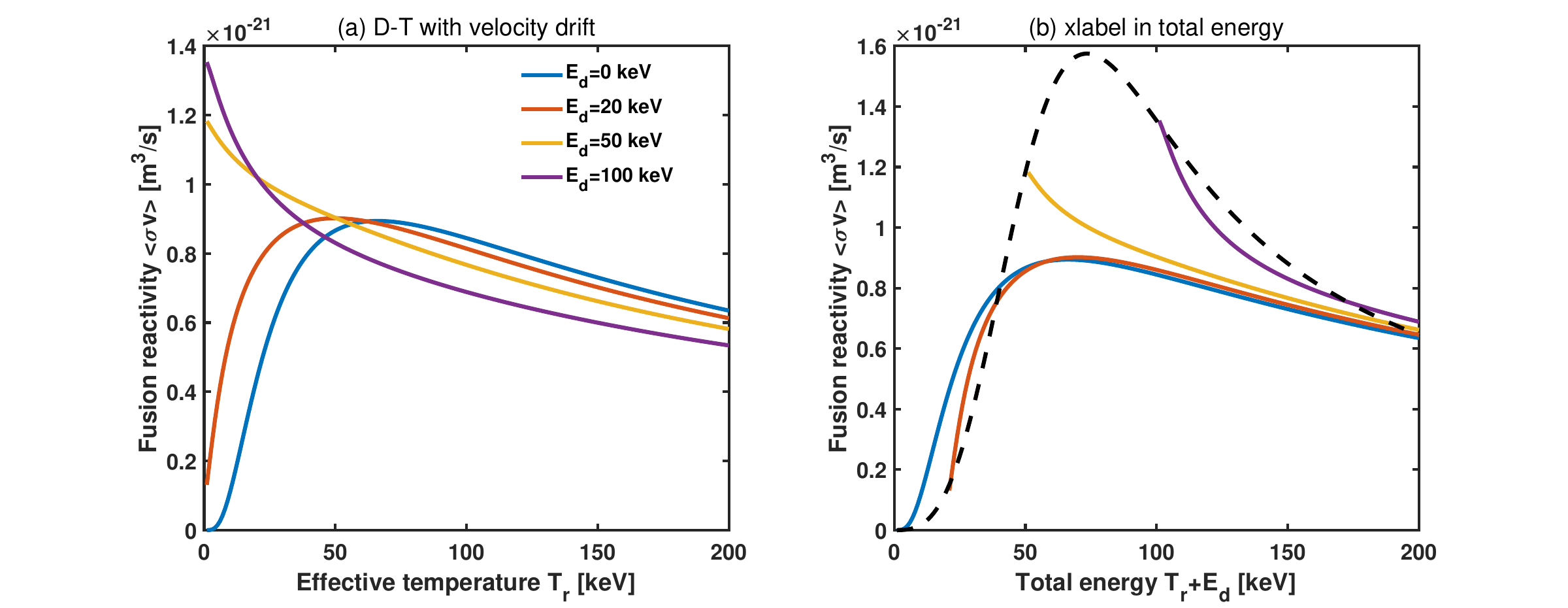}\\
\caption{The impact of the relative drift velocity difference between deuterium and tritium on the fusion reaction rate, where the dashed line indicates the reaction rate when the beam is injected into a low-temperature $T_r=1$keV target plasma.}\label{fig:sgmvdmdt}
\end{center}
\end{figure}Reactivity of fusion reactions
\begin{equation}
\langle\sigma v\rangle=\int\int d{\bm v}_1d{\bm v}_2\sigma(|{\bm v}_1-{\bm v}_2|)|{\bm v}_1-{\bm v}_2|f_1({\bm v}_1)f_2({\bm v}_2),
\end{equation}
where $f_1$ and $f_2$ are the normalized distribution functions of the two ions, that is, $\int f_{1,2}d{\bm v}=1$.
For two ions with unequal temperatures but both following beam Maxwellian distribution,
\begin{equation}
f_j(v)=\Big(\frac{m_j}{2\pi k_BT_j}\Big)^{3/2}\exp\Big[-\frac{m_j({\bm v}-{\bm v}_{dj})^2}{2k_BT_j}\Big],
\end{equation}
by performing a change of variables, the integration yields [Xie (2023)]
\begin{eqnarray}
\langle\sigma v\rangle_{DM}&=&\frac{2}{\sqrt{\pi}v_{tr}v_{d0}}\int_0^{\infty}\sigma(v)v^2\exp\Big(-\frac{v^2+v_{d0}^2}{v_{tr}^2}\Big)\sinh\Big(2\frac{vv_{d0}}{v_{tr}^2}\Big)dv\\\nonumber
&=&\sqrt{\frac{2}{\pi m_rk_B^2T_rT_d}}\int_0^{\infty}\sigma(E)\sqrt{E}\exp\Big(-\frac{E+E_{d}}{k_BT_r}\Big)\sinh\Big(\frac{2\sqrt{EE_d}}{k_BT_r}\Big)dE,
\end{eqnarray}
where $\sinh(x)=(e^x-e^{-x})/2$, and the reduced mass, effective temperature, effective thermal velocity, relative drift velocity, and relative drift energy are defined as
\begin{eqnarray}\nonumber
m_r=\frac{m_1m_2}{m_1+m_2},~~T_{r}=\frac{m_1T_2+m_2T_1}{m_1+m_2},\\
v_{tr}=\sqrt{\frac{2k_BT_r}{m_r}},~~
v_{d0}=|{\bm v}_{d2}-{\bm v}_{d1}|, ~~E_d=k_BT_d=\frac{m_rv_{d0}^2}{2}.
\end{eqnarray}
The above results can be reduced to the expression without beam in the appendix in the degenerate case, and can also be reduced to the expression of $\delta$-beam in Miley (1974, 1975) and Morse (2018). Figure \ref{fig:sgmvdmpb} shows the influence of relative drift velocity on the reactivity of hydrogen-boron fusion at different energies. When the effective temperature is the same, the beam effect usually enhances the reactivity. However, from the perspective of total energy\footnote{Here, for simplicity, the x-axis is taken as $T_r+E_d$, for a more reasonable choice, see Xie (2023).}, the reactivity does not increase significantly, but rather decreases at low energies.This also reflects that for the same amount of energy, the fusion reaction rate of thermonuclear fusion is more favorable compared to that of beam-driven fusion. This is due to the enhanced effect of the high-energy tail of the Maxwellian distribution function, which significantly increases the fusion reaction rate. Figure 1.5 presents this effect well. Beam-driven fusion is only more favorable when it can effectively utilize the resonance peak, which occurs in the energy range greater than 400 keV for the hydrogen-boron reaction. Figure 1.6 shows the influence of the relative drift velocity at different energies on the fusion reaction rate of deuterium-tritium fusion. It can be seen that when the beam energy exceeds 40 keV, its reaction rate can surpass that of the Maxwellian distribution with the same energy. This is also one of the reasons why some neutral beam heating methods (typically using injection energies above 50 keV) in current tokamaks can enhance fusion yield.

For arbitrary distribution functions $f_1$ and $f_2$, calculating the reaction rate $\langle\sigma v\rangle$ usually requires integration over the high-dimensional velocity space. Xie (2023) derived the two- and one-dimensional integral forms of the fusion reaction rate for drift bi-Maxwellian distribution and calculated their influences on the fusion reaction rates of deuterium-tritium, deuterium-deuterium, deuterium-helium, and hydrogen-boron.

\section{Summary of this chapter}

\begin{figure}[htbp]
\begin{center}
\includegraphics[width=12cm]{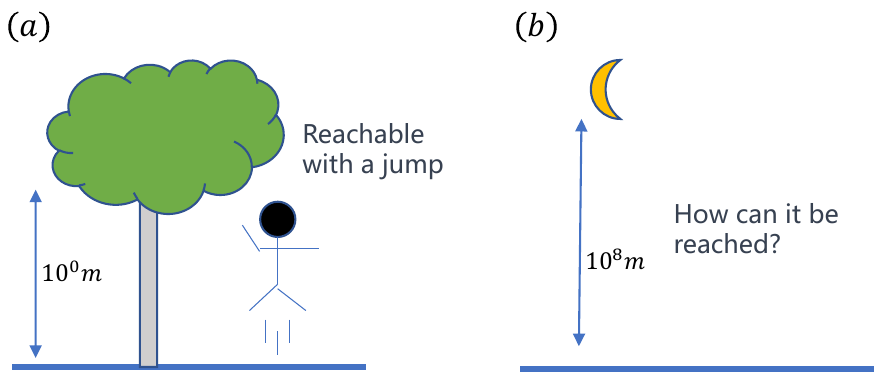}\\
\caption{Fusion energy is not a goal that can be reached with a simple jump, so most methods for optimizing "high jump" techniques are not applicable to fusion.}\label{fig:howfar}
\end{center}
\end{figure}

Due to the fact that the Coulomb cross section is much larger than the fusion cross section (the physical meaning of which will be clearer in subsequent chapters), the fusion fuel must be confined to achieve fusion energy. Moreover, due to the extremely low fusion reaction cross section, the vast majority of fusion reactions are not suitable as raw materials for controlled fusion energy, and only a few types of reactions are possible. Each of these fusion reactions has its own disadvantages, such as high-energy neutrons, scarce raw materials, and excessively high reaction conditions. This makes fusion energy extremely difficult to achieve. Based on energy balance, the minimum requirements for achieving fusion gain in terms of temperature, density, and confinement time can be deduced. These conditions are already close to being achievable for deuterium-tritium fusion, but are at least tens or even hundreds of times more difficult for deuterium-deuterium, deuterium-helium, and hydrogen-boron fusion.

When considering radiation and the actual situation of fusion reactors in later sections, more difficulties will be encountered. As a simple analogy, for many other technological research and development fields, the level of difficulty may only be from jumping to a height of 1 meter to reaching for a branch 10 meters high; but for controlled fusion energy research, it is like when jumping to a height of 1 meter, one must be able to reach the height of the moon, which is 400,000 kilometers away, a difference of eight orders of magnitude. No matter how much you optimize the high jump technique, it will not be possible to achieve this. A completely new way of thinking is necessary. In this chapter, we have seen that nuclear fusion reactions are the main limiting factor. To break through, we need to ask one by one: is it possible to increase the reaction cross section? Can the reaction rate be increased and maintained? Can the non-thermalized distribution that can increase the reaction rate be sustained? Can issues such as material damage caused by high-energy neutrons and tritium breeding be resolved? Some of these questions are still being explored by nuclear physics or other fields internationally, and some will be discussed later in this paper.

\vspace{30pt}
Key points of this chapter:
\begin{itemize}
\item The main limitations of fusion energy come from the reaction cross section and the scarcity of raw materials. Only a few reactions, such as ${\rm D-{}T}$, ${\rm D-{}D}$, ${\rm D-{}^{3}He}$, and ${\rm p-{}^{11}B}$, are suitable for fusion energy.
\item If the fusion reaction rate can be increased by 100 times, fusion energy may have already been realized.
\item The scientific feasibility of deuterium-tritium fusion has already been achieved. The zeroth-order challenge for fusion energy lies in tritium breeding and high-energy neutron protection.
\item Non-deuterium-tritium fusion, from the perspective of the fusion temperature, density, and energy confinement time as aspects of scientific feasibility, is at least 10 times more difficult than deuterium-tritium fusion.
\end{itemize} 
\chapter{Range of Optional Parameters and Lawson Criterion}\label{chap:lawson}

In the previous chapter, we discussed the basic fusion reactions and identified four fusion reactions that are worth considering. We also plotted their reaction cross sections and reaction rates, and provided preliminary parameter requirements for fusion temperature, density, and confinement time. In this chapter, we will further explore the physical implications of these data in fusion energy research, and clarify the range of optional parameters, particularly the subtle differences in the implications behind different assumptions. Due to our discussion, we usually ignore some more complex practical factors and adopt optimistic assumptions. Therefore, the conclusions obtained are necessary conditions, but not necessarily sufficient conditions. Through the calculations in this chapter, quantitative criteria that can serve as references for the theoretical feasibility of various fusion fuels will be obtained, and some directions for breakthrough or reduction of these theoretical conditions can be indicated. Concrete implementation plans will be discussed in subsequent chapters.

\section{Fusion Mean Free Path}

For collisions, we are familiar with the mean free path $\lambda_m$, which refers to the average distance traveled before a collision occurs, the mean collision time $\tau_m$, which refers to the average time before a collision occurs, and the mean collision frequency $\nu_m$, which refers to the number of collisions occurring per unit time (per second). These data can all be calculated based on the collision cross section.Similarly, for the fusion reaction, the number of nuclear reactions per unit volume per unit time is
\begin{equation}
R_{12}=\frac{n_1n_2}{1+\delta_{12}}\langle\sigma v\rangle,
\end{equation}
where $n_1$ and $n_2$ are the number densities of the two nuclei, $\sigma$ is the fusion reaction cross section, and $\langle\sigma v\rangle$ is the velocity-averaged fusion reaction rate. If the two nuclei are different, then $\delta_{12}=0$; if they are the same, then $\delta_{12}=1$. The reaction frequency representing the average number of fusion reactions occurring for particle 1 per second can be defined as[Dolan (1981)]
\begin{equation}
\nu_{m1}\equiv\frac{R_{12}}{n_1}=\frac{n_2}{1+\delta_{12}}\langle\sigma v\rangle.
\end{equation}
and the average time for one fusion to occur can be defined as
\begin{equation}
\tau_{m1}=\frac{1}{\nu_{m1}}=(1+\delta_{12})\frac{1}{n_2\langle\sigma v\rangle},
\end{equation}
and the average free path representing the average distance traveled by a particle for one fusion event can be defined as
\begin{equation}
\lambda_{m1}=\langle v\rangle_1\cdot\tau_{m1}=(1+\delta_{12})\frac{\langle v\rangle_1}{n_2\langle\sigma v\rangle},
\end{equation}
where
\begin{equation*}
\langle v\rangle_1=\frac{\int vf_1({\bm v})d{\bm v}}{\int f_1({\bm v})d{\bm v}},
\end{equation*}
is the average velocity of particle 1. For a Maxwellian distribution, $\langle v\rangle_1=\sqrt{\frac{8k_BT_1}{\pi m_1}}$. For a monoenergetic beam and a stationary target, it can be simplified as $\lambda_{m1}=\frac{1+\delta_{12}}{n_2\sigma}$.  Similar definitions can be made for particle 2. Usually, $\lambda_{m1}\neq\lambda_{m2}$, $\tau_{m1}\neq\tau_{m2}$, $\nu_{m1}\neq\nu_{m2}$.\begin{figure}[htbp]
\begin{center}
\includegraphics[width=15cm]{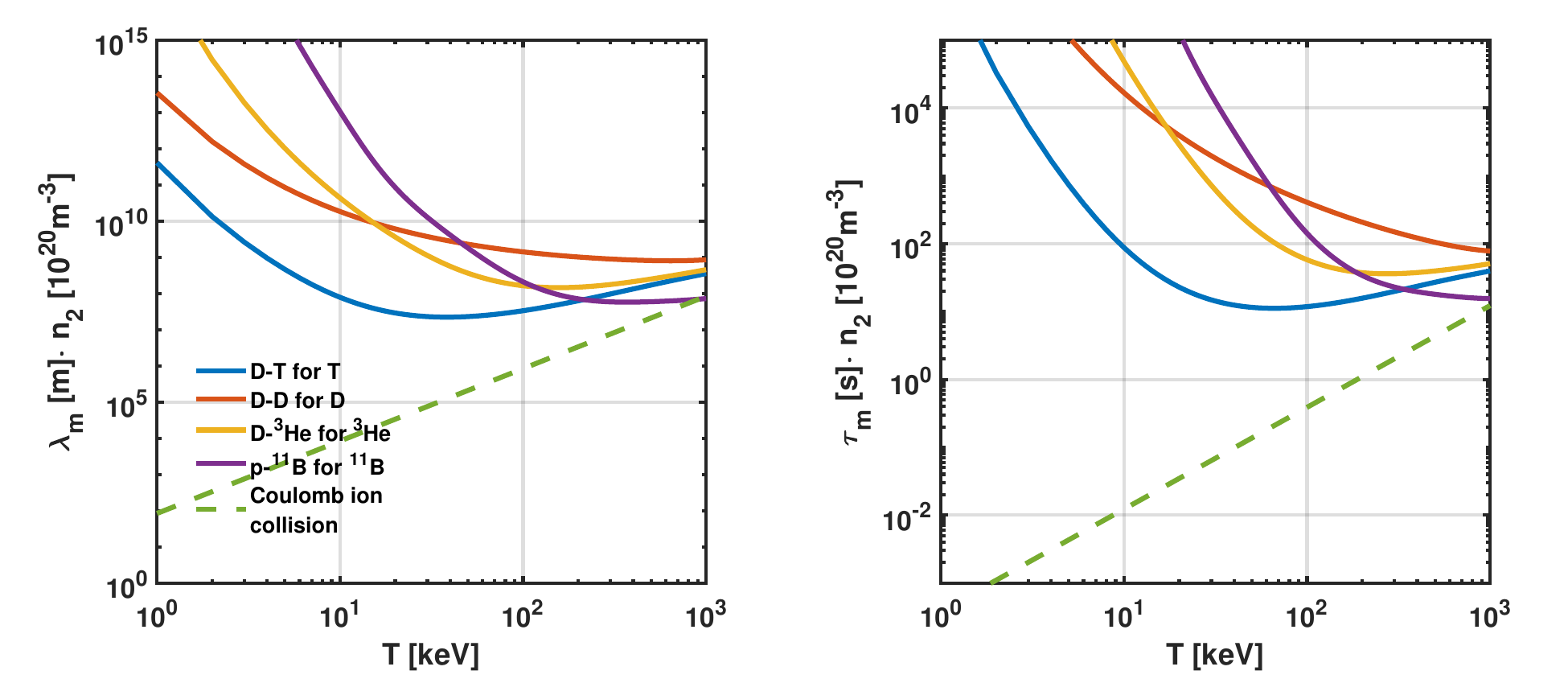}\\
\caption{Average fusion reaction mean free path and average reaction time, compared with the Coulomb collision time of ions (protons).}\label{fig:fusionmfp}
\end{center}
\end{figure}

Figure \ref{fig:fusionmfp} shows the variation of the average fusion reaction mean free path $\lambda_{m1}$ and average reaction time $\tau_{m1}$ with temperature $T$, assuming the same temperature for both ions $T_{eff}=T_1=T_2=T$ when the target plasma density is $n_2=10^{20}\,{\rm m^{-3}}$. Some typical values are listed in Table \ref{tab:fusionmfp}. The figure also shows the Coulomb collision time of protons [Wesson11], which is found to be much smaller than the average time for fusion reactions. It is observed that both the mean free path and the collision time are inversely proportional to the density, that is, $n_2\tau_{m1}$ and $n_2\lambda_{m1}$ remain constant at the same temperature. Therefore, the average reaction (collision) time is small and the average fusion (collision) mean free path is short at high density.

\begin{figure}[htbp]
\begin{center}
\includegraphics[width=15cm]{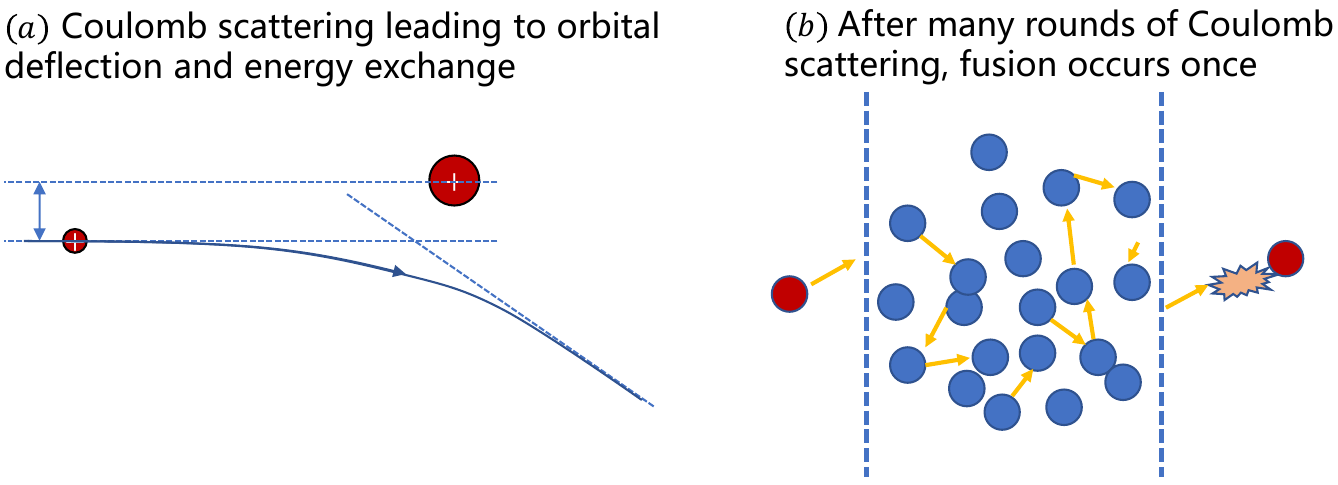}\\
\caption{Illustration of Coulomb collision scattering, where many Coulomb collisions must occur for 1 fusion reaction to occur due to the larger Coulomb collision cross section compared to the fusion reaction cross section.}\label{fig:scattervsfusion}
\end{center}
\end{figure}\begin{table}[htp]
\caption{Average mean free paths and average reaction times for fusion nuclear reactions under typical temperature parameters, with a density of $n_2=10^{20}{\rm m^{-3}}$, where the incursion ions are indicated within parentheses.}
\begin{center}
\footnotesize
\begin{tabular}{c|c|c|c|c|c|c}
\hline\hline
Fusion Reaction & $\lambda_{10keV}$[m] & $\lambda_{100keV}$[m] & $\lambda_{200keV}$[m] & $\tau_{10keV}$[s] &  $\tau_{100keV}$[s] &  $\tau_{200keV}$[s] \\\hline
DT (T) &  7.9e7  & 3.4e7 &  6.4e7 &  87 & 12 & 16 \\
DD (D) & 1.9e10 &  1.4e9 &  1.0e9 &  1.7e4 & 4.1e2 & 2.1e2 \\
DH (He) & 4.4e10 & 1.7e8  &  1.5e8 &  4.9e4 & 60 & 37 \\
HB (B) & 1.1e13 &  2.1e8 & 7.2e7  & 2.3e7  & 1.4e2 & 34 \\
Coulomb (Proton) & 8.5e3 & 8.5e5  &  3.4e6 & 1.2e-2  & 3.8e-1 & 1.1 \\
\hline\hline
\end{tabular}
\end{center}
\label{tab:fusionmfp}
\end{table}

From this perspective, the previously mentioned Coulomb scattering cross section is much larger than the fusion reaction cross section, such as being 1000 times larger. This means that before one fusion reaction occurs, there have already been 1000 Coulomb collisions, as shown in Figure \ref{fig:scattervsfusion}. For a cold target plasma, each Coulomb collision causes the incursion ions to lose a certain amount of energy, which means that the energy of the incursion ions is primarily used to heat the target plasma rather than to induce fusion. Unless the background plasma is already in a thermalized state with energy comparable to or higher than the incursion ions, i.e., achieving fusion gain through the thermonuclear mode. This also suggests that in order to accurately measure the fusion reaction cross section using the beam-target method, the target needs to be sufficiently thin, i.e., much smaller than the average Coulomb collision mean free path. Note that the collision cross section between charged ions and neutral atoms is even larger than the Coulomb cross section, resulting in an ionization cross section of approximately $10^7{\rm b}$, which means a shorter average mean free path. This also indicates that wall confinement is not feasible and achieving fusion energy gain is impossible.

From the figure, it can be seen that within the typical fusion temperature range of 10-200 keV, $\tau_m\cdot n_2\sim10^{21}-10^{22}{\rm s\cdot m^{-3}}$. If the actual confinement parameter $n\tau_E$ is smaller than this value, the fuel is not fully burned. If it is larger than this value, the reaction products have already dominated in the fusion reactor. Therefore, the above data has another physical meaning: if it is required that most ions undergo fusion reactions fully, then the average fusion collision time is equivalent to the required confinement time; the average fusion mean free path is equivalent to the system size required for beam reaction, or the distance ions have traveled in the case of thermonuclear reaction. If the confinement time is shorter than the average fusion reaction time here, it means that the fuel is not fully undergoing fusion reactions and the combustion rate is not high. The subsequent Lawson criterion is essentially limited by these two parameters, resulting in a narrow optimal range of fusion temperature and a minimum requirement for density and confinement time.

\section{Fusion Power Density}

Fusion power density refers to the fusion power released per unit volume. It is directly proportional to the square of the density of ions undergoing fusion reactions. The power density limits the selectable range of density. If the density is too high, the energy released per unit volume is too large and the pressure is too high, making it uncontrollable. Or, the heat load scattered by the surface is too large for the wall to withstand. Therefore, for fusion energy research, the fusion reaction region needs to either have a low density or an extremely small volume. Otherwise, like a hydrogen bomb, the energy released in a single event exceeds a kiloton of TNT. On the other hand, if the density is too low, the fusion power is too low and cannot meet the economic requirements. Therefore, the plasma density in fusion energy research has an optimal range.

\begin{figure}[htbp]
\begin{center}
\includegraphics[width=15cm]{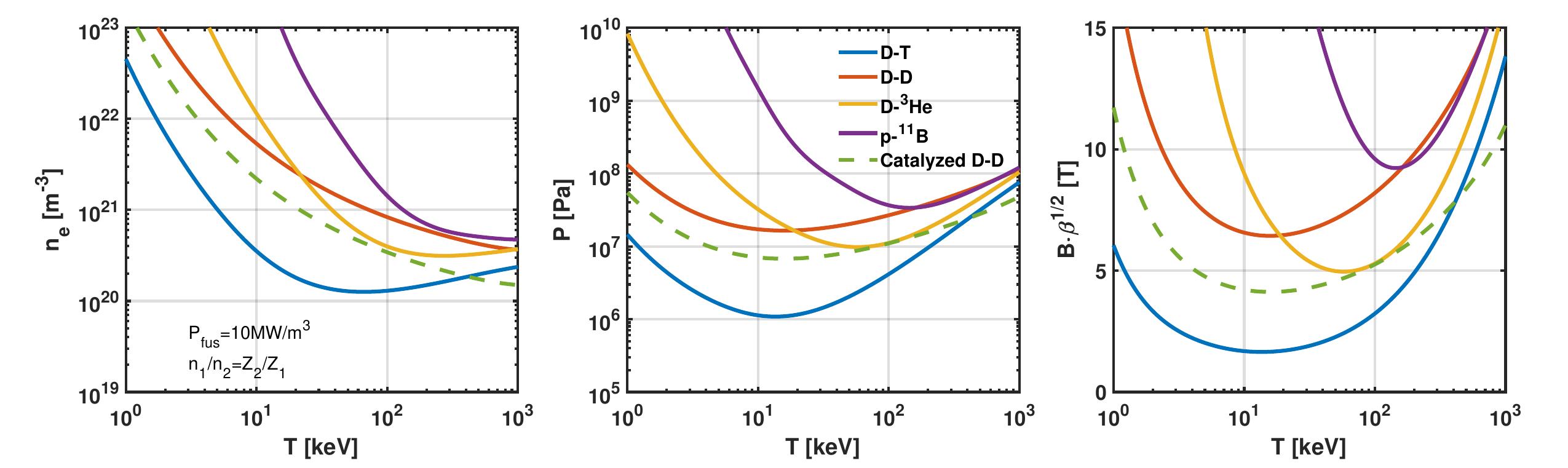}\\
\caption{Plasma density and corresponding plasma thermal pressure and required magnetic field when the fusion power density is set to $P_{fus}=10{\rm MW/m^3}$, with all particle temperatures set to the same value.}\label{fig:fusionpfus}
\end{center}
\end{figure}The volumetric fusion power, $P_{fus}$, is the product of the reaction rate $R_{12}$ and the energy released in a single reaction, $Y$, as follows:
\begin{eqnarray}
    P_{fus}=\frac{1}{1+\delta_{12}}n_1n_2\langle\sigma v\rangle Y.
\end{eqnarray}
Assuming that based on economic requirements, the fusion power density cannot be less than $P_{min}$, and based on thermal load requirements, it cannot exceed $P_{max}$. For example, if $P_{min}=0.1{\rm MW/m^3}$ and $P_{max}=100{\rm MW/m^3}$, the density can only vary by a factor of $\sqrt{P_{max}/P_{max}}\simeq32$ at the same temperature. In reality, there is usually an optimal $P_{fus}$, such as $P_{fus}\simeq10{\rm MW/m^3}$, which limits the density. It is also worth noting that the fusion reaction rate is sensitive to temperature, and the optimal density can vary by orders of magnitude for different temperatures. Therefore, depending on the situation, an optimized density can be selected, such as optimizing for the lowest pressure or the lowest density. For magnetic confinement fusion, the density calculated in this way is close to the plasma density in the actual device. For inertial confinement or magnetically inertial confinement fusion, the density here can be considered as the average density of the fusion region over the entire cavity volume. Therefore, although the density in the target pellet area is extremely high, the volume of the cavity is large, resulting in a relatively low average density\footnote{In early literature before 1960, such as Bishop (1958), Glasstone (1960), Lawson (1955), it was considered unrealistic to have excessively high densities based on power density. For example, Lawson (1955) discussed the assumption of radiation transparency due to bremsstrahlung and concluded that the fusion power with opaque parameters should be at least $5\times10^{10}V^{1/3}~{\rm W/m^3}$, stating that "This is certainly greater than could be handled in a controlled reactor". Looking at the research history of inertial confinement fusion, it is evident that this "impossibility" is not really impossible, as long as the cavity is large and the target pellet is small. This also tells us that it is necessary to distinguish the premises of conclusions in order to differentiate between what is truly impossible in principle and what is limited by our concepts. This is also what this book aims to clarify.}.By using the quasi-neutrality condition, we maximize the fusion power $P_{fus}$ with the electron density $n_e=Z_1n_1+Z_2n_2$ being a constant. In this case, the ratio of ion densities is given by $n_1/n_2=Z_2/Z_1$. Figure \ref{fig:fusionpfus} shows the plasma density and the corresponding plasma thermal pressure $p$ for different fusion reactions when the fusion power density is $P_{fus}=10{\rm MW/m^3}$. It can be seen that the optimal density range is between $10^{20}-10^{22}{\rm m^{-3}}$ and the corresponding pressure is approximately 10-1000 times atmospheric pressure. For comparison, atmospheric pressure is approximately $10^5{\rm Pa}$ and atmospheric density is about $10^{25}{\rm m^{-3}}$. In the case of magnetic confinement fusion, due to the constraint of the beta parameter $\beta\propto p/B^2$, the required magnetic field $B$ also has a minimum value. A detailed discussion on this is presented in subsequent chapters. As a quantitative comparison, according to the data in the appendix, 1 atmosphere corresponds to a magnetic field of about 0.5T, and 1000 atmospheres correspond to a magnetic field of about 15.8T. In other words, if $\beta=1$, this magnetic field is the required field for confinement corresponding to the pressure. If the power density is set to $P_{fus}=0.1{\rm MW/m^3}$, then the above density decreases by only a factor of 10, and the required magnetic field decreases by approximately $\sqrt{10}\simeq3.16$ times. $0.1{\rm MW/m^3}$ represents the fusion reactor region of one cubic meter, which can produce energy equivalent to that used by 1000 100W appliances, that is, the daily electricity consumption of a small residential area.

Since the energy of the fusion reactor can only be exported from the surface, that is, the surface material has a maximum surface energy limit $P_S=P_{fus}V/S\leq P_{S,max}$, the power density $P_{fus}$ of the fusion reactor cannot be too high, or only a smaller volume can be used, minimizing the ratio of the volume to the surface area $V/S\propto r$. In comparison, the energy of a fission reactor can be exported from within the reactor, with a typical power density of 50-1000${\rm MW/m^3}$. From this perspective, the size of a fusion reactor is usually much larger than that of a fission reactor, which to some extent affects the economy of the fusion reactor. The power density of coal-fired power plants is about 0.2${\rm MW/m^3}$, and its auxiliary facilities are not as complex as those in a fusion reactor. As a quantitative calculation, we can assume a spherical plasma and set the maximum average surface heat load as $P_{S,max}=10{\rm MW/m^2}$. Then the maximum radius of the device is given by
\begin{eqnarray}
    R_{max}=\frac{4\pi P_{S,max}}{\frac{4}{3}\pi P_{fus}}=\frac{3P_{S,max}}{P_{fus}}.
\end{eqnarray}
Setting $P_{fus}=10{\rm MW/m^3}$, $R_{max}=3{\rm m}$, and the total power of the device is given by $P=4\pi R_{max}^2P_{S,max}=1.13{\rm GW}$. It can be seen that the requirements of the wall load and economy (the size cannot be infinitely large) of the device actually limit the total power range of a single fusion reactor.

\section{Ideal Ignition Conditions}Let us discuss some fusion conditions in ideal situations, especially the minimum temperature required for fusion ignition.

\subsection{Ideal Ignition Temperature}
In addition to energy release through fusion reactions, the confined fusion plasma also has various energy loss mechanisms. Assuming no consideration of transport losses and only considering radiation losses, specifically bremsstrahlung losses, and assuming that all the energy of charged products can deposit into the plasma, we can calculate the so-called "ideal ignition condition," which represents the temperature at which the power from charged products exceeds the power from bremsstrahlung radiation.

In the weak relativistic case, the Bremsstrahlung power is given by [Nevins (1998)]:
\begin{eqnarray}\nonumber
P_{brem}&=&C_Bn_e^2\sqrt{k_BT_e}\Big\{Z_{eff}\Big[1+0.7936\frac{k_BT_e}{m_ec^2}+1.874\Big(\frac{k_BT_e}{m_ec^2}\Big)^2\Big]\\&&
+\frac{3}{\sqrt{2}}\frac{k_BT_e}{m_ec^2}\Big\} ~{\rm [W\cdot m^{-3}]}.
\end{eqnarray}
Here, $C_B=1.69\times10^{-38}\times(1000)^{1/2}=5.34\times10^{-37}$, the temperature $k_BT_e$ and energy $m_ec^2$ are in units of keV, the density $n_e$ is in the unit of ${\rm m^{-3}}$, and the effective charge number $Z_{eff}=\sum(n_iZ_i^2)/n_e$. Removing the terms related to $m_ec^2$ reduces the equation to the non-relativistic case. The last term is from electron-electron bremsstrahlung.

\begin{figure}[htbp]
\begin{center}
\includegraphics[width=14cm]{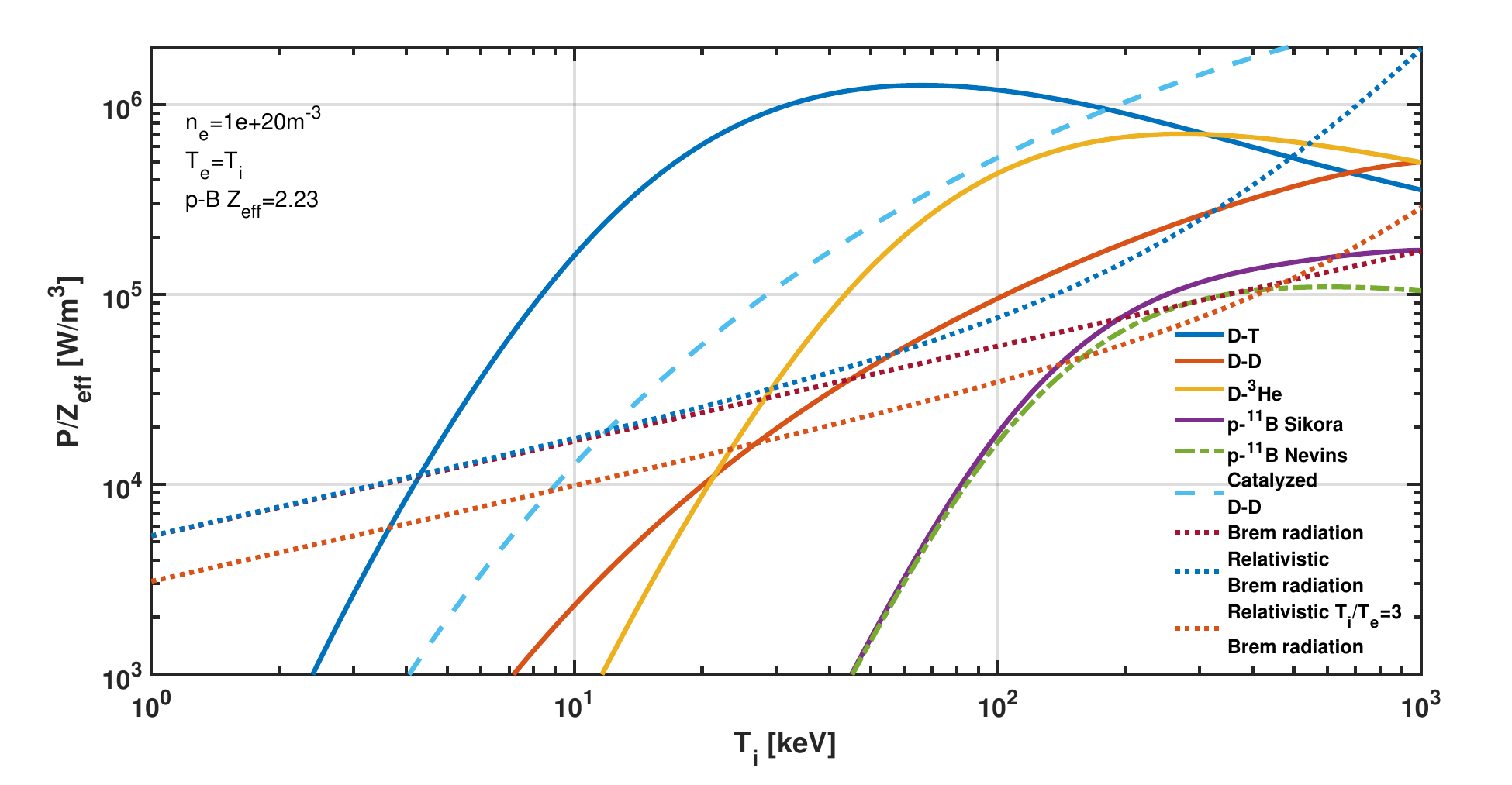}\\
\caption{Comparison between fusion power (from charged products) and bremsstrahlung power for different fusion reactions. The temperature corresponding to the intersection of the two curves on the left is the ideal ignition temperature $T_i$. The calculation of relativistic bremsstrahlung uses $Z_{eff}=2.23$ for the p-${\rm ^{11}}$B reaction.}\label{fig:idealTi}
\end{center}
\end{figure}\begin{table}[htp]
\caption{Ideal ignition temperature $T_i$ when the electron temperature is equal to the ion temperature. }
\begin{center}
\begin{tabular}{c|c|c|c|c|c|c}
\hline\hline
Fusion Reaction & D-T & D-D &  {D-$\rm^3He$} & \thead {\normalsize p-B \\ \normalsize (Nevins)} &  \thead {\normalsize p-B \\ \normalsize (Sikora)} & \thead {\normalsize D-D \\ \normalsize Catalyzed} \\\hline
$n_1/n_i$ &  0.5  & 1.0 &  0.739 &  0.918 & 0.918 & 1.0 \\
$Y_+$ [MeV] &  3.52  & 2.43 &  18.4 &  8.68 & 8.68 & 13.4 \\
$Z_{i}$ &  1.0  & 1.0 &  1.261 &  1.328 & 1.328 & 1.0 \\
$Z_{eff}$ &  1.0  & 1.0 &  1.414 &  2.235 & 2.235 & 1.0 \\
Non-relativistic $T_i$ [keV] &  4.3  & 45 &  28 &  280 & 193 & 12 \\
Relativistic $T_i$ [keV] & 4.3 &  72 &  30 &  - & - & 12 \\
\hline\hline
\end{tabular}
\end{center}
\label{tab:idealTi}
\end{table}

Since the fusion power and bremsstrahlung power are both proportional to the square of the density, their ratio is only related to the temperature and the density ratio of the two ions. When the two ions undergoing fusion are different, the charge numbers of the two ions are $Z_1$ and $Z_2$ respectively, and the density ratio is $x:(1-x)$. According to the quasi-neutrality condition, the electron density is $n_e=Z_in_i$, and the ion densities are $n_1=xn_i$ and $n_2=(1-x)n_i$, with average charge number $Z_i=xZ_1+(1-x)Z_2$ and effective charge number $Z_{eff}=[xZ_1^2+(1-x)Z_2^2]/[xZ_1+(1-x)Z_2]$.

For a given electron and ion temperature, in the non-relativistic case, to maximize $P_{fus}/P_{brem}$, we need to calculate
\begin{eqnarray}\nonumber
max\Big\{\frac{x(1-x)/[xZ_1+(1-x)Z_2]^2}{Z_{eff}}\Big\}=max\Big\{\frac{x(1-x)}{[x(Z_1^2-Z_2^2)+Z_2^2][x(Z_1-Z_2)+Z_2]}\Big\}.
\end{eqnarray}
From this, we can determine the optimal density ratio $x$, for Deuterium-Tritium fusion, $x=0.5$, $Z_i=1.0$, $Z_{eff}=1.0$; for Deuterium-Helium fusion, $x=0.739$, $Z_i=1.261$, $Z_{eff}=1.414$; for Hydrogen-Boron, $x=0.918$, $Z_i=1.328$, $Z_{eff}=2.235$.According to the optimized density ratio mentioned above, Figure \ref{fig:idealTi} shows the curves of fusion power ($P_{fus}$) and bremsstrahlung radiation power ($P_{brem}$), both normalized by their respective $Z_{eff}$ values, where the ion and electron temperatures are assumed to be equal ($T_i=T_e$). Table \ref{tab:idealTi} provides the specific values of the ideal ignition temperature ($T_i$)\footnote{For the ideal ignition temperature data for DD reactions, Freidberg (2007) reports 30 keV, Gross (1984) reports 48 keV, and Glasstone (1960) reports 36 keV. The differences may arise from the different treatment of the output energy of the two reaction channels. Dawson (1981) assumes the combustion of T and He3, resulting in 35 keV.}. From the graph, it can be seen that the ideal ignition temperature for deuterium-tritium fusion is the lowest, at about 4.3 keV, and is almost unaffected by relativistic effects. For hydrogen-boron fusion, if the relativistic bremsstrahlung radiation formula is used, there is no ignition point, i.e., the radiation loss exceeds the fusion power at any temperature. For this reason, hydrogen-boron fusion is usually considered infeasible or extremely difficult. To overcome this condition, the plasma typically needs to be in the regime of thermal ions, i.e., $T_e/T_i<1$. For example, from the graph, it can be seen that the bremsstrahlung radiation at an electron temperature of 100 keV is smaller than the hydrogen-boron fusion power at 200 keV, or when $T_e/T_i=1/3$, the hydrogen-boron fusion power is significantly greater than the bremsstrahlung radiation power. Various possible breakthroughs will be discussed later in the book.

\begin{figure}[htbp]
\begin{center}
\includegraphics[width=14cm]{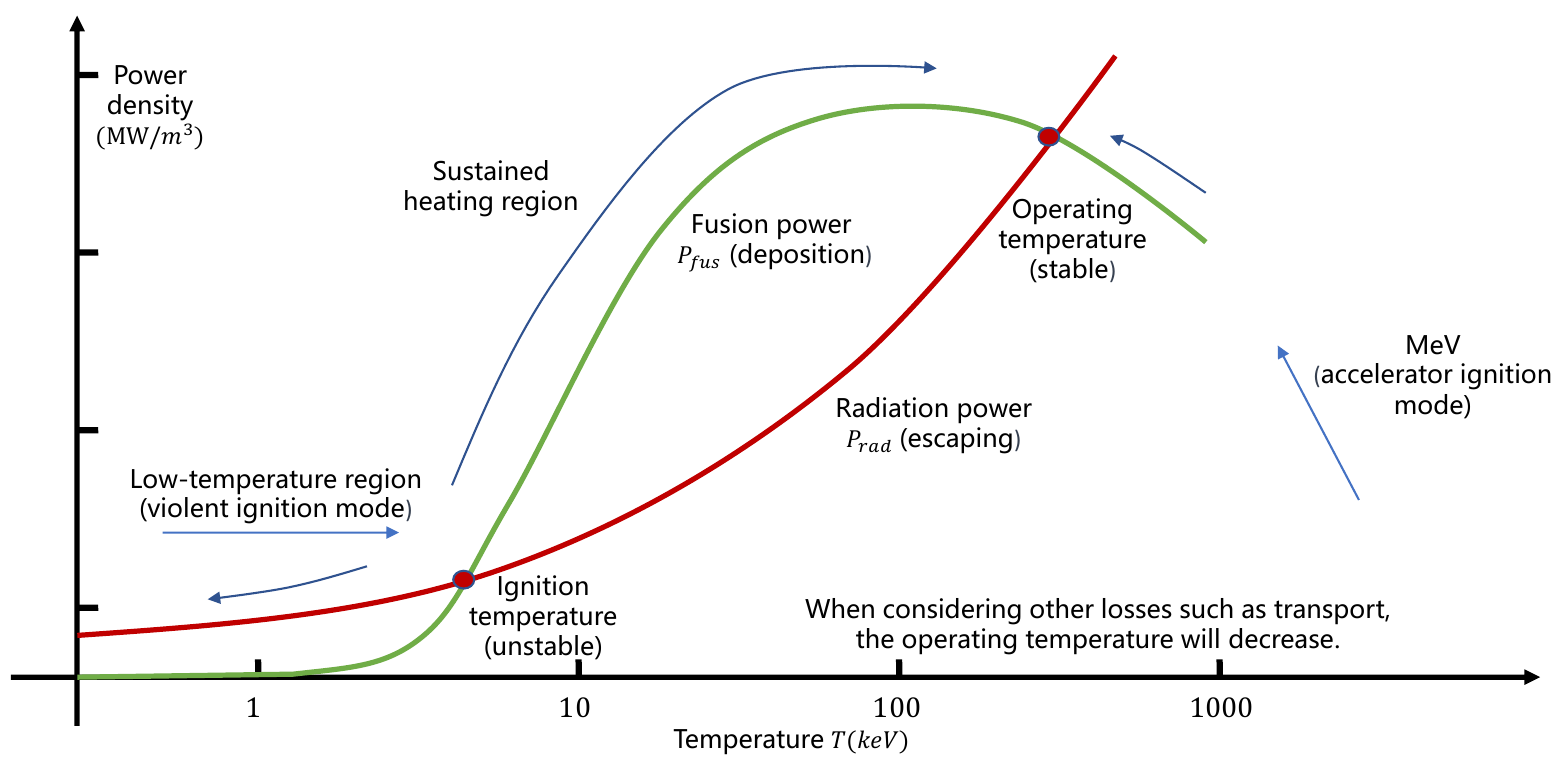}\\
\caption{Operating temperature range for fusion. When the temperature is lower than the first point where radiation and fusion power are equal, ignition can only be maintained with strong external input power; when it exceeds the first point, fusion power exceeds the loss power, causing the temperature to continue to rise until the second equilibrium point; and the higher energy range is typically the operation mode of accelerator fusion schemes.}\label{fig:fusionTi}
\end{center}
\end{figure}

Taking deuterium-tritium fusion as an example, Figure \ref{fig:idealTi} also shows that when the temperature is too high (above 440 keV), the bremsstrahlung radiation will exceed the fusion power, resulting in plasma cooling and temperature reduction. In other words, the optimal fusion temperature lies between the temperatures at which bremsstrahlung radiation and fusion power intersect, enabling dynamic equilibrium and avoiding a divergent increase in temperature. This is one of the reasons why fusion reactors are considered to be inherently safe. Figure \ref{fig:fusionTi} illustrates this dynamic equilibrium process and the operating modes in different temperature ranges [McNally (1982)].

\subsection{Maximum Allowable Impurity Content}

The previous discussions did not consider the effects of fusion products and impurities. If these effects are taken into account, $Z_{eff}$ will be greater than the values mentioned above, resulting in significant changes to the ideal ignition conditions. Let's calculate the maximum allowable impurity content.

\begin{figure}[htbp]
\begin{center}
\includegraphics[width=15cm]{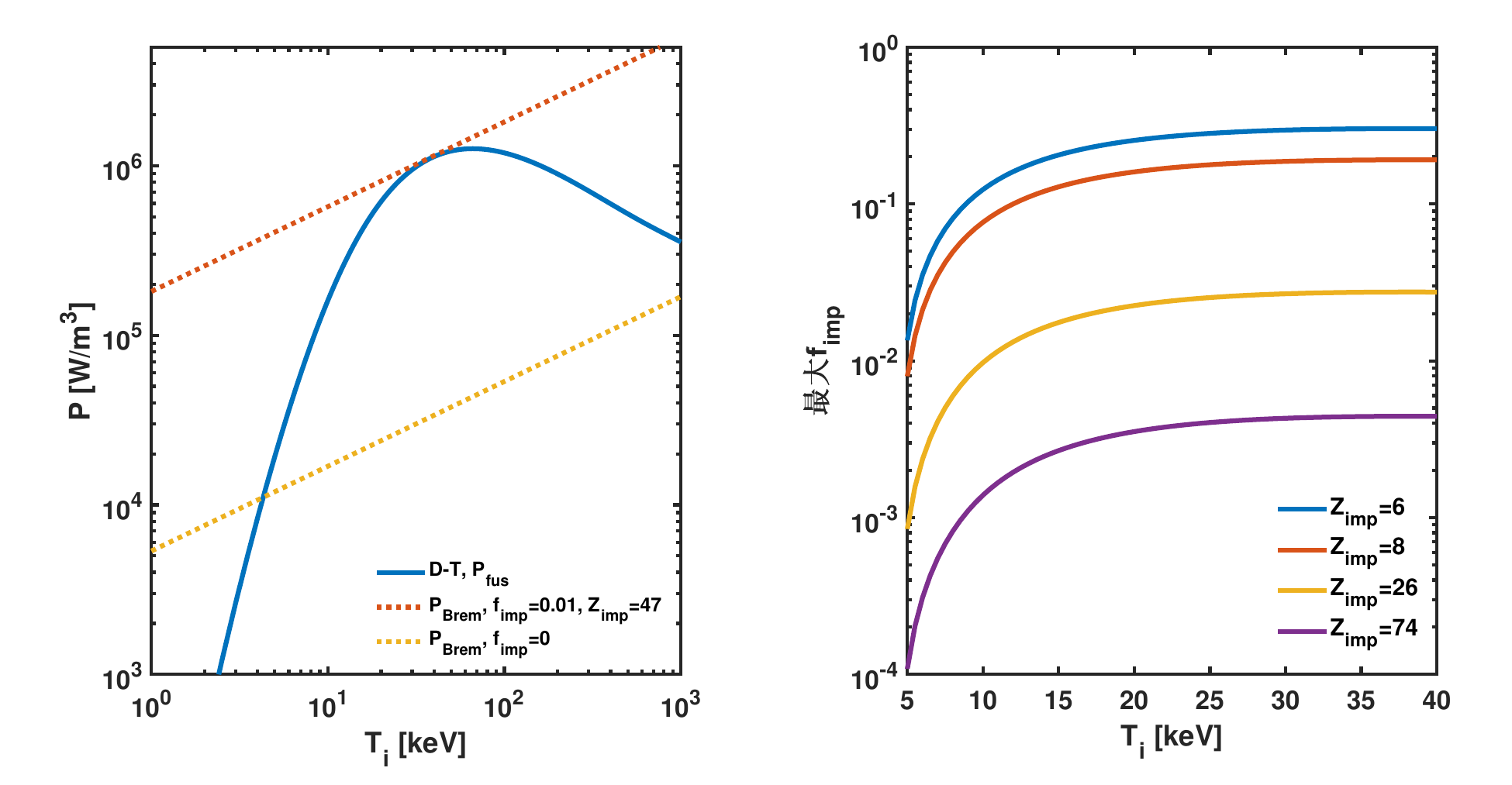}\\
\caption{The maximum allowable level of impurities in deuterium-tritium fusion.}\label{fig:fimpTi}
\end{center}
\end{figure}

Let's first consider the D-T fusion reaction, where the reaction ions have $Z_1 = Z_2 = 1$. To keep the fusion power constant, we assume an initial ion density of $n_{i0}$, corresponding to an electron density of $n_{e0}=n_{i0}$. By introducing an impurity with a charge number of $Z_{imp}$ and an impurity fraction of $f_{imp}=n_{imp}/n_{i0}$, we have
\begin{eqnarray}
n_e = n_{i0} + f_{imp}n_{i0}Z_{imp},\\
Z_{eff} = \frac{1+f_{imp}Z_{imp}^2}{1+f_{imp}Z_{imp}}.
\end{eqnarray}
Therefore, the multiplicative factor for the non-relativistic bremsstrahlung power loss is
\begin{eqnarray}
g_{imp} = \frac{n_e^2Z_{eff}}{n_{e0}^2} = (1+f_{imp}Z_{imp}^2)(1+f_{imp}Z_{imp}).
\end{eqnarray}
It can be seen that the radiation multiplication factor $g_{imp}$ is proportional to $Z_{imp}^3$. Assuming an impurity fraction of $f_{imp}=0.01$, for oxygen impurity with $Z_{imp}=8$, the corresponding radiation multiplication factor is $g_{imp}=1.77$; for iron impurity with $Z_{imp}=26$, the corresponding radiation multiplication factor is $g_{imp}=9.78$; for tungsten impurity with $Z_{imp}=74$, the corresponding radiation multiplication factor is $g_{imp}=97.0$. That is, for heavy impurities, even at very low concentrations, they can greatly increase the power loss due to the bremsstrahlung radiation, making it difficult or even infeasible to achieve ignition. In the case of D-T fusion, the maximum ratio of fusion power to non-relativistic bremsstrahlung power is about 33, corresponding to a temperature of about 39 keV. Therefore, the content of heavy impurities should not be too high. Figure \ref{fig:fimpTi} shows the maximum allowable content of different impurities in D-T fusion. It can be seen that the content of heavy impurities should be controlled to be below 0.01 or even 0.001.


For D-D fusion, D-He fusion, and H-B fusion, the allowable content of impurities is even more stringent. In particular, in H-B fusion, the accumulation of helium ash increases $Z_{eff}$, further enlarges the ratio of bremsstrahlung power to fusion power, and makes ignition more difficult to achieve. This also reflects the difficulty of achieving fusion in high-$Z$ plasmas, as high-$Z$ impurities lead to greater power loss due to radiation. Therefore, it is necessary to avoid high-$Z$ impurities and control the impurities. Additionally, even at temperatures above 10 keV, high-$Z$ atoms may not be fully ionized, resulting in additional radiation losses.

\section{The Main Radiations and Characteristics}


There are multiple radiation loss mechanisms in a plasma. Blackbody radiation comes from the equilibrium radiation. The main radiation in a fusion plasma comes from the acceleration and deceleration of charged particles in electromagnetic fields. Bremsstrahlung radiation is mainly produced by the collision process of the interaction between electrons and ions in Coulomb fields. Synchrotron radiation, on the other hand, is produced by the circular motion of charged particles in a magnetic field. Here, we mainly discuss these three mechanisms, neglecting other radiations such as line radiation and impurity radiation, as they are not dominant in high-temperature fully ionized fusion plasmas. For a more detailed discussion on radiation in fusion plasmas, refer to Bekefi (1966).

\subsection{Blackbody Radiation}

First, let us consider a case where there is no other radiation except for blackbody radiation. In other words, the plasma is in an optically thick state, where all other radiations, such as bremsstrahlung radiation, are absorbed, and energy is emitted only through blackbody radiation due to thermal equilibrium.

From the knowledge of nuclear reaction cross-sections in the appendix, we know that even at very low temperatures, there is a certain probability of fusion reactions occurring, although the reaction cross-section is extremely low. The energy produced from these fusion reactions can heat the plasma, causing the temperature to increase and further increasing the reaction rate until the energy released from fusion reactions becomes smaller than the energy emitted outward through blackbody radiation, reaching an equilibrium state.

\begin{figure}[htbp]
\begin{center}
\includegraphics[width=14cm]{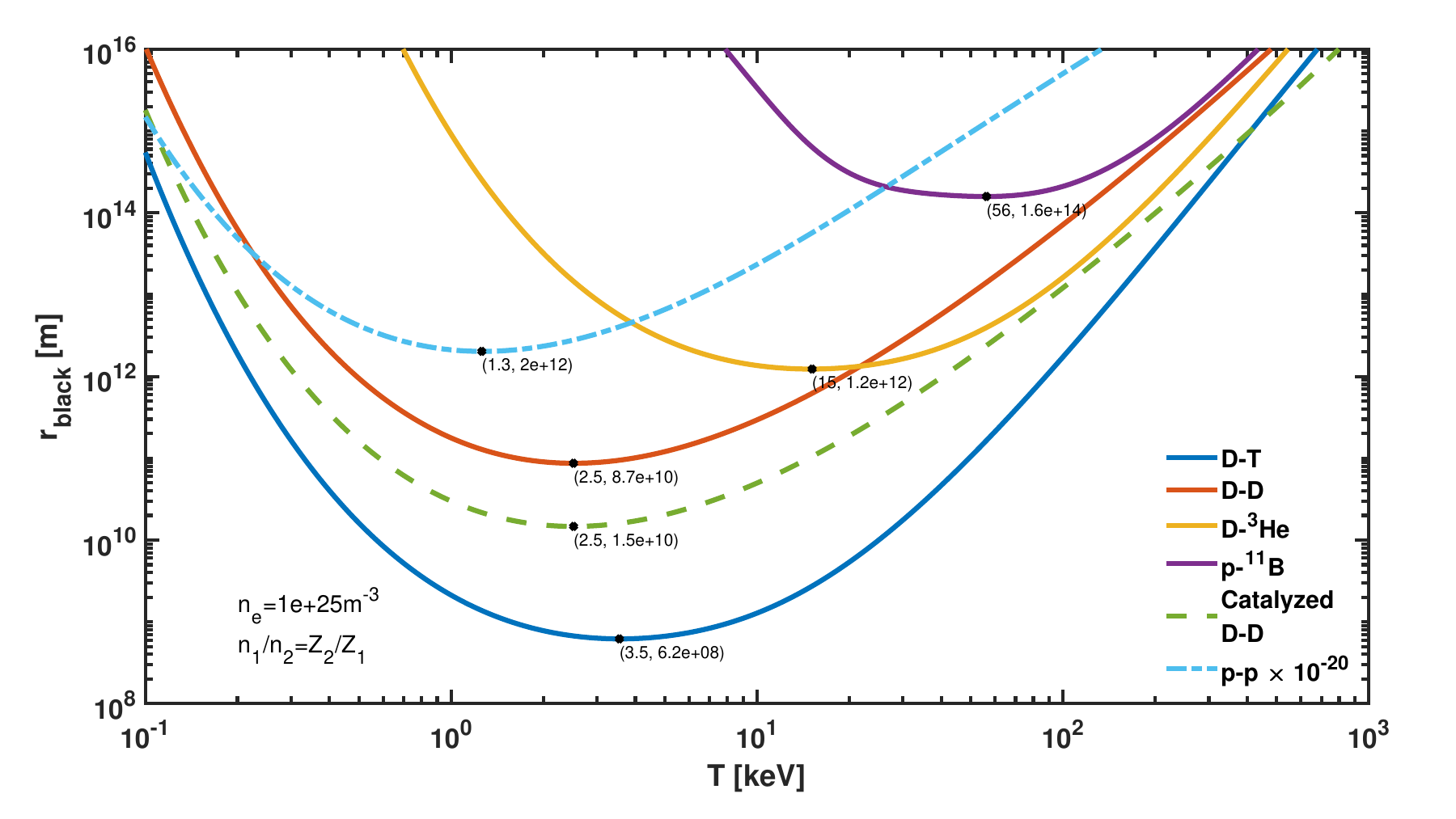}\\
\caption{Minimum plasma radius at the balance between blackbody radiation power and fusion power.}\label{fig:rblack}
\end{center}
\end{figure}The formula for black body radiation power is
\begin{eqnarray}
P_{black}=\alpha T^4\cdot S,
\end{eqnarray}
where
\begin{eqnarray}
\alpha=\frac{2\pi^5k_B^4}{15c^2h^3}=5.67\times10^{-8}{\rm W\cdot m^{-2}\cdot K^{-4}},
\end{eqnarray}
which is the Stefan-Boltzmann constant. Here, $h$ is the Planck constant, $T$ is the thermal temperature, and $S$ is the surface area. Assuming that the plasma is a sphere with a radius of $r$, the surface area is given by $S=4\pi r^2$ and the volume is given by $V=\frac{4}{3}\pi r^3$. When the fusion power is equal to the black body radiation power, we have
\begin{eqnarray}
P_{black}=P_{fus},
\end{eqnarray}
which yields the critical radius
\begin{eqnarray}
r_{black}=3\alpha T^4\frac{1+\delta_{12}}{n_1n_2\langle\sigma v\rangle Y},
\end{eqnarray}
where $Y$ represents the total energy released per fusion reaction, including both charged and uncharged products. Since $r$ is usually large enough for the neutron energy to remain in the plasma, the minimum value of $r$ can be obtained for different nuclear reactions. For a given density, optimizing the ion density ratio $n_1/n_2=Z_2/Z_1$, the minimum value of $r$ is determined by the minimum value of $T^4/\langle\sigma v\rangle$. Figure \ref{fig:rblack} shows the results for several fusion reactions. It can be seen that for a density of $n_e=10^{25}{\rm m^{-3}}$, the minimum radius $r_{black}$ required for deuterium-tritium fusion is $6.2\times10^{8}{\rm m}$, while for hydrogen-boron fusion, it is $1.6\times10^{14}{\rm m}$. From the formula, it can also be seen that the critical radius is inversely proportional to the square of the density. Therefore, when the density reaches $10^{31}{\rm m^{-3}}$, the minimum radius $r_{black}$ for deuterium-tritium fusion is $6.2\times10^{-4}{\rm m}$, which is close to the size of an inertially confined compressed target. For catalyzed deuterium-deuterium fusion, it is $1.5\times10^{-2}{\rm m}$, while for hydrogen-boron fusion, it is $1.6\times10^{2}{\rm m}$.

The calculated $r_{black}$ above can be considered as the typical size required for confinement radiation. As can be seen, $r_{black}$ is usually very large. At the same time, for typical fusion temperatures, such as $T=10$keV, in a plasma with a size less than $100$m, the black body radiation power is much larger than the fusion power. Therefore, fusion reactors on the ground are not in radiation equilibrium (where the electrons, ions, and photons reach the same temperature) and can be considered transparent to radiation, neither reflecting nor absorbing it. This also means that the thermal temperature of photons from black body radiation is lower than the electron and ion kinetic temperature of the plasma. An exception occurs in the case of extremely high density inertial confinement, where the deuterium-tritium target may approach radiation thermal equilibrium.For plasma systems whose size can reach the critical size mentioned above, radiation can be somewhat confined, leading to self-sustaining fusion reactions and forming a chain reaction similar to fission reactors [Glasstone (1960)]. From the data of the pp reaction in the figure, it can be seen that the optimal temperature is 1.3 keV, which happens to be the temperature at the center of the sun. However, according to the central density of the sun $10^{32} {\rm m^{-3}}$, the calculated $r_{black}=2.0\times10^{18} {\rm m}$ is much larger than the actual radius of the sun $r_{black}=6.96\times10^{8} {\rm m}$, with a difference of $\frac{2.0\times10^{18}}{6.96\times10^{8}}=2.9\times10^{9}$ times. This discrepancy comes from the uneven distribution of density and temperature inside the sun. On the one hand, blackbody radiation is emitted at a temperature of 6000K on the surface of the sun. On the other hand, fusion mainly occurs in the high-temperature central region with a radius smaller than 0.1-0.2$r$. This deviation causes a multiplier of $\simeq(\frac{1.3\times10^{3}}{0.6/1.16})^4(0.1)^3=4.0\times10^{10}$, which is very similar to the previous value. The remaining deviation can also be understood because as the temperature in the fusion center decreases with increasing radius, the reaction rate also rapidly decreases, resulting in an overestimate of fusion power. In other words, central fusion can greatly influence the critical radius $r_{black}$.

Based on the above analysis, we can conclude that for ground-based fusion reactors, both low-density magnetic confinement and magnetically driven inertial confinement devices based on magnetic confinement schemes are transparent to bremsstrahlung radiation and inevitable energy loss terms. For high-density inertial confinement deuterium-tritium fusion, radiation is not completely transparent and can be retained to some extent in the plasma. It should also be noted that at extremely high densities, the fusion power density is also extremely high, and the corresponding plasma pressure is extremely high. Therefore, fusion should only occur in a very small region to avoid destructive losses to the device.

This can also be estimated from another perspective. The radiation pressure of blackbody radiation is $\alpha T^4/c$, where $c$ is the vacuum speed of light. For $T=10$ keV, it is roughly $10^{11}$ times atmospheric pressure. In stars, this can be balanced by gravity, while in inertial confinement fusion, it is temporarily maintained by strong external drivers.

\begin{figure}[htbp]
\begin{center}
\includegraphics[width=15cm]{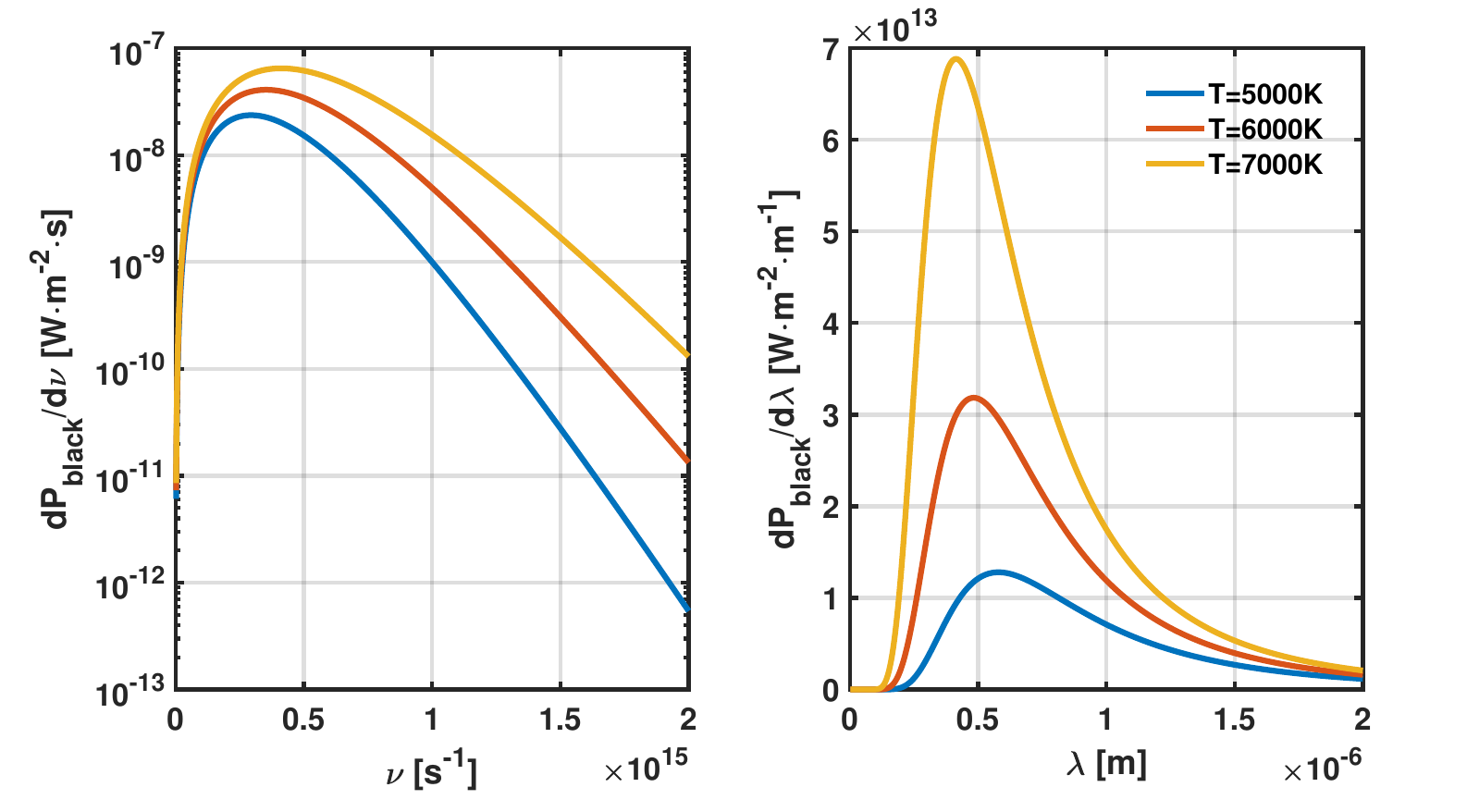}\\
\caption{Blackbody radiation spectrum distribution at different temperatures.}\label{fig:blacknu}
\end{center}
\end{figure}The spectral distribution of the blackbody radiation intensity per unit area and per unit solid angle is given by
\begin{eqnarray}
\frac{dP_{black}}{d\nu}\equiv B(\nu,T)=\frac{2\nu^2}{c^2}\frac{h\nu}{e^{h\nu/k_BT}-1}.
\end{eqnarray}
For the surface temperature of the Sun, $T=6000$K, the peak spectrum of the radiation lies in the visible light range, i.e., 400-760nm, as can also be seen from Figure \ref{fig:blacknu}. Based on the blackbody radiation formula, we can obtain many useful quantitative information. For example, we can estimate the normal radiation of a human body to be about 100W, which also represents the power consumption of normal metabolism.

\subsection{Bremsstrahlung Radiation}

In the previous section, we mentioned the total power of bremsstrahlung radiation. Here, we discuss its spectral distribution. In a plasma with Maxwellian distribution, the bremsstrahlung power as a function of frequency $\nu$ is given by [Gross (1984), Glasstone (1960)]: 
\begin{eqnarray}
j_{brem}(\nu)\equiv\frac{dP_{brem}}{d\nu}=g\frac{32\pi}{3^{3/2}}\Big(\frac{2\pi}{k_BT_e}\Big)^{1/2}\frac{e^6}{m_e^{3/2}c^3(4\pi\epsilon_0)^3}n_e\sum(n_iZ_i^2)e^{-\frac{h\nu}{k_BT_e}},
\end{eqnarray}
The power distribution as a function of wavelength $\lambda$ is given by:
\begin{eqnarray}
\frac{dP_{brem}}{d\lambda}=g\frac{32\pi}{3^{3/2}}\Big(\frac{2\pi}{k_BT_e}\Big)^{1/2}\frac{e^6}{m_e^{3/2}c^2(4\pi\epsilon_0)^3}n_e\sum(n_iZ_i^2)\lambda^{-2}e^{-\frac{hc}{\lambda k_BT_e}},
\end{eqnarray}
where the frequency-wavelength relationship is $\nu=c/\lambda$, and the Gaunt factor $g\simeq1.11$ is approximately constant. The integral of the power over frequency $\nu$ gives the total bremsstrahlung power per unit volume:
\begin{eqnarray}
P_{brem}=\int_{0}^{\infty} \frac{dP_{brem}}{d\nu} d\nu=g\frac{32\pi}{3^{3/2}}\frac{(2\pi k_BT_e)^{1/2}e^6}{m_e^{3/2}c^3h(4\pi\epsilon_0)^3}n_e^2Z_{eff},
\end{eqnarray}
or
\begin{eqnarray}
P_{brem}&=&C_Bn_e^2(k_BT_e)^{1/2}Z_{eff}=5.39\times10^{-37}n_e^2T_e^{1/2}Z_{eff}~{\rm [W/m^3]},\\
C_B&=&g\frac{32\pi}{3^{3/2}}\frac{(2\pi )^{1/2}e^6}{m_e^{3/2}c^3h(4\pi\epsilon_0)^3}.
\end{eqnarray}
Here, $Z_{eff}=\sum(n_iZ_i^2)/n_e$ is the effective charge number, and $n_e=\sum n_iZ_i$ is the electron density. In the equation with specific numerical values, the temperature $T_e$ is in units of keV, and the density $n_e$ is in units of ${\rm m^{-3}}$. The above results are consistent with the weakly relativistic bremsstrahlung radiation formula used in the previous section under the reduced condition, with the slight difference in coefficients arising from the approximate values of the Gaunt factor.

\begin{figure}[htbp]
\begin{center}
\includegraphics[width=15cm]{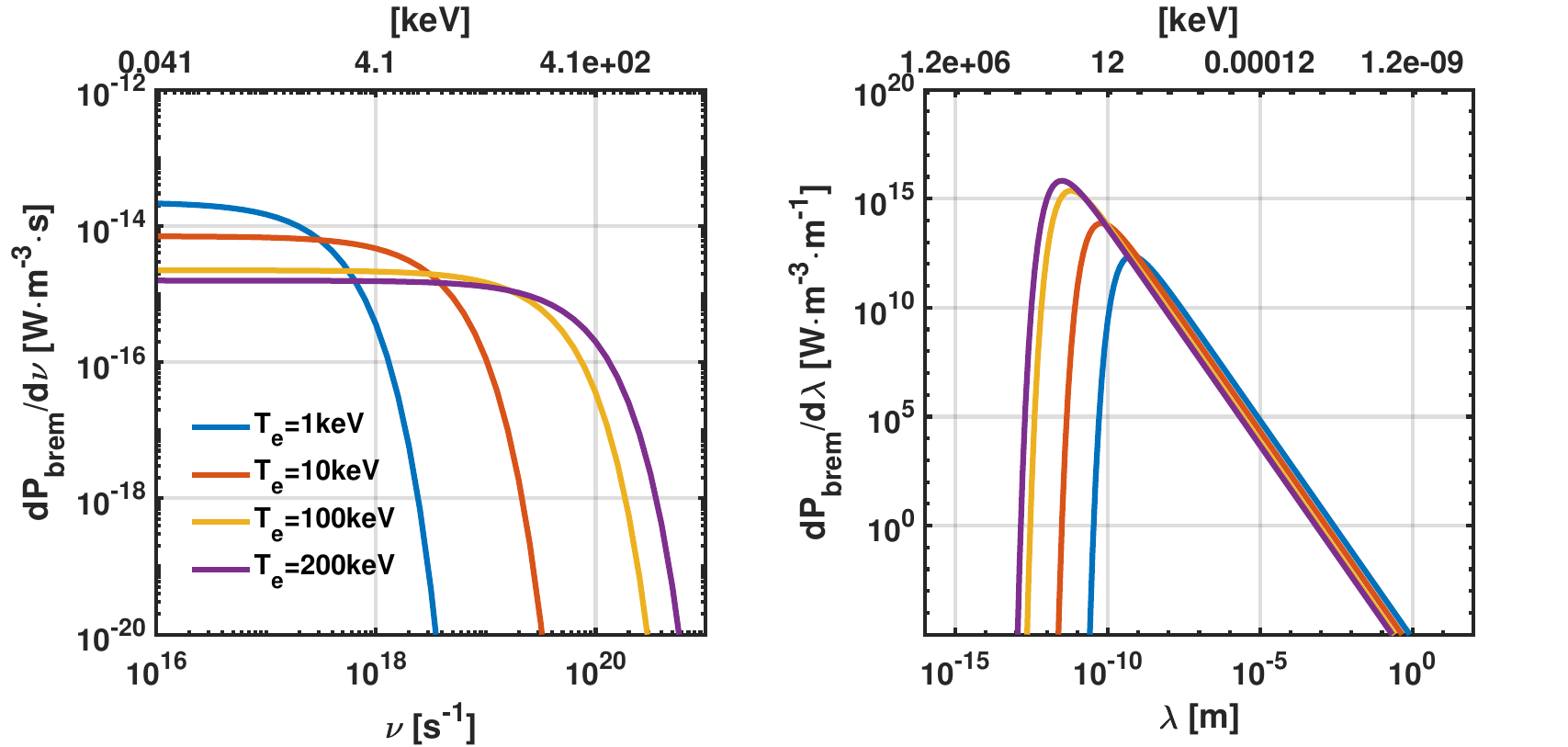}\\
\caption{Spectral distribution of bremsstrahlung radiation at different temperatures.}\label{fig:bremnu}
\end{center}
\end{figure}Figure 1 displays the bremsstrahlung spectrum distribution at different temperatures. It can be seen that bremsstrahlung is a broad continuous spectrum, mainly concentrated in the energy region close to the electron temperature. At typical fusion temperatures, it corresponds mainly to soft X-rays and ultraviolet radiation, which can be absorbed by materials and converted into heat, i.e., can be recovered and reused.

Fortunately, bremsstrahlung radiation varies with $T^{1/2}$, while fusion power in the region of interest varies with $T^{5/2}$, making fusion gain possible. The discussion above mainly focuses on radiation from electrons in the Coulomb field of ions. When the temperature is higher than 50keV, bremsstrahlung from electron-electron interactions will also account for a certain proportion and cannot be ignored.

In a hot plasma, according to Kirchhoff's law, the absorption coefficient for photons through inverse bremsstrahlung is [Spitzer (1956) p89, Hutchinson (2002) p204]
\begin{eqnarray}\nonumber
\alpha_\nu=\frac{j_{brem}(\nu)}{4\pi B(\nu)}&=&\frac{g\frac{32\pi}{3^{3/2}}\Big(\frac{2\pi}{k_BT_e}\Big)^{1/2}\frac{e^6}{m_e^{3/2}c^3(4\pi\epsilon_0)^3}n_e^2Z_{eff}e^{-\frac{h\nu}{k_BT_e}}}{4\pi \frac{2\nu^2}{c^2}\frac{h\nu}{e^{h\nu/k_BT}-1}}\\&\simeq&g{\frac{4}{3^{3/2}}\Big(\frac{2\pi}{k_BT_e}\Big)^{1/2}\frac{n_e^2Z_{eff}e^6}{m_e^{3/2}hc\nu^3(4\pi\epsilon_0)^3}}.
\end{eqnarray}
where $B(\nu)$ is the blackbody radiation intensity, the coefficient $4\pi$ comes from the solid angle of a sphere, and it reduces into the Rayleigh-Jeans limit when the frequency is high and the wavelength is short, i.e., $h\nu>k_BT$. Since $B(\nu)$ is proportional to $\nu^2$ when $h\nu\ll k_BT$, the absorption coefficient $\alpha_\nu$ increases as $\nu$ decreases. Therefore, absorption is more important for longer wavelengths. However, due to the lower total energy at longer wavelengths, we focus on estimating absorption under the short wavelength approximation. The mean absorption free path is given by\footnote{Hutchinson (2002) p166, the coefficient has slight differences. Glasstone (1960) and Gross (1984) calculate a coefficient of $7.0\times10^5$. These differences are of the same order of magnitude and do not fundamentally affect the conclusions of this book.}
\begin{eqnarray}
\lambda_{\nu}=\frac{1}{\alpha_\nu}=\frac{3}{4}\Big(\frac{3k_BT}{2\pi}\Big)^{1/2}\frac{hcm_e^{3/2}(4\pi\epsilon_0)^3\nu^3}{gn_e^2Z_{eff}e^6}=9.23\times10^{4}\frac{T_e^{1/2}\nu^3}{Z_{eff}n_e^2} [m].
\end{eqnarray}
In the latter equality, the temperature $T_e$ is in units of keV, and the density $n_e$ is in units of ${\rm m^{-3}}$.Therefore, for typical parameters of magnetic confinement fusion, density $n_e=10^{20} \rm m^{-3}$, effective charge $Z_{eff}=1$, $T=10$ keV, bremsstrahlung radiation photon frequency $\nu=10^{18} \rm s^{-1}$, the mean absorption free path is $\lambda_{\nu}\approx3\times10^{19}$ m, much larger than the size of fusion plasma on the ground. In other words, bremsstrahlung radiation is optically transparent in magnetic confinement fusion. For parameters at the center of the sun, density $n_e\approx10^{31}\rm m^{-3}$, $T\approx1.3$ keV, bremsstrahlung radiation photon frequency $\nu\approx10^{18} \rm s^{-1}$, we have $\lambda_{\nu}\approx1\times10^{-3}$ m, very short, therefore it is optically thick. This is also the key reason why solar energy can confine radiation for millions of years. The density and temperature parameters of inertial confinement fusion are of the same order of magnitude as those at the center of the sun, and the mean free path of bremsstrahlung radiation is similar to the size of the compressed plasma target, so bremsstrahlung radiation is not completely transparent in inertial confinement fusion.

\subsection{Synchrotron Radiation}

When there is a magnetic field, charged particles will perform circular motion around the magnetic field lines, and the change in direction corresponds to acceleration, so radiation will be emitted. This kind of radiation is called cyclotron radiation, or synchrotron radiation. These two names are generally not strictly distinguished in fusion literature. In the usual literature, synchrotron radiation refers to the cyclotron radiation of charged particles in the relativistic energy range. Similarly, because of the large mass and slow motion of ions, cyclotron radiation mainly comes from electron radiation.

\begin{figure}[htbp]
\begin{center}
\includegraphics[width=15cm]{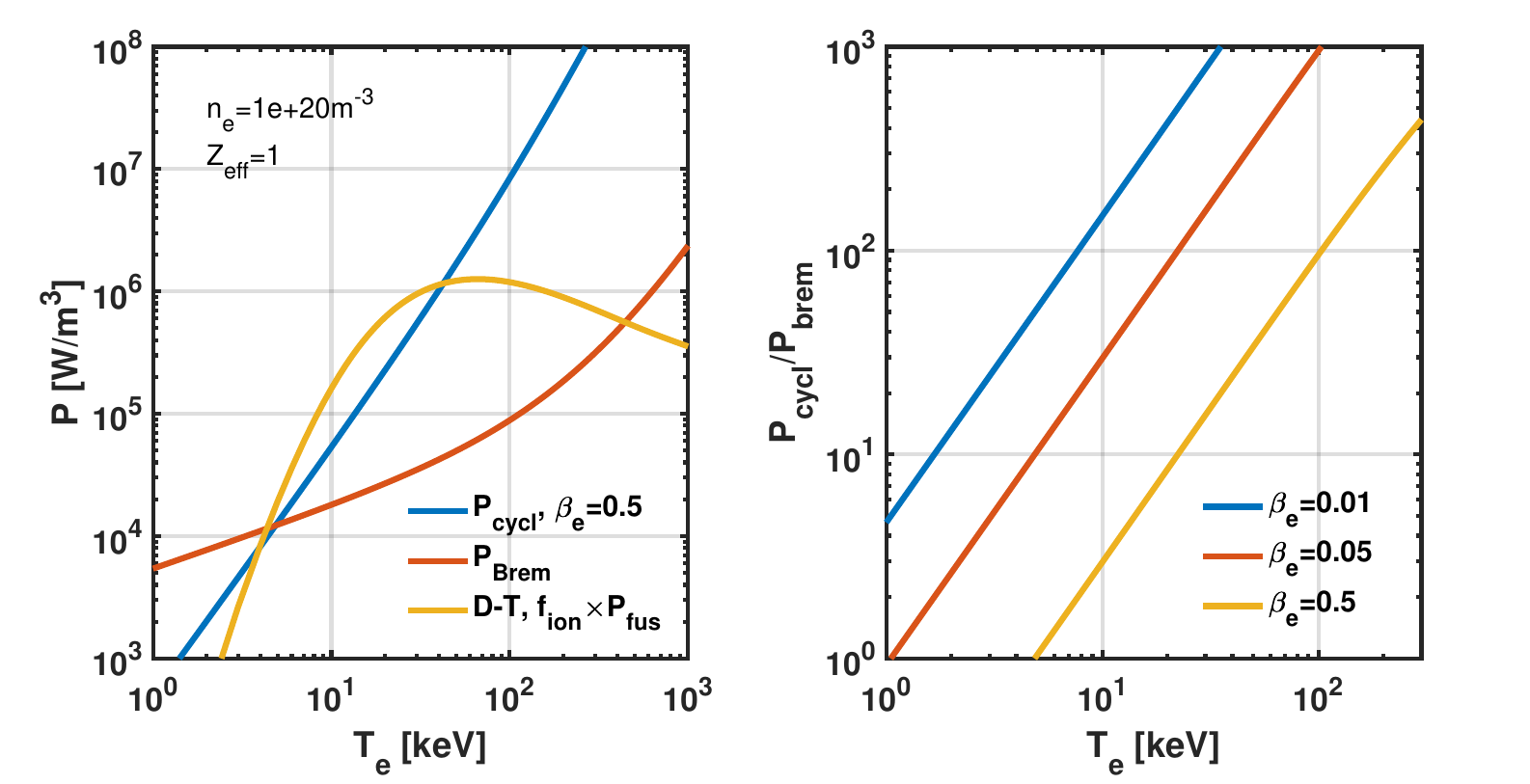}\\
\caption{Relative sizes of synchrotron radiation, bremsstrahlung radiation, and deuterium-tritium fusion.}\label{fig:Pcyclbrem}
\end{center}
\end{figure}

Different from bremsstrahlung radiation, cyclotron radiation mainly consists of integer multiples of the cyclotron frequency. The cyclotron frequency is given by
\begin{eqnarray}
\nu_c=\frac{eB}{2\pi m_e}=2.80\times10^{10}B ~~{\rm [Hz]},
\end{eqnarray}
where the magnetic field $B$ is measured in Tesla. Therefore, in typical magnetic confinement devices, it mainly corresponds to the microwave frequency range. In principle, this frequency range can be absorbed by the plasma and efficiently reflected by the device walls. We will ignore these absorption and reflection effects for now. 

For an individual electron, the power of cyclotron radiation is given by
\begin{eqnarray}
P_{cycl}=\frac{e^4B^2}{6\pi\epsilon_0 m_e^2c^3}\frac{v_\perp^2}{1-v^2/c^2}.
\end{eqnarray}
For a Maxwellian distribution of electrons, taking the average $\langle v_\perp^2\rangle=2k_BT_e/m_e$, in the weakly relativistic case, the non-relativistic cyclotron radiation power per unit volume is given by [Dolan (1981) p68]
\begin{eqnarray}
P_{cycl}&=&\frac{e^4B^2n_e}{3\pi\epsilon_0m_e^2c}\Big(\frac{k_BT_e}{m_ec^2}\Big)\Big[1+\frac{3.5k_BT_e}{m_ec^2}+\cdots\Big]\\&\simeq&6.21\times10^{-21}B^2n_eT_e\Big[1+\frac{T_e}{146}\Big]~~{\rm [W/m^3]},\end{eqnarray}
The latter equation, where the magnetic field $B$ is in units of tesla (T), the electron temperature $T_e$ is in units of kilo-electron volts (keV), and the electron density $n_e$ is in units of ${\rm m^{-3}}$. If we assume that the plasma beta $\beta_e=2\mu_0n_ek_BT_e/B^2$ is a constant and remove the relativistic term, then  
\begin{eqnarray}
P_{cycl}\simeq\frac{2\mu_0e^4}{3\pi\epsilon_0m_e^3c^3\beta_e}n_e^2k_B^2T_e^2=2.50\times10^{-38}\frac{n_e^2T_e^2}{\beta_e}~{\rm [W/m^3]},
\end{eqnarray}
where $T_e$ is in units of keV and $n_e$ is in units of ${\rm m^{-3}}$. From the equation above, it can be seen that the cyclotron radiation is proportional to $T_e^2$. Comparing this expression with the expression for bremsstrahlung radiation mentioned earlier, the ratio between the two can be written as 
\begin{eqnarray}
\frac{P_{cycl}}{P_{brem}}=\frac{2.50\times10^{-38}T_e^{3/2}}{5.39\times10^{-37}\beta_eZ_{eff}}=4.64\times10^{-2}\frac{T_e^{3/2}}{\beta_eZ_{eff}}.
\end{eqnarray}
From the above equation, it can be seen that besides temperature, a strong magnetic field is also a detrimental factor for cyclotron radiation. From this perspective, we can also see that a high plasma beta $\beta$ is advantageous for reducing cyclotron radiation. For a 5 keV temperature of deuterium-tritium fusion, cyclotron radiation may not be significant, but for advanced fuels such as hydrogen-boron, a strong magnetic field and high temperature cannot be avoided, hence cyclotron radiation will be larger than bremsstrahlung radiation. For the 100 keV parameter of advanced fuels, cyclotron radiation will increase by nearly a hundred times, making it the main source of radiation losses. Figure \ref{fig:Pcyclbrem} illustrates the relative values of cyclotron radiation, bremsstrahlung radiation, and deuterium-tritium fusion.

At 50 keV, 94\% of the energy is emitted by cyclotron frequencies of second order or higher. Rose (1961) provides a detailed discussion on the spectrum of cyclotron radiation, which can be referred to. Considering that cyclotron radiation can be absorbed and effectively reflected, in the case of a fusion reactor, whether it can be effectively confined needs a more detailed assessment. There is currently no consensus in the literature. The calculation of cyclotron radiation produced by a single particle is enormous, but taking into account the resonance cavity and wall reflection effect, it can be significantly weakened. In the subsequent evaluation of the parameter range for the magnetic confinement configuration, we will use the formula for cyclotron radiation with reflection.

\subsection{Summary of Radiation}

In conclusion, radiation is a major mechanism for plasma energy loss, and the continuous spectrum of Bremsstrahlung radiation is inevitable. Except under extremely high density conditions, it is transparent to plasma, meaning a 100\% loss. If effective methods can be found to avoid Bremsstrahlung radiation or to allow its deposition within the plasma, it would lower the ignition conditions. When a magnetic field is present, synchrotron radiation becomes another important loss mechanism, especially at high temperatures and strong magnetic fields, which will become the most critical limiting factor in the subsequent discussion of magnetically confined advanced fuel fusion. Therefore, its reflection and absorption are crucial for the feasibility or difficulty assessment of fusion energy implementation.

If radiation loss cannot be avoided, as discussed later in the Lawson criterion for engineering fusion gain conditions, if radiation energy can be efficiently utilized or generated, it can also reduce the parameter conditions. Whether radiation can be effectively reflected mainly depends on the wavelength and atomic spacing in the material. For X-rays, the wavelength is usually close to or smaller than the atomic spacing, so it is difficult to reflect. For the microwave range, it is easier to reflect, and the key is the magnitude of the reflectivity. For example, we expect a reflectivity of 95\% or even more for synchrotron radiation.

\section{Consideration of radiation in the Lawson criterion}
The Lawson diagram can provide a more systematic representation of parameter requirements. In the previous chapters, we showed the simplest Lawson diagram where radiation was not separately taken into account. Here, we calculate radiation separately, taking into account the unequal temperatures of electrons and ions and the efficiency of power generation, and analyze the corresponding three-parameter Lawson diagram. In the original Lawson paper and subsequent literature, there are various assumptions regarding the criteria for the triple products, which may vary in different publications. The definition in this section is a relatively standard one. The model and results in this section can be considered as the most central content of this book.

\subsection{Fusion gain factor}

Assuming that all the charged ions produced in the fusion process are used to heat the fuel, radiation is considered as a direct loss term, and transport losses are measured by the energy confinement time $\tau_E$. The following equation is obtained:
\begin{eqnarray}
\frac{dE_{th}}{dt}=P_{ext}-\frac{E_{th}}{\tau_E}+f_{ion}P_{fus}-P_{rad},
\end{eqnarray}
where the volume-specific thermal energy $E_{th}$ and fusion power $P_{fus}$ are defined as:
\begin{eqnarray}
E_{th}=\frac{3}{2}k_B\sum_jn_jT_j=\frac{3}{2}k_B(n_eT_e+n_iT_i),\\
P_{fus}=\frac{1}{1+\delta_{12}}n_1n_2\langle\sigma v\rangle Y.
\end{eqnarray}
Here, $Y$ represents the energy released per single nuclear reaction, $Y_+$ represents the energy of the charged ions produced per single nuclear reaction, $f_{ion}=Y_{+}/Y$ is the proportion of energy released that is carried by the charged ions, $n_1$ and $n_2$ are the number densities of the two types of ions, and $T_j$ represents the temperature of each component (including electrons and ions). Impurity effects are neglected, and it is assumed that fusion products are quickly removed from the plasma after depositing their energy (compared to $\tau_E$), while the fusion fuel is quickly replenished. Note that $n_i=n_1+n_2$ and under quasi-neutrality condition, $n_e=Z_1n_1+Z_2n_2=Z_in_i$.

The energy gain factor, $Q$, is defined as:
\begin{eqnarray}
Q\equiv\frac{P_{out}-P_{in}}{P_{in}}.
\end{eqnarray}
When $Q=1$, the energy gain is equivalent to the energy input (scientific breakeven), meaning that the energy produced by fusion is equal to the energy input. The condition $Q=\infty$ is defined as ignition, where the fusion reactor operates in a steady state without any external input power. The output power is given by:
\begin{eqnarray}
P_{out}=\frac{E_{th}}{\tau_E}+(1-f_{ion})P_{fus}+P_{rad}.
\end{eqnarray}
The input power is given by:
\begin{eqnarray}
P_{in}=P_{ext}.
\end{eqnarray}At steady state, $dE_{th}/dt=0$, we have
\begin{eqnarray}
   P_{ext}=\frac{E_{th}}{\tau_E}-f_{ion}P_{fus}+P_{rad},
\end{eqnarray}
thus obtaining $P_{out}-P_{in}=P_{fus}$, that is,
\begin{eqnarray}
Q=\frac{P_{fus}}{\frac{E_{th}}{\tau_E}-f_{ion}P_{fus}+P_{rad}}.
\end{eqnarray}

We still use the Maxwellian distribution for the fusion reaction rate, that is, $\langle\sigma v\rangle=\langle\sigma v\rangle_M$. The main radiation term is
\begin{eqnarray}
P_{rad}=P_{brem}+P_{cycl}.
\end{eqnarray}
We temporarily ignore cyclotron (synchrotron) radiation $P_{cycl}$, only consider bremsstrahlung radiation $P_{brem}$, and use the previously weak relativistic formula
\begin{eqnarray}
P_{brem}&=&C_Bn_e^2\sqrt{k_BT_e}Z_{eff}g_{eff}~{\rm [MW\cdot m^{-3}]},\\
g_{eff}(T_e,Z_{eff})&=&\Big[1+0.7936\frac{k_BT_e}{m_ec^2}+1.874\Big(\frac{k_BT_e}{m_ec^2}\Big)^2\Big]
+\frac{1}{Z_{eff}}\frac{3}{\sqrt{2}}\frac{k_BT_e}{m_ec^2} .
\end{eqnarray}We assume that the temperatures of the two ions are equal, $T_1=T_2=T_i$, but the electron temperature may be different. We define $f_T=T_e/T_i$, and also define $x_1=x=n_1/n_i$ and $x_2=n_2/n_i$. For the same ion, $x_2=x_1$; for different ions, $x_2=1-x_1$.
From the expression of $Q$, we obtain
\begin{eqnarray}
P_{fus}=\frac{Q}{(1+Qf_{ion})}\Big(\frac{E_{th}}{\tau_E}+P_{brem}\Big).
\end{eqnarray}
Explicitly writing it out, we have
\begin{eqnarray}
\frac{1}{1+\delta_{12}}\frac{x_1x_2}{Z_i^2}\langle\sigma v\rangle Y=\frac{Q}{(1+Qf_{ion})}\Big[\frac{\frac{3}{2}k_B(T_e+T_i/Z_i)}{n_e\tau_E}+C_B\sqrt{k_BT_e}Z_{eff}g_{eff}\Big],
\end{eqnarray}
meaning
\begin{eqnarray}
n_e\tau_E=\frac{\frac{3}{2}k_B(T_e+T_i/Z_i)}{\frac{(1/Q+f_{ion})}{1+\delta_{12}}\frac{x_1x_2}{Z_i^2}\langle\sigma v\rangle Y -C_B\sqrt{k_BT_e}Z_{eff}g_{eff}}.
\end{eqnarray}
Removing the radiation term and assuming $Q=\infty$ and $f_T=1$, we can simplify it to the "ignition" form in Chapter \ref{chap:fuels}
\begin{eqnarray}
 n_e\tau_E=\frac{3}{2}k_BZ_i(1+Z_i)({1+\delta_{12}})\frac{T}{\langle\sigma v\rangle Y_+}.
\end{eqnarray}

\subsection{Lawson ideal fusion reactor}

\begin{figure}[htbp]
\begin{center}
\includegraphics[width=14cm]{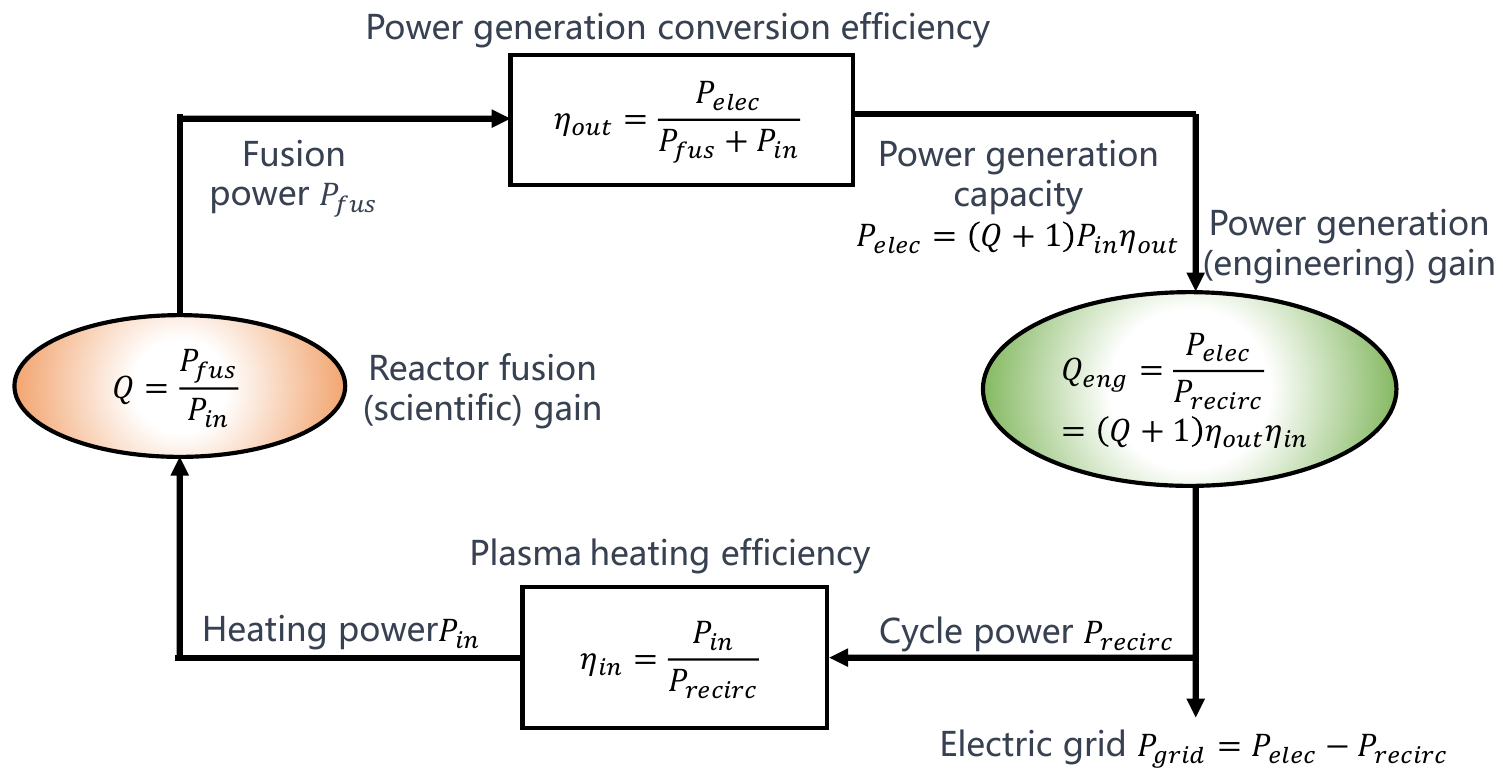}\\
\caption{Energy conversion process diagram of fusion power reactor.}\label{fig:lawsonrelation}
\end{center}
\end{figure}
\begin{figure}[htbp]
\begin{center}
\includegraphics[width=14cm]{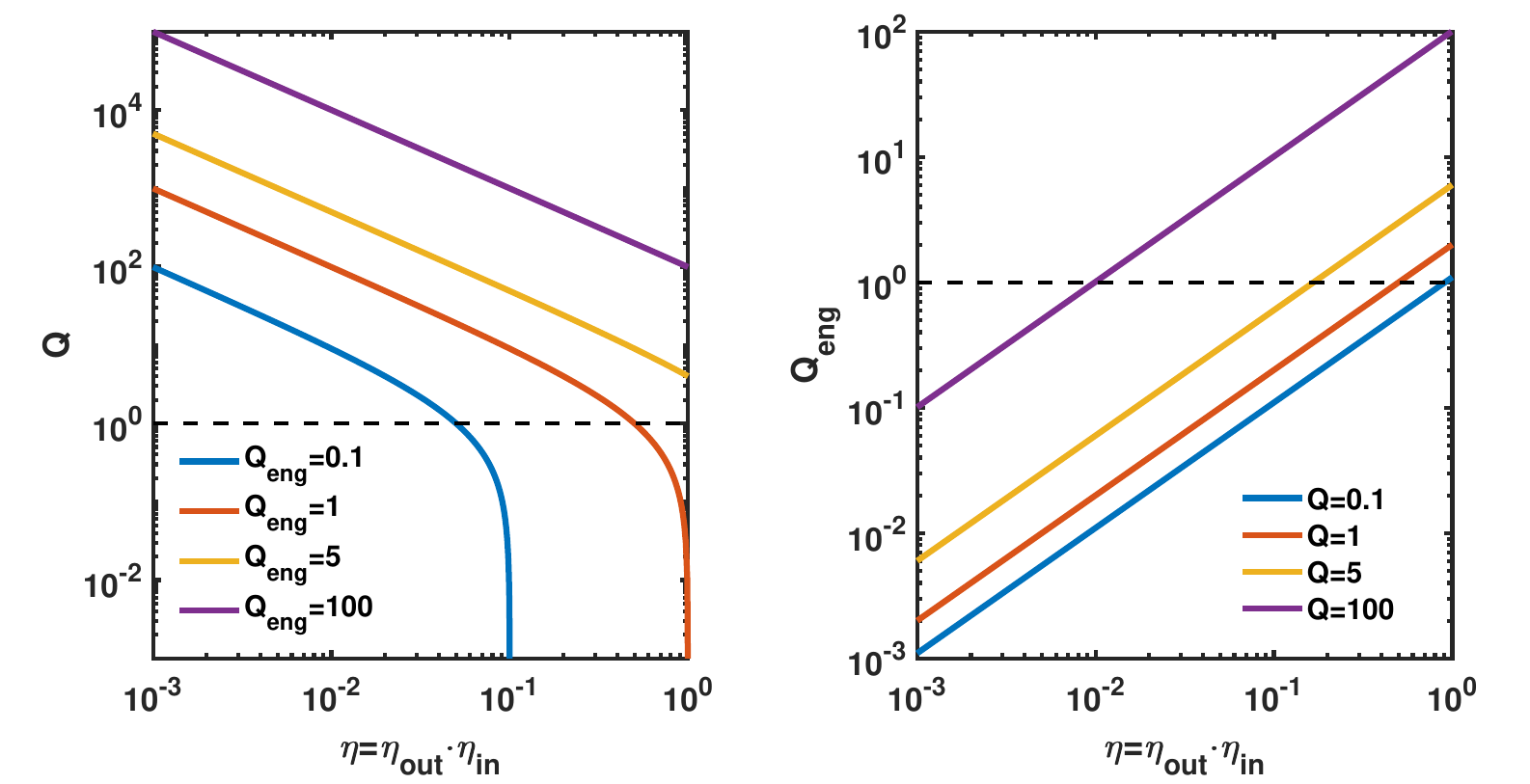}\\
\caption{Relationship between the scientific gain factor $Q$ and the engineering gain factor $Q_{eng}$ with the energy conversion efficiency $\eta=\eta_{out}\cdot\eta_{in}$ in fusion reactors.}\label{fig:Qeta}
\end{center}
\end{figure}

Let us consider the more practical process of fusion power generation, as an ideal fusion power reactor, the energy conversion process is shown in Figure \ref{fig:lawsonrelation} [Morse (2018)]. The energy conversion efficiency is defined, with the efficiency of converting energy from the grid to plasma heating denoted as $\eta_{in}$. This process includes both the energy conversion efficiency of the driver/heating device itself and the heating efficiency of converting the output energy to plasma internal energy. The efficiency of converting energy released from the fusion reactor to electricity is denoted as $\eta_{out}$. In this process, the total energy in the fusion reactor consists of the input energy $P_{in}$ and fusion-produced energy $P_{fus}$, which goes through various processes such as radiation, transport, ionization, and ultimately dissipates outward through the boundaries. Some of this energy can be converted to electricity, while some is lost. If using a thermal engine for power generation, typically $\eta_{out}\leq40\%$, the original paper by Lawson (1955) takes $\eta_{out}=1/3$; if using direct power generation with charged particles, it may be possible to achieve $\eta_{out}\geq80\%$. The relationship between the scientific gain factor $Q$ and the engineering gain factor $Q_{eng}$ is given by
\begin{eqnarray}
 Q_{eng}\equiv\frac{P_{elec}}{P_{recirc}}=(Q+1)\eta_{out}\eta_{in}.
\end{eqnarray}
There is also the definition of the engineering gain factor as [Freidberg07]
\begin{eqnarray}
 Q_{eng,2}\equiv\frac{P_{grid}}{P_{recirc}}=(Q+1)\eta_{out}\eta_{in}-1.
\end{eqnarray}
These two definitions are essentially the same, only differing in their physical meanings. The former represents net electrical gain when $Q_{eng}>1$, while the latter represents net electrical gain when $Q_{eng,2}>0$. Figure \ref{fig:Qeta} shows the relationship between $Q$ and $Q_{eng}$ with the energy conversion efficiency $\eta=\eta_{out}\cdot\eta_{in}$.

Under optimistic conditions, taking $\eta_{in}=1$, $\eta_{out}=1/3$, achieving a balance between the gained and lost electrical energy, that is, the net electrical energy grid is zero ($P_{grid}=0$), results in an engineering gain factor of $Q_{eng}=1$, and in this case, a scientific gain factor of $Q=2$ is required. For inertial confinement fusion, due to the usually low efficiency of the driver, for example, $\eta_{in}\leq0.02$, to achieve an engineering gain of $Q_{eng}\geq1$, typically $Q\geq50-150$ is required. We also note that as long as the energy conversion efficiency is high enough, $\eta_{out}\eta_{in}\simeq1$, even if $Q<1$, net electrical gain can still be achieved, that is, $Q_{eng}>1$.In energy balance, there still exists \footnote{In some literature and in Lawson's original work, the heating term of the fusion charged particles is not subtracted from $P_{loss}$, that is, $P_{loss}=\frac{E_{th}}{\tau_E}+P_{rad}$. In this case, the meanings of $Q$ and $\tau_E$ are slightly different from those in this book, so they should be distinguished. Currently, the definition used in this book is more commonly used in the literature, such as Wurzel (2022) and Morse (2018).}
\begin{eqnarray}
 P_{in}=P_{loss}=\frac{E_{th}}{\tau_E}-f_{ion}P_{fus}+P_{rad}.
\end{eqnarray}
Thus, the $Q$ defined in this section is equivalent to the $Q$ defined in the previous section. For the calculation of the Lawson criteria, it is only necessary to replace $Q$ with the engineering gain factor $Q_{eng}$, in order to calculate the parameter conditions required when $Q_{eng}$ is a specific value.

\subsection{Computational Results}

Based on the above model, we can calculate the Lawson diagrams for different values of the electron-to-ion temperature ratio $f_T=T_e/T_i$, gain factor $Q$, and ion density ratio $x$, and plot the curves of $n_e\tau_E$ or $n_e\tau_ET_i$ as a function of temperature $T_i$.

\begin{figure}[htbp]
\begin{center}
\includegraphics[width=15cm]{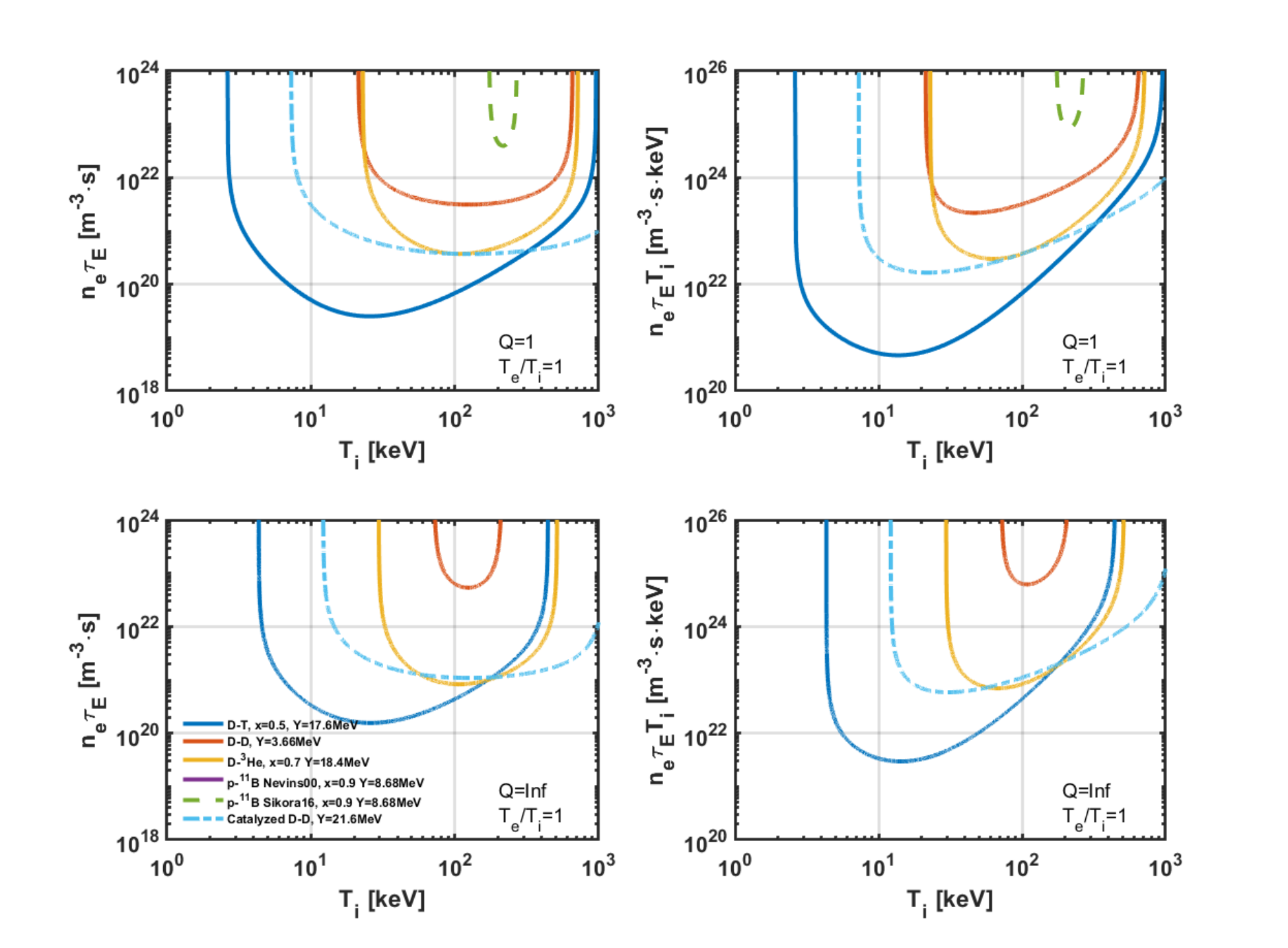}\\
\caption{Lawson diagrams for $T_e/T_i=1$, with $Q=1$ and $Q=\infty$.}\label{fig:Lawson_ntauE_fT=1}
\end{center}
\end{figure}

\begin{figure}[htbp]
\begin{center}
\includegraphics[width=15cm]{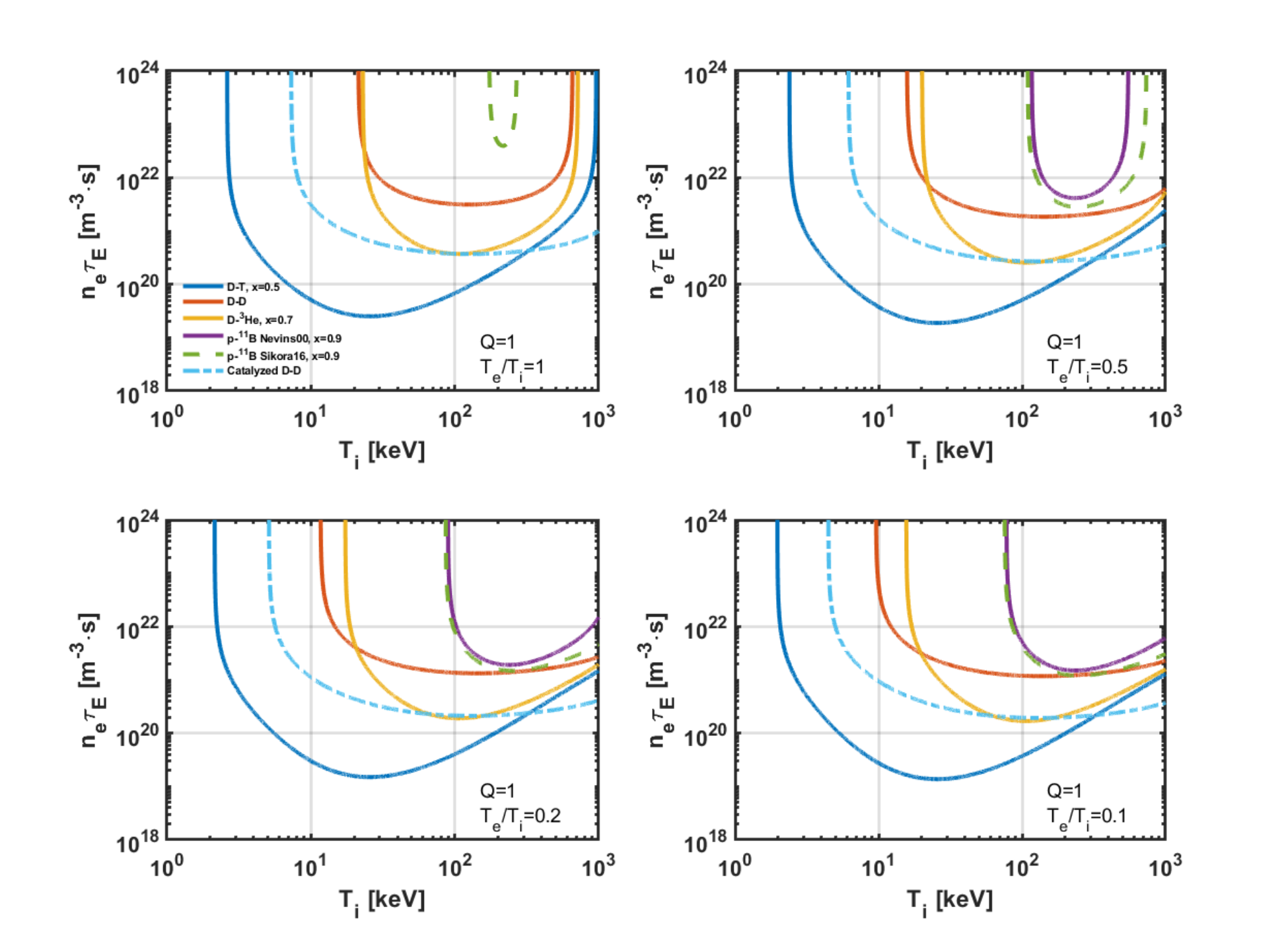}\\
\caption{Lawson diagrams for $Q=1$, with temperature ratio $T_e/T_i=1,0.5,0.2,0.1$.}\label{fig:Lawson_ntauE_Q1_scanfT}
\end{center}
\end{figure}

\begin{figure}[htbp]
\begin{center}
\includegraphics[width=15cm]{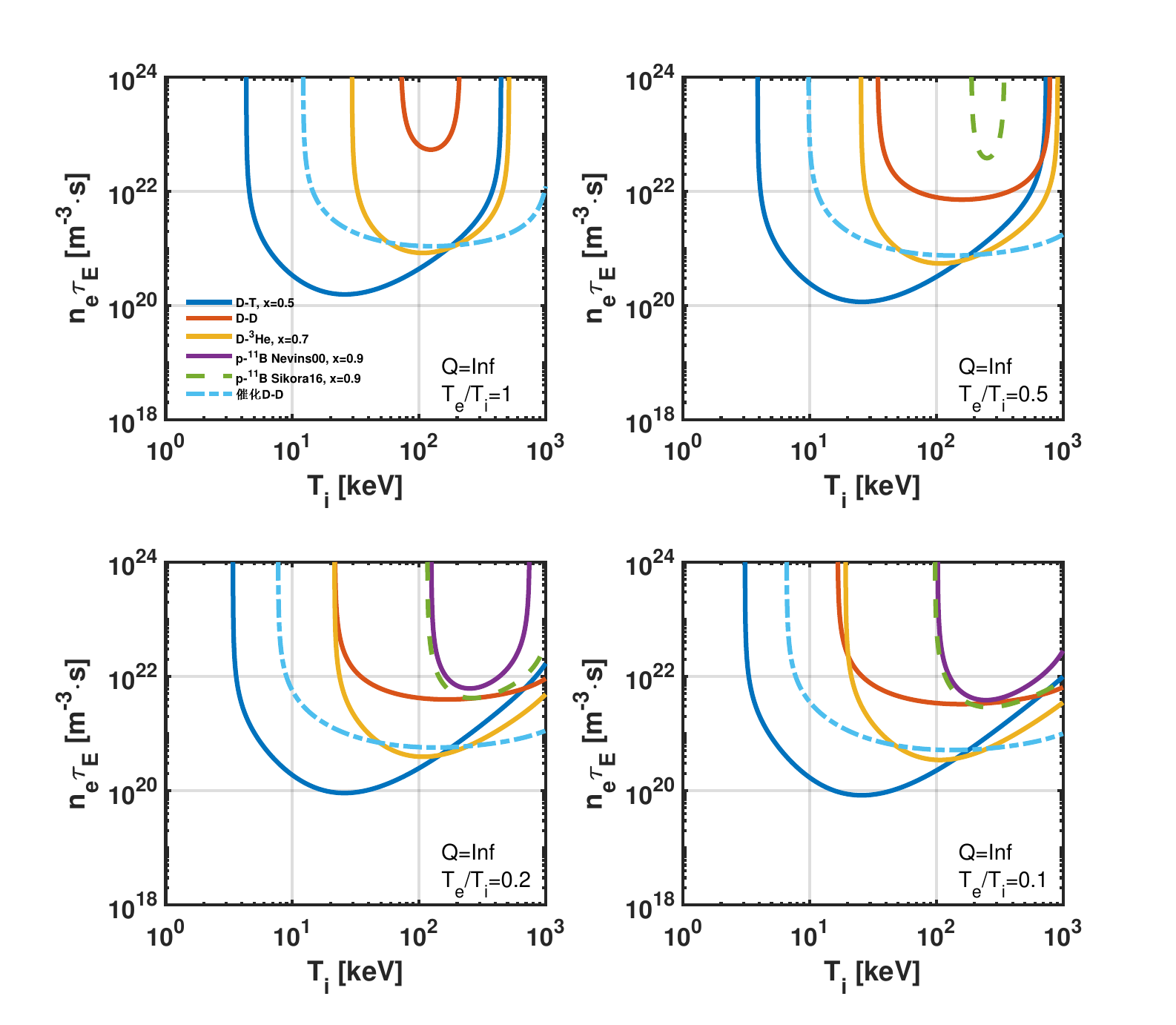}\\
\caption{Lawson diagram with scanning temperature ratios $T_e/T_i=1,0.5,0.2,0.1$ for $Q=\infty$.}\label{fig:Lawson_ntauE_QInf_scanfT}
\end{center}
\end{figure}

In order to obtain the lowest $n_e\tau_E$, the density ratio $x$ needs to be optimized. First, we do not optimize separately, but choose similar values of $x$ based on the ideal ignition conditions mentioned earlier. Figure \ref{fig:Lawson_ntauE_fT=1} shows the triple product requirements for equally gain and loss conditions and ignition conditions when $f_T=T_e/T_i$. It can be seen from the figure that only with the larger Sikora (2016) cross-section data, can the equal gain and loss condition $Q=1$ be achieved for hydrogen-boron fusion, but the ignition condition $Q=\infty$ (i.e., the parameter range where the fusion power of charged products exceeds the radiation loss power) cannot be achieved. Using Nevins (2000) data is even more difficult, as there is no parameter range with $Q\geq1$ for equally gain and loss conditions. Figure \ref{fig:Lawson_ntauE_Q1_scanfT} shows the Lawson diagram for the fixed gain factor $Q=1$, with scanning temperature ratios $T_e/T_i=1,0.5,0.2,0.1$. It can be seen that when $T_e/T_i=0.5$, there is a gain interval for hydrogen-boron fusion. Figure \ref{fig:Lawson_ntauE_QInf_scanfT} shows the Lawson diagram for the fixed gain factor $Q=\infty$, with scanning temperature ratios $T_e/T_i=1,0.5,0.2,0.1$. It can be seen that at $T_e/T_i=0.5$ or $T_e/T_i=0.2$, hydrogen-boron fusion can reach ignition conditions. From these two figures, it can be observed that the hot ion mode with $T_e/T_i<1$ is crucial for achieving hydrogen-boron fusion, but it has a minor impact on deuterium-tritium and catalyzed deuterium-deuterium fusion, and has a certain influence on deuterium-deuterium and deuterium-helium fusion.

These results indicate that besides hydrogen-boron fusion, the other three fusion approaches (deuterium-tritium, deuterium-deuterium, and deuterium-helium) are scientifically feasible, with only differences in the required parameter values. For hydrogen-boron fusion, in addition to the normal thermonuclear fusion mode, the hot ion operational mode must also be realized. It should also be noted that the above discussion did not consider cyclotron radiation, which will impose more stringent conditions on the scientific feasibility of fusion. We will further analyze this in subsequent chapters. Furthermore, it should be noted that in the previous discussion on the average fusion reaction time $\tau_m$, within the typical fusion temperature range of 10-200 keV, $\tau_m\cdot n_2\sim10^{21}-10^{22}{\rm s\cdot m^{-3}}$. Comparing this with the requirement of $n\tau_E$ in the Lawson criterion, it can be seen that for deuterium-tritium fusion, $n\tau_E\simeq10^{19}-10^{20}{\rm s\cdot m^{-3}}$. Therefore, even if a high gain $Q$ is achieved, the fuel is usually not fully burned. For deuterium-deuterium, catalyzed deuterium-deuterium, and deuterium-helium fusion, a higher $n\tau_E\simeq10^{21}-10^{22}{\rm s\cdot m^{-3}}$ is required to achieve a gain, indicating a higher burning rate. For hydrogen-boron fusion, $n\tau_E\geq10^{22}{\rm s\cdot m^{-3}}$ is required when $T_e/T_i=1$, and an extremely high and rapid refueling rate is needed. In the hot ion mode of hydrogen-boron fusion, when $n\tau_E\sim10^{22}{\rm s\cdot m^{-3}}$, a higher burning rate is also needed, and the accumulation of products cannot be ignored. This actually means that a too long confinement time may not necessarily be advantageous, but may be detrimental, leading to the accumulation of fusion products and preventing further effective fusion and cooling of the fusion reactor. Subsequent sections will further evaluate this process, emphasizing the importance and quantitative conditions for effective removal of fusion products. From this perspective, the pursuit of confinement time $\tau_E$ is also limited. As long as the value required by the Lawson criterion is met, there is no need to further pursue an increase in the order of magnitude of the confinement time. However, in current fusion research, the confinement time is still much lower than or only close to the required value, so it is necessary to focus on confinement.

At the same time, it should be noted that for hydrogen-boron fusion, the temperature itself is already one-tenth of the energy of fusion products, so the energy of fusion products will also change. This was not taken into account in the zero-order approximation in this section. However, this also indicates that to achieve gain in hydrogen-boron fusion, the burning rate needs to exceed 10\%. This is a high requirement for confinement. It will also cause the problem of helium ash accumulation, which means that the actual situation is more difficult than the conditions assumed by the Lawson criterion. This problem can be understood from the perspective of the average fusion free path mentioned earlier, especially the parameter $n_2\tau_m$. In subsequent sections, we will discuss more advanced models that include the effect of burning rate.

\section{Energy exchange and hot ion mode}
In the previous sections, we have emphasized the main focus of fusion energy research, which is thermonuclear fusion. We have also pointed out that hydrogen-boron fusion requires a hot ion mode. Studying these issues, such as the heat exchange process between particles and whether the hot ion mode can be maintained, mainly involves Coulomb collision processes between charged particles. Compared to collisions between individual particles, it requires consideration of collective effects in plasma. The relevant theory has been established at the early stage of fusion research [Spitzer (1956)]. This part of the content is essentially related to the second law of thermodynamics. In this section, we will discuss this topic briefly.

\subsection{Basic Processes and Parameters of Coulomb Collisions}

For two types of particles $i$ and $j$, which have no drift and follow the Maxwellian distribution, when their temperatures $T_i$ and $T_j$ are different, they will undergo heat exchange through collisions to make the temperatures approach each other [Wesson (2011)]:
\begin{eqnarray}
 \frac{dT_i}{dt}=\frac{T_j-T_i}{\tau_{ij}}.
\end{eqnarray}
Here, the heat exchange time is given by
\begin{eqnarray}\label{eq:tauij}
 \tau_{ij}=\frac{3\sqrt{2}\pi^{3/2}\epsilon_0^2m_im_j}{n_je^4Z_i^2Z_j^2\ln\Lambda}\Big(\frac{k_BT_i}{m_i}+\frac{k_BT_j}{m_j}\Big)^{3/2}.
\end{eqnarray}
From the perspective of heat exchange power (ignoring particle density changes), the power transferred from particle $j$ to particle $i$ is given by
\begin{eqnarray}
 P_{ij}=\frac{d(\frac{3}{2}k_Bn_iT_i)}{dt}=\frac{3}{2}k_Bn_i(T_j-T_i)\nu_{ij}=-P_{ji}.
\end{eqnarray}
Here, the heat exchange frequency is defined as
\begin{eqnarray}
 \nu_{ij}=\frac{1}{\tau_{ij}}.
\end{eqnarray}For electrons and singly charged ions with $Z=1$, the thermal exchange times are given by
\begin{eqnarray}
 \tau_{ei}=\tau_{ie}=\frac{m_i}{2m_e}\tau_e,
\end{eqnarray}
where $\tau_e$ is the electron collision time (electron-ion collisions)
\begin{eqnarray}
 \tau_{e}=3(2\pi)^{3/2}\frac{\epsilon_0^2m_e^{1/2}(k_BT_e)^{3/2}}{n_iZ^2e^4\ln\Lambda}=1.09\times10^{16}\frac{T_e^{3/2}}{n_iZ^2\ln\Lambda}.
\end{eqnarray}
The ion collision time is given by
\begin{eqnarray}
 \tau_{i}=12\pi^{3/2}\frac{\epsilon_0^2m_i^{1/2}(k_BT_i)^{3/2}}{n_iZ^4e^4\ln\Lambda}=6.60\times10^{17}\frac{{(m_i/m_p)}^{1/2}T_i^{3/2}}{n_iZ^4\ln\Lambda}.
\end{eqnarray}
In the latter equation, the temperatures $T_e$ and $T_i$ are in units of keV, and the density is in units of ${\rm m^{-3}}$. In this process, the collision frequency is mainly determined by the fast component, that is, the electrons. However, due to the large difference in mass between electrons and ions, only a fraction of energy transfer equal to $m_e/m_i$ occurs each time.

The Coulomb logarithm $\ln\Lambda$ is given by [Gross (1984) p50]
\begin{eqnarray}
\Lambda=\frac{\lambda_D}{\bar{b}_0}=(12\pi n\lambda_D^3)=\frac{12\pi}{Z^2n_e^{1/2}}\Big(\frac{\epsilon_0k_BT}{e^2}\Big)^{3/2}=1.24\times10^7\frac{T^{3/2}}{Z^2n_e^{1/2}}.
\end{eqnarray}
Here, the temperature $T$ is in units of K, and the electron density $n_e$ is in units of ${\rm m^{-3}}$.

We perform some specific calculations for typical parameters. For simplicity, following [Wesson (2011), sec.14.5], we assume $\ln\Lambda\simeq17$ and $\ln\Lambda_i\simeq1.1\ln\Lambda$. Using the above formulas, we assume ions to be hydrogen and boron (with atomic mass numbers $A_i=1$ and 11, and charge numbers $Z=1$ and 5, respectively). To simplify the calculation, we do not consider the hydrogen-boron mixture ratio and only calculate the thermal exchange time for a single type of ion, namely either protons or borons, where $Z_in_i=n_e$. For electrons with an energy of 100 keV, we find that at a density of $10^{20}{\rm m}^{-3}$, the thermal exchange times with protons and borons are approximately 5.88 seconds and 12.9 seconds, respectively. This result indicates that under typical hydrogen-boron fusion parameters, the temperature of electron ions will approach each other on a timescale of a few seconds.

For collisional relaxation processes with non-Maxwellian distribution functions, it is necessary to solve the complex Fokker-Planck equation, which we will not discuss here.

\subsection{Fast particle energy deposition process}\begin{figure}[htbp]
\begin{center}
\includegraphics[width=15cm]{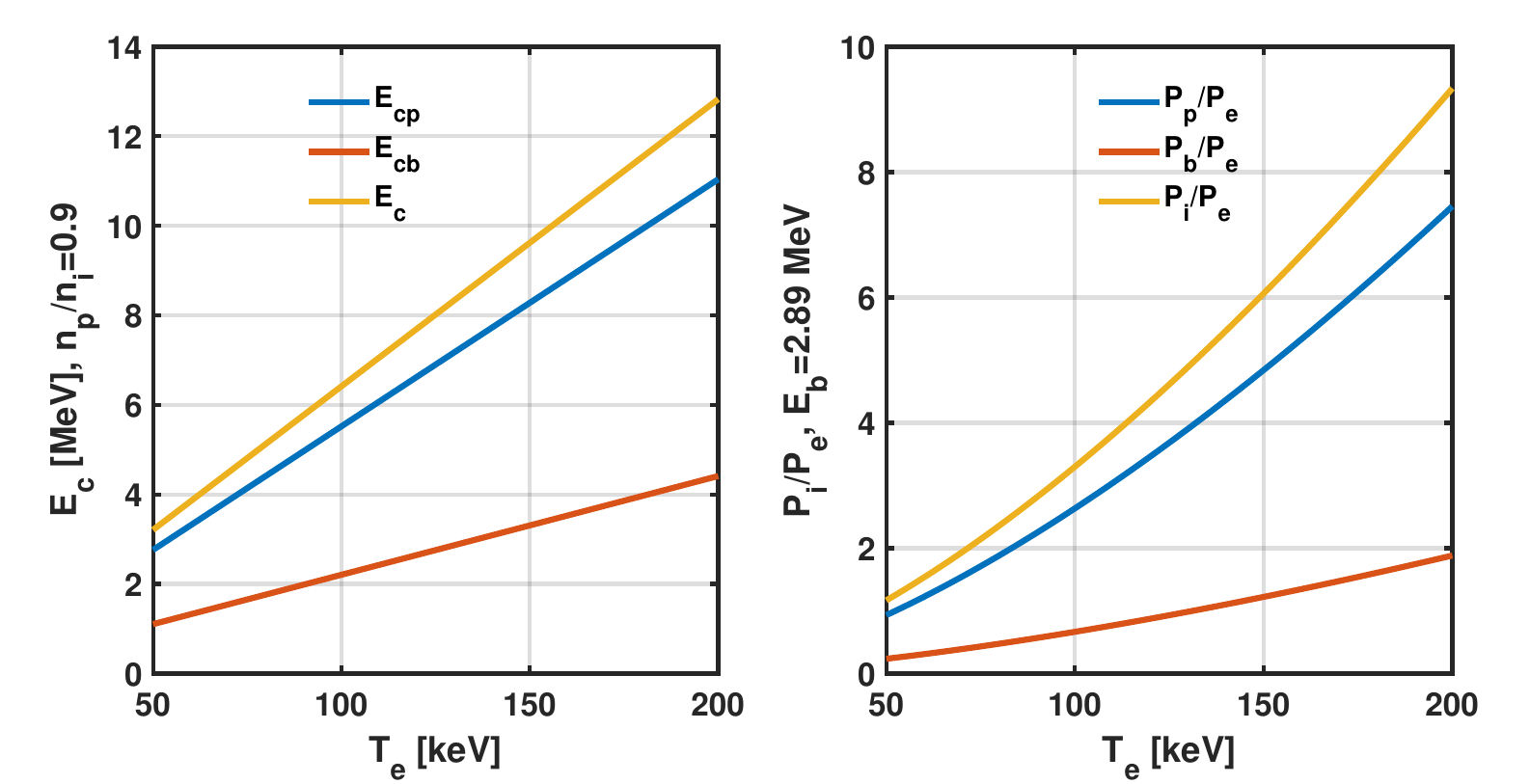}\\
\caption{The ratio of heating electrons and ions to alpha particle products of hydrogen-boron fusion as a function of electron temperature $T_e$.}\label{fig:fastionEc}
\end{center}
\end{figure}For the slowing down process of mono-energetic high-energy ions (neutral beam heating or fusion product heating), the electron mass is small, and when $v_b$ is less than the electron thermal velocity, the main resistance to high-energy ions is in the parallel direction. Therefore, the electron beam $v_b$ contributes to the electron heating power as follows:

\begin{eqnarray}
P_{be}=F_{be}v_b=\frac{m_bv_b^2}{\tau_{se}}=\frac{2E_b}{\tau_{se}}=\frac{2m_e^{1/2}m_bA_{De}E_b}{3(2\pi)^{1/2}(k_BT_e)^{3/2}},
\end{eqnarray}

where $m_b$ is the mass of the beam ions, $E_b=\frac{1}{2}m_bv_b^2$ is the beam energy, and $\tau_{se}$ is the slowing down time caused by electrons, which is expressed as:

\begin{eqnarray}
\tau_{se}=\frac{3(2\pi)^{1/2}(k_BT_e)^{3/2}}{m_e^{1/2}m_bA_{De}},~~A_{De}=\frac{n_ee^4Z_b^2\ln\Lambda_e}{2\pi\epsilon_0^2m_b^2}.
\end{eqnarray}

For background ions, due to their mass being close to that of the beam ions, the vertical scattering process is also important, and the heating power is given by:

\begin{eqnarray}
P_{bi}=F_{bi}v_b+\frac{1}{2}m_b\langle v_\perp^2\rangle=\frac{m_bv_b^2}{\tau_{si}}=\frac{2E_b}{\tau_{si}}=\frac{m_b^{5/2}A_{Di}}{2^{3/2}m_iE_b^{1/2}},
\end{eqnarray}

where

\begin{eqnarray}
\tau_{si}=\frac{m_i}{m_b}\frac{2v_b^3}{A_{Di}},~~A_{Di}=\frac{n_ie^4Z_i^2Z_b^2\ln\Lambda_i}{2\pi\epsilon_0^2m_b^2}.
\end{eqnarray}

Therefore, the ratio between the heating of ions and electrons by high-energy ions is:

\begin{eqnarray}
\frac{P_{bi}}{P_{be}}=\frac{3(2\pi)^{1/2}(k_BT_e)^{3/2}}{2m_e^{1/2}m_bA_{De}E_b}\frac{m_b^{5/2}A_{Di}}{2^{3/2}m_iE_b^{1/2}}\simeq\frac{3\pi^{1/2}(k_BT_e)^{3/2}}{4m_e^{1/2}n_e}\frac{m_b^{3/2}n_iZ_i^2}{m_iE_b^{3/2}}=\Big(\frac{E_{ci}}{E_b}\Big)^{3/2},
\end{eqnarray}

where the critical energy is:

\begin{eqnarray}
E_{ci}=\Big(\frac{3\sqrt{\pi}}{4}\Big)^{2/3}\Big(\frac{n_iZ_i^2}{n_e}\Big)^{2/3}\Big(\frac{m_i}{m_e}\Big)^{1/3}\frac{m_b}{m_i}k_BT_e.
\end{eqnarray}

Here, we assume $\ln\Lambda_e\simeq\ln\Lambda_i$. If we consider the combined effect of ions, then:

\begin{eqnarray}
E_c=\Big(\frac{3\sqrt{\pi}}{4}\Big)^{2/3}\Big(\sum_i\frac{n_iZ_i^2}{n_e}\frac{m_e}{m_i}\Big)^{2/3}\frac{m_b}{m_e}k_BT_e.
\end{eqnarray}Figure \ref{fig:fastionEc} shows the proportion of heating between alpha particles and electrons and ions in hydrogen-boron fusion. It can be seen that the main heating is done by ions, which provides the possibility of achieving the hot ion mode in hydrogen-boron fusion. Note that the energy distribution of alpha particles is not monoenergetic, which will have some impact on the results. Here, an average energy of $E_\alpha = 2.89$ MeV is taken.

The situation of deuterium-tritium fusion is different from hydrogen-boron. The energy of charged product alpha particles is 3.52 MeV, while the background temperature is only around 10 keV, so alpha particles mainly heat electrons rather than ions. In simple terms, it can be understood that if the velocity of fast ions is closer to the thermal velocity of electrons, they will preferentially heat electrons, while if it is closer to the thermal velocity of ions, they will preferentially heat ions. The content of this section is discussed in detail in textbooks such as Wesson (2011) and Freidberg (2007).

\subsection{Hot ion mode}

Let us use the simplest calculation to estimate the electron temperature in the hot ion mode of hydrogen-boron fusion.

\begin{figure}[htbp]
\begin{center}
\includegraphics[width=15cm]{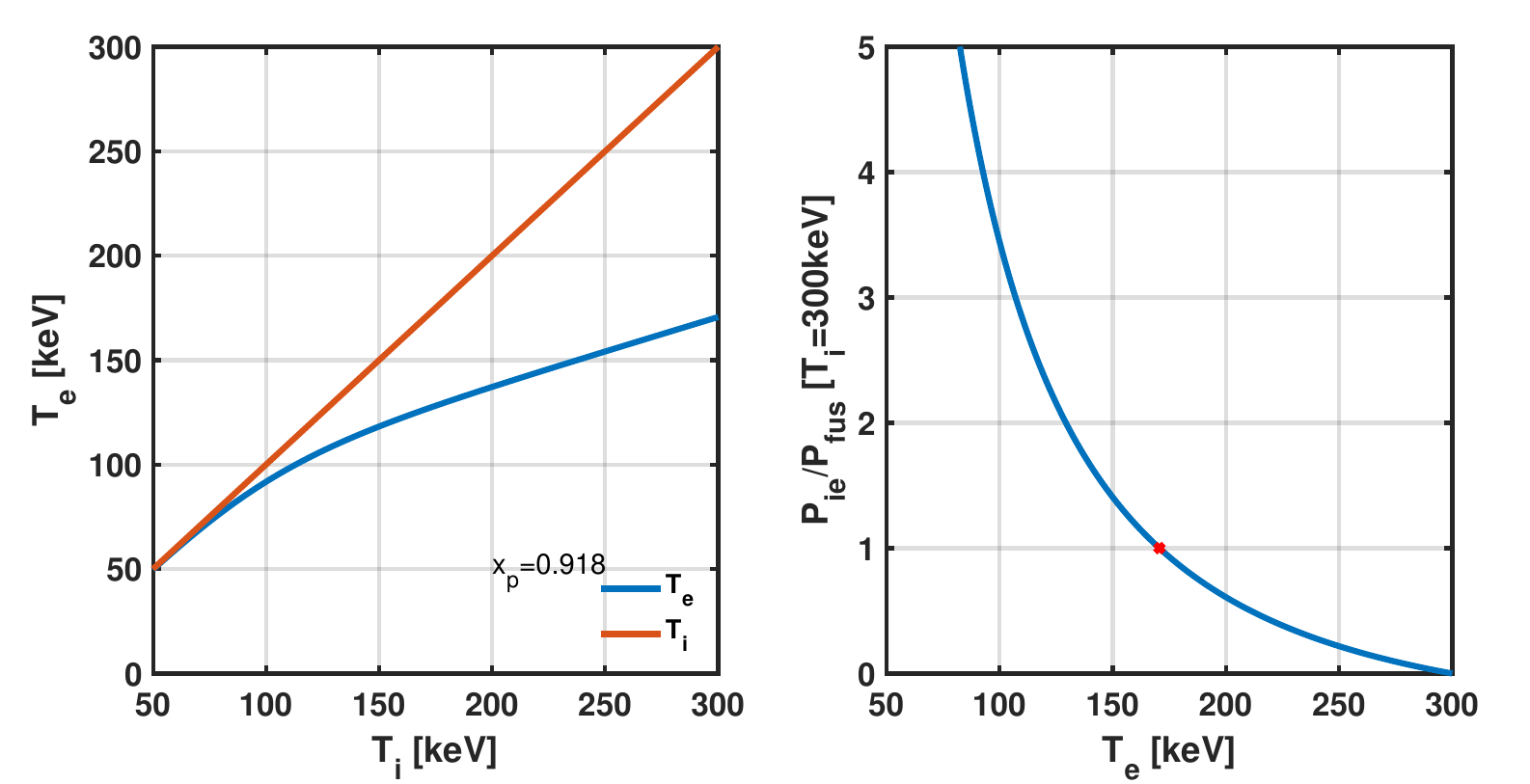}\\
\caption{The relationship between the electron temperature $T_e$ and the ion temperature $T_i$ in the hot ion mode of hydrogen-boron fusion.}\label{fig:Tehotion} 
\end{center}
\end{figure}

Assuming that all the power generated in hydrogen-boron fusion is used to heat ions, electrons maintain their energy by exchanging heat with ions. Neglecting radiation and transport losses and only considering the dominant terms, we have
\begin{eqnarray}
-\frac{dW_i}{dt}=\frac{3}{2}k_B\Big(\frac{n_p}{\tau_{p,e}}+\frac{n_B}{\tau_{B,e}}\Big)(T_i-T_e)=f_{ion}P_{fus}.
\end{eqnarray}
Where $\tau_{p,e}$ and $\tau_{B,e}$ are the thermal exchange times between protons and electrons and between boron ions and electrons, which can be calculated using Equation (\ref{eq:tauij}). The results are shown in Figure \ref{fig:Tehotion}. The electron temperature calculated using this method is the lowest possible steady-state temperature, and the actual value will be higher than this value considering other effects. It can be seen that when $T_i=300$ keV, $T_e=170$ keV, i.e. $T_e/T_i=0.57>0.5$. For smaller $T_i$, the difference between $T_e$ and $T_i$ is even smaller. This indicates that it is more difficult to maintain a high temperature difference in the hot ion mode. This conclusion is similar to Moreau (1977) and Dawson (1981).\begin{figure}[htbp]
\begin{center}
\includegraphics[width=15cm]{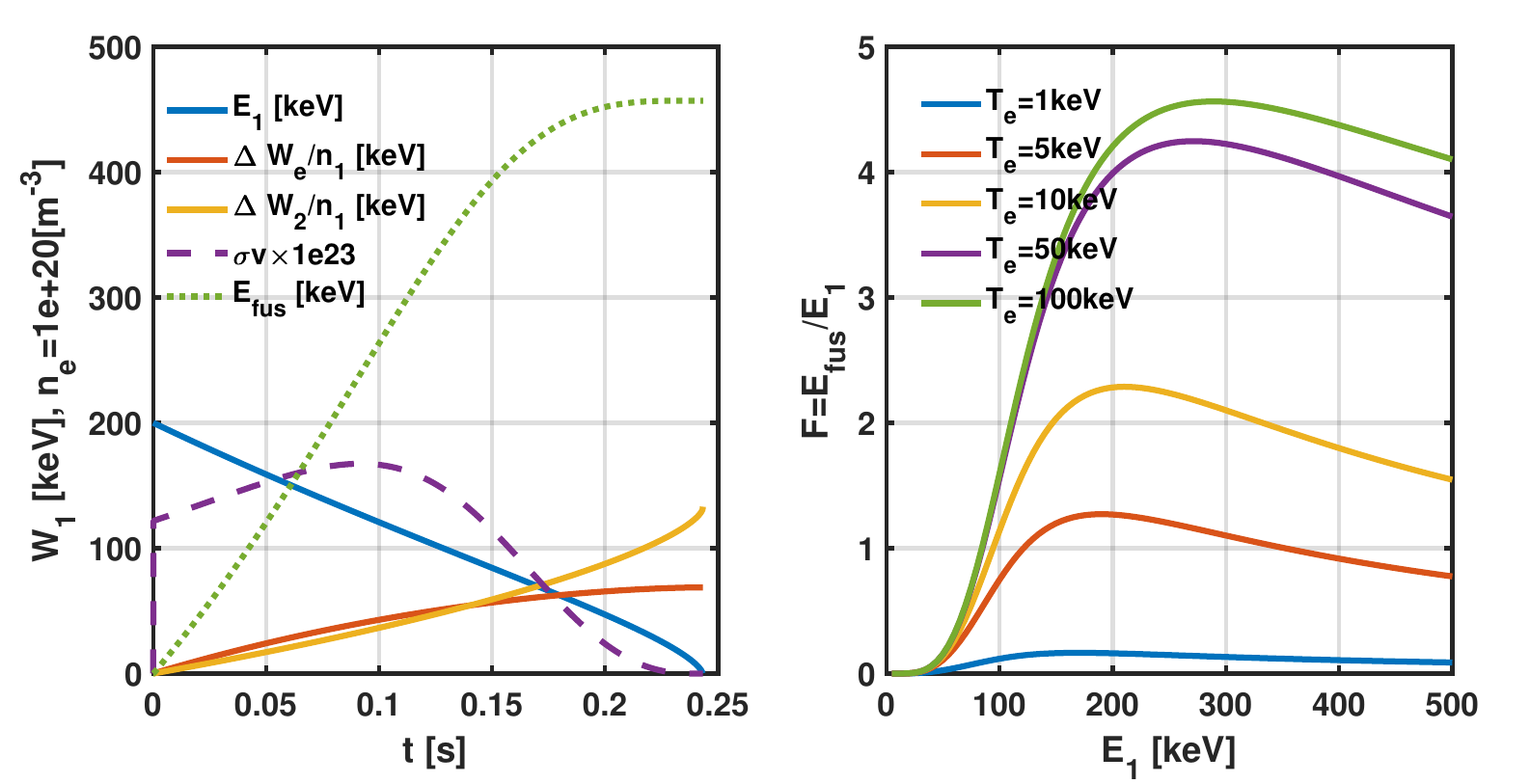}\\
\caption{The gain factor $F=E_{fus}/E_1$ of beam-driven fusion, showing the fusion gain caused by injecting high-energy deuterium ions into a tritium background plasma, for different background electron temperatures $T_e$.}\label{fig:Epfusiondt}
\end{center}
\end{figure}

\begin{figure}[htbp]
\begin{center}
\includegraphics[width=15cm]{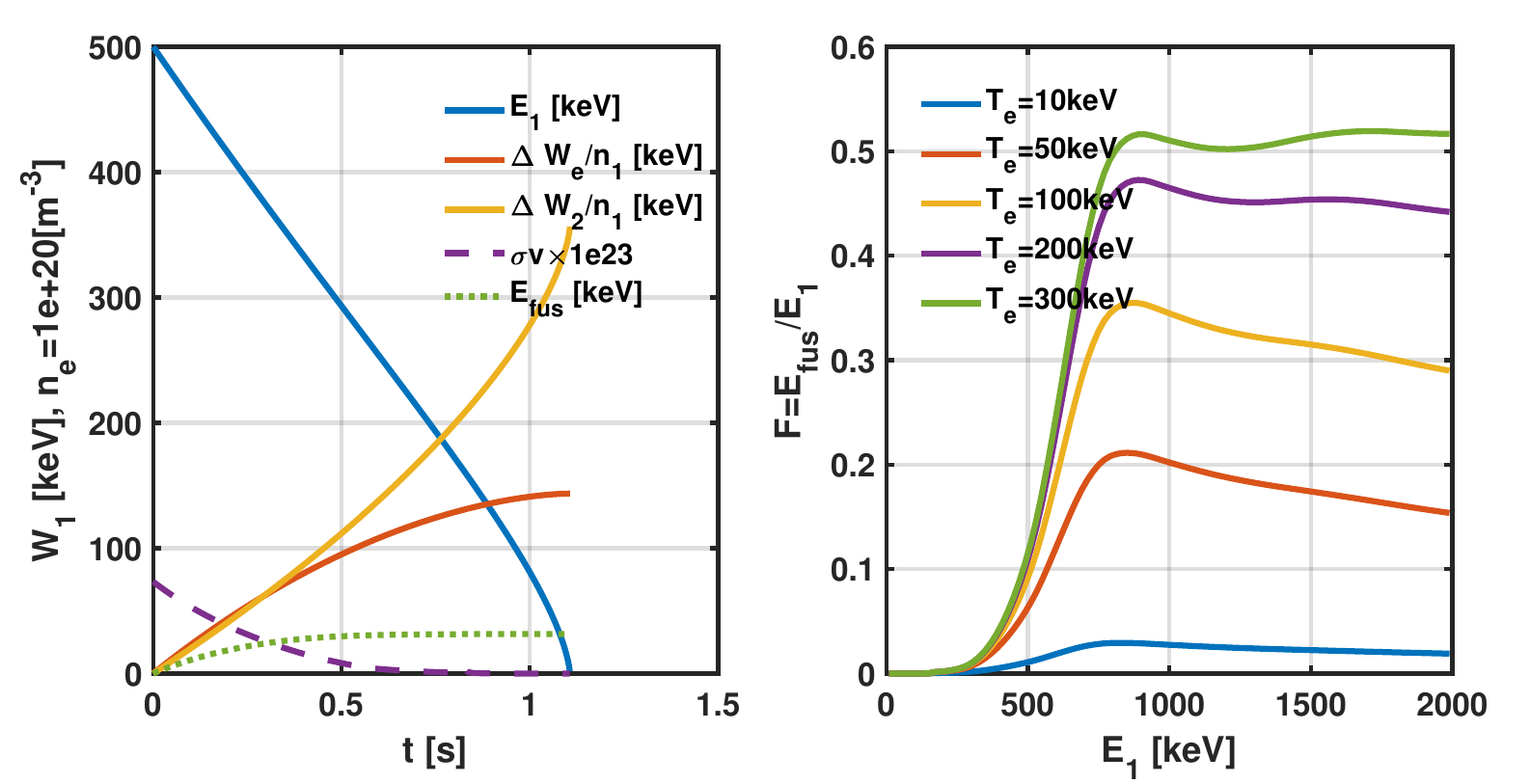}\\
\caption{The gain factor $F=E_{fus}/E_1$ of beam-driven fusion, showing the fusion gain caused by injecting high-energy proton ions into a boron background plasma, for different background electron temperatures $T_e$.}\label{fig:Epfusionpb}
\end{center}
\end{figure}

\subsection{Beam-Driven Fusion}

With the formula derived for high-energy particle slowing down, we can now estimate the fusion energy gain obtained by injecting high-energy ions into a cold background plasma, such as deuterium injection into a tritium background or proton injection into a boron background. Assuming the high-energy ions are denoted as $m_1$ and the background ions as $m_2$, with $n_1\ll n_2$, the time evolution equation for the energy $W_1=\frac{1}{2}n_1m_1v_1^2=n_1E_1$ of the high-energy ions is given by
\begin{eqnarray}
 \frac{dW_1}{dt}&=&-P_{12}-P_{1e},\\
 P_{1e}&=&\frac{2m_e^{1/2}}{3(2\pi)^{1/2}m_1(k_BT_e)^{3/2}}\frac{n_ee^4Z_1^2\ln\Lambda_e}{2\pi\epsilon_0^2}W_1,\\
 P_{12}&=&\frac{m_1^{1/2}}{2^{3/2}m_2E_1^{3/2}}\frac{n_2e^4Z_2^2Z_1^2\ln\Lambda_i}{2\pi\epsilon_0^2}W_1.
\end{eqnarray}
By solving this equation, the time variation of the beam energy can be obtained. The beam ions undergo fusion reactions with the background ions, and the instantaneous fusion power is given by
\begin{eqnarray}
P_{fus}=n_1n_2\frac{1}{1+\delta_{12}}\langle\sigma v\rangle Y.
\end{eqnarray}
Assuming $n_1\ll n_2$, we have $n_2=n_e/Z_2$, and the densities $n_1$ and $n_e$ can be normalized out without directly affecting the conclusions. At the same time, we assume that the background ions are cold. Note that in this case, the reaction rate $\langle\sigma v\rangle$ is given for the center-of-mass energy $E_c=E_1m_2/(m_1+m_2)$ and the center-of-mass velocity $v=\sqrt{2E_1/m_1}$.Figure \ref{fig:Epfusiondt} shows the gain factor $F=E_{fus}/E_1$ of beam fusion. It calculates the fusion gain caused by the injection of high-energy deuterium ions into the background tritium plasma. For different background electron temperatures $T_e$, it can be seen that although this beam fusion method can achieve a gain of $F>1$, the maximum gain is only around $F\simeq4$. Considering the energy required to sustain the background plasma and the energy loss in beam generation, it is difficult to achieve an engineering fusion gain $Q_{eng}>1$ overall. The above scheme was first analyzed in detail by Dawson (1971). However, if the number of beam particles is large enough, the energy can be deposited into the background plasma and heat it up to the fusion temperature, and confine it for a certain period of time, which may achieve a gain through the thermal nuclear reaction. This is the goal of neutral beam heating. Figure \ref{fig:Epfusionpb} shows the fusion gain factor $F=E_{fus}/E_1$ when high-energy protons are injected into a boron background plasma. It can be seen that the maximum $F\simeq0.5<1$, which means that fusion gain for hydrogen-boron cannot be achieved through this method, and the energy gain and loss are not balanced. This conclusion is similar to Moreau (1977). Santini (2006) also discussed the energy gain of injecting non-thermalized beams into a background plasma and concluded that achieving gain is very difficult.

In addition, in the two beam fusion examples mentioned above, we used a lower mass number for the beam and a higher mass number for the background. This is because for the fusion reaction, the relative velocity or the energy in the center of mass frame is meaningful. For higher mass numbers, a larger beam energy is required to achieve the same relative velocity, which means higher cost and a more difficult gain factor.

\section{Parameter Ranges for Different Confinement Methods}

According to the Lawson criterion's triple product requirements, controlled fusion confinement methods are generally classified into magnetic confinement fusion, inertial confinement fusion, and magneto-inertial confinement fusion. For ease of subsequent discussion, we roughly list the parameter ranges for the main types of confinement methods in Table \ref{tab:cftype}. The literature mainly discusses deuterium-tritium fusion and the given ranges may be narrower than those listed here. Here, we slightly expand the parameter range for advanced fuels for later discussion.
\begin{table}[htp]
\footnotesize
\caption{Approximate parameter range for different fusion confinement methods.}
\begin{center}
\begin{tabular}{c|c|c|c|c}
\hline\hline
Type & \thead {Magnetic \\ Confinement}  & \thead {Magnetic \\ Inertial Confinement} & \thead {Inertial \\ Confinement} & \thead {Gravitational \\Confinement}  \\\hline
Density $\rm [m^{-3}]$ &  $10^{19}-10^{22}$  & $10^{24}-10^{30}$ &  $>10^{30}$ &  $>10^{30}$  \\
Confinement Time $[s]$  &  $>0.1$  & $10^{-8}-10^{-3}$ &  $<10^{-8}$ &  $>10^{10}$\\
\thead {Ignition \\ Temperature Required} &  Yes  & Yes &  Yes &  No\\
\thead {Magnetic \\ Field Required} &  Yes  & Yes &  No &  No  \\\hline\hline
\end{tabular}
\end{center}
\label{tab:cftype}
\end{table}

The parameter range for gravitational confinement is usually characterized by extremely high density and long energy confinement time, and it can effectively confine radiation. Therefore, the parameter range for gravitational confinement is different from the other methods. Inertial confinement also involves high density, but with extremely short confinement time, making it a pulsed scheme. In magnetic confinement, the main goal is to achieve steady-state or long-pulse power generation, which is closer to classical plasma physics, and it is limited by most of the physical factors discussed in this chapter. The objective of magnetic inertial confinement is to take advantage of the strengths of both magnetic confinement and inertial confinement, and realize fusion energy in the intermediate density and confinement time range. It also belongs to the pulsed scheme.

\section{Summary of this Chapter}

This chapter discusses the basic parameter relationships and requirements for fusion reactors. Firstly, we point out the meaning of average fusion reaction time and average fusion mean free path, and indicate their parameter ranges. In particular, we explain why fusion energy research mainly focuses on thermonuclear fusion by comparing it with Coulomb collisions. At the same time, due to the requirements for power density economy and controllability, the plasma density for fusion can only be selected within a narrow range. By comparing fusion power with bremsstrahlung radiation power, we obtain the optimum ignition temperature and indicate that the optimal fusion temperature is within a narrow range. Based on a comprehensive energy balance analysis, we derive the Lawson criterion, pointing out that the fusion triple product of temperature, density, and confinement time needs to exceed a certain value in order to achieve fusion gain. In the cases where temperature and density have optimal ranges, the minimum requirement for confinement time is also determined. Due to excessive bremsstrahlung radiation, we discuss the feasibility of the hot ion mode. Additionally, we explore the energy gain of non-thermal fusion through beam-driven fusion, indicating the significant difficulties in achieving fusion energy with this method. In order to break through the conclusions discussed in this chapter, there are some possible but extremely difficult ideas: (1) Are there any fusion schemes that are exceptions and not within the scope of the Lawson criterion? If so, it may be possible to surpass the parameter requirements here. For example, cold fusion. (2) At energy levels much higher than MeV, if the ion-ion Coulomb cross section is smaller than the fusion cross section, can it be used to achieve fusion? It requires comparing the nuclear ion-electron collision cross section and the energy scale. (3) Is it possible to have economic feasibility even without reaching the ignition condition? (4) Is it possible to have a chain reaction of fusion that can significantly reduce the fusion conditions, as discussed in McNally (1973) and Lu Hefu (1960) in IAEA 1973?

\vspace{30pt}
Key points of this chapter:
\begin{itemize}
\item Due to the limitations of power generation and driver energy conversion efficiency, in order to achieve positive energy output from fusion, it is necessary to first achieve energy balance at a scientific level.
\item The ratio of the power due to bremsstrahlung radiation to the fusion power determines the minimum ignition temperature and the optimal temperature range for fusion.
\item The limitation of power density determines that the average density of the fusion plasma can only be in a narrow range, around $10^{19}-10^{22}{\rm m^{-3}}$, for magnetic confinement fusion.
\item The average reaction time for fusion determines the lower limit of the required confinement time.
\item Energy balance provides the minimum conditions for the scientific feasibility of fusion, and it has scientific feasibility for deuterium-tritium, deuterium-deuterium, and deuterium-helium fusion.
\item For hydrogen-boron fusion, there is a parameter range that can be operated in the hot ion mode or when the fusion reaction rate is significantly enhanced, but the required conditions are extremely harsh given the current technological capabilities.
\end{itemize} 
\chapter{Parameter Range for Magnetic Confinement Fusion}\label{chap:mcf}

Magnetic confinement is achieved by using magnetic fields to confine fusion plasma for an extended period of time, during which the plasma can be considered to be in quasi-equilibrium, with a basic balance between thermal pressure and electromagnetic forces. Therefore, the two zeroth-order effects brought about by magnetic confinement fusion are the upper limit of beta ($\beta$) and the synchrotron (cyclotron) radiation limit. In addition, there are engineering constraints that limit the maximum achievable magnetic field.

The scientific feasibility of magnetic confinement for deuterium-tritium fusion has been basically verified. Therefore, from a zeroth-order perspective, there is no need to further discuss its scientific feasibility. The main challenges for its use as an energy source lie in engineering and economic factors such as neutron shielding, tritium breeding, and commercialization costs. Lidsky (1983) published a famous critical article from an engineering perspective, arguing that fusion energy from deuterium-tritium reactions is not worth pursuing and is far from comparable to fission reactors. Reinders (2021) comprehensively reviewed the history of fusion and expressed disappointment, stating that fusion energy is not as promising as it is advertised and collected some major criticisms. Of course, not everyone agrees with these criticisms, for example, Stacey (1999) responded to some of the criticisms.

This chapter explores the scientific feasibility of magnetic confinement for non-deuterium-tritium fusion from a zeroth-order perspective. The results presented here represent the minimum conditions required for advanced fuels, focusing mainly on hydrogen-boron fusion, deuterium-helium fusion, and also discussing deuterium-deuterium fusion. Dawson (1981) and McNally (1982) have discussed advanced fuel fusion in detail in the early days. However, due to the consideration of multiple factors in the models involved, it is not easy to determine which factors can be overcome. In this chapter, we mainly consider the most basic energy balance and a few necessary limiting factors to explore the difficulties of magnetic confinement for non-deuterium-tritium fusion. This way, we can more easily determine whether these few limiting factors can be overcome and assess the difficulty of overcoming them, thus obtaining a clearer understanding of the difficulties associated with non-deuterium-tritium fusion and identifying potential breakthrough directions. We will also use slightly more complex models for certain discussions.

\section{Zeroth-order Parameter Evaluation Model}\label{sec:mcfmodel1}

In the previous chapter, we have outlined the ranges of some key parameters, especially for magnetic confinement fusion, whose parameters can be roughly determined through the following process:
\begin{itemize}
\item Temperature: The minimization of the ratio of Bremsstrahlung radiation to fusion power determines the range of temperature $T_i$. In the case where the fusion reaction rate cannot be increased, a hot ion mode with $T_e/T_i<1$ is required for hydrogen-boron reactions. When the fusion reaction rate can be increased, the requirements for temperature ratio can be relaxed.
\item Density: The economic and controllable requirements limit the fusion power density within certain maximum and minimum values, so the density cannot be too low or too high. Typically, the range is $10^{19}-10^{22}{\rm m^{-3}}$.
\item Magnetic field: Given the temperature and density, the pressure is determined, and thus the value of $\beta B^2$ is determined. The magnetic constraint $\beta$ has an upper limit, which implies a lower limit for the magnetic field $B$.
\item Energy confinement time: Based on energy balance, the requirements for the energy confinement time $\tau_E$ and the wall reflection rate $R_w$ can be calculated.
\end{itemize}

These zeroth-order factors limit the parameter range of fusion energy and cannot be freely chosen, but can only vary within a narrow range. Based on the previous discussion, we will further analyze the parameter space.

\subsection{Model and Constraint Factors}

We calculate the energy balance between fusion reactions, radiation and confinement losses, and limit the parameter range through physical and economic requirements. We neglect the effects of fusion products such as helium ash.
The energy balance in a steady state is given by
\begin{eqnarray}
\frac{dE_{th}}{dt}=-\frac{E_{th}}{\tau_E}+f_{ion}P_{fus}+P_{heat}-P_{rad}=0,
\end{eqnarray}
Here, we adopt the usual approach from system codes [Costley (2015)], where all the energy of charged products in the fusion power $P_{fus}$ is treated as input energy. For example, for deuterium-tritium fusion, the energy carried by charged products accounts for only 1/5 of the total fusion energy ($f_{ion}=0.2$), while for hydrogen-boron fusion, all the energy of the fusion products is carried by charged alpha particles ($f_{ion}=1$). $P_{heat}$ represents the external heating power, and we obtain
\begin{eqnarray}
P_{heat}=P_{rad}+\frac{E_{th}}{\tau_E}-f_{ion}P_{fus}.
\end{eqnarray}
The fusion gain factor $Q_{fus}$ is defined as
\begin{eqnarray}
Q_{fus}\equiv\frac{P_{fus}}{P_{heat}}.
\end{eqnarray}
The fusion power per unit volume is given by
\begin{eqnarray}
P_{fus}=\frac{1}{1+\delta_{12}}n_1n_2\langle\sigma v\rangle Y,
\end{eqnarray}
Here, $n_1$ and $n_2$ are the number densities of two types of fusion ions. For the same ion, $\delta_{12}=1$; for different ions, $\delta_{12}=0$. $Y$ represents the energy release from a single nuclear reaction.Considering non-thermalization effects (non-Maxwellian distribution, such as the beam distribution using resonance peaks), the unequal temperatures of hydrogen boron ions, and possible increases in reaction cross section, we attribute all these effects to an amplification factor $f_\sigma$, that is, the reaction rate is assumed to be
\begin{eqnarray}
\langle\sigma v\rangle =f_\sigma \langle\sigma v\rangle_M,
\end{eqnarray}
where $\langle\sigma v\rangle_M$ is the reaction rate with the Maxwellian distribution as described in the appendix.

Plasma energy storage
\begin{eqnarray}
E_{th}= \frac{3}{2}k_B(n_iT_i+n_eT_e).
\end{eqnarray}

For the radiation term, we only consider bremsstrahlung and cyclotron (synchrotron) radiation
\begin{eqnarray}
P_{rad}= P_{brem}+P_{cycl}.
\end{eqnarray}
In this section, we consider the minimum conditions for magnetically confined fusion. Since the radiation from relativistic effects is greater than that from non-relativistic effects, we prioritize calculations based on non-relativistic or weakly relativistic formulas. The bremsstrahlung radiation is the same as in Nevins (1998)
\begin{eqnarray}\nonumber
P_{brem}&=&C_Bn_e^2\sqrt{k_BT_e}\Big\{Z_{eff}\Big[1+0.7936\frac{k_BT_e}{m_ec^2}+1.874\Big(\frac{k_BT_e}{m_ec^2}\Big)^2\Big]\\&&
+\frac{3}{\sqrt{2}}\frac{k_BT_e}{m_ec^2}\Big\} ~{\rm [MW\cdot m^{-3}]}.
\end{eqnarray}
where $C_B=5.34\times10^{-37}\times10^{-6}$, temperature $k_BT_e$ and energy $m_ec^2$ are in keV, and density $n_e$ is in ${\rm m^{-3}}$.
For cyclotron radiation, we use the same formula as in Costley (2015), but neglect the volume-averaging effect
\begin{eqnarray}
P_{cycl}=4.14\times10^{-7}n_e^{0.5}T_e^{2.5}B^{2.5}(1-R_w)^{0.5}\Big(1+2.5\frac{T_e}{511}\Big)\cdot\frac{1}{a^{0.5}}~{\rm [MW\cdot m^{-3}]},
\end{eqnarray}
where $R_w$ is the wall reflection coefficient, temperature $T_e$ is in keV, density is in $10^{20}{\rm m^{-3}}$, magnetic field $B$ is in T, and minor radius $a$ is in m. McNally (1982) provides several early derived formulas for cyclotron radiation, and the results differ. It is not easy to accurately assess the magnitude of cyclotron radiation in practical devices. Here, we temporarily use the above formula for calculation, while attributing any deviations from reality to the requirement for the reflection coefficient $R_w$. For example, if the actual value is smaller than that calculated by this formula, the requirement for the reflection coefficient can be lower.The plasma beta is given by
\begin{eqnarray}
\beta=\frac{2\mu_0k_B(n_iT_i+k_Bn_eT_e)}{B^2},
\end{eqnarray}
where $\mu_0$ is the permeability of vacuum. In order for fusion to be economical, the required conditions are
\begin{eqnarray}
Q_{fus}\geq Q_{min},\\
P_{fus}\geq P_{min},
\end{eqnarray}
for example, taking $Q_{min}=1$ and $P_{min}=1{\rm MW\cdot m^{-3}}$. For magnetic confinement equilibrium, it is physically required that
\begin{eqnarray}
\beta\leq\beta_{max},
\end{eqnarray}
for magnetic confinement, the mechancial equilibrium generally requires a volume-averaged $\beta_{max}=1$; when considering instability, $\beta_{max}$ becomes smaller, for example, a spherical tokamak can achieve $\beta=0.4$, and a field reversed configuration can achieve $\beta\simeq1$.

Other parameters and symbols: two ion charge numbers $Z_1$, $Z_2$, ion temperature $T_1=T_2=T_i$, electron temperature $T_e=f_TT_i$, electron density $n_e=Z_in_i$, ion density $n_i=n_1+n_2$, average ion charge number $Z_i=(Z_1n_1+Z_2n_2)/(n_1+n_2)$. Density of the first ion species $n_1=f_1n_i$, density of the second ion species $n_2=f_2n_i$, when both ion species are the same, $f_1=f_2=1$, otherwise $f_2=1-f_1$. Average charge number $Z_i=f_1Z_1+f_2Z_2$, effective charge number $Z_{eff}=\sum(n_iZ_i^2)/n_e$.

There are 8 input parameters in this model: the proportion of the first ion species $f_1$, the ratio of electron and ion temperatures $f_T$, reaction rate factor $f_{\sigma}$, magnetic field $B$, energy confinement time $\tau_E$, electron density $n_e$, ion temperature $T_i$, and the reflectivity of synchrotron radiation $R_w$. For simplicity of discussion, the effect of the minor radius $a$ in synchrotron radiation is also included in $R_w$, and $a=1$m is set.

According to the above model, we can scan these 8 parameters and then examine whether there exists a suitable parameter range or search for the optimal parameter range through the posterior constraints of $Q_{min}$, $P_{min}$, $\beta_{max}$ and other output parameters.

\begin{figure}[htbp]
\begin{center}
\includegraphics[width=13cm]{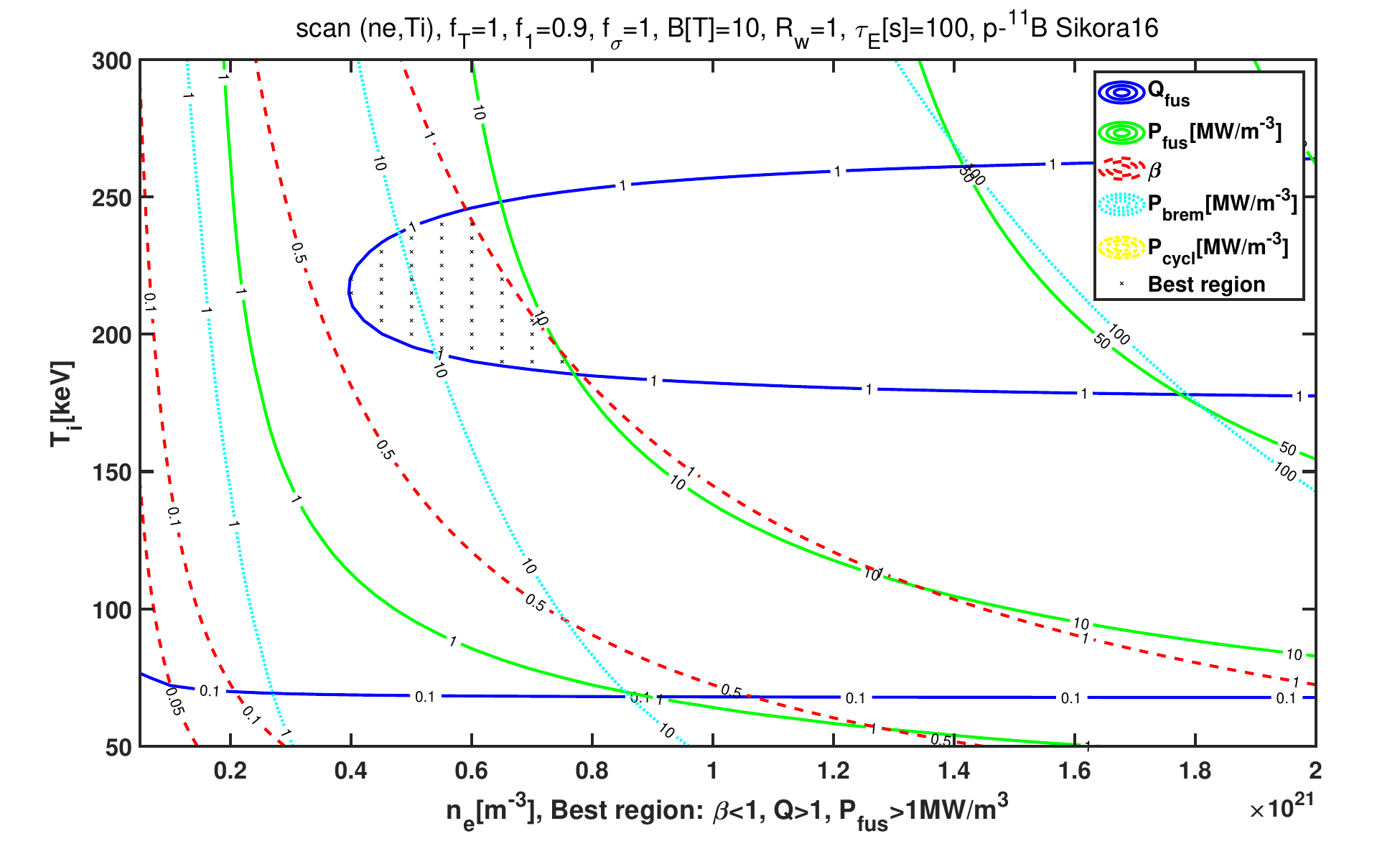}\\
\caption{Parameter scan for hydrogen-boron (${p-{}^{11}B}$), assuming equal electron and ion temperatures and that synchrotron radiation can be completely reflected. In this case, there may exist a parameter range, but it is extremely narrow.}\label{fig:mcfscan2dpb1} 
\end{center}
\end{figure}

\begin{figure}[htbp]
\begin{center}
\includegraphics[width=13cm]{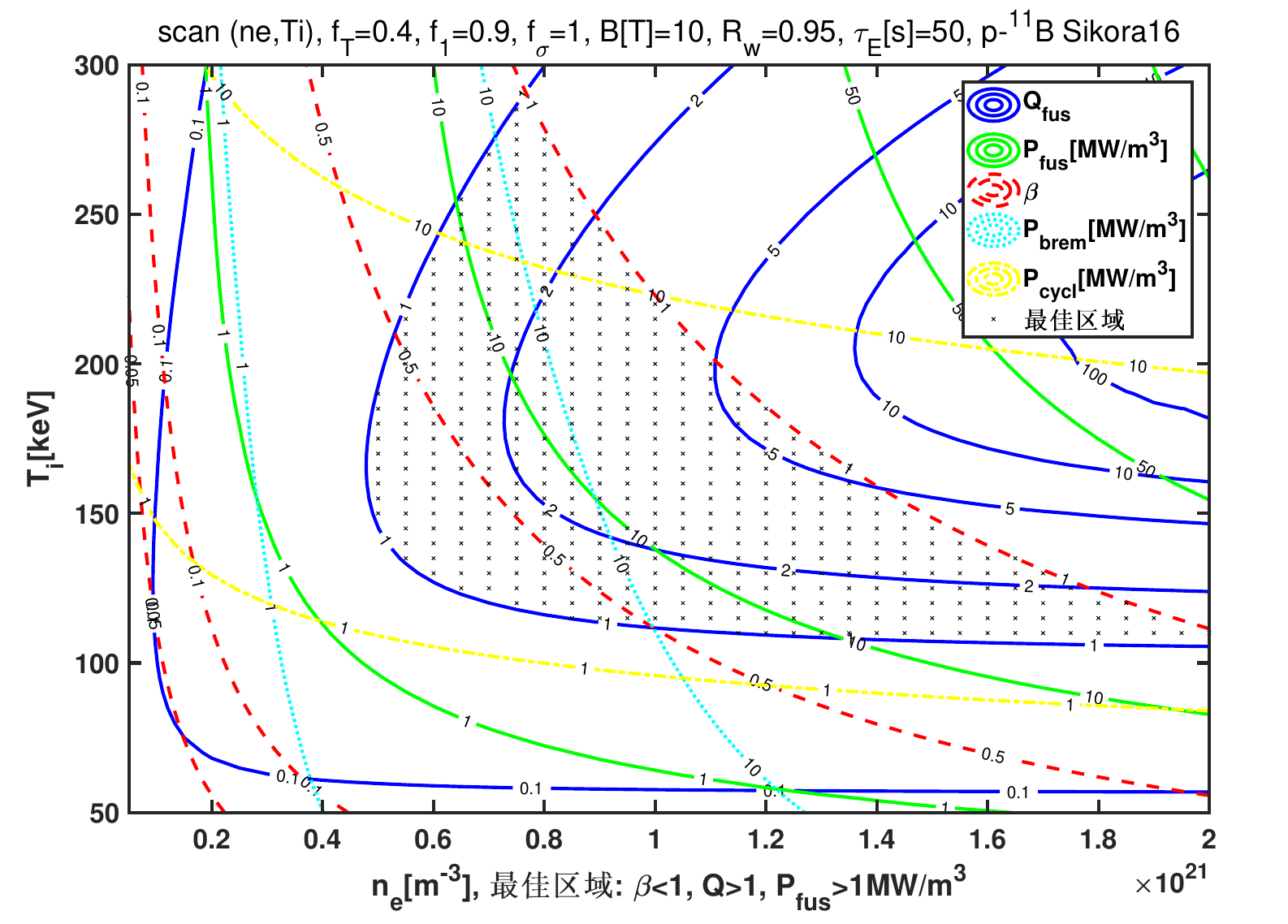}\\
\caption{Parameter scan for hydrogen-boron (${p-{}^{11}B}$), assuming a ratio of electron and ion temperatures of $f_T=0.4$ and a synchrotron radiation reflectivity coefficient of $R_w=0.95$. In this case, there may exist a parameter range, but it is still demanding.}\label{fig:mcfscan2dpb2}
\end{center}
\end{figure}\subsection{Hydrogen-Boron Fusion Parameters}

Hydrogen-boron fusion, characterized by abundant raw materials and no neutron byproducts, is a more ideal fusion energy source. However, due to its low reaction cross-section, achieving energy net gain from thermonuclear reactions requires conditions far more demanding than deuterium-tritium fusion. We will quantitatively calculate the key conditions required for a magnetic confinement device to achieve net gain from hydrogen-boron fusion using the above simple zero-dimensional model. The latest experimental data is used for the hydrogen-boron fusion reaction cross-section. The first ion species is assumed to be a proton. The constrained parameters are set as follows: $Q_{min}=1$, $P_{min}=1 \, \text{MW} \cdot \text{m}^{-3}$, and $\beta_{max}=1$.

Figure \ref{fig:mcfscan2dpb1} shows that when cyclotron radiation is not considered and the electron and ion temperatures are equal, there is a parameter range where the gain is greater than 1, but this range is very narrow. However, once cyclotron radiation is taken into account, for example, by setting a reflectivity $R_w=0.99$, this gain range disappears.

Figure \ref{fig:mcfscan2dpb2} shows that by reducing the electron temperature, there is a positive gain range. It should be noted that when the magnetic field is 10 T, the positive gain range is greater than 0.4 in terms of beta. In other words, if the magnetic field is lower, the beta will exceed the limit value of 1. At this point, the energy confinement time is also not low, $\tau_E=50$ s.

By scanning a broader range of parameters using this simplest model, we can conclude that hydrogen-boron fusion may have a feasible parameter range under stringent optimization conditions, but the range is extremely narrow and the difficulty is very high. The relative importance of the required conditions can roughly be ranked as follows: hot ion mode, reaction rate improvement, cyclotron radiation reflectivity, high beta, magnetic field that is neither too high nor too low, and high confinement time. These conditions are almost beyond the conventional range of current technological capabilities. This is also the reason why Nevins (2000) and early literature considered hydrogen-boron fusion to be almost infeasible. This is not even considering how to raise the temperature to above 150 keV and achieve energy confinement times of tens of seconds on an economically viable device.

The limiting factors here can also be understood from Chapter \ref{chap:lawson}: the pressure limit restricts the magnetic field for hydrogen-boron fusion from being too low, not lower than 6 T, while cyclotron radiation restricts the field from being too high. This limits the available magnetic field values, which in turn requires a $\beta$ as close to 1 as possible. Due to the small range of variable temperature, this further limits the density to a narrow range. Even if the energy confinement time can be very long and the density requirements are lowered, too low a density results in a fusion power per unit volume that is too low to meet economic requirements. Hence, the parameter range for magnetic confinement hydrogen-boron fusion is very narrow. Even with only these three set limiting factors, the parameters constrain each other. When considering more practical factors, the difficulty increases further. Figures \ref{fig:mcfscan2dpb1} and \ref{fig:mcfscan2dpb2} are essentially more complex versions of the Lawson diagrams in Chapter \ref{chap:lawson}. To overcome the above constrains, the following questions can be considered: Can bremsstrahlung and cyclotron radiation be recycled? Can β be exceeded, such as wall confinement? However, almost all of these are not feasible.

\begin{figure}[htbp]
\begin{center}
\includegraphics[width=13cm]{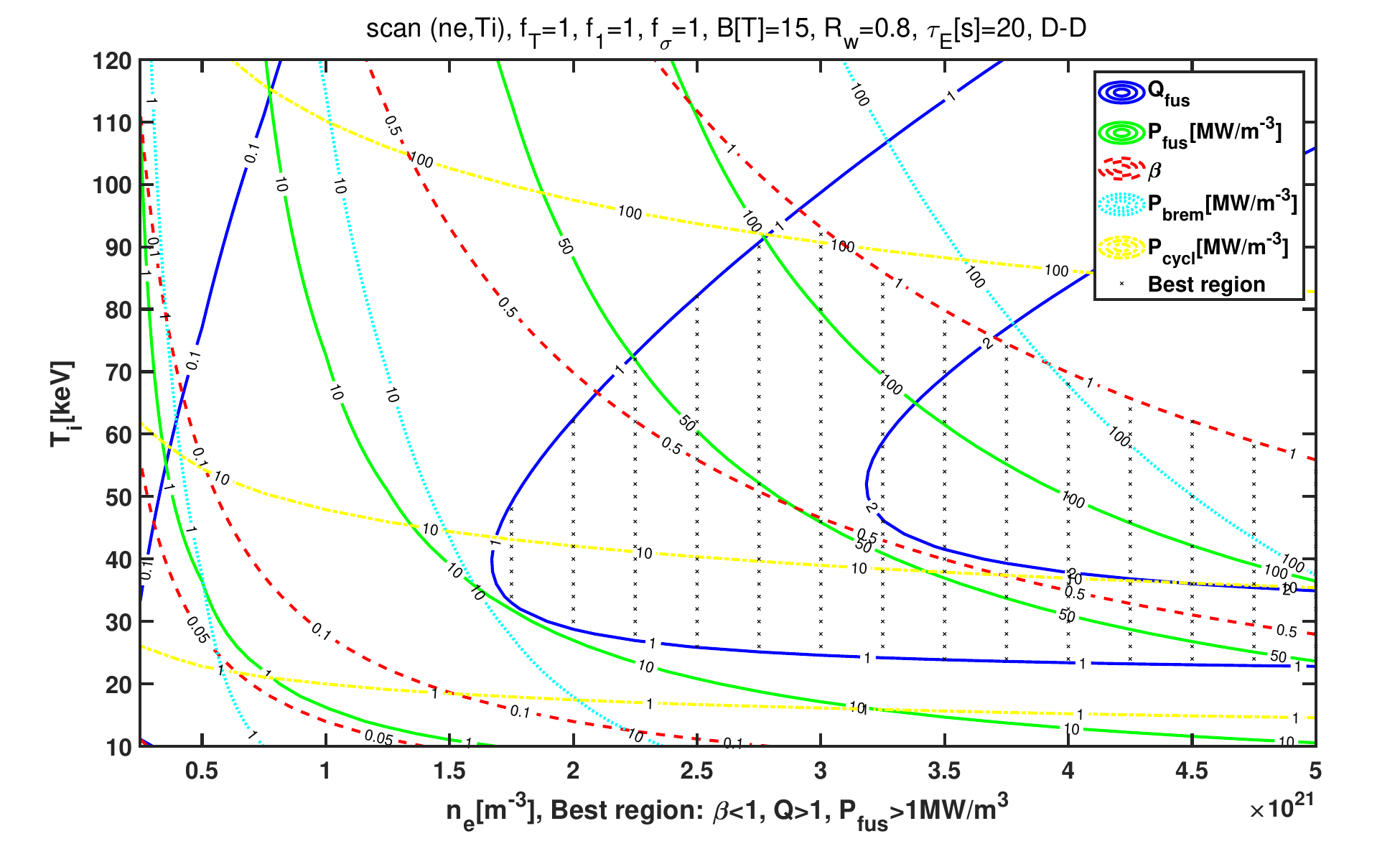}\\
\caption{Typical parameter range for magnetic confinement deuterium-deuterium (D-D) fusion.}\label{fig:mcfscan2ddd1}
\end{center}
\end{figure}

\begin{figure}[htbp]
\begin{center}
\includegraphics[width=13cm]{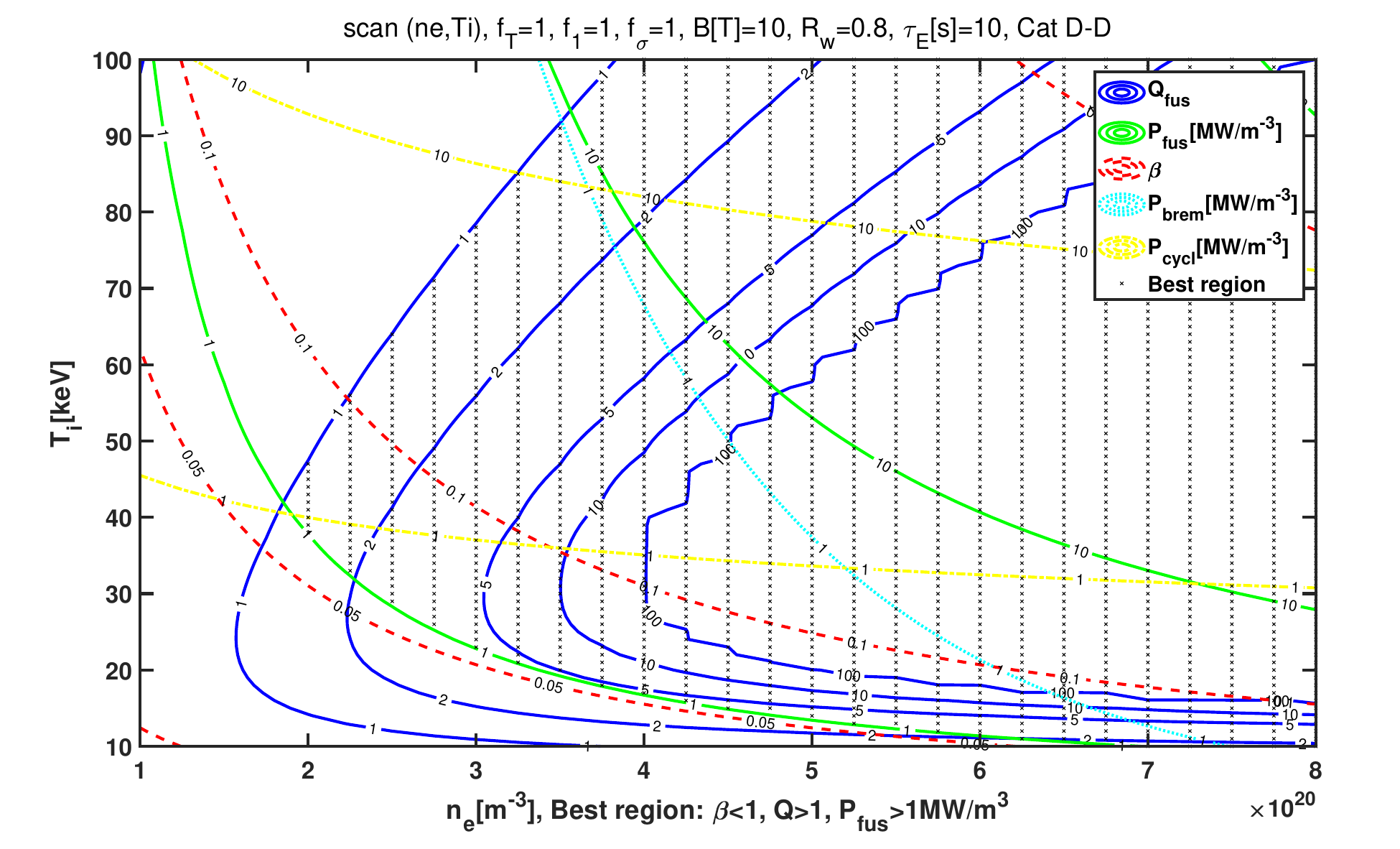}\\
\caption{Typical parameter range for magnetic confinement catalyzed deuterium-deuterium (D-D) fusion.}\label{fig:mcfscan2dcatdd1}
\end{center}
\end{figure}

\begin{figure}[htbp]
\begin{center}
\includegraphics[width=13cm]{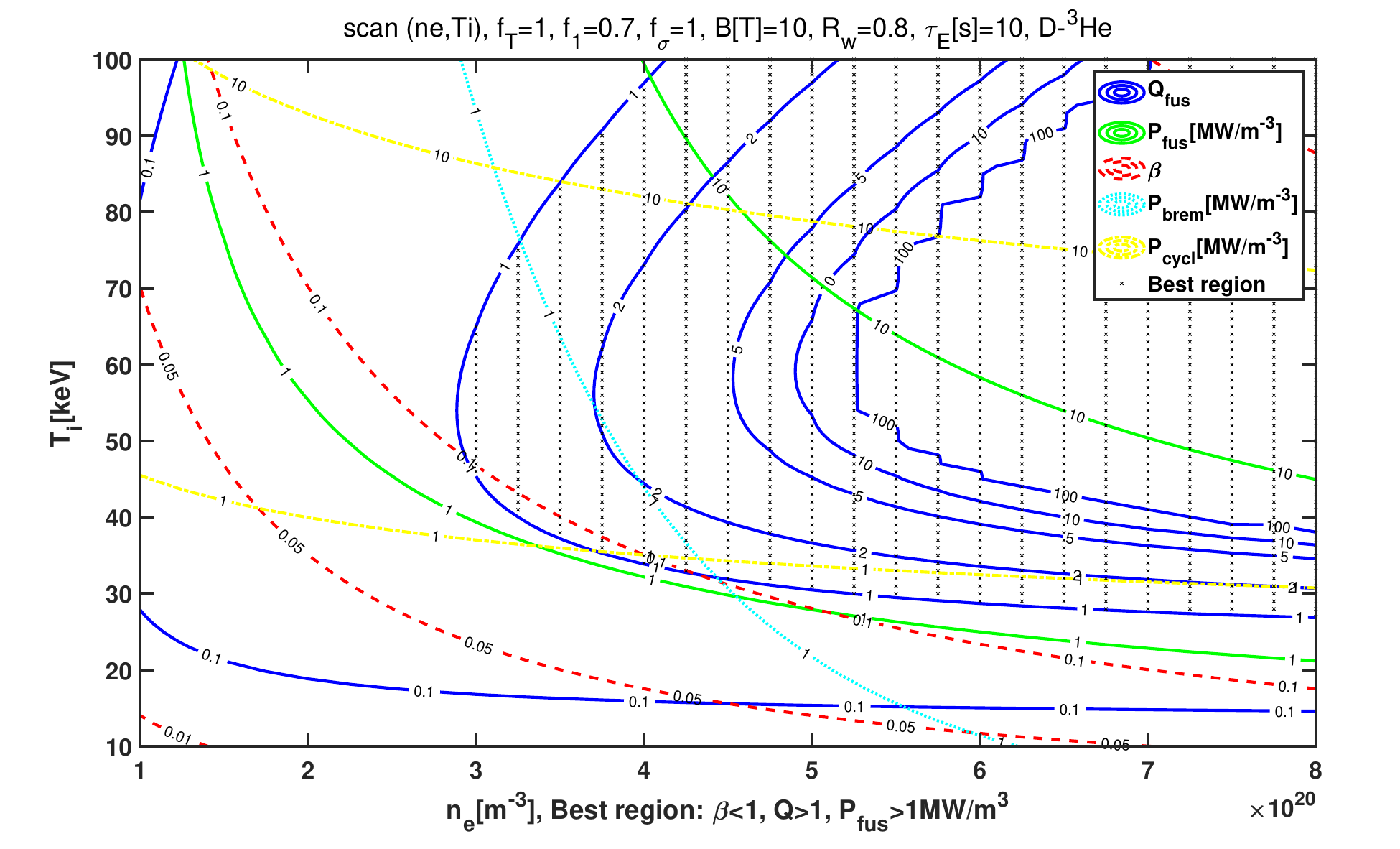}\\
\caption{Typical parameter range for magnetic confinement deuterium-helium (${ \rm D-{}^3He }$) fusion.}\label{fig:mcfscan2ddhe1}
\end{center}
\end{figure}

\begin{figure}[htbp]
\begin{center}
\includegraphics[width=13cm]{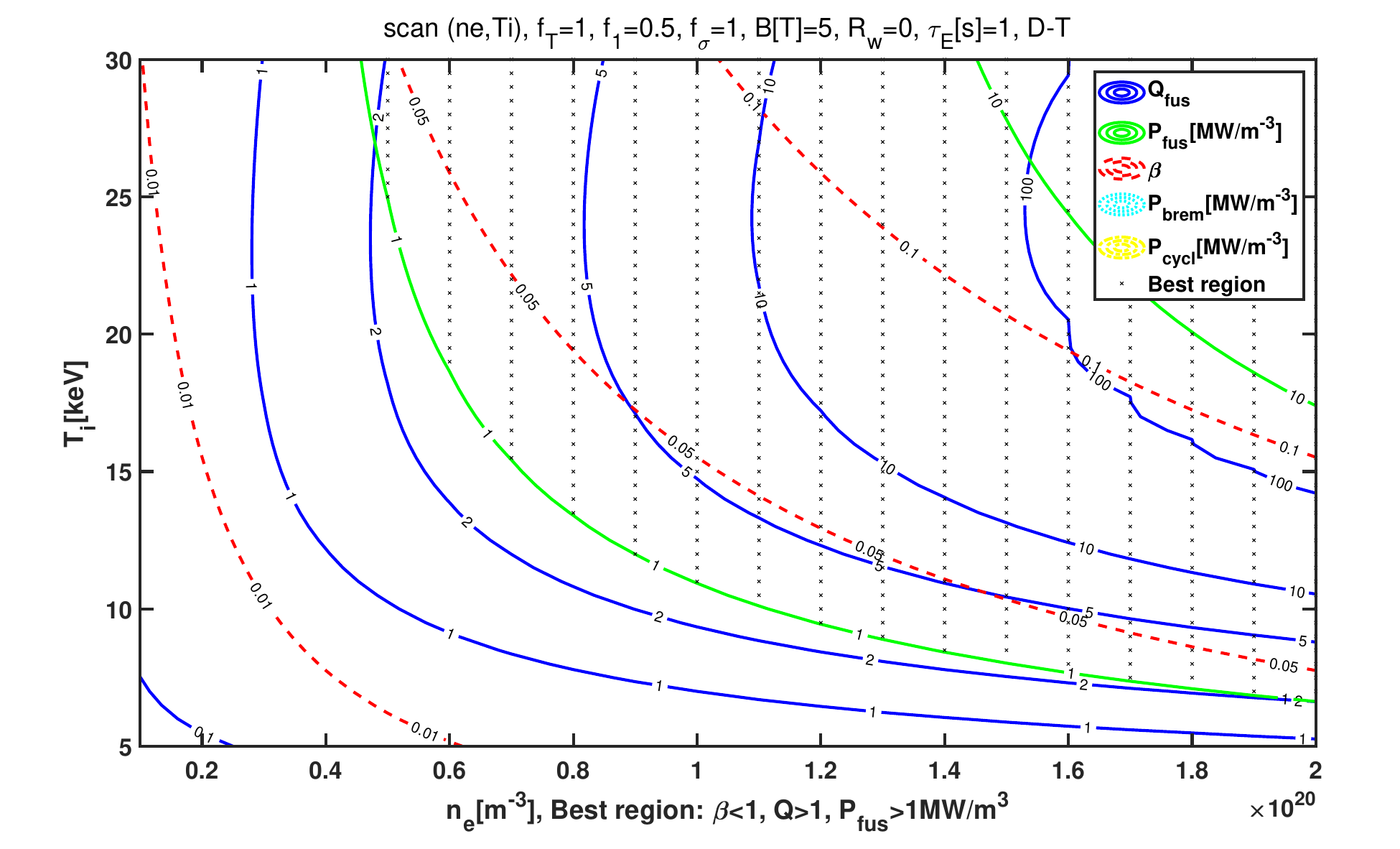}\\
\caption{Typical parameter range for magnetic confinement deuterium-tritium (D-T) fusion.}\label{fig:mcfscan2ddt1}
\end{center}
\end{figure}

\subsection{D-T, D-D, and ${\rm \bf D-{}^{3}He}$ Fusion Parameters}

In Chapter \ref{chap:lawson}, we pointed out the scientific feasibility of D-T, D-D, and ${\rm D-{}^{3}He}$ fusion through the analysis of the Lawson criterion. Here, we can further analyze their parameter ranges using the above model.

Figure \ref{fig:mcfscan2ddd1} shows a typical parameter range for D-D fusion, which is also quite challenging due to the high density required, the higher β requirement, and the problem of excessive fusion power per unit volume at high density. Figure \ref{fig:mcfscan2dcatdd1} shows a typical parameter range for catalyzed D-D fusion, which is relatively easier compared to non-catalyzed D-D fusion. The requirements for density, magnetic field, confinement time, temperature, reflection from cyclotron radiation walls, and other factors are within the reach of existing technology. Figure \ref{fig:mcfscan2ddhe1} shows a typical parameter range for ${\rm D-{}^{3}He}$ fusion, indicating that ${\rm D-{}^{3}He}$ fusion is also within the reach of existing technology. Figure \ref{fig:mcfscan2ddt1} shows a typical parameter range for D-T fusion, demonstrating that the requirements for D-T fusion are the lowest and no additional treatment for cyclotron radiation is needed. From the analysis here, if hydrogen-boron fusion does indeed exceed current technological capabilities, and if the issue of difficult tritium and neutron resolution for D-T fusion cannot be resolved, ${\rm D-{}^{3}He}$ fusion can be prioritized for experimental research. ${\rm D-{}^{3}He}$ can be used to test the scientific and engineering feasibility of non-deuterium-tritium fusion, and if it is only used for experimental reactors, the required 3He can be resolved. Overall, catalytic D-D fusion may be the most promising, with relatively lower required conditions except for the need to resolve a certain proportion of neutrons. Whether catalytic D-D fusion can be fully realized in practice still requires further evaluation, including whether secondary reactions can have sufficient burn rates. We will discuss this further in the next section.

Preliminary conclusions can be reached that catalytic deuterium-deuterium (D-D) fusion is most likely to achieve commercialization of fusion energy through magnetic confinement. This is similar to the conclusion of McNally82. There are various variants of catalytic D-D, such as the addition of a small amount of 3He to D-D, which enables more efficient realization of D-D fusion reactors with a small amount of 3He as catalyst.

The above model can be used for a wider range of parameter scans. Here we only present the main conclusions and do not discuss them in detail.

\section{Analysis of Several Fusion Reactions with Deuterium}

Based on the analysis in the previous section, which was based on the Lawson criterion and the parameter interval graph for magnetic confinement, we can see that three fusion reactions involving deuterium have scientific feasibility as fusion energy sources. Their main reactions and side reactions overlap with each other. Therefore, here we consider several steady-state fusion reactions involving deuterium, so that the aforementioned D-T, D-D, D-3He fusion reactions, and their mixtures, can be calculated under the same model. Here, different particle compositions are distinguished, as well as particle confinement time $\tau_N$ and energy confinement time $\tau_E$, but we still assume that the ion temperatures are the same.

\subsection{Model}
Consider the following four fusion reactions:
\begin{eqnarray*}
{\rm D+T} &\to& {\rm n (14.07MeV) +{}^4He (3.52MeV)},\\
    {\rm D+D} &\to& {\rm n (2.45MeV)+{}^3He (0.82MeV)}  (50\%),\\\nonumber
    {\rm D+D }&\to& {\rm p (3.03MeV) +T (1.01MeV)} (50\%),\\
    {\rm D+{}^3He} &\to& {\rm p (14.68MeV) +{}^4He (3.67MeV)}.\end{eqnarray*}
The corresponding reaction cross sections are denoted as $\langle \sigma v\rangle_1$, $\langle \sigma v\rangle_2$, $\langle \sigma v\rangle_3$, and $\langle \sigma v\rangle_4$; the released energy is denoted as $Y_1=17.59$MeV, $Y_2=3.27$MeV, $Y_3=4.04$MeV, and $Y_4=18.35$MeV, respectively; the energies of the charged products are denoted as $Y_{1+}=3.52$MeV, $Y_{2+}=0.82$MeV, $Y_{3+}=4.04$MeV, and $Y_{4+}=18.35$MeV, respectively; the energies of the neutral neutron products are denoted as $Y_{1n}=14.07$MeV, $Y_{2n}=2.45$MeV, and $Y_{3n}=Y_{4n}=0$; the ion species involved are ${\rm p}$, ${\rm D}$, ${\rm T}$, ${\rm {}^3He}$, and ${\rm {}^4He}$, with respective densities assumed as $n_p$, $n_d$, $n_t$, $n_h$, and $n_\alpha$; the corresponding charges are $Z_p=Z_d=Z_t=1$ and $Z_h=Z_\alpha=2$. Other secondary reactions such as ${\rm T-T}$ and ${\rm {}^3He-{}^3He}$ have much lower reaction rates and can be ignored.

The particle balance equations in steady state are [Nakao (1979), Khvesyuk (2000)]:
\begin{eqnarray}
S_p+\frac{1}{2}n_d^2\langle \sigma v\rangle_3+n_dn_h\langle \sigma v\rangle_4-\frac{n_p}{\tau_N}=0,\\
S_d-n_dn_t\langle \sigma v\rangle_1-2\cdot\frac{1}{2}n_d^2\langle \sigma v\rangle_2-2\cdot\frac{1}{2}n_d^2\langle \sigma v\rangle_3-n_dn_h\langle \sigma v\rangle_4-\frac{n_d}{\tau_N}=0,\\
S_t-n_dn_t\langle \sigma v\rangle_1+\frac{1}{2}n_d^2\langle \sigma v\rangle_3-\frac{n_t}{\tau_N}=0,\\
S_h+\frac{1}{2}n_d^2\langle \sigma v\rangle_2-n_dn_h\langle \sigma v\rangle_4-\frac{n_h}{\tau_N}=0,\\
S_\alpha+n_dn_t\langle \sigma v\rangle_1+n_dn_h\langle \sigma v\rangle_4-\frac{n_\alpha}{\tau_N}=0.
\end{eqnarray}
For simplicity, we assume that the particle confinement time $\tau_N$ and the energy confinement time $\tau_E$ are the same for all components, and $\tau_N=2\tau_E$.The energy balance equation in steady state is
\begin{eqnarray}
P_{heat}+P_{fus,+}=P_{rad}+\frac{E_{th}}{\tau_E},
\end{eqnarray}
where the total fusion power is
\begin{eqnarray}
P_{fus}=n_dn_t\langle \sigma v\rangle_1Y_1+\frac{1}{2}n_d^2\langle \sigma v\rangle_2Y_2+\frac{1}{2}n_d^2\langle \sigma v\rangle_3Y_3+n_dn_h\langle \sigma v\rangle_4Y_4,
\end{eqnarray}
and the power of charged products is
\begin{eqnarray}
P_{fus,+}=n_dn_t\langle \sigma v\rangle_1Y_{1+}+\frac{1}{2}n_d^2\langle \sigma v\rangle_2Y_{2+}+\frac{1}{2}n_d^2\langle \sigma v\rangle_3Y_{3+}+n_dn_h\langle \sigma v\rangle_4Y_{4+},
\end{eqnarray}
and the power of neutral product neutrons is
\begin{eqnarray}
P_n=P_{fus}-P_{fus,+}=n_dn_t\langle \sigma v\rangle_1Y_{1n}+\frac{1}{2}n_d^2\langle \sigma v\rangle_2Y_{2n}.
\end{eqnarray}

Other parameter relationships remain the same as in the previous section. For example, the plasma stored energy is given by
\begin{eqnarray}
E_{th}= \frac{3}{2}k_B(n_iT_i+n_eT_e).
\end{eqnarray}
The radiation term still includes two parts: bremsstrahlung radiation and synchrotron radiation
\begin{eqnarray}
P_{rad}= P_{brem}+P_{cycl}.
\end{eqnarray}
The reaction rates are still assumed to follow the Maxwellian distribution. The ion density, electron density (due to quasi-neutrality), and effective charge are given by
\begin{eqnarray}
n_i=n_p+n_d+n_t+n_h+n_\alpha,\\
n_e=Z_pn_p+Z_dn_d+Z_tn_t+Z_hn_h+Z_\alpha n_\alpha,\\
Z_{eff}=(n_pZ_p^2+n_dZ_d^2+n_tZ_t^2+n_hZ_h^2+n_\alpha Z_\alpha^2)/n_e.
\end{eqnarray}
The fusion gain factor is defined as
\begin{eqnarray}
Q_{fus}\equiv \frac{P_{fus}}{P_{heat}}.
\end{eqnarray}
The external heating power $P_{heat}$ can still be obtained from the aforementioned energy balance equation
\begin{eqnarray}
P_{heat}=P_{rad}+\frac{E_{th}}{\tau_E}-P_{fus,+}.
\end{eqnarray}

\subsection{Results}The above models can be used to analyze ${\rm D-T}$, ${\rm D-{}^3He}$, catalyzed ${\rm D-D-{}^3He}$, catalyzed ${\rm D-D-T}$, catalyzed ${\rm D-D-{}^3He-T}$, and pure ${\rm D-D}$ fusion.

\begin{itemize}
  \item For ${\rm D-T}$ fusion, the particle sources $S_d$ and $S_t$ are taken such that $n_d=n_t$ and $S_h=0$.
  \item For ${\rm D-{}^3He}$ fusion, the particle sources $S_d$ and $S_h$ are taken such that $n_d=n_h$ and $S_t=0$.
  \item For pure ${\rm D-D}$ fusion, there is only the particle source $S_d$.
  \item For catalyzed ${\rm D-D-{}^3He}$ fusion, $S_t=0$ and the lost ${\rm {}^3He}$ is recycled, i.e., $S_h=n_h/\tau_N$.
  \item For catalyzed ${\rm D-D-T}$ fusion, $S_h=0$ and the lost ${\rm T}$ is recycled, i.e., $S_t=n_t/\tau_N$.
  \item For catalyzed ${\rm D-D-{}^3He-T}$ fusion, both the lost ${\rm {}^3He}$ and ${\rm T}$ are recycled, i.e., $S_h=n_h/\tau_N$ and $S_t=n_t/\tau_N$.
\end{itemize}

In all the above cycles, we assume that the source terms for protons and $\alpha$ particles are $S_p=S_\alpha=0$.

\begin{figure}[htbp]
\begin{center}
\includegraphics[width=15cm]{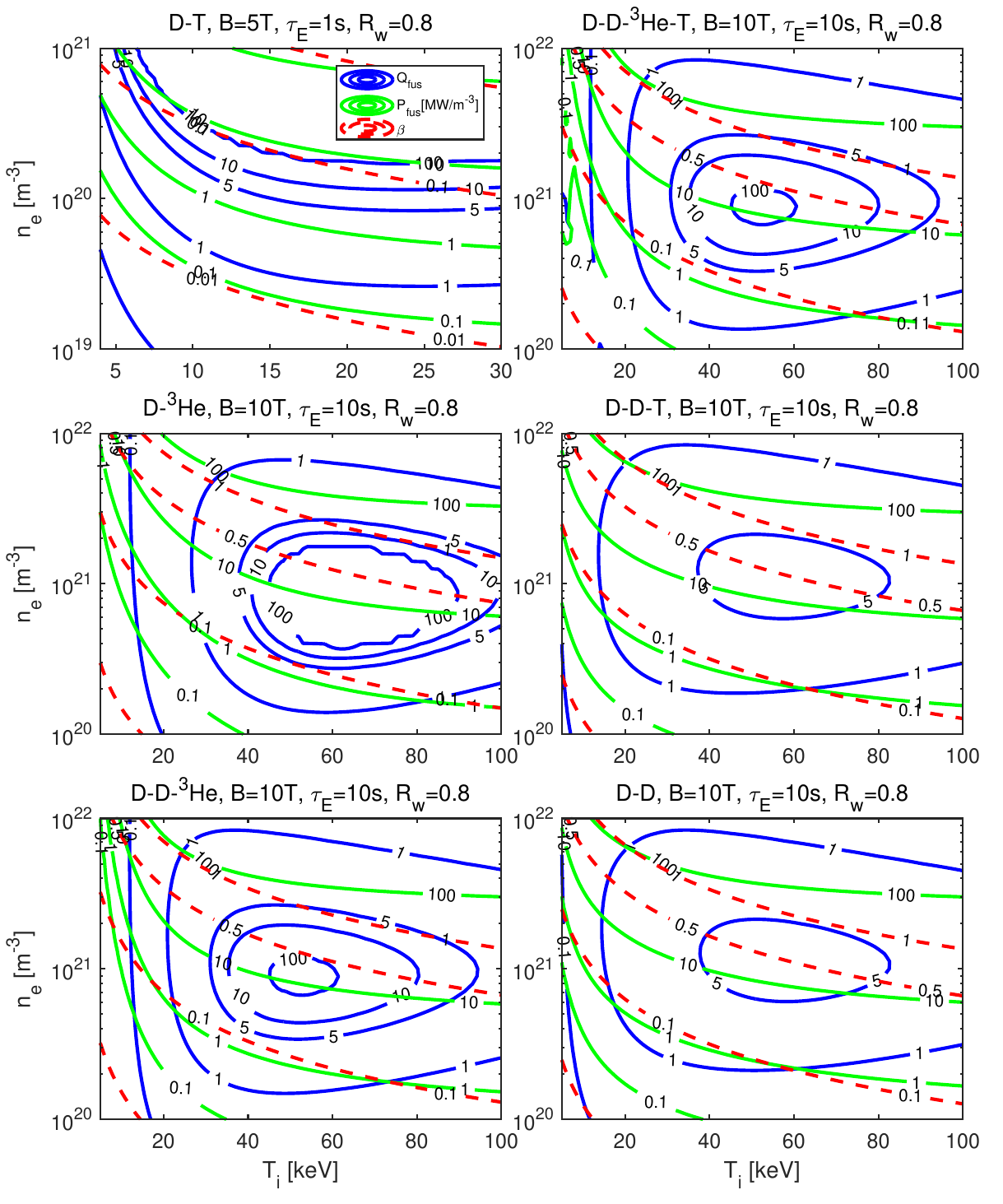}
\caption{Parameter ranges for different minor reactions in magnetic confinement Deuterium (${\rm D}$) fusion.}\label{fig:dfuelscan2d}
\end{center}
\end{figure}

The independent variables of the above models are: electron density $n_e$, energy confinement time $\tau_E$, ion temperature $T_i$, and the magnetic field $B$ and wall reflection coefficient $R_w$ related to the synchrotron radiation. There are a total of 5 variables. Other quantities are output variables calculated according to the equations, such as fusion power $P_{fus}$, plasma pressure $\beta$, and fusion gain $Q_{fus}$.

Figure \ref{fig:dfuelscan2d} shows the typical calculation results for the above six cases. It can be seen that ${\rm D-T}$ fusion is the easiest, followed by ${\rm D-{}^3He}$, then ${\rm D-D-{}^3He-T}$ and ${\rm D-D-{}^3He}$, followed by ${\rm D-D-T}$ and pure ${\rm D-D}$. In terms of reaction ease and availability of raw materials, the optimal reactions are ${\rm D-D-{}^3He-T}$ and ${\rm D-D-{}^3He}$. However, the gain space for pure ${\rm D-D}$ fusion is limited. Compared to the Lawson diagram mentioned earlier, we can also see here that a higher density does not necessarily lead to a higher fusion gain $Q_{fus}$, but there is an optimal range. This is closely related to the accumulation of products and dilution of fusion fuel. That is, at the same confinement time $\tau_E$, a higher density $n_e$ leads to a higher fusion burn rate and more product accumulation. Similarly, with a fixed density $n_e$, increasing $\tau_E$ will also result in product accumulation and a decrease in fusion gain. That is, a larger $n_e\tau_E$ is not necessarily better. This is why helium ash removal technology is needed in magnetic confinement fusion.To make these various fusion schemes comparable, we can also set $\beta$, $P_{fus}$, and $Q_{fus}$ as given values to inversely calculate other quantities and compare the parameter requirements of these schemes. This part can be left for readers to try.

\section{Parameters of Hydrogen-Boron Fusion with Heat Exchange and Burn Rate}

Similar to the discussion on deuterium fusion in the previous section, here we focus on hydrogen-boron fusion, distinguishing different particle components and considering energy conversion between them. We also differentiate particle confinement time and energy confinement time.

The model includes three types of ions: protons, boron, and helium ash, with densities denoted as $n_p$, $n_b$, and $n_\alpha$, respectively. It is assumed that the ions follow the Maxwellian distribution and have the same temperature $T_i$. The balance equations for the particle numbers of ions in the steady state are:
\begin{eqnarray}
\frac{dn_p}{dt}&=&S_p-n_pn_b\langle\sigma v\rangle-\frac{n_p}{\tau_{Np}}=0,\\
\frac{dn_b}{dt}&=&S_b-n_pn_b\langle\sigma v\rangle-\frac{n_b}{\tau_{Nb}}=0,\\
\frac{dn_\alpha}{dt}&=&3n_pn_b\langle\sigma v\rangle-\frac{n_\alpha}{\tau_{N\alpha}}=0,
\end{eqnarray}
where the source terms $S_p$ and $S_b$ are used to maintain the required proton and boron densities $n_p$ and $n_b$. The electron density is determined by quasi-neutrality condition, $n_e=Z_pn_p+Z_bn_b+Z_\alpha n_\alpha$. Assuming $n_p$ and $n_b$ are input quantities, we can omit the need to solve the first two equations. The third equation for helium ash density becomes:
\begin{eqnarray}
n_\alpha=3n_pn_b\langle\sigma v\rangle \tau_{N\alpha}=3n\tau n_b\langle\sigma v\rangle h,
\end{eqnarray}
where $h=\tau_{N\alpha}/\tau_E$, and $n\tau=n_p\tau_E$. It is assumed that $f_{\alpha}=1$ by default. If there is no effective helium ash removal technique, then $h\simeq1$ because $\tau_{N\alpha}\simeq\tau_{Np}\simeq\tau_E$, which would result in a large accumulation of helium ash. For example, if we take $h=1$, $n_b/n_p=0.15$, $\langle\sigma v\rangle=4.4\times10^{-22}{\rm m^3/s}$ ($T_i=300keV$), and $n\tau=10^{22}{\rm m^{-3}\cdot s}$, we obtain $n_{\alpha}/n_p\simeq1.8$. This is essentially limited by the fusion average reaction time discussed in Chapter \ref{chap:lawson}. If we choose a smaller $n\tau$, the accumulation of helium ash would be reduced, but this would also result in failure to meet the Lawson criterion for gain.The energy equation per unit volume in steady state is given by
\begin{eqnarray}
\frac{dE_e}{dt}&=&P_{xe}+P_{\alpha e}+P_{ie}-P_r-\frac{E_e}{\tau_{Ee}}=0,\\
\frac{dE_i}{dt}&=&P_{xi}+P_{\alpha i}-P_{ie}-\frac{E_i}{\tau_{Ei}}=0,
\end{eqnarray}
where $P_{xi}$ and $P_{xe}$ are the heating powers for ions and electrons from external sources, $P_{\alpha e}$ and $P_{\alpha i}$ are the heating powers from fusion $\alpha$ particles, $P_{ie}$ is the power exchanged between electrons and ions, and $P_r$ is the radiative term. $\tau_{Ee}$ and $\tau_{Ei}$ represent the energy confinement time for electrons and ions. The energy equations can be summed up into one equation
\begin{eqnarray}
P_{x}+P_{\alpha }=P_{r}+\frac{E}{\tau_{E}},
\end{eqnarray}
where $E=E_{e}+E_i$ is the total thermal energy, and $E/\tau_E=E_e/\tau_{Ee}+E_i/\tau_{Ei}$.

\begin{figure}[htbp]
\begin{center}
\includegraphics[width=15cm]{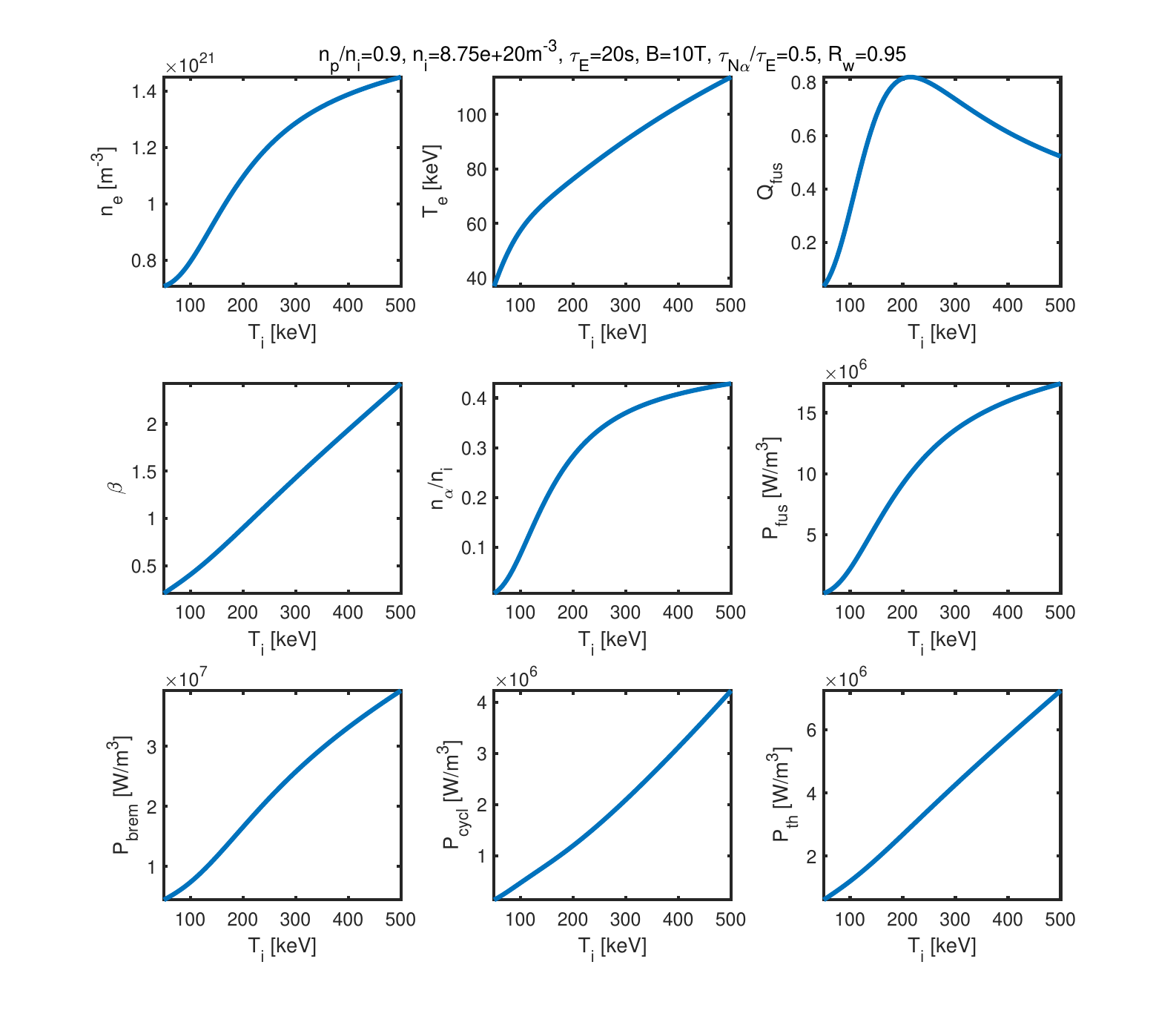}\\
\caption{Typical example of magnetically confined hydrogen-boron fusion, where fusion gain region is difficult to exist when helium ash cannot be effectively removed.}\label{fig:mcfpb1}
\end{center}
\end{figure}

We adopt an optimistic estimation and for simplification, we assume $P_{xe}=P_{\alpha e}=0$, meaning that external sources and fusion $\alpha$ particles only heat the ions and not the electrons. This allows the electron temperature to be as low as possible, thus reducing radiation loss. This assumption slightly underestimates the electron temperature, but it is reasonable because based on the calculations in Chapter \ref{chap:lawson}, hydrogen-boron fusion products, $\alpha$ particles mainly heat the ions, and in the future, neutral beams and ion cyclotron resonance heating may be predominantly used for ion heating in hydrogen-boron reactors. The heat exchange power is given by
\begin{eqnarray}
P_{ie}=\frac{3}{2}k_B\Big(\frac{n_p}{\tau_{p,e}}+\frac{n_b}{\tau_{b,e}}+\frac{n_\alpha}{\tau_{\alpha,e}}\Big)(T_i-T_e),
\end{eqnarray}
where $\tau_{p,e}$, $\tau_{b,e}$, and $\tau_{\alpha,e}$ are the exchange times for protons, boron ions, and $\alpha$ particles with electrons and can be calculated by formula (\ref{eq:tauij}).

Fusion power is given by
\begin{eqnarray}
P_{fus}=n_pn_b\langle\sigma v\rangle Y,
\end{eqnarray}
where $Y=8.68$MeV. Fusion gain factor is given by
\begin{eqnarray}
Q_{fus}=\frac{P_{out}-P_{in}}{P_{in}}=\frac{P_{fus}}{P_{x}},
\end{eqnarray}
where $P_{out}=P_r+E/\tau_E$, $P_{in}=P_{x}$, and $P_\alpha=P_{fus}$.The input parameters in the model are $n_p, n_b, h, \tau_{E}, \tau_{Ei}/\tau_{Ee}, T_i$. The electron temperature $T_e$ is obtained through the thermal exchange equation. At the same time, $P_r, P_\alpha, P_{fus}, P_{x}$, and $Q_{fus}$ can be calculated. The difference between this model and the one in Section \ref{sec:mcfmodel1} is that here we consider helium ash $n_\alpha$ and the electron temperature $T_e$ is obtained through thermal exchange instead of being arbitrarily given. This model is similar to the one discussed in Dawson (1981).

Figure \ref{fig:mcfpb1} shows a set of typical examples, where the results indicate that the hot ion mode is achievable, but the beta ratio and helium ash issues remain limiting factors.

\section{Energy Confinement Time Limit}
The classical collision confinement determines the energy confinement time limit of magnetic confinement fusion. Although turbulent transport and neo-classical transport can reduce the actual confined energy time magnitude, the latter two can be optimized and improved, while the classical collision confinement cannot be further improved. The transport coefficient is given by
\begin{eqnarray}
D=\frac{\Delta x^2}{\Delta t}\propto \frac{n}{T^{1/2}B^2}, ~~\Delta x=\rho_{cs},~~\Delta t=\tau_c.
\end{eqnarray}
where $\rho_{cs}$ is the gyroradius and $\tau_c$ is the collision time. The energy confinement time is
\begin{eqnarray}
\tau_E\sim \frac{a^2}{D}\sim \frac{a^2B^2T^{1/2}}{n}.
\end{eqnarray}
Thus, the larger the magnetic field $B$, the better the confinement; and the larger the device size $a_{plasma}$, the better the confinement. However, both parameters will lead to increased device costs. Here, we see that $n\tau_E$ appears as a whole, which means that when the density is high, $\tau_E$ decreases, but it has no significant effect on $n\tau_E$. The higher the temperature, the better the confinement, but the cost of high temperature also increases the cost, and the increase in plasma pressure makes it more difficult to achieve controllability.

Let's do some specific calculations. The formulas for classical and neo-classical transport coefficients can be expressed as
\begin{eqnarray}
\xi_{i,c}\sim D_{i,c}=\frac{\rho_i^2}{\tau_{ei}},\\
\xi_{i,neo}\sim D_{i,neo}=\frac{\epsilon^{-3/2}q^2\rho_i^2}{\tau_{ei}}=\epsilon^{-3/2}q^2\cdot D_{i,c},
\end{eqnarray}
where $\epsilon=r/R$ is the inverse aspect ratio and $q$ is the safety factor. The collision frequency $\tau_{ei}$ is used instead of $\tau_i$ because the plasma is maintained quasi-neutral, which is reflected in the bipolar diffusion [Freidberg (2007) chap14]. Taking typical values of $\epsilon=0.5$ and $q=2$, we have $D_{i,neo}/D_{i,c}=11.3$, which means that the neo-classical transport coefficient is one order of magnitude larger than the classical transport coefficient. Here, the neo-classical transport mainly refers to tokamaks, where the characteristic scale of collisions is the banana orbit width, rather than the gyroradius $\rho_i$, and the difference in size between the banana orbit width and the gyroradius is reflected in the correction coefficient of the neo-classical to classical transport. The specific formula for the gyroradius calculation is given in Appendix \ref{chap:appendix}.\begin{table}[htp]
\caption{The gyroradius and confinement times under typical fusion parameters, with $\epsilon=0.5$, $q=2$, $a=1$ m, and $B=1$ T.}
\begin{center}
\begin{tabular}{c|c|c|c}
\hline\hline
 & \textbf{Parameter 1} & \textbf{Parameter 2} & \textbf{Parameter 3} \\\hline
Temperature (keV) & 1 & 20 & 200 \\
Density (${\rm m}^{-3}$) & $1 \times 10^{19}$ & $1 \times 10^{20}$ & $1 \times 10^{20}$ \\
Proton gyroradius (m) & $0.00457$ & $0.0204$ & $0.0646$ \\
Electron gyroradius (m) & $1.07 \times 10^{-4}$ & $4.77 \times 10^{-4}$ & $1.51 \times 10^{-3}$ \\
Electron collision time (s) & $6.4 \times 10^{-5}$ & $5.72 \times 10^{-4}$ & $1.81 \times 10^{-2}$ \\
Proton collision time (s) & $3.53 \times 10^{-3}$ & $3.15 \times 10^{-2}$ & $0.997$ \\
\thead {\normalsize Classical \\ \normalsize transport coefficient (${\rm m^2/s}$)} & $3.55 \times 10^{-4}$ & $7.95 \times 10^{-4}$ & $2.51 \times 10^{-4}$ \\
\thead {\normalsize Neoclassical \\ \normalsize transport coefficient (${\rm m^2/s}$)} & $4.02 \times 10^{-3}$ & $8.99 \times 10^{-3}$ & $2.84 \times 10^{-3}$ \\
\thead {\normalsize Classical \\ \normalsize confinement time (s)} & $489$ & $218$ & $691$ \\
\thead {\normalsize Neoclassical \\ \normalsize confinement time (s)} & $43.2$ & $19.3$ & $61.1$ \\
\hline\hline
\end{tabular}
\end{center}
\label{tab:tauE}
\end{table}
Due to the transport diffusion equation in cylindrical coordinates:
\begin{eqnarray}
\frac{\partial n}{\partial t} = D\nabla^2n = D\frac{1}{r}\frac{\partial}{\partial r}\Big(r\frac{\partial n}{\partial r}\Big),~n(r=a)=0,
\end{eqnarray}
the zero-order solution is given by:
\begin{eqnarray}
n(r,t) = n_0J_0\Big(\frac{r}{\sqrt{D\tau_D}}\Big)e^{-t/\tau_D},
\end{eqnarray}
where $J_0(x_0\simeq2.4)=0$ represents the first zero point of the Bessel function. Thus, the characteristic time of transport diffusion is:
\begin{eqnarray}
\tau_D = \frac{a^2}{Dx_0^2}\simeq0.1736\frac{a^2}{D},
\end{eqnarray}
which represents the time confinement under a cylindrical configuration with transport coefficient $D$ and radius $a$. For simplicity, we also estimate the neoclassical confinement time using this formula.

Based on the above formulas, several sets of typical results are calculated and shown in Table \ref{tab:tauE}. It can be observed that the classical confinement time is on the order of 100 seconds, which meets the requirements for energy confinement time. However, the neoclassical transport's confinement time will pose significant challenges for fusion reactions with higher constraints, such as hydrogen-boron fusion. Nevertheless, we also note that in the collision transport model mentioned above, $D$ is proportional to the square of the gyro radius $\rho_i$, and $\rho_i$ is inversely proportional to the magnetic field $B$. Therefore, $\tau_D\propto B^2$, meaning that a magnetic field that is 10 times larger results in a confinement time 100 times longer. In practice, it is usually not collision transport but rather an anomalous transport, such as Bohm transport where $D\propto T/B$. On the other hand, $\tau_D\propto a^2$, meaning that if the device is doubled in size, the confinement time becomes four times longer. This is also the reason why larger devices and stronger magnetic fields are more likely to achieve fusion constraints.

These estimates provide quantitative references and directions for improving constraints and achieving the desired energy confinement time.

\section{Summary of this Chapter}

Based on the Lawson criterion discussed in the previous chapter, this chapter further analyzes the parameter space of magnetic constrained fusion, taking into account both the effects of cyclotron radiation and the beta limit. Using the first model, we point out the stringent conditions for hydrogen-boron fusion and highlight the feasibility of deuterium-tritium, deuterium-deuterium, and deuterium-helium three fusion methods in theory. Furthermore, we establish particle balance equations for the six fusion methods involving deuterium and analyze their parameter space. We emphasize the complexity of fuel scarcity (difficulty in breeding), gain condition feasibility, and neutron shielding difficulty, and propose that D-D-He3 fusion is the most likely candidate for commercial fusion. Finally, we analyze a more detailed model for hydrogen-boron fusion, which includes helium ash and energy exchange, and further identify the challenges and key breakthrough conditions for hydrogen-boron fusion.

On the whole, for magnetic confinement fusion, a strong magnetic field can improve the requirement for the beta, thus compactly achieving high parameters. However, for advanced fuels, a strong magnetic field will result in excessive cyclotron radiation, requiring efficient cyclotron radiation recovery technology. Due to the magnitude of the difficulties in non-deuterium-tritium fusion compared to deuterium-tritium fusion, many fusion scientists assume that fusion energy is obtained through deuterium-tritium fusion. This is particularly evident in Wesson's (2011) classic tokamak book, which discusses fusion-related issues solely in the context of deuterium-tritium fusion. However, economic requirements prohibit the use of low density, and if extremely low density is used, longer confinement time requirements are needed. It is possible to significantly reduce cyclotron radiation by optimizing the magnetic field configuration, such that the central magnetic field is low and the boundary magnetic field is high. This is because the limit of magnetic confinement is based on average $\beta \leq 1$, not local $\beta$. In Appendix \ref{chap:appendix}, we discuss the equilibrium of field reversal configuration. If the fusion reaction rate can be greatly increased, it is possible to reduce the conditions for magnetic confinement fusion. However, calculations based on the complex Fokker-Planck collision model indicate that the majority of proposed non-thermalized distributions cannot sustain fusion gain. Rider (1995) discusses this in detail, pointing out that the energy required to sustain a non-thermalized distribution exceeds the energy output of fusion. It is worth exploring whether viable non-thermalized distributions can be constructed in practice.

This chapter mainly discusses the case of a uniform distribution. Appendix \ref{chap:tokamak} establishes a model that is more closely related to specific experimental devices with respect to the effects of non-uniform temperature and density. There, the effects of non-uniformity on fusion power and radiation have certain differences, resulting in some changes in the relative values and some optimization possibilities compared to the uniform situation.

\vspace{30pt}
Key points of this chapter:
\begin{itemize}
\item For magnetic confinement fusion, D-T, D-D, and D-He3 are all feasible.
\item The conditions for p-11B are extremely strict and difficulty is extremely high. Significantly advancing technology is required across multiple aspects.
\item Taking into account fuel scarcity (difficulty of breeding), ease of achieving gain conditions, and neutron shielding difficulty, the most likely fusion approach to achieve commercialization is D-D-He3.
\item If low neutron content is required, only p-11B can be chosen. At that time, breakthroughs are needed in multiple aspects, such as hot ion mode, reaction rate improvement, radiation recovery, and plasma confinement.
\end{itemize} 
\chapter{Parameter Range for Inertial Confinement Fusion}\label{chap:icf}

Inertial confinement refers to the process of achieving confinement by relying on the inertia of mass without any additional constraints, allowing the fusion reaction to occur within a short period of time, releasing energy. Due to the short confinement time, a high density is usually required to meet the fusion gain conditions. In this chapter, we establish the simplest model to explore the minimum conditions for inertial confinement. From a physical model perspective, inertial confinement usually falls into the high energy density range, where quantum and relativistic effects are more significant. The radiation, state equation, energy deposition, and reaction rate in this range may be more complex compared to the classical physical range of magnetic confinement. However, we do not consider these complex mechanisms here, rather we incorporate these potential factors into a few key factors.

Compared to magnetic confinement, the key difference of inertial confinement at the zero order is the absence of $\beta$ constraint, which allows for lower magnetic fields and the neglect of synchrotron radiation. Although magnetic fields may be generated in practice, synchrotron radiation may not necessarily be negligible. Due to the high density, the bremsstrahlung radiation has a certain optical depth, which may result in some radiation being absorbed. However, we do not consider radiation effects here.

Although there were earlier proposals for inertial confinement, Nuckolls (1972) is generally considered the originator of laser inertial confinement. The physics of hydrogen bombs also falls under inertial confinement, which is more complex than the discussion in this chapter and will be addressed in subsequent chapters. For a detailed and comprehensive discussion of inertial confinement, refer to works such as Atzeni (2004), Wang Ganchang (2005), Pfalzner (2006), etc.

\section{Criterion of the Simplest Model}

\begin{figure}[htbp]
\begin{center}
\includegraphics[width=11cm]{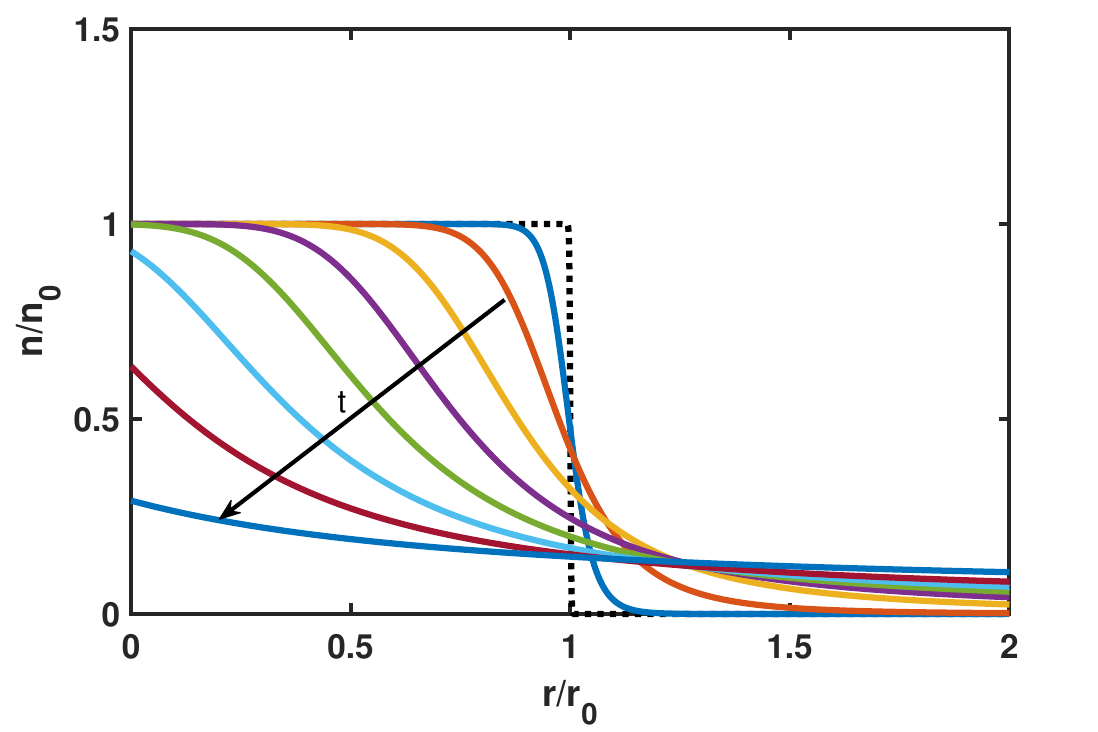}\\
\caption{A schematic representation of the density profile $n(r)$ evolving with time $t$ in inertial confinement fusion, with temperature and pressure profiles behaving similarly.}\label{fig:icfrhoRt}
\end{center}
\end{figure}

In the inertial confinement scheme, the main constraint is on the mass of the fuel, and the confinement time is approximately the time it takes for the plasma with ion sound velocity to propagate from the boundary to the center, which is usually very short. Because of the extremely short confinement time, a very high density is required to achieve fusion gain conditions. The ion sound velocity is given by\footnote{Some literature takes $c_s=\sqrt{{2k_BT_e}/{m_i}}$ without affecting the order of magnitude estimation.}
\begin{eqnarray}
c_s=\sqrt{\frac{k_BT_e}{m_i}}=3.10\times10^5\sqrt{\frac{T_e}{A_i}}~{\rm [m/s]},
\end{eqnarray}
where $m_i$ is the average ion mass, and in the later equation, the temperature $T_e$ is given in keV. Strictly speaking, the average method for $m_i$ is relatively complex. We adopt a simplified approach, taking $m_i=\frac{\sum_j n_jA_j}{\sum_j n_j}m_p=A_im_p$, where $m_p$ is the mass of a proton, $n_j$ is the density of each ion component, and $A_j$ is the mass number of each ion component. For an equal proportion of deuterium and tritium, $m_i=2.5m_p$.Figure \ref{fig:icfrhoRt} illustrates the initial plasma density distribution $n(r)$ of compressed target at time $t=0$, with a uniform distribution and a radius of $R$. Then, a rarefaction wave propagates inward, causing the gradual disintegration of the target. The instantaneous radius of the target front is given by
\begin{eqnarray}
r(t)=R-c_st,
\end{eqnarray}
where $\tau_c$ represents the time of mass constraint, given by
\begin{eqnarray}
\tau_{c}=\frac{R}{c_s}.
\end{eqnarray}
For a deuterium-tritium target with $R=1{\rm mm}$ and $T=10{\rm keV}$, we have $c_s=6.19\times10^5{\rm m/s}$ and $t_c=1.6\times10^{-9}{\rm s}$, indicating that the typical time for inertial confinement fusion is on the order of nanoseconds (${\rm 1ns=10^{-9}s}$). The relationship between number density $n_i$ and mass density $\rho$ is given by $n_i=\rho/m_i$, so we have
\begin{eqnarray}
n_i\tau_{c}=\frac{1}{m_ic_s}\rho R.
\end{eqnarray}

Energy gain $G$ is defined as
\begin{eqnarray}
G\equiv\frac{P_{fus}\tau_{c}}{E_{th}},
\end{eqnarray}
where the volumetric energy $E_{th}$ and fusion power $P_{fus}$ are given by
\begin{eqnarray}
    E_{th}=\frac{3}{2}k_B\sum_jn_jT_j,\\
    P_{fus}=\frac{1}{1+\delta_{12}}n_1n_2\langle\sigma v\rangle Y.
\end{eqnarray}
Here, $Y$ represents the energy released from a single nuclear reaction, $n_1$ and $n_2$ are the number densities of the two ions, and $T_j$ is the temperature of each component (including electrons and ions). If the two ions are different, $\delta_{12}=0$; if they are the same, $\delta_{12}=1$.

Assuming that all components have the same temperature, and letting $x_1=n_1/n_i$ and $x_2=n_2/n_i$, we obtain
\begin{eqnarray}
n_i\tau_{c}=G\frac{\frac{3}{2}(x_1+x_2+Z_i)k_BT}{\frac{1}{1+\delta_{12}}x_1x_2\langle\sigma v\rangle Y},
\end{eqnarray}
which can be rewritten as
\begin{eqnarray}
\rho R=G\frac{\frac{3}{2}(x_1+x_2+Z_i)}{\frac{1}{1+\delta_{12}}x_1x_2\langle\sigma v\rangle Y}m_i^{1/2}(k_BT)^{3/2}.
\end{eqnarray}\begin{figure}[htbp]
\begin{center}
\includegraphics[width=15cm]{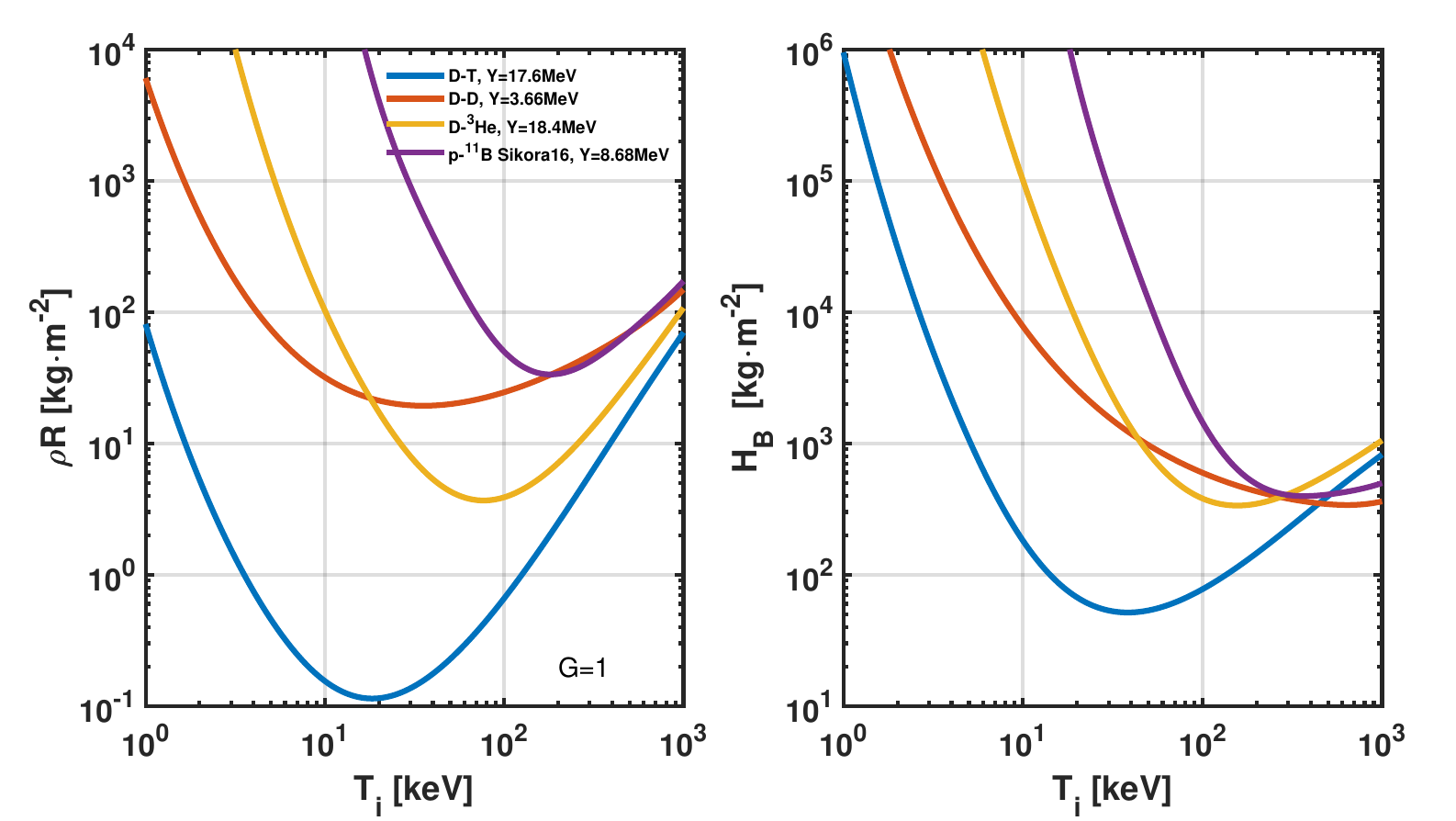}\\
\caption{Requirements for the $\rho R$ parameter in inertial confinement fusion and the $H_B$ parameter, assuming equal proportions of ions.}\label{fig:rhoR_icf} 
\end{center}
\end{figure}

Figure \ref{fig:rhoR_icf} (left) shows the variation of different fusion reactions' $\rho R$ with temperature when $G=1$. It can be seen that for a target pellet with post-compression radius $R=0.1{\rm mm}$, a D-T density of at least $10^3{\rm kg/m^3}$, i.e. solid density, is required. Considering the efficiency of the inertial confinement driver, in order to have positive energy gain, $G$ needs to be at least $100-1000$, therefore the typical density for inertial confinement fusion is $10^5-10^6{\rm kg/m^3}$. If considering hydrogen-boron fusion, the $\rho R$ value needs to be about 200 times higher.

For inertial confinement fusion, the ratio of $\tau_c$ to the characteristic fusion time $\tau_{fus}$ is significant, where
\begin{eqnarray}
\tau_{fus}=\frac{1}{\langle\sigma v\rangle n}.
\end{eqnarray}
Thus,
\begin{eqnarray}
\frac{\tau_c}{\tau_{fus}}=\langle\sigma v\rangle n\frac{R}{c_s}=\langle\sigma v\rangle \frac{\rho R}{m_ic_s}.
\end{eqnarray}
It characterizes the combustion proportion of the fuel.

\section{Model considering burn rate}

For magnetic confinement fusion, fusion power density is a meaningful measure. For inertial confinement fusion, "burn rate" is relatively more important and is defined as
\begin{eqnarray}
f_b=\frac{N_{fus}}{N_0}=\frac{N_0-N_{t}}{N_0},
\end{eqnarray}
where $N_0$ and $N_{t}$ are the remaining particle numbers at the initial time and at time $t$, respectively. After central ignition of the target pellet, the fusion burn wave quickly propagates through the remaining fuel and ignites the external plasma, where not all plasma can undergo fusion reactions. The disassembly time of the target pellet is
\begin{eqnarray}
\tau_s=\frac{R}{c_s},
\end{eqnarray}
where $R$ is the final compressed target pellet radius. As shown in Figure \ref{fig:icfrhoRt}, we assume that only the region where density and temperature are close to the initial values can undergo fusion burn reaction. Before the sparse wave stops the burn, the average time for fuel to react is 
\begin{eqnarray}
\tau_b=\frac{\int_0^{\tau_s}dt m(t)}{m_0}=\frac{\int_0^{\tau_s}(4\pi\rho/3) (R-c_st)^3dt}{(4\pi\rho/3)R^3}=\frac{R}{4c_s},
\end{eqnarray}
where $m_0$ is the initial fuel mass and $m(t)$ is the remaining compressed mass. Here, we can simplify the density ratio by setting $x_1=x_2=\frac{1+\delta_{12}}{2}$ and using $\rho=nm_i$ to approximate,
\begin{eqnarray}
\rho R=\frac{8m_ic_s}{(1+\delta_{12})\langle\sigma v\rangle}\frac{f_b}{1-f_b},
\end{eqnarray}
or 
\begin{eqnarray}
f_b=\frac{\rho R}{H_B+\rho R},~~H_B=\frac{8m_ic_s}{(1+\delta_{12})\langle\sigma v\rangle}.
\end{eqnarray}
The above equations are accurate when $\rho R\ll1$ and $\rho R\simeq 1$. Here, $m_i=A_im_p, A_i=\frac{A_1+A_2}{2}$ for D-T fusion. For 15-100keV,
\begin{eqnarray}
H_B\simeq6-10 ~~[{\rm g/cm^2}].
\end{eqnarray}
Figure \ref{fig:rhoR_icf} on the right shows the variation in fusion reaction $H_B$ with temperature.

Now let's look at the requirements for advanced fuels. From Figure \ref{fig:rhoR_icf} on the right, it can be seen that to achieve the same burn rate, the $\rho R$ for advanced fuels only needs to be one order of magnitude higher than that for deuterium-tritium fusion. However, as shown in Figure \ref{fig:rhoR_icf} on the left, to achieve the same energy gain $G$, the $\rho R$ required for advanced fuels needs to be one to three orders of magnitude higher than that for deuterium-tritium fusion. At the same time, the required temperature for advanced fusion is also one order of magnitude higher than that for deuterium-tritium fusion.\begin{table}[htp]
\caption{Inertial Confinement Fusion (ICF) burn rate $H_B$ parameters for different fusion fuels [Atzeni (2004) p46].} 
\begin{center}
\small
\begin{tabular}{c|c|c|c|c}
\hline\hline
Fusion Reaction & \thead {\normalsize Ideal Ignition \\ \normalsize Temperature [keV]}  & $H_B^{min}$ [${\rm g/cm^2}$] & $T (H_B^{min})$ [keV] & \thead {Energy \\ Yield [GJ/mg]}  \\\hline
D-T &  4.3  & 7.3 &  40 &  0.337  \\
D-D  &  35  & 52 &  500 & 0.0885 \\
D-3He  &  28  & 51 &  38 &  0.0357\\
p-11B  &  -  & 73 &  250 &  0.0697 \\\hline\hline
\end{tabular}
\end{center}
\label{tab:icfHB}
\end{table}

From the right plot in Fig. \ref{fig:rhoR_icf}, it can be seen that for a 50\% burn rate of hydrogen boron fuel, a $\rho R\simeq500{\rm kg/m^2}$ is required at 200keV, which means that for a compressed plasma pellet with $R=0.1{\rm mm}$, the required density is $\rho\simeq5\times10^6{\rm kg/m^3}$, with a total ion mass of $M=\rho \frac{4}{3}\pi R^3={\rm2.1\times10^{-5}kg=21mg}$, an ion number of $N_i=M/m_i=M/(6m_p)={\rm 2.1\times10^{21}}$, and a released energy of $E_{fus}=\frac{1}{2}N_iY={\rm 1.45\times10^{9}J=1.45GJ}$. In comparison, the density of solid boron is $2.3\times10^3{\rm kg/m^3}$, which means that the density of the pellet needs to be compressed to about 1000 times its solid density. It should also be noted that $t_c=R/c_s=0.06{\rm ns}$, the released fusion energy is already equivalent to the energy of a TNT explosion weighing about 350 kg or 400 kWh of electricity. Whether such an energy release rate still falls into the category of controllable fusion energy needs further evaluation. Increasing the pellet radius can reduce the compression rate requirement, but it will further increase the energy released in a single event. Therefore, in order to keep the energy release within a controllable range, the compression rate can only be further increased.

Table \ref{tab:icfHB} lists the inertial confinement fusion burn rate $H_B$ parameters for four typical controlled fusion fuels.[Atzeni (2004) chap2] provides a more detailed discussion of the above discussed inertial confinement models and calculations, including a discussion on advanced fuels, for reference.

\section{Energy Efficiency}

\begin{figure}[htbp]
\begin{center}
\includegraphics[width=14cm]{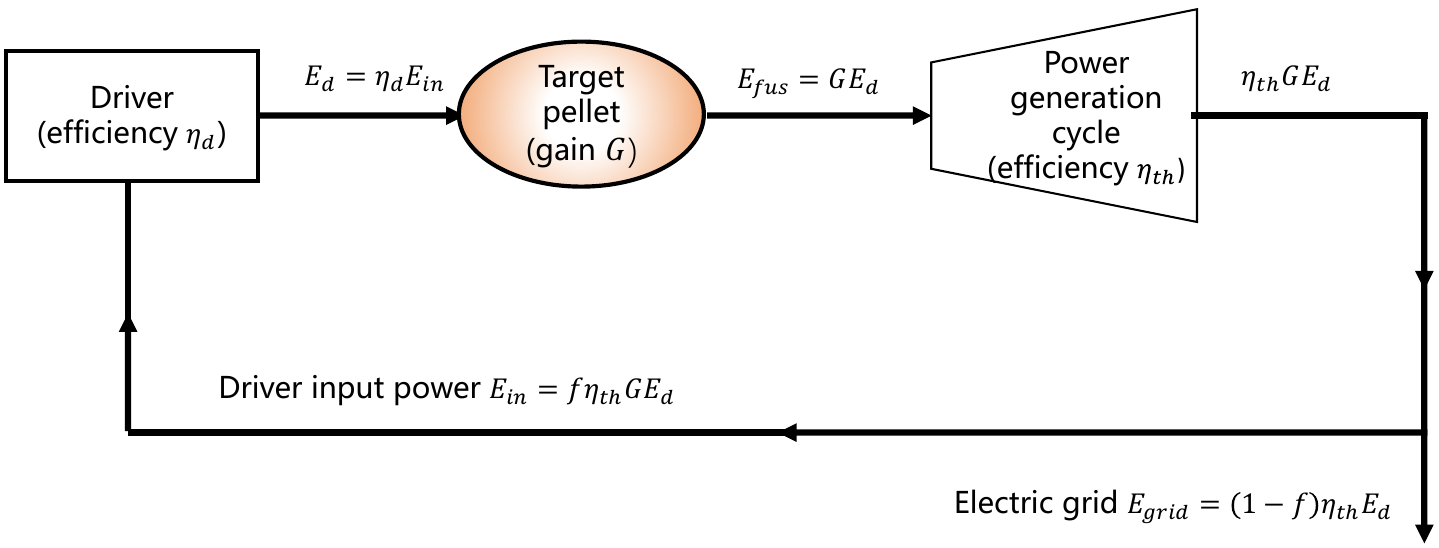}\\
\caption{Energy conversion diagram for inertial confinement fusion power generation.}\label{fig:icfgrid}
\end{center}
\end{figure}

One major advantage of using ICF for energy is that the driver can be located far away from the reactor, providing operational and maintenance flexibility. However, a significant drawback is the relatively low energy conversion efficiency. Similar to Figure \ref{fig:lawsonrelation}, we can refine the energy conversion process for fusion reactors based on inertial confinement, as shown in Figure \ref{fig:icfgrid}. This includes the efficiency from the energy source to the driver, and the efficiency from the driver to the heated plasma.

Let $\eta_d$ represent the efficiency of the driver, then $E_d=\eta_dE_{in}$. Let $G$ represent the gain from fusion targets, then $E_{fus}=GE_d$. Let $\eta_{th}$ represent the efficiency of the power plant, then the energy $E_{th}=\eta_{th}GE_d$. A fraction of the electrical energy ($f$) is provided to the driver, and the actual energy incorporated into the grid is $E_{grid}=(1-f)\eta_{th}GE_d$. This results in an energy balance for the cycle, $f\eta_{th}GE_d =E_{in}$. From this, we obtain $f\eta_{th}\eta_dG=1$. Typically, $f$ is around 50\%, $\eta_d$ is around 10\%, and $\eta_{th}$ is 40\%, so $G>50$. In other words, in order to achieve power generation with inertial confinement, the $\rho R$ parameter calculated in the left graph of Figure \ref{fig:rhoR_icf} needs to be increased by at least a factor of 50.

\begin{figure}[htbp]
\begin{center}
\includegraphics[width=14cm]{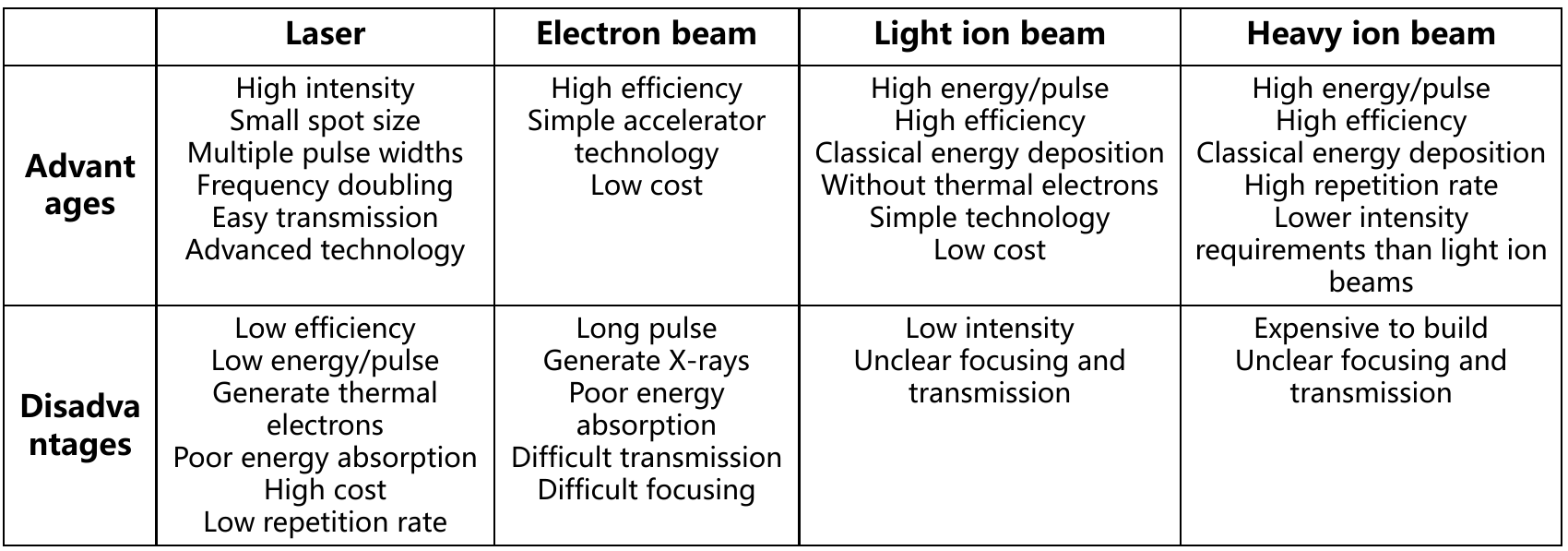}\\
\caption{Comparison of different drivers for inertial confinement fusion [Wang, G. (2005)].}\label{fig:icfdriver}
\end{center}
\end{figure}

Different driving methods for inertial confinement primarily manifest the energy conversion efficiency of the driver, such as lasers typically achieving 4-8\%, while heavy ions can achieve 20-30\%. For a single pulse release of 100 MJ, 10 pulses per second would be required to achieve a fusion power station of 1 GW. The low energy conversion efficiency in this case would result in significant energy waste. Furthermore, achieving economically and efficiently changing fusion targets for pulse power is also a major challenge. The feasibility of inertial confinement fusion energy to a large extent depends on the driver. Figure \ref{fig:icfdriver} provides a comparison of the characteristics of four common drivers [Wang, G. (2005)]. For fast ignition and central ignition, the requirements of the driver can be slightly reduced. In December 2022, the experimental results of the National Ignition Facility (NIF) at Lawrence Livermore National Laboratory in the United States achieved for the first time an energy output from fusion greater than the energy output from lasers, with a laser energy output of 2.05 MJ and a fusion energy output of 3.15 MJ. The 2.05 MJ of laser energy is converted from about 300 MJ of electrical energy, indicating that the energy conversion efficiency of the laser is less than 1\%.

\section{Summary of this chapter}

This chapter discusses the quantitative criteria for inertial confinement fusion, which is a variant of the Lawson criteria, changing the requirement of the product of density and confinement time $n\tau_E$ to the requirement of the product of density and target capsule radius $\rho R$. The key difficulty it faces is that if the energy output from a single fusion event needs to be low and controllable, such as in the range of 1-100 MJ, the target capsule radius $R$ needs to be small. In this case, the density $\rho$ needs to be very high, i.e., the compression ratio needs to be more than 1000 times the solid density, or even higher. This puts high demands on the intensity, focusing, and energy efficiency of the drivers.

There is currently no complete and feasible plan for economically generating power using inertial confinement. In addition to the zeroth-order quantity mentioned here, the instability in inertial confinement is also a problem to be solved.

\vspace{30pt}
Key points of this chapter:
\begin{itemize}
\item Inertial confinement fusion has no beta or synchrotron radiation restrictions, and even has opacity to bremsstrahlung, making it possible to ignore radiation losses, resulting in fewer theoretical restrictions compared to magnetic confinement.
\item The confinement time in inertial confinement fusion is limited by the ion acoustic velocity, which imposes a minimum requirement on $\rho R$.
\item The efficiency of the driver in inertial confinement is still relatively low, which increases the requirements on $\rho R$ by tens or even hundreds of times compared to the theoretical lower limit.
\item The typical target density for inertial confinement D-T fusion is 1000 times the solid density, and for hydrogen-boron fusion it needs to be about $10^5$ times.
\item Achieving economic viability in pulsed power generation using inertial confinement is also a difficult issue.
\end{itemize}

\chapter{Parameter Range for Magneto-inertial Confinement Fusion}\label{chap:mif}

Magneto-inertial confinement fusion is expected to combine the advantages of high density in inertial confinement and long confinement time in magnetic confinement, in order to achieve fusion energy more economically. The basic process is to first generate an initial plasma in some way, and then rapidly compress the plasma to achieve fusion conditions within a short time. During this process, there is a certain level of magnetic confinement which may result in longer confinement time compared to conventional inertial confinement.

In this chapter, we establish the simplest model to explore the minimum conditions for magnetic inertial confinement.

\section{Estimation of Parameters in a Simple Model}

\begin{figure}[htbp]
\begin{center}
\includegraphics[width=15cm]{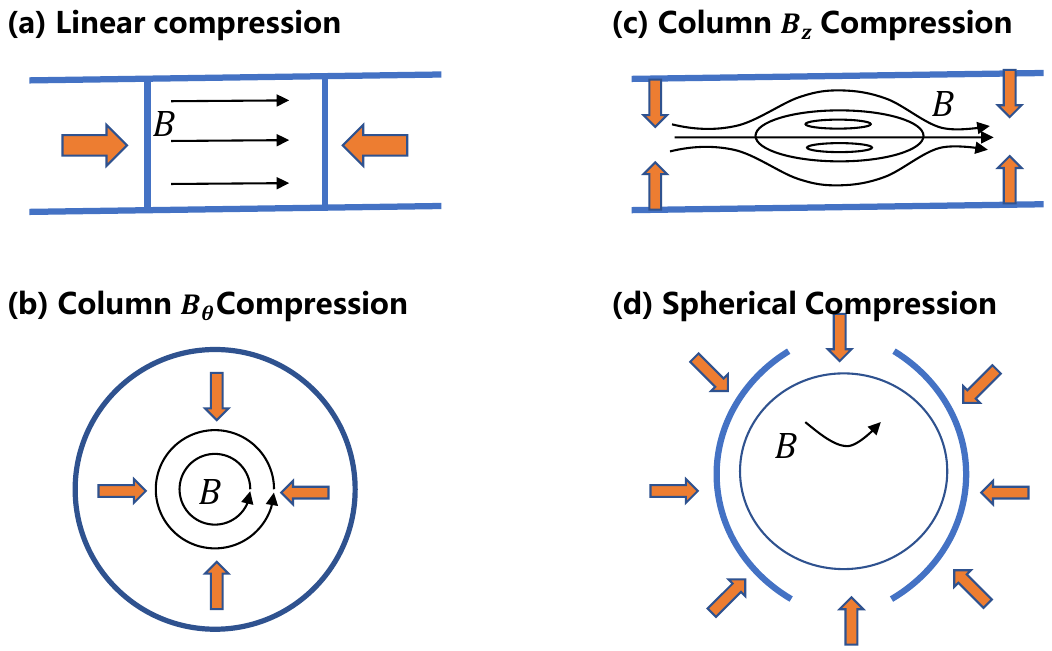}\\
\caption{Schematic diagram of the four compression methods for magnetized target fusion.}\label{fig:mifcartoon}
\end{center}
\end{figure}

Let's consider the simplest compression model based on ideal conservation relationships, in order to have a more quantitative understanding of the parameter range. The number of particles $N=nV$ is conserved, where $n$ is the number density of particles, including electrons and ions, in the plasma, and $V$ is the volume; the magnetic flux $\psi=BS$ is conserved, where $B$ is the magnetic field and $S$ is the area through which the magnetic field passes; the adiabatic compression $pV^{\gamma}$ is conserved, where the pressure $p=k_BnT$ and $T$ is the temperature. The adiabatic index is taken as $\gamma=5/3$. The compression ratio $C=R_0/R_f$, where $R_0$ and $R_f$ are the sizes of the plasma along the compression direction before and after compression, respectively. Based on these three conservation relationships, we can compile a table of the parameter relationships before and after compression for four compression methods ranging from one dimension to three dimensions, as shown in Table \ref{tab:mifC}, namely, one-dimensional linear compression, such as colliding beams; two-dimensional cylindrical compression in the $\theta$ direction of a magnetic field, such as z-pinch; two-dimensional cylindrical compression in the $z$ direction of a magnetic field, such as magnetic mirror or field-reversed configuration; and three-dimensional compression in a spherical shape, such as spherical tokamak or spheromak. Figure \ref{fig:mifcartoon} shows the corresponding compression diagrams.
\begin{table}[htp]
\footnotesize
\caption{The ratio of compressed parameters to the pre-compressed parameters in magnetically inertial fusion, where the compression ratio $C=R_0/R_f$.}
\begin{center}
\begin{tabular}{c|c|c|c|c|c|c}
\hline\hline
Parameter & \thead {Pre-\\Compression} & \thead {Post-\\Compression} & Linear & Cylinder $B_\theta$ & Cylinder $B_z$ & Sphere \\ \hline
Volume & $V_0$ & $V_f$ & $C^{-1}$ & $C^{-2}$ & $C^{-2}$ & $C^{-3}$  \\
Area & $S_0$ & $S_f$ & $C^{0}$ & $C^{-1}$ & $C^{-2}$ & $C^{-2}$  \\
Density & $n_0$ & $n_f$ & $C^{1}$ & $C^{2}$ & $C^{2}$ & $C^{3}$  \\
Temperature & $T_0$ & $T_f$ & $C^{\gamma-1}$ & $C^{2(\gamma-1)}$ & $C^{2(\gamma-1)}$ & $C^{3(\gamma-1)}$  \\
Magnetic Field & $B_0$ & $B_f$ & $C^{0}$ & $C^{1}$ & $C^{2}$ & $C^{2}$  \\
Beta & $\beta_0$ & $\beta_f$ & $C^{\gamma}$ & $C^{2\gamma-2}$ & $C^{2\gamma-4}$ & $C^{3\gamma-4}$  \\
\hline\hline
\end{tabular}
\end{center}
\label{tab:mifC}
\end{table}

To complete the model, we need several additional parameter relationships. For the input driver energy $E_L$, assuming an ideal lossless situation, it can be estimated as the sum of plasma internal energy $E_{th}$ and magnetic energy $E_B$:
\begin{eqnarray}
E_L&=&E_{th}+E_B=E_{th}[1+2/(3\beta_f)],\\
E_{th}&=&\frac{3}{2}k_Bn_fT_fV_f,
\end{eqnarray}
The actual value of $E_L$ will be larger than the above value. The difference between this model and the magnetic confinement and inertial confinement models lies in the addition of the $E_B$ term. For magnetic confinement, as it approaches quasi-steady state, the power consumption of $E_B$ within the confinement time $\tau_E$ can be ignored; for inertial confinement, the magnetic field is low, so $E_B$ can also be ignored. However, for magnetically inertial confinement, on one hand, there is a higher magnetic field, and on the other hand, the magnetic field for each shot needs to be provided by the driver, so the $E_B$ term needs to be included. In this section, we estimate the difficulty of magnetic confinement and focus solely on deuterium-tritium (DT) fusion. The peak compression (dwell) time is denoted as $\tau_{dw}$, and we assume that fusion occurs within this time. The fusion energy released is given by
\begin{eqnarray}
E_{fus}=\frac{1}{4}n_f^2\langle\sigma v\rangle \cdot Y\cdot \tau_{dw}\cdot V_f,
\end{eqnarray}
where $Y=17.6$ MeV is the energy released per fusion event. For simplicity, we fix the post-compression temperature at the optimum DT fusion temperature, $T_f=10$ keV, at which the fusion reaction rate is approximately $\langle\sigma v\rangle\simeq1.1\times10^{-22}$ m$^3$/s. The fusion gain is defined as
\begin{eqnarray}
Q_{fus}=\frac{E_{fus}}{E_L},
\end{eqnarray}
where we can normalize the plasma volume $V_f$. Therefore, the input parameters for the above model are only the initial density $n_0$, the final magnetic field $B_f$, the peak compression time $\tau_{dw}$, and the compression ratio $C$, while the others are output parameters.

\begin{figure}[htbp]
\begin{center}
\includegraphics[width=15cm]{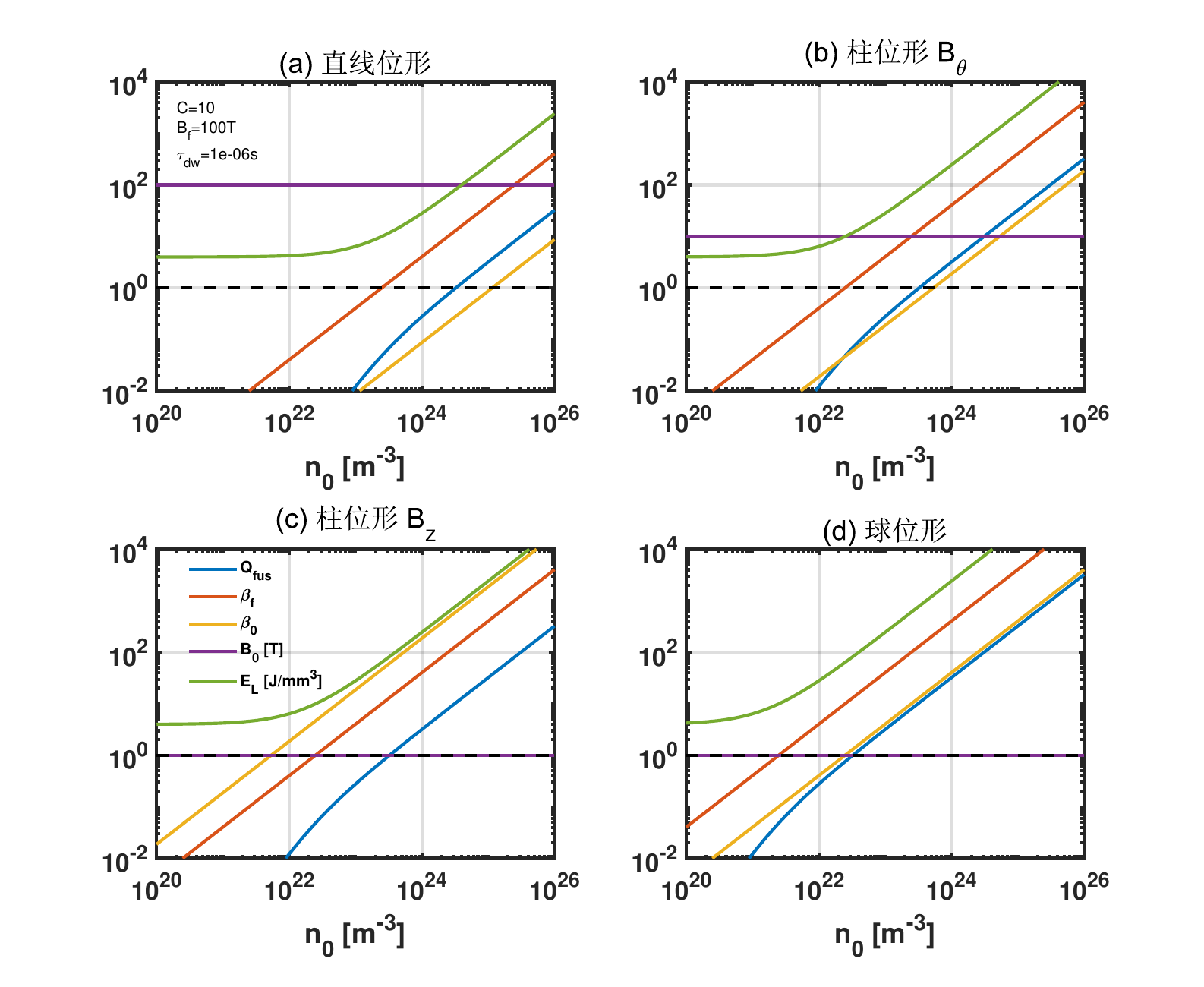}\\
\caption{When the compression parameter is $B_f=100{\rm T}$, $\tau_{dw}=10^{-6}{\rm s}$, and $C=10$.}\label{fig:mif1}
\end{center}
\end{figure}

Figure \ref{fig:mif1} shows the compression parameters for $B_f=100$ T, $\tau_{dw}=10^{-6}$ s, and $C=10$. Considering that a very high magnetic field leads to significant synchrotron radiation losses and can cause equipment walls to break or melt, while a very low magnetic field cannot achieve high density due to the beta limit, which requires a high confinement time. Therefore, we set $B_f=100$ T. If we intend to take full advantage of the long confinement time provided by magnetic confinement, we need $\beta<1$. We will also discuss the case of $\beta\geq1$.

In the figure \ref{fig:mif1}, a set of typical compression parameters is shown for $\tau_{dw}=10^{-6}$ s and $C=10$. These parameters are often used in the design of the liner compression scheme based on Field-reversed configuration (FRC). This is because typical driver power is below 100 MJ, which limits the size and parameter range of the device. For example, the compression process is definitely very fast, exceeding 10 km/s. For a device with a radius of 10 cm, $\tau_{dw}<10^{-5}$ s, so it is reasonable to assume a peak time of $\tau_{dw}=10^{-6}$ s. Under these parameters, it can be seen that all four typical configurations require an initial density $n_0$ greater than $10^{22}$ m$^{-3}$ to achieve a gain $Q_{fus}>1$. Moreover, at the corresponding density, it is necessary to have a beta greater than 1, or even a final beta greater than 1. Such a target plasma cannot be provided by magnetic confinement alone but can only be provided by wall or inertial confinement schemes, such as the MAGO (Garanin, 2015) scheme and MagLIF scheme.\begin{figure}[htbp]
\begin{center}
\includegraphics[width=15cm]{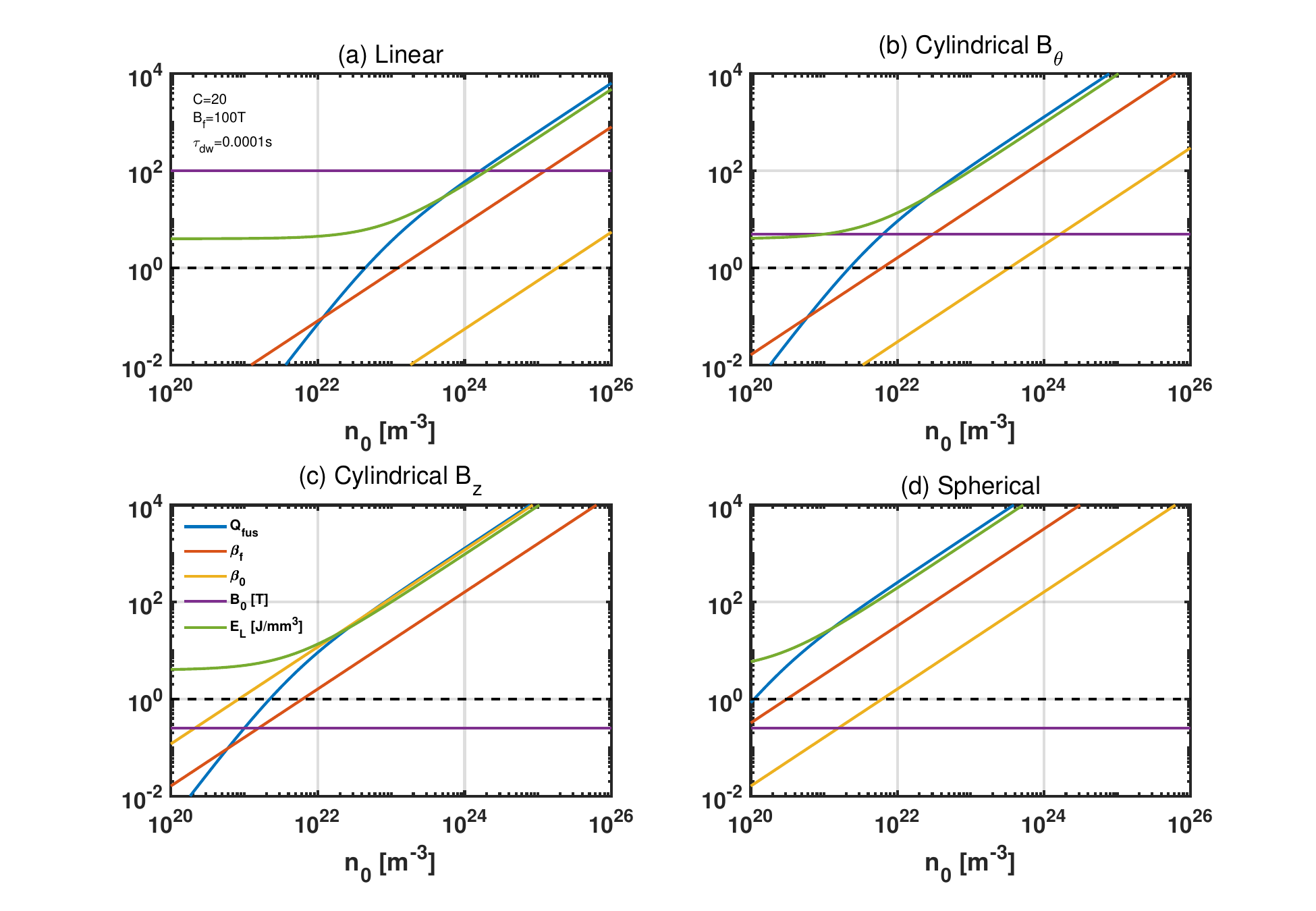}\\
\caption{Compression parameters: $B_f=100{\rm T}$, $\tau_{dw}=10^{-4}{\rm s}$, $C=20$.}\label{fig:mif2}
\end{center}
\end{figure}

In order to overcome the limitations mentioned above, we consider larger peak time $\tau_{dw}=10^{-4}{\rm s}$ and a higher compression rate $C=20$, and the results are shown in Figure \ref{fig:mif2}. It can be seen that the situation has improved compared to Figure \ref{fig:mif1}. For compression of the column $B_{\theta}$, there is a parameter range near the initial density $n_{0}\simeq4\times10^{21}{\rm m^{-3}}$ that satisfies $Q_{fus}>1$ and $\beta<1$. For spherical compression, the appropriate parameter range is $n_{0}\simeq2\times10^{20}{\rm m^{-3}}$. We also note that the appropriate parameter range is very narrow, and the calculation of $Q_{fus}$ in the above model is too ideal and definitely higher than the actual value.

By scanning more parameters, the following conclusions can be drawn: (1) In magnetic-confinement target plasma compression, fusion gain can only be achieved if the compression peak time is sufficiently large and the initial density is sufficiently high. For example, $\tau_{dw}\geq 10^{-4}{\rm s}$ is required, which can only be achieved by slow compression. However, slow compression will result in significant transport losses that violate the conservation assumptions in the model. (2) If it is desired to use fast compression ($\tau_{dw}\leq10^{-5}{\rm s}$) to achieve magnetic-inertial fusion, $\beta>1$ is necessary. This can only be achieved by wall confinement or inertia confinement in the initial plasma, and additional models need to be evaluated for their feasibility. (3) Non-deuterium-tritium fusion requires even more stringent parameter conditions than the calculations above. Although the results of Lindemuth (2009) are often used to illustrate the potential low-cost advantages of magnetized target fusion (MTF) schemes, Lindemuth (2017) also highlights the difficulties, and its conclusions are similar to those mentioned above.

In summary, the parameter range for magnetic-inertial confinement is very narrow, and it is almost impossible to use magnetic-confinement target plasma. If inertial-confinement target plasma is used, it needs to be evaluated using the inertial confinement model. The above results can also be simplified by replacing them with a single question: Is it possible to achieve plasma with fusion temperature ($10-100{\rm keV}$) and a confinement time of milliseconds ($10^{-3}{\rm s}$) at atmospheric density ($10^{25}{\rm m^{-3}}$)? If so, non-deuterium-tritium magnetized inertial confinement fusion may be possible. Currently, many fusion schemes around the world expect to use magnetic-inertial compression, and they all need to verify the feasibility of zero-order factor through the above model, explaining which condition their scheme can overcome. Otherwise, no matter how the parameters are improved, fusion energy cannot be realized.

\section{Model with Compression Circuit}

The previous section discussed the possible parameter range of magnetic inertia constraints under the simplest model. Here we consider a more realistic situation, mainly referring to the model proposed by Dahlin (2004). For other magnetized target schemes, similar modeling analysis can be conducted. Here we consider the scheme of compressing FRC (Field Reversed Configuration) plasma by a metal liner driven by a capacitor circuit in the z-direction. This is because FRC is a very representative magnetically confinement target plasma, which is highly compressible and transferable.

\subsection{Model Equations}

We adopt the cylindrical coordinates $(r, \theta, z)$. The thickness of the liner is given by
\begin{eqnarray}
d(R) = -R + \sqrt{R^2 + d_0^2 + 2R_0d_0},
\end{eqnarray}
where $d_0$ is the initial thickness of the liner, $R$ is the inner radius of the liner, and $R_0$ is the initial inner radius. 

The motion equation of the liner is given by
\begin{eqnarray}
\rho\frac{d{\bm v}}{dt} = {\bm j} \times {\bm B} - \nabla p,
\end{eqnarray}
where $\rho$ is the mass density of the liner, ${\bm v}$ is the radial velocity, ${\bm j}$ is the current density inside the liner, $p$ is the pressure (including the internal plasma pressure and magnetic pressure), and ${\bm B}$ is the magnetic field given by
\begin{eqnarray}
{\bm B} = \frac{\mu_0 I(r)}{2\pi r}{\hat{\bm \theta}},
\end{eqnarray}
where $\mu_0$ is the vacuum permeability and $I(r)$ is the current in the liner. Since the current in the liner is in the $\hat{\bm z}$ direction, the magnetic field is in the ${\hat{\bm \theta}}$ direction. Here we ignore the component of the magnetic field in the $\theta$ direction caused by FRC itself.

Assuming ${\bm v}$ is uniform and zero-dimensional, the motion equation of the liner can be simplified as
\begin{eqnarray}
\rho d\frac{d^2 R^*}{dt^2} = -\frac{\mu_0 I^2}{8\pi^2(R+d)^2} + \Big(2neT+\frac{B_z^2}{2\mu_0}\Big),
\end{eqnarray}
where $R^*$ is the center radius of the radial mass distribution, given by
\begin{eqnarray}
R^* = \frac{\int_R^{R+d}(r2\pi r)dr}{\int_R^{R+d}(2\pi r)dr} = \frac{2}{3}\cdot\frac{3R^2+3Rd+d^2}{2R+d}.
\end{eqnarray}
It is possible that the liner may partially or completely melt or even vaporize, but we temporarily ignore this effect.

Assuming the plasma distribution is uniform, the volume of the plasma satisfies
\begin{eqnarray}
\frac{V(R)}{V_0} = \frac{R^\delta}{R_0^\delta},
\end{eqnarray}
where the scaling factor $\delta$ can be chosen according to the specific situation. For example, for pure radial compression, $\delta=2$, for FRC it is approximately $\delta=2.4$, and for other configurations it can be adjusted accordingly. Temperature can be determined by energy balance:
\begin{eqnarray}
3ne\frac{dT}{dt}=-p\cdot\nabla{\bm v}+P_{\alpha}-P_{rad},
\end{eqnarray}
where $n$ is the plasma number density, $e$ is the unit charge, and $T$ is the temperature in eV units. We assume that all the energy of fusion $\alpha$ products is deposited, and only consider bremsstrahlung radiation:
\begin{eqnarray}
P_{rad}=1.692\times10^{-38}n_e(n_H+4n_{He})\sqrt{T},
\end{eqnarray}
where $n_e$ is the electron density and $n_{He}$ is the $\alpha$ particle density, and impurities are neglected. The energy deposition and energy loss mentioned above represent the most optimistic case, so the calculated energy gain in the following calculations will be larger than the actual achievable value. By averaging the equation, we have
\begin{eqnarray}
\langle p\nabla\cdot{\bm v}\rangle=\frac{1}{V}\int\int\int(p\nabla\cdot{\bm v})rdrd\theta dz=\frac{p}{V}\frac{dV}{dt}.
\end{eqnarray}
Using $p=2neT$, we obtain
\begin{eqnarray}
\frac{dT}{dt}=-\frac{2T\delta}{2}\frac{\dot{R}}{R}+\frac{\delta(P_{\alpha}-P_{rad})}{3ne}.
\end{eqnarray}

Plasma density is given by
\begin{eqnarray}
\frac{dn_H}{dt}=\frac{d}{dt}\Big(\frac{n_H}{V}\Big)=\frac{1}{V}\frac{dN_H}{dt}+N_H\frac{dV^{-1}}{dt},
\end{eqnarray}
where $N_H$ is the total number of hydrogen ions, and we have
\begin{eqnarray}
\frac{1}{V}\frac{dN_H}{dt}=-n_Dn_T\langle\sigma v\rangle_{DT},
\end{eqnarray}
where $n_D$ and $n_T$ are the densities of deuterium and tritium, and here we assume $2n_D=2n_T=n_H$. Also, we have
\begin{eqnarray}
N_H\frac{dV^{-1}}{dt}=-\frac{N_H}{V^2}\frac{dV}{dt}=-\frac{\delta n_H}{R}\frac{dR}{dt}.
\end{eqnarray}
Therefore, we have
\begin{eqnarray}
\frac{dn_H}{dt}=-n_Dn_T\langle\sigma v\rangle_{DT}-\frac{\delta n_H}{R}\frac{dR}{dt}.
\end{eqnarray}
The plasma density $n=n_H+n_{He}$, where
\begin{eqnarray}
n_{He}=\frac{1}{2}\Big(n_0\frac{R_0^\delta}{R^\delta}-n_H\Big).
\end{eqnarray}Let us consider the equations for the driving circuit. Assuming that the circuit has capacitance C and total resistance $R_C$, for the overall circuit, the inductance is given by
\begin{eqnarray}
L=\frac{\mu_0 l}{2\pi}\ln\Big(\frac{a}{R+d}\Big),
\end{eqnarray}
where l is the length of the liner, and a is the radius of the loop. Since R is variable during the compression process, L also changes with time. Kirchhoff's law gives
\begin{eqnarray}
u_C+u_L+u_R=0.
\end{eqnarray}
The voltage across the inductor is given by
\begin{eqnarray}
u_R=\frac{d}{dt}(LI)=L\frac{dI}{dt}+I\frac{dL}{dt},
\end{eqnarray}
and the current is
\begin{eqnarray}
I(t)=C\frac{du_C}{dt}.
\end{eqnarray}

Adding Ohm's law $u_R=R_CI(t)$, we finally obtain the circuit equation
\begin{eqnarray}
\frac{d^2u_C}{dt^2}=\Big(\frac{\mu_0l}{2\pi L(R+d)}\frac{dR}{dt}-\frac{R_C}{L}\Big)\frac{du_C}{dt}-\frac{1}{LC}u_C.
\end{eqnarray}

Combining the above equations, the complete self-consistent ordinary differential equations can be obtained, consisting of six first-order ordinary differential equations:
\begin{eqnarray}
\frac{dq_1}{dt}&=&q_2,\\
\frac{dq_2}{dt}&=&-\frac{\mu_0 I^2}{8\pi^2\rho d(R+d)^2}+\frac{2neT}{\rho d}+\frac{B^2}{2\mu_0\rho d},\\
\frac{dq_3}{dt}&=&(P_{\alpha}-P_{rad})\frac{R^{2\delta/3}}{3ne},\\
\frac{dq_4}{dt}&=&-\frac{1}{4}R^{\delta}n_H^2\langle\sigma v\rangle_{DT},\\
\frac{dq_5}{dt}&=&q_6,\\
\frac{dq_6}{dt}&=&\Big(\frac{\mu_0l}{2\pi L(R+d)}\frac{dR}{dt}-\frac{R_C}{L}\Big)q_6-\frac{1}{LC}q_5,
\end{eqnarray}
where the variables $q_i$ are defined as
\begin{eqnarray}
q_1=R^*,~~q_2=\frac{dR^*}{dt},~~q_3=R^{2\delta/3}T,~~q_4=n_HR^{\delta},~~q_5=u_C,~~q_6=\frac{I}{C}.
\end{eqnarray}The energy gain of each shot is proportional to the burn rate $f_b$:
\begin{eqnarray}
f_b=\frac{N_0-N_H}{N_0}=\frac{n_0V_0-N_H}{n_0V_0}.
\end{eqnarray}
The output fusion energy is given by:
\begin{eqnarray}
E_{fus}=\frac{1}{2}f_bn_0V_0Y,
\end{eqnarray}
where $Y=17.6$ MeV represents the energy release from a single deuterium-tritium fusion. The main energy input comes from the capacitor:
\begin{eqnarray}
E_C=\frac{1}{2}Cu_0^2.
\end{eqnarray}
Thus, the energy gain factor is defined as:
\begin{eqnarray}
Q_{fus}=\frac{E_{fus}}{E_C}.
\end{eqnarray}

This completes the entire calculation model, and we can optimize it for different parameters to see how much energy gain $Q_{fus}$ can be achieved.



\begin{figure}[htbp]
\begin{center}
\includegraphics[width=15cm]{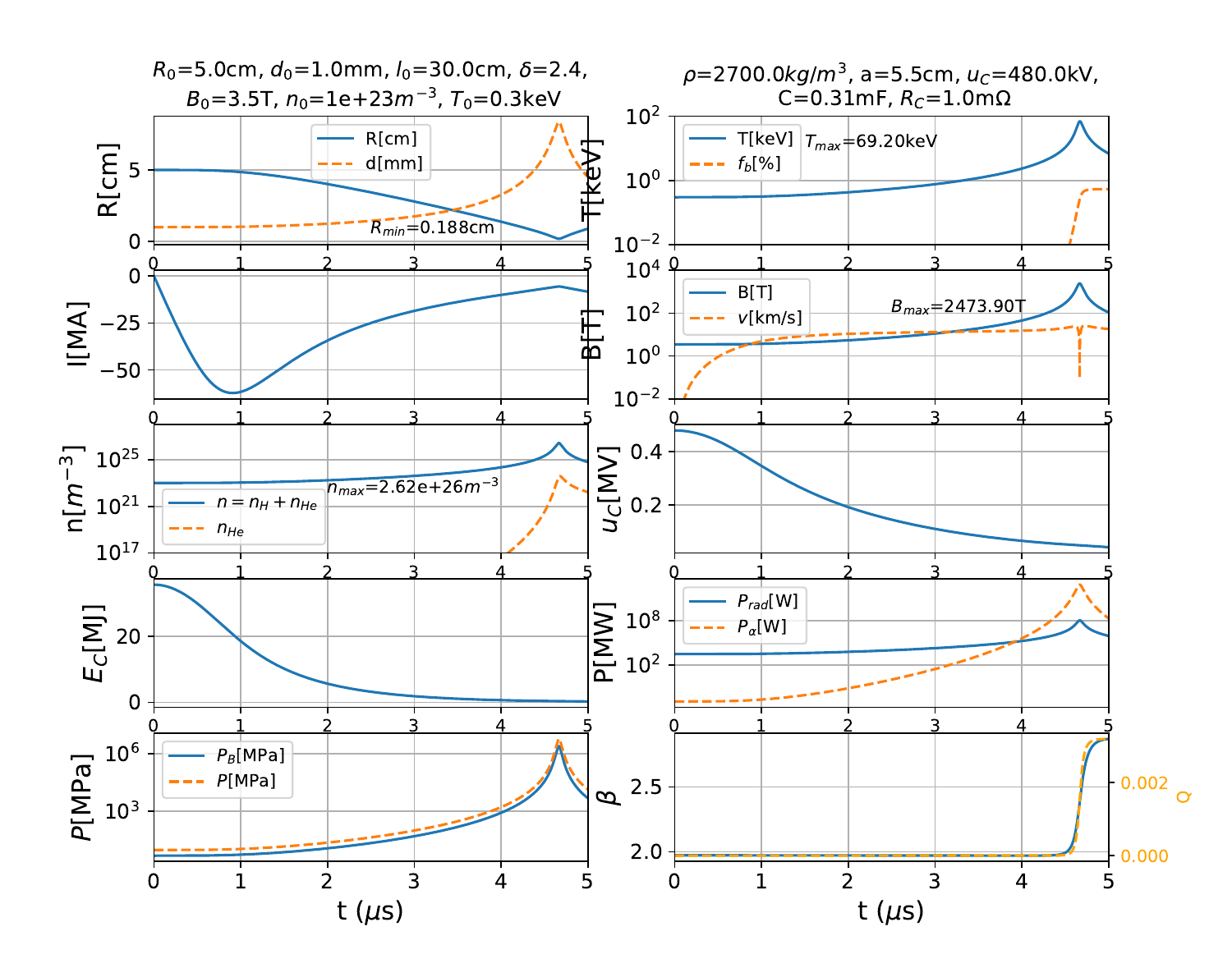}\\
\caption{Typical Example 1 of Magnetic Inertial Confinement Fusion Compression Process.}\label{fig:dahlin04R0=005}
\end{center}
\end{figure}

\begin{figure}[htbp]
\begin{center}
\includegraphics[width=15cm]{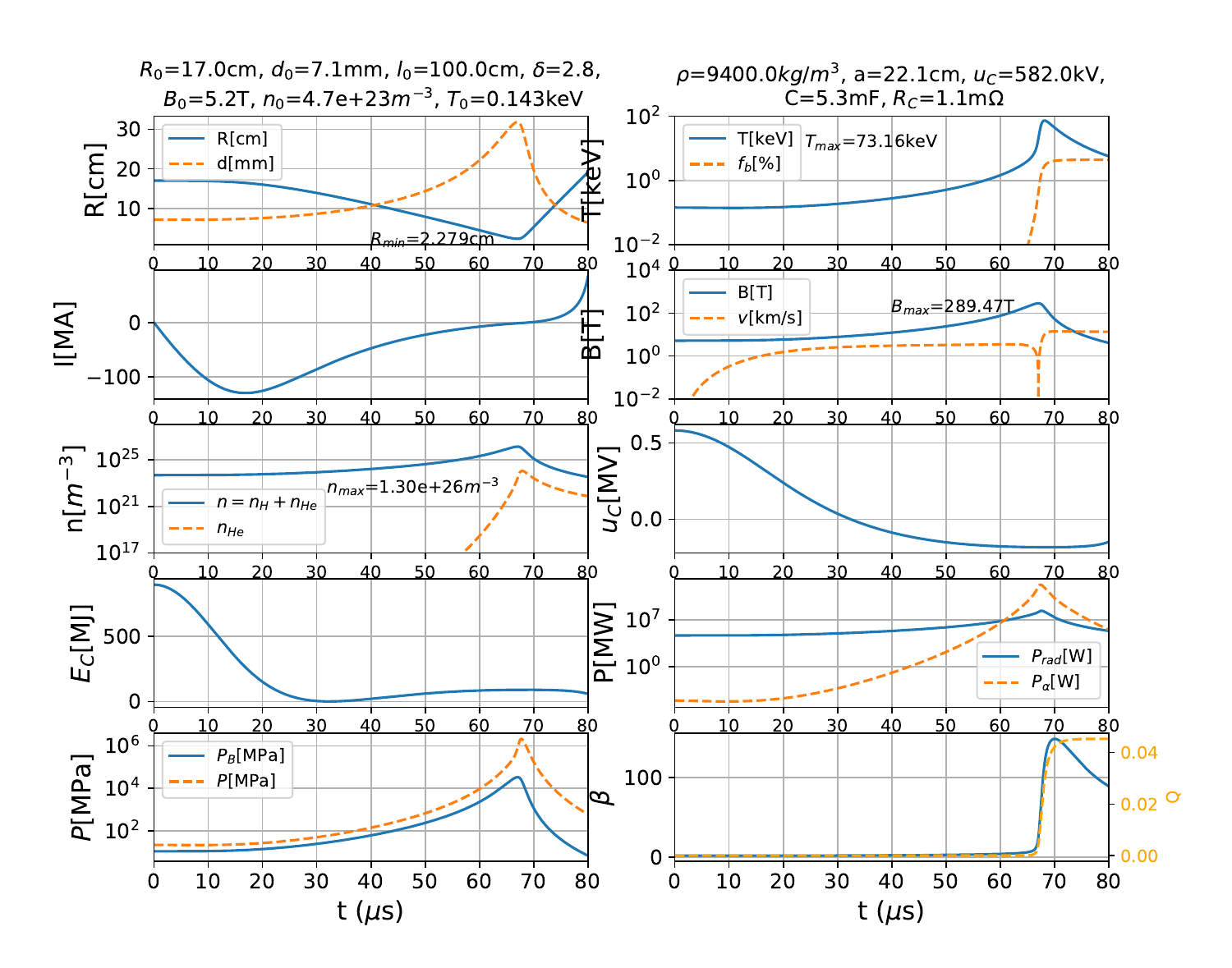}\\
\caption{Typical Example 2 of Magnetic Inertial Confinement Fusion Compression Process.}\label{fig:dahlin04R0=017}
\end{center}
\end{figure}

\subsection{Typical Calculation Results}
Figures \ref{fig:dahlin04R0=005} and \ref{fig:dahlin04R0=017} show the calculation results for two typical parameter sets. We can see that the fusion gain is less than 1, which means it does not meet the condition for scientific feasibility. The key reason for this is the short dwell time, only about 1 microsecond. A larger driver would make this time even shorter, while a smaller driver would not be sufficient for effective compression.

These results are consistent with the results of the simplified model in the previous section, and further demonstrate the difficulties faced in achieving high fusion gain for a more realistic MTF fusion device. Dahlin (2004) conducted more parameter scans and ultimately found only a few parameters that yielded $Q_{fus}\simeq1$. In other words, achieving fusion gain based on the FRC MTF scheme is challenging at the fundamental level, mainly because a way to overcome the limitation of dwell time needs to be found. In the models above, we have assumed that the energy confinement time is longer than the peak compression time. If the energy confinement time is shorter than this time, or if a significant loss of plasma occurs due to instability, the difficulty of achieving fusion gain for this scheme will be further increased.

\section{Summary of this Chapter}
From the analysis of the two models in this chapter, it can be seen that magnetic inertial confinement faces challenges similar to magnetic confinement and inertial confinement, as the compression time is extremely short and difficult to increase by orders of magnitude. Especially when using a magnetically confined target plasma, even under ideal conditions, it can only barely achieve the conditions for deuterium-tritium fusion and almost cannot achieve conditions for non-deuterium-tritium fusion. Ultimately, if only deuterium-tritium can be used, the problem is related to materials and engineering. If this limitation is to be overcome, wall confinement or inertial confinement targets should be considered, which have a higher theoretical limit compared to magnetically confined targets. The difficulty lies in how to confine a high-temperature plasma for a long time in wall confinement, as rapid energy loss occurs, or in the intensity and efficiency of the driver for inertial confinement.

In fact, the main reason for not using wall confinement is not because the wall cannot withstand high temperatures (although the temperature is high, the density is extremely low, so the heat is also low), but because the wall temperature is low and it will quickly cool the main plasma. It is similar to a fluorescent lamp, which has a temperature of several thousand degrees or even tens of thousands of degrees but does not melt the tube wall. Compared to the mean free path, collision occurs many times, so even with a very low loss rate, it will quickly lose plasma and introduce impurities.From the above analysis, we can see that it is necessary for fusion to have ultra-high pressure in order to increase density. So, can the high-pressure environment naturally present on the Earth be used to provide conditions for fusion? Naturally, we think of the high pressure in the deep sea. By simple calculation, the depth of the Mariana Trench is approximately 10,000 meters, which is equivalent to approximately 1,000 atmospheres, or 100 MPa, while the mechanical structural materials currently used in existing magnetic confinement fusion devices are close to the material pressure limit, at approximately 1,000 MPa.

Therefore, it can be seen that, in addition to building fusion reactors on land, there is no advantage to relying on high-pressure environments in the deep sea to build fusion reactors. It is better to further increase the material pressure limit.

\vspace{30pt}
Key points of this chapter:
\begin{itemize}
\item The combination of magnetic and inertial confinement aims to achieve fusion energy gain more economically by taking advantage of the long confinement time of magnetic confinement and the high confinement density of inertial confinement.
\item Due to the limitations of the beta ratio and compression time, the gain of magnetic and inertial confinement is not optimistic and can only achieve weak gain in deuterium-tritium fusion under ideal conditions, unable to achieve gain in advanced fuels fusion.
\item The analysis of zero order quantities shows that the feasibility of magneto-inertial  confinement fusion is no higher than that of magnetic or inertial confinement alone.
\end{itemize}
\chapter{Evaluation of Fusion Schemes}\label{chap:approach}

\begin{figure}[htbp]
\begin{center}
\includegraphics[width=15cm]{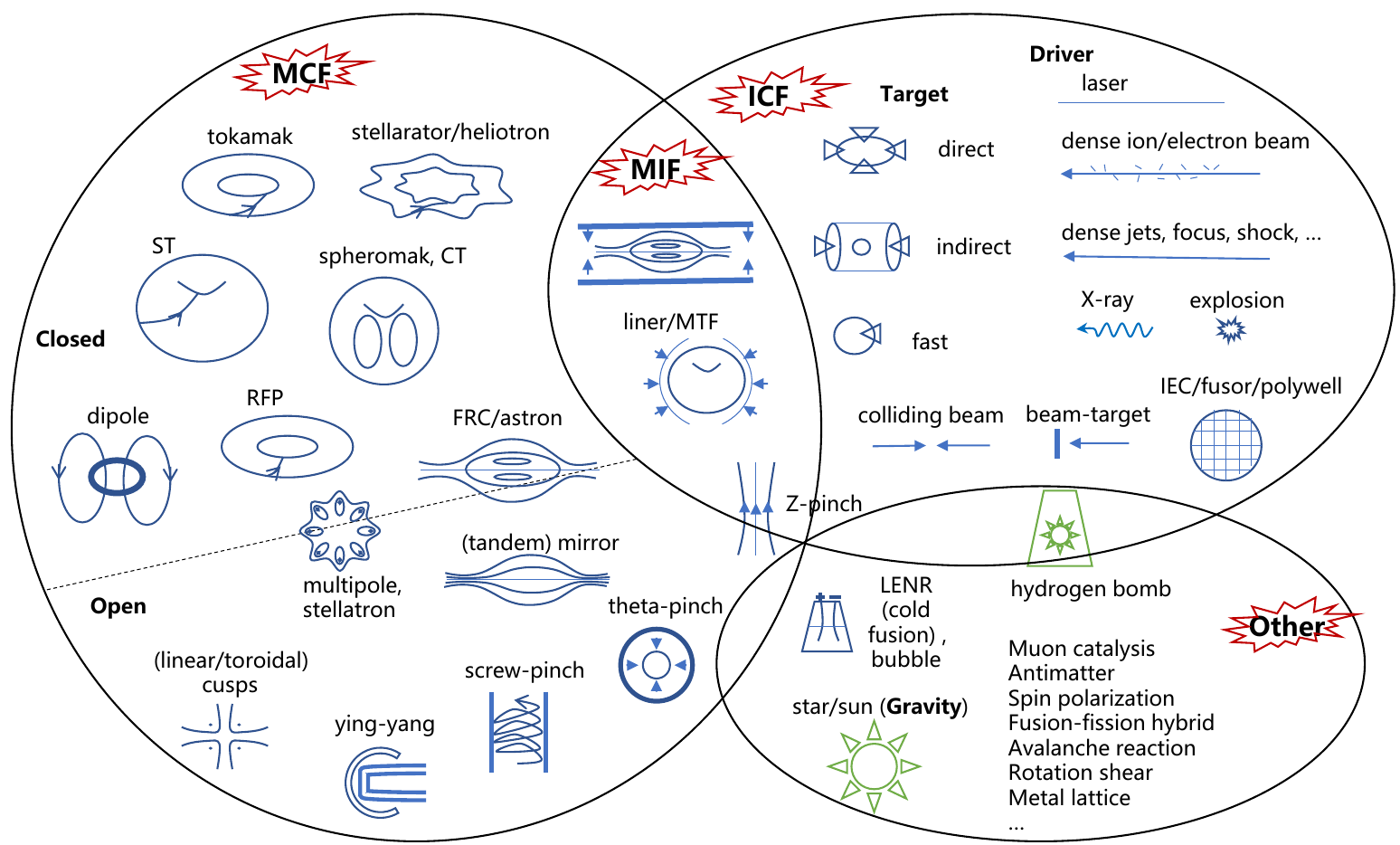}\\
\caption{Fusion scheme zoo, in which stars and hydrogen bombs belong to schemes that have already achieved fusion energy gain.}\label{fig:fusionzoo}
\end{center}
\end{figure}

This chapter discusses the basic methods, achievable parameters, and key challenges of various fusion schemes. Figure \ref{fig:fusionzoo} summarizes the main fusion schemes that have been proposed and potential ones. The schemes not listed in the figure are usually derived from the schemes listed in the figure. Among them, magnetic confinement fusion (MCF) refers to the confinement by magnetic fields, which usually has a long confinement time but low density. Inertial confinement fusion (ICF) achieves fusion reactions within a short time of inertial scattering, usually with extremely short confinement time but high density. Gravity confinement involves confining fusion fuel through gravity, such as in stars. The combination of magnetic confinement and inertial confinement is referred to as magneto-inertial confinement (MIF), or magnetized target fusion (MTF). Other methods include wall confinement, which includes cold fusion and lattice fusion. Muon catalysis, spin polarization, avalanche reactions, rotation, and hybrid reactors are methods that improve reaction rates, enhance confinement, or improve energy utilization in the aforementioned schemes.

\begin{figure}[htbp]
\begin{center}
\includegraphics[width=15cm]{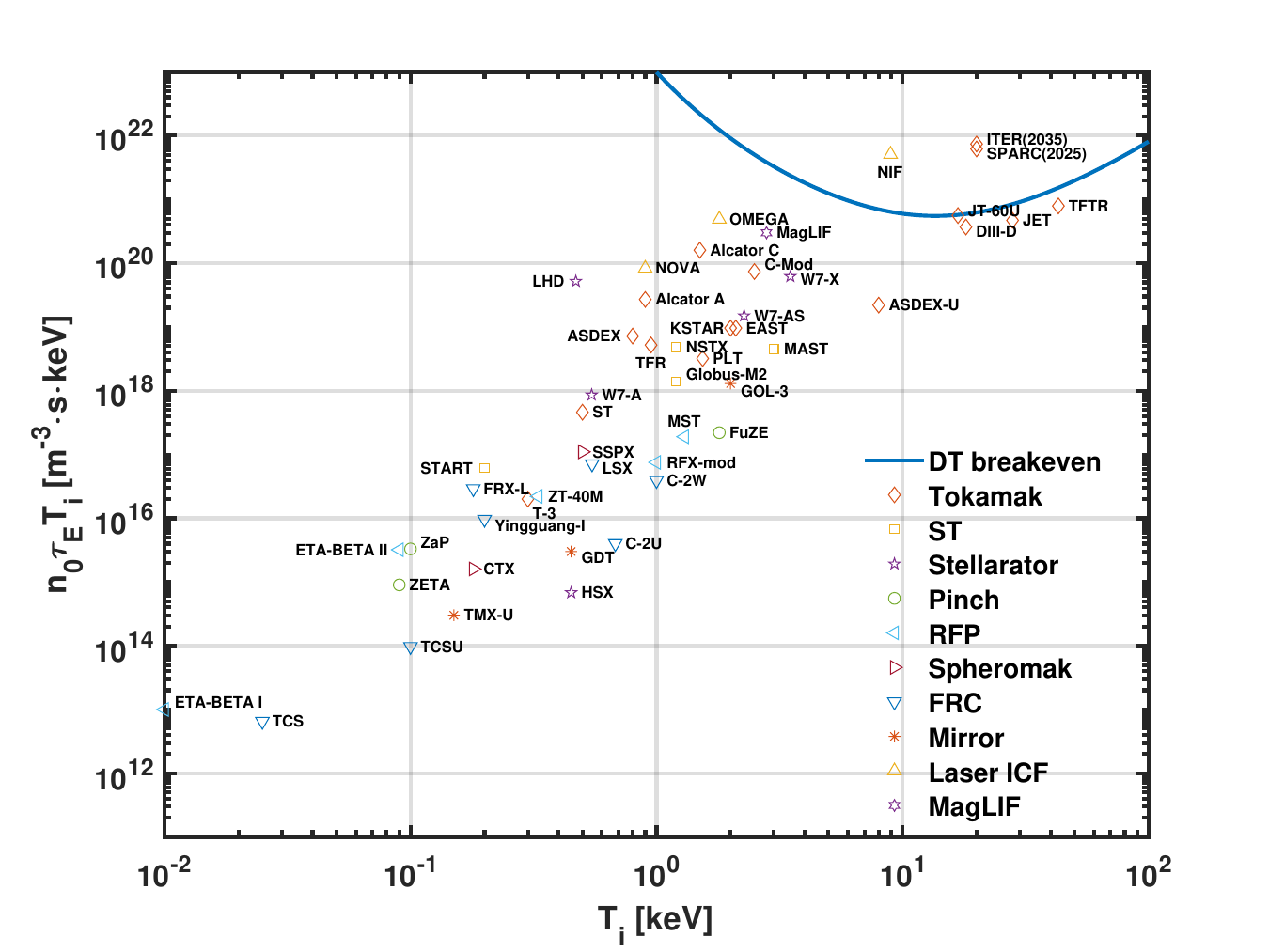}\\
\caption{Parameters achieved in various experimental devices as given in the literature [Wurzel (2022)].}\label{fig:lawsondat}
\end{center}
\end{figure}

The achieved parameters and planned parameters for various experimental schemes are shown in Figure \ref{fig:lawsondat}. As can be seen, some devices are approaching or surpassing the critical line for the scientific feasibility of deuterium-tritium fusion.

\section{Implemented Fusion Energy Devices}

Currently, only stars and hydrogen bombs belong to fusion schemes that have successfully achieved energy gain, and theoretically can be considered as energy devices. The conditions required for the former cannot be realized on Earth, and the energy release of the latter is still uncontrollable. Here, we conduct a quantitative analysis of why they can achieve fusion energy gain, in order to understand their differences from other fusion schemes.

\subsection{Stars and Sun}
\begin{figure}[htbp]
\begin{center}
\includegraphics[width=15cm]{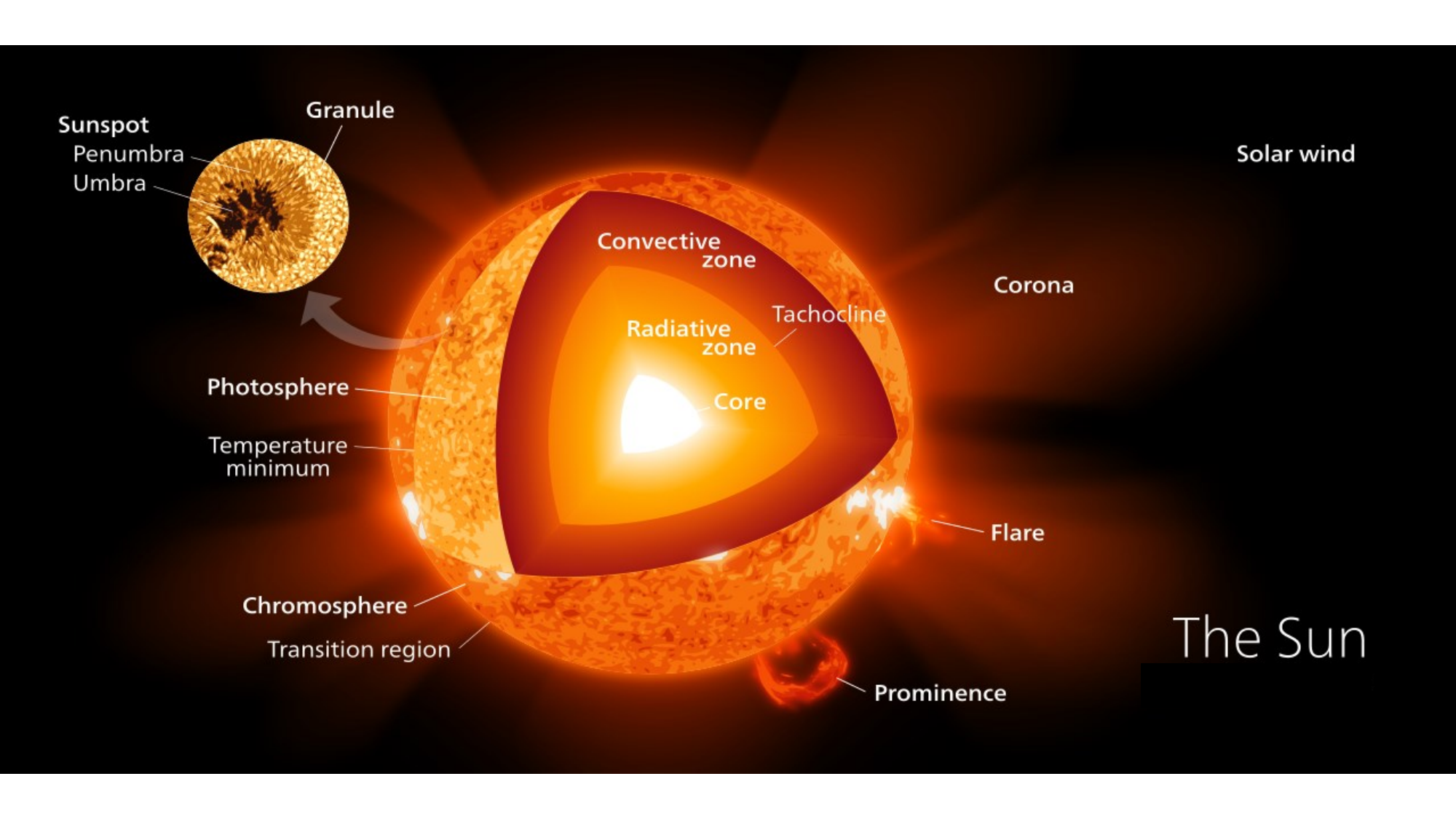}\\
\caption{Solar Model (Wikipedia).}\label{fig:Sunposter}
\end{center}
\end{figure}

Stars maintain continuous fusion reactions by the gravitational confinement. Taking the Sun as an example, as shown in Figure \ref{fig:Sunposter}. We will perform some quantitative calculations.

The radius of the Sun is $r=6.96\times10^8$m, which is about 109 times that of the Earth. The mass is $m=1.99\times10^{30}$kg, accounting for 99.86\% of the solar system. By mass, the composition is 73\% hydrogen, 25\% helium, and a small amount of carbon, oxygen, neon, iron, etc. The average density is $1.41\times10^3{\rm kg/m^3}$, and the estimated central density is $1.62\times10^5{\rm kg/m^3}$.

Studies have shown that 99\% of the fusion energy is generated in the core region of the Sun, with a radius of about 24\% (almost zero fusion at radius 30\%). The average density is $1.5\times10^5{\rm kg/m^3}$, and the temperature is 1.3 keV (15 million K). The surface temperature is 6000K. Most of the Sun's energy comes from the p-p (proton-proton) chain fusion reaction, with only 0.8\% coming from the secondary CNO cycle reaction.
The energy generated by the fusion in the core region is transferred to various outer layers, and finally emitted mainly in the form of radiation on the surface, with a small amount escaping as the energy of solar wind particles, etc.The measured value of the solar radiation reaching the outer atmosphere of the Earth is approximately $1.36\times10^3{\rm W/m^2}$, with a distance to the Earth of $R=1.5\times10^{11}$m and a surface area of the sphere being $4\pi R^2$. Therefore, the total radiation power of the Sun is approximately $3.84\times10^{26}{\rm W}$.

The pp reaction chain can be summarized as a reaction of ${\rm 4p+2e^{-}\to \alpha+2\nu_e+26.73MeV}$, where 0.7\% of the total mass is converted to energy, with a small amount carried by neutrinos. Assuming that all of the radiated energy comes from the pp fusion reaction, it can be deduced that this reaction occurs every $9.0\times10^{37}$ times per second, converting a total of $3.6\times10^{38}$ protons into helium nuclei. The total number of protons in the Sun is approximately $8.9\times10^{56}$, so the proportion of reactions occurring per second is only $4.0\times10^{-19}$, with a total conversion of $6.0\times10^{11}{\rm kg}$ of protons per second. Assuming that these fusion reactions occur in the core area, which is 0.24 times the radius of the Sun, the average unit volume power is $20{\rm W/m^{-3}}$. As a comparison, the normal metabolic power of a human body is about 80W. Therefore, the fusion reaction rate of the Sun is extremely low but due to its large volume and mass, the total power generated is enormous.

The fusion rate in the core area is self-regulating: when the fusion rate increases, the core area expands due to heating, resulting in a decrease in density and a decrease in fusion rate; when the fusion rate decreases, the core area cools and contracts, resulting in an increase in density and an increase in fusion rate. Therefore, it maintains a disturbed equilibrium state, maintaining the current fusion reaction rate.

Considering that the pp reaction cross-section data is temperature-sensitive [Adelberger (2011)], and the temperature in the central region of the Sun varies with radius (the temperature and density profiles are also given by theoretical models), there are difficulties in accurately calculating the fusion power of the Sun based on reaction rates. However, this does not prevent us from making estimates. The reaction rate is mainly limited by the first pp reaction, which is approximately $10^{-51}-10^{-49}{\rm m^3/s}$ in the core area. The energy released is the total energy of one pp cycle, $0.5\times26.73$MeV.
Based on the previous density data, the proton density in the core area is calculated to be $n_p=6.7\times10^{31}{\rm m^{-3}}$, with a volume of $V_c=2.0\times10^{25}{\rm m^{3}}$, and the number of fusion reactions is $2N=2n_p^2\langle\sigma v\rangle_{pp} V_c$. Substituting $N=9.0\times10^{37}$, the required reaction rate is calculated to be $\langle\sigma v\rangle_{pp}\simeq4\times10^{-51}{\rm m^3/s}$. For temperatures of 0.5, 1.0, and 1.3keV, the corresponding pp fusion reaction rate data is approximately $1.4\times10^{-51}$, $4.5\times10^{-50}$, and $1.3\times10^{-49}{\rm m^3/s}$, respectively, within the estimated range. In the previous section, we discussed the Lawson criterion and pointed out that the bremsstrahlung power of the hydrogen-boron reaction is greater than the fusion power, making it difficult to achieve fusion energy. So why is the power of the pp reaction in the sun lower and not limited by the Lawson criterion? This is mainly due to the enormous volume and strong gravitational field of the star, which results in a very high density and allows radiation to be confined and reused. Radiation is opaque and has a very short mean free path for bremsstrahlung (Chapter \ref{chap:lawson}). The appropriate Lawson criterion for the sun is derived in Chapter \ref{chap:fuels}, which incorporates radiation losses and reuse into the energy confinement time, $\tau_E$. However, even so, a tremendous $\tau_E$ is required, exceeding millions of years, making it uncontrollable on Earth.

In other words, when humans want to realize fusion energy like an "artificial sun", it obviously does not mean using the pp reaction chain of the sun to achieve fusion energy, but only refers to achieving controlled fusion energy. The fuel used must necessarily have high reaction cross-sections.

\subsection{Hydrogen Bomb}

The hydrogen bomb, also known as a thermonuclear weapon, belongs to the second-generation nuclear weapons. It mainly utilizes the energy released from the nuclear fusion reactions of hydrogen isotopes (deuterium and tritium) to cause destruction and casualties. It is a powerful and large-scale destructive weapon. As a comparison, the first-generation nuclear weapon, the atomic bomb "Little Boy" in 1945, had an equivalent yield of about 13 kilotons of TNT, while the largest hydrogen bomb detonation in history, the Tsar Bomba in 1961, had an equivalent yield of about 50 million tons of TNT, and the resulting mushroom cloud reached a height of about 60 kilometers, causing damage to buildings up to 160 kilometers away.

The successful research of the hydrogen bomb took place after World War II. In 1952, the United States conducted the first experimental demonstration of the hydrogen bomb's principle at the Eniwetok Atoll in the Pacific Ocean, using liquid deuterium as the nuclear fuel, which was not practical. In 1953, the Soviet Union conducted the first practical hydrogen bomb test. In 1954, the United States conducted the first practical hydrogen bomb test at Bikini Atoll, using lithium deuteride as the fusion fuel. In 1957, with the assistance of the United States, the United Kingdom conducted a hydrogen bomb test. In 1966, China conducted the first experimental demonstration of the hydrogen bomb's principle and completed an airburst test the following year. In 1968, France conducted its first hydrogen bomb test.Early phase fusion-fission two-phase hydrogen bombs are gradually being retired due to their large size and excessive power. The current total number of hydrogen bombs exceeds 10,000, with the majority being fusion-fission-fusion three-phase bombs. In the three-phase bombs, a layer of uranium-238 material is wrapped around the shell of the hydrogen bomb to induce uranium-238 fission with high-energy neutrons generated from fusion, thereby producing more energy. Neutron bombs and shock wave bombs in the third generation of nuclear weapons also belong to hydrogen bombs, but they have undergone specialized treatments. Currently, the testing of hydrogen bombs has been banned, and the main direction of development is miniaturization and cleanliness to reduce radiation pollution, in order to avoid nuclear non-proliferation treaties through fusion technology research. 

To begin with, we need to understand the principles of atomic bombs or fission nuclear power plants. Atomic bombs use fissionable radioactive fuels, such as uranium-235, which generates more neutrons during the fission process. These neutrons can further trigger additional fissions. As long as the fission material exceeds a certain critical mass, where the generated neutrons are faster than the escaping neutrons during transportation, a chain reaction will be initiated, resulting in sustained release of energy, even explosion. The critical mass of an atomic bomb is inversely proportional to the square of its density. Therefore, control of the chain reaction can be achieved through the following methods: bringing together two subcritical mass fission materials to exceed the critical mass, compressing subcritical mass materials through explosions, etc., to increase density and reduce critical mass. The addition of a neutron source or neutron reflector can also reduce the critical mass.

The most efficient shape to achieve a critical mass with the fewest materials is a sphere. If neutron reflecting materials are added around it, the critical mass can be further reduced. The critical point for a neutron-reflecting spherical uranium-235 is about 15 kg, while for plutonium it is about 10 kg. Without neutron reflection, the critical mass of uranium-235 is approximately 52 kg, with a critical diameter of 17 cm. When fully fissioned, 1 kg of uranium-235 releases an energy equivalent to about 17,000 tons of TNT. In typical atomic bombs, due to the instantaneous nature of the explosion, the chain reaction is far from complete before the explosive material disintegrates. The fragmented uranium or plutonium falls below critical mass again, while the voids increase and neutrons escape, resulting in lower utilization efficiency. Taking the examples of "Fat Man" and "Little Boy," their efficiency is approximately 17\% and 1.4\%, respectively.

A hydrogen bomb is ignited by an atomic bomb. As atomic bombs are limited by the critical mass and cannot be made too small, even small atomic bomb-driven hydrogen bombs can cause large-scale explosions. Therefore, hydrogen bombs cannot be made too small, and the release of energy cannot be controlled. According to publicly available information, the smallest hydrogen bomb can only reach an equivalent of 1-10 kilotons of TNT.

\begin{figure}[htbp]
\begin{center}
\includegraphics[width=15cm]{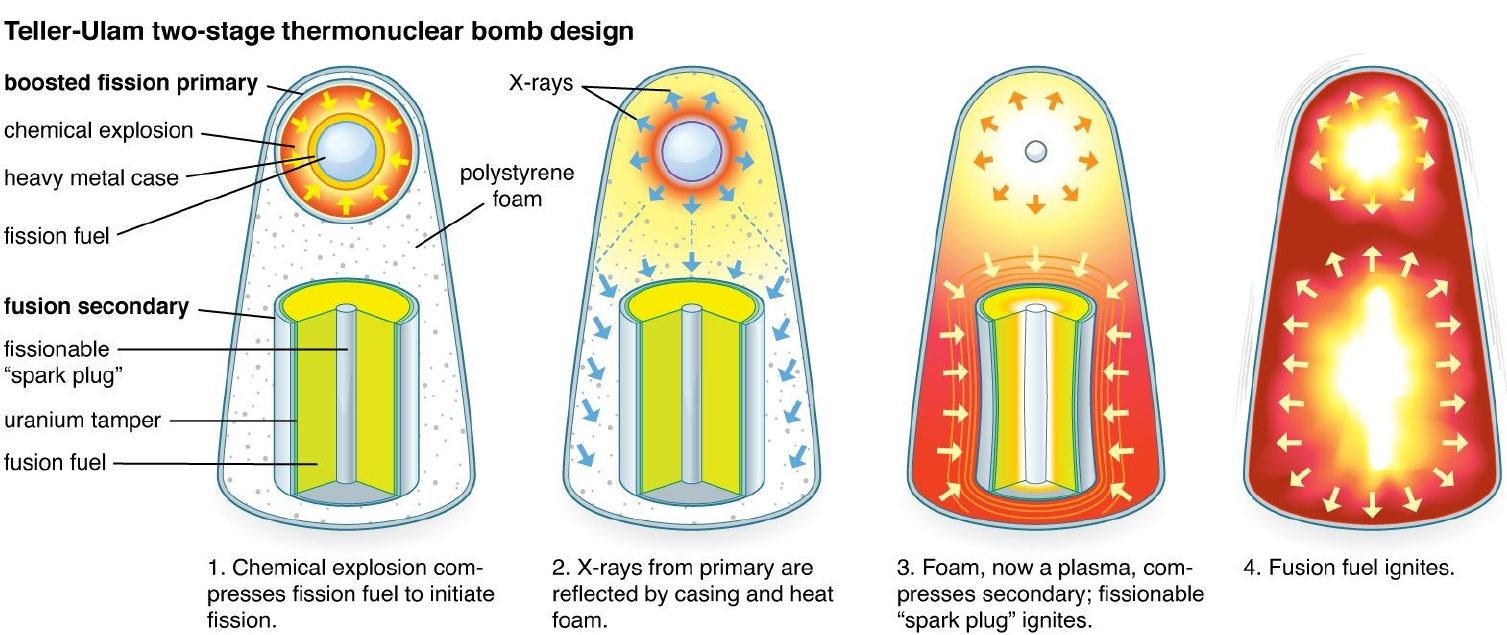}\\
\caption{Diagram illustrating the principle of a hydrogen bomb (Encyclopaedia).}\label{fig:HBomb1}
\end{center}
\end{figure}\begin{figure}[htbp]
\begin{center}
\includegraphics[width=10cm]{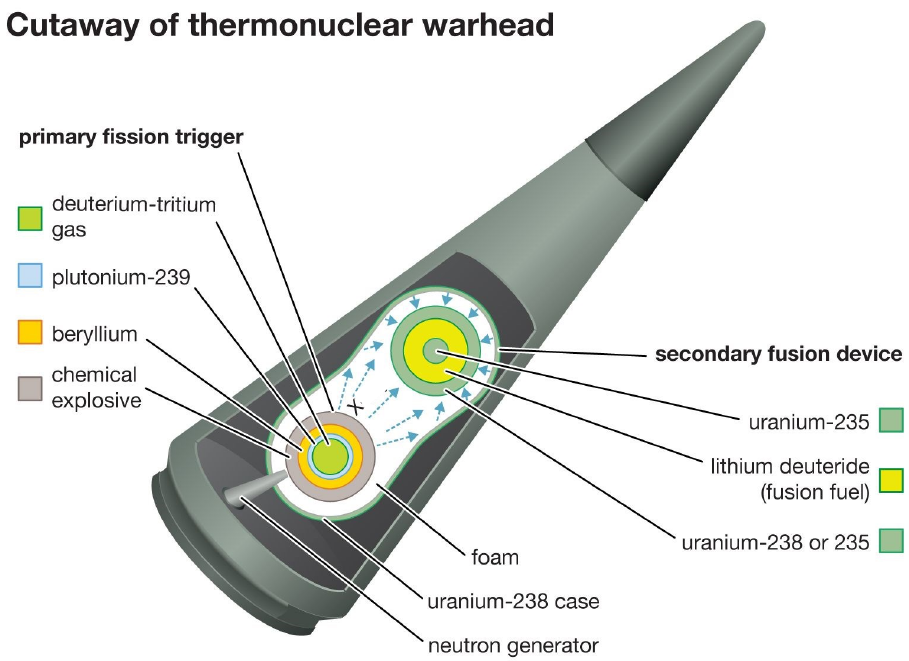}\\
\caption{Schematic diagram of a typical hydrogen bomb structure (Encyclopaedia).}\label{fig:HBomb2}
\end{center}
\end{figure}

The complete design details of a hydrogen bomb are still strictly confidential. We will discuss it quantitatively based on publicly available information. The process diagram is shown in Figure \ref{fig:HBomb1} and Figure \ref{fig:HBomb2}. The structure of the hydrogen bomb generally consists of two stages. The primary stage is for fission initiation, with uranium ${\rm{}^{235}_{~92}U}$ or plutonium ${\rm{}^{239}_{~94}Pu}$ as the raw materials. The secondary stage is for fusion with greater explosive power, with deuterium and tritium as the raw materials. The nuclear reaction is ${\rm D+T} \to {\rm n (14.07MeV) +{}^4He (3.52MeV)}$, or in more modern cases, using lithium ${\rm{}^{6}_{3}Li}$ deuteroxide, which utilizes the nuclear reaction ${\rm ^6Li+n} \to {\rm T+{}^4He +4.7MeV}$ to produce tritium. The thermal neutrons are provided by the fission reaction, with a reaction cross-section of up to 942b, resulting in high tritium efficiency. Lithium deuteroxide is a solid and does not require cooling compression, making it low in production cost, small in size, light in weight, and easy to transport. Additionally, it can avoid the problem of storing tritium due to its short half-life.

In the "Teller-Ulam configuration", the compression of the secondary stage to detonate the hydrogen bomb mainly relies on the ablative mechanism of the reflector/pusher layer. The X-rays generated by the fission in the primary nuclear bomb compress the secondary nuclear bomb, which is known as the radiation implosion of the secondary nuclear bomb. After internal compression, the secondary nuclear bomb is heated by the fission explosion inside it. The brief complete process is as follows: the high explosive initiates the fission reaction → the primary fission generates X-rays and $\gamma$-rays that heat the fusion bomb → the primary fission ends and the pusher layer surface ablates, pushing the compressed fusion bomb → the fusion material inside the fusion bomb reacts → the fusion reaction occurs.

In Figure \ref{fig:HBomb1}, for the pre-detonation hydrogen bomb, the spherical object on the top is the primary fission bomb, and the cylindrical body below is the secondary fusion bomb. After the detonation of the high explosive in the primary fission bomb, it compresses the primary fission bomb beyond its critical state. The primary fission bomb undergoes a fission reaction, reaching temperatures ranging from millions to billions of degrees, radiating gamma rays and X-rays, heating the inner wall and outer shell of the hydrogen bomb, as well as the reflective layer of the secondary fusion bomb, transforming the polymer foam layer into a plasma state. After the completion of the fission reaction in the primary fission bomb, it begins to expand. The rapid surface heating of the pusher layer of the secondary fusion bomb leads to surface ablation and ejection, pushing the remaining components of the secondary fusion bomb, including the pusher layer, fusion fuel, and fissionable materials, inward. The compressed secondary fissionable material reaches its critical mass, initiating fission and compressing the fusion material outward. At the same time, neutrons generated in the process react with lithium deuteride to produce tritium. The fuel in the secondary fusion bomb reaches a temperature of 300 million degrees, initiating a deuterium-tritium fusion reaction, which quickly burns out, generating additional neutrons that further induce uranium-238 in the outer wall to undergo fission. This forms a fireball, resulting in a massive explosion. The entire process occurs in less than 1 second. From the above process, it can be seen that the hydrogen bomb does not solely rely on fusion energy, and the proportion of fusion to fission energy release may vary depending on different design approaches. Its superiority over atomic bombs lies in the enhancement of the fission process by the fusion process, allowing for more complete combustion of the fissionable materials.

According to publicly available information, the typical peak time of a hydrogen bomb nuclear explosion is about 20ns, and the temperature can reach $10^2-10^3$keV. In order to prevent the temperature from becoming too high, adjustments will be made to the design details. Under high density conditions, the radiation is optically thick, so the radiation can be reabsorbed through inverse bremsstrahlung and the radiation absorption can be enhanced by a reflecting layer, thereby heating the fusion fuel. The radiation field can even reach thermal equilibrium [Atzeni (2004) Chap2]. According to information from Wikipedia, the radiation pressure of the first Ivy Mike hydrogen bomb was about 7.3TPa, and the plasma pressure was about 35TPa; the more modern W-80 hydrogen bomb has a radiation pressure of about 140TPa, and a plasma pressure of about 750TPa. The compression speed of the implosion is about 400km/s, assuming that the thickness of the fusion material before compression is 1 centimeter, then the compression time is about 25ns, which is close to the aforementioned 20ns. With these data, assuming that the explosion temperature of the hydrogen bomb is about $T\simeq100$keV, the plasma pressure is $P\simeq500$TPa, and assuming that the electron and ion temperatures are the same, the ion number density is estimated to be $n\simeq 1.6e28{\rm m^{-3}}$. Therefore, the corresponding $n_e\tau\simeq8e20{\rm m^{-3}\cdot s}$, which exceeds the ignition condition of deuterium-tritium fusion $n_e\tau\simeq2e20{\rm m^{-3}\cdot s}$, indicating the feasibility of this principle.

Based on the hydrogen bomb, people have also proposed a fusion energy scheme for underground cavern nuclear melting-reheat conversion power generation, and a conceptual design is described in the literature by Peng Xianjue (1997). There are two main obstacles to the development of fusion energy using this scheme: (1) In 1996, nuclear testing was fully banned internationally, requiring "non-proliferation" of nuclear weapons. (2) It is estimated that the scheme is still lacking in economic viability, mainly due to the cost of raw materials.Let's calculate the economics of using a hydrogen bomb. Assuming each bomb has an equivalent of 10,000 tons of TNT, its energy is $10^7\times4.18{\rm MJ}\simeq40{\rm TJ}$, which is approximately equivalent to 11.6 million kilowatt-hours of electricity. Assuming a competitive electricity price of about 0.2 RMB per kilowatt-hour, the price of converting it to electricity would be about 2.3 million RMB. In other words, for the power generation of a hydrogen bomb with an equivalent of 10,000 tons of TNT to be economical, its manufacturing cost must be much lower than 2.3 million RMB, considering energy conversion efficiency and other costs. As we are unable to obtain the true cost of a hydrogen bomb, it is impossible to accurately evaluate the actual feasibility of this scheme. If a hydrogen bomb with a larger TNT equivalent is used, the economics may improve, but it will also further complicate the control of the released energy.

Let's quantitatively calculate the $\rho R$ parameter corresponding to the hydrogen bomb as an inertial confinement fusion method. For a hydrogen bomb with an equivalent of 10,000 tons of TNT, if all of it comes from deuterium-tritium fusion, it would require about 0.1 kg of fuel. Assuming the density is only compressed to 100 times that of solid state, i.e., $\rho=10^5{\rm kg/m^3}$, at this point, the radius $R\simeq6.2$ mm, corresponding to $\rho R\simeq620{\rm kg\cdot m^{-2}}$. In Chapter \ref{chap:icf}, we calculated the $H_B$ parameter for the inertial confinement burning rate, which corresponds to a burning rate of 50\%, and at 30 keV, deuterium-tritium fusion $H_B\simeq50{\rm kg\cdot m^{-2}}$, which is an order of magnitude lower than the previously calculated $\rho R\simeq620{\rm kg\cdot m^{-2}}$. This means that the hydrogen bomb has enough margin in principle, so it is feasible. For hydrogen bombs with TNT equivalent of ten thousand tons, fusion gain can be achieved by compressing the density by 10-100 times. The feasibility of the hydrogen bomb scheme is mainly due to its large $R$ and high radiation optical thickness. The disadvantage is that this method is difficult to miniaturize because once the driver is smaller, the compression rate becomes smaller. And with a smaller compression rate, the theoretically required $R$ becomes larger, resulting in a larger TNT equivalent. According to available information\footnote{http://nuclearweaponarchive.org/Nwfaq/Nfaq4-4.html}, the W80 hydrogen bomb with a 15,000 ton equivalent can be compressed to a density of $\rho=720 {\rm g/cm^3}$ at a compression ratio of 878 times.

According to calculations, conventional chemical explosives cannot drive the fusion of hydrogen bombs to achieve gain. It can be seen that due to factors such as the inability to reduce the TNT equivalent, the need for a powerful atomic bomb as a driver, and the requirement of high radiation optical thickness, it is difficult to replicate the hydrogen bomb method in a controlled fusion reactor scheme. Based on its technical principles, the most critical limitation is that a hydrogen bomb can be made large but difficult to make small. Theoretically, advanced fuels such as deuterium-deuterium, deuterium-helium, and hydrogen-boron can be achieved, but the cost is that the lower limit of energy release in a single release is larger than that of a deuterium-tritium hydrogen bomb, making it uncontrollable.

\section{Magnetic Confinement Fusion}

The advantage of magnetic confinement lies in its potential to achieve steady-state confinement, thus enabling stable and continuous energy release for power generation. It is considered to be the mainstream direction in fusion energy research, especially represented by the tokamak. The parameters of tokamaks are close to the Lawson criterion, making it the most recognized approach to achieving fusion energy. A simple straight magnetic field can only confine the motion of charged electrons and ions in the plasma in the perpendicular direction, and cannot confine the motion in the direction parallel to the magnetic field lines. Therefore, for open magnetic field configurations, efforts need to be made to minimize the loss at the endpoints. For example, for a magnetic mirror configuration, the mirror ratio needs to be increased by enhancing the magnetic field at the endpoints to minimize the loss. From this perspective, a commonly well-confinement configuration is a closed magnetic field line arrangement. Among various options, a torus configuration is the most direct closed magnetic field line approach. Most high-parameter magnetic confinement devices can be regarded as torus devices, such as tokamaks, stellarators, reversed field pinches (RFPs), dipoles, etc.

Here, we need to pay special attention to the difference between the confinement time of energy and particles and the discharge time of plasma. The former is usually not long, with durations exceeding 1 second considered high, and can usually only be indirectly measured. In contrast, the discharge time can be significantly longer, for example, exceeding 1 hour, and its value can be directly obtained from the discharge waveform. The energy confinement time and discharge time can be analogized using a simple analogy: continuously dripping ink into a cup of hot water, the time it takes for each ink drop to spread from the center to the cup's rim represents the energy confinement time, which is usually short. As long as the cup and its water remain intact, and the action of dropping ink continues, it represents a sufficiently long discharge time (the single operation time of the device).

All magnetic confinement devices face the problem that, if advanced fuel is to be achieved, the magnetic field cannot be too small. The challenges then become how to address radiation issues, followed by confinement problems, and then impurity and heating issues. If deuterium-tritium fusion is pursued, there are additional challenges such as tritium breeding and high-energy neutrons. The quantitative conditions required are discussed in detail in Chapter \ref{chap:mcf}. Therein, it is also pointed out that the temperature and density ranges for magnetic confinement fusion energy systems are very narrow, with the only adjustable parameter being the energy confinement time.

For magnetic confinement devices, based on nearly 70 years of research, there are three approaches to improving the confinement capability and achieving high fusion parameters: increasing the device size, increasing the magnetic field, and improving the confinement through new physical mechanisms. Increasing the device size deviates from economic considerations to some extent, while increasing the magnetic field mainly presents engineering difficulties. New physical mechanisms are continuously being explored.

\subsection{Tokamak and its derivatives}
\begin{figure}[htbp]
\begin{center}
\includegraphics[width=12cm]{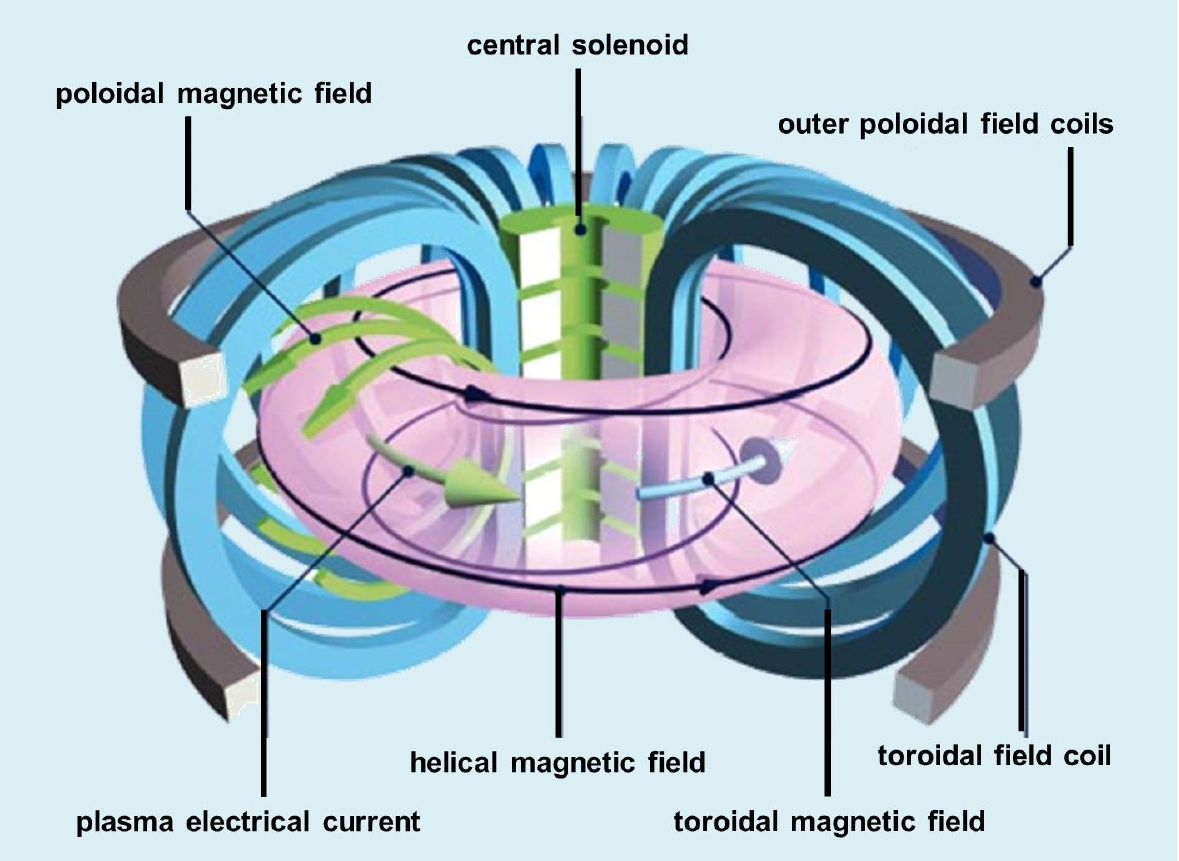}\\
\caption{Schematic diagram of tokamak magnetic field structure (from Wikipedia)\label{fig:tokamak}}
\end{center}
\end{figure}

The tokamak, with its magnetic field configuration composed of poloidal field and toroidal field, as shown in Figure \ref{fig:tokamak}, can be regarded as axisymmetric. The toroidal field is mainly provided by external toroidal field coils, while the poloidal field is balanced by the magnetic field generated by the internal plasma current and the magnetic field generated by the external poloidal field coils. It is currently the most successful scheme in magnetic confinement fusion, with the highest plasma parameters, and is approaching the scientific feasibility threshold for deuterium-tritium fusion. The highest ion temperature is around 46 keV, achieved by the US TFTR and Japan JT60U using neutral beam heating, which is about 500 million degrees; the highest plasma pressure is achieved by the US C-Mod device in 2016, reaching 2 atmospheres, corresponding to a density close to $2.5\times10^{20}{\rm m^{-3}}$; the longest energy confinement time is achieved by the UK JET, reaching nearly 1 second; the highest deuterium-tritium fusion power is achieved by JET in 1997, reaching 16.1 MW; the largest fusion energy release is achieved by JET in 2021, reaching 59 MJ; and the longest high-confinement mode operation is achieved by China's EAST in 2017, lasting 101 seconds. Currently, the world's largest fusion device under construction is the International Thermonuclear Experimental Reactor (ITER), which is based on the tokamak scheme.

There have been several important milestones in the development history of tokamaks. These include the confirmation in 1968 that the T-3 tokamak in the former Soviet Union achieved excellent performance with plasma temperatures reaching tens of millions of degrees, the discovery of high-confinement mode by Germany's ASDEX in 1982, and the parameter verification of close to deuterium-tritium fusion gain conditions achieved by the US TFTR, UK JET, and Japan JT60U around 1997, which confirmed the scientific feasibility of magnetic confinement fusion.The conventional tokamak has a lower beta compared to advanced fuels, with the highest beta achieved by the US DIIID reaching instantaneous values of over 10\%. Tokamaks have many derivatives, such as the spherical tokamak with a low aspect ratio. The aspect ratio $A = R/a$, where $R$ is the major radius of the plasma and $a$ is the minor radius. The beta value of a spherical tokamak can be higher, with typical values reaching up to 40\%. The main limitation lies in the engineering aspect, specifically when the aspect ratio is small, the space for the central column is limited, making it unable to accommodate large toroidal field coils. As a result, the toroidal field in a spherical tokamak is usually significantly lower than that of a conventional tokamak. Currently, the two largest spherical tokamaks, MAST-U and NSTX-U, only have a central magnetic field of 0.5-1T. 

Tokamaks can optimize their parameters to a certain extent by adjusting their shape, such as creating a negative triangular shape or elongating the plasma. For example, the recently achieved super H mode has higher boundary pedestal than the conventional H mode. With the development of low-temperature superconducting and high-temperature superconducting magnets, long-pulse high-field tokamak devices have become a key focus of recent years. The central magnetic field in these devices can reach above 10T.

Another major physical challenge of the tokamak is its requirement for high current drive. For a fusion reactor, the plasma current typically needs to be 5-10MA. High current can easily lead to instabilities and large disruptions, causing significant damage to the device.

In summary, the tokamak and its derivatives are currently at the forefront of fusion energy candidate solutions. The difficulty lies in further optimization and addressing issues related to power generation and economics. In order to study advanced fuel fusion based on the tokamak, there still needs to be a certain level of breakthrough in physics or the need for larger-sized and stronger-magnetic-field devices. The specific conditions that need to be achieved are detailed in Chapter \ref{chap:mcf}.

\subsection{Stellarator}

\begin{figure}[htbp]
\begin{center}
\includegraphics[width=11cm]{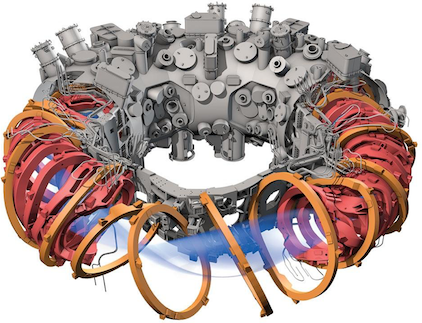}\\
\caption{Schematic diagram of a stellarator (source: wikipedia).}\label{fig:stellarator}
\end{center}
\end{figure}The stellarator, also known as a toroidal magnetic confinement fusion device, is shown in Figure \ref{fig:stellarator}. It uses complex three-dimensional coils to generate a twisted three-dimensional magnetic field structure to confine the plasma. The plasma current in the stellarator can be very small. Currently, the parameters of the stellarator are second only to those of the tokamak in magnetic confinement fusion and are approaching those of the tokamak. Through optimization, people have no doubt that it can achieve deuterium-tritium fusion conditions. Compared with the tokamak, its biggest advantage is that it does not have a major disruptive problem caused by plasma current. Its disadvantage is that the magnetic field is three-dimensional and requires extremely high precision. Achieving a strong magnetic field of 5T or even 10T is extremely challenging. Therefore, based on current technology, it is speculated that the stellarator is further away from achieving non-deuterium-tritium fusion compared to the tokamak-based approach. However, the advantage of not having a major disruptive problem could make the stellarator more competitive than the tokamak in terms of the overall fusion reactor in the future.

\subsection{Field-Reversed Configuration}

\begin{figure}[htbp]
\begin{center}
\includegraphics[width=12cm]{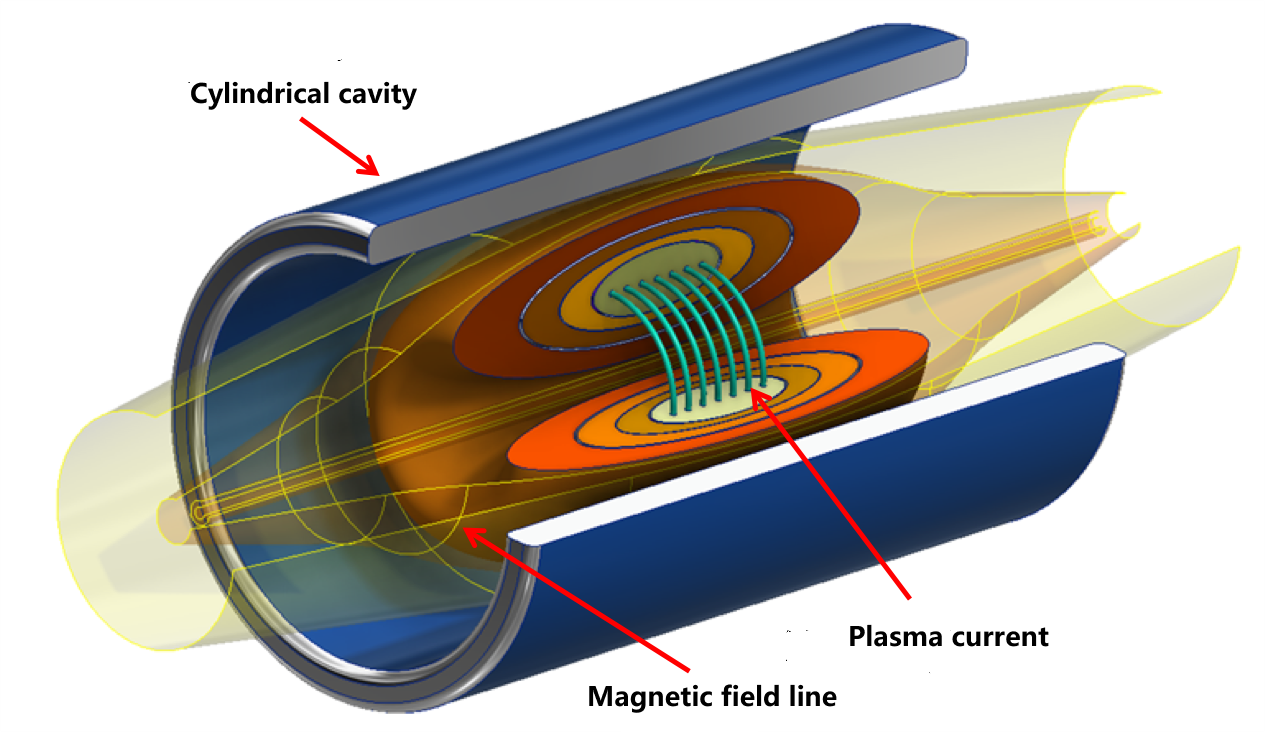}\\
\caption{Schematic diagram of the field-reversed configuration (Figure by Yang Yuanming).}\label{fig:frc}
\end{center}
\end{figure}

The field-reversed configuration (FRC), as shown in Figure \ref{fig:frc}, is a linear device. Its background magnetic field is provided by a magnetic mirror or a $\theta$-pinch, which is an open-ended system of magnetic field lines. However, by using an internal reversed plasma current, a reversed magnetic field is generated, which creates a closed magnetic surface in the central region, providing better plasma confinement than a magnetic mirror. Therefore, it can also be regarded as a closed magnetic field configuration. FRC has obvious advantages such as high beta, easy transfer, and direct energy conversion, making it a popular potential fusion scheme for research. The FRC can be considered as a type of compact torus. Other types of compact torus include spheromak and spherical tokamak, which are also important potential fusion schemes under investigation. The spherical tokamak has a central column and a toroidal magnetic field, while the spheromak does not have a central column but has a toroidal magnetic field. The FRC does not have a central column and has a weak or zero toroidal magnetic field.

The main difficulty of the field-reversed configuration (FRC) at present is that there has not been a significant breakthrough in improving the energy confinement, which means that although this approach has many advantages in theory, it is still unable to reach the level of fusion reactors in terms of its parameters. Some famous devices in the history of fusion research can be classified as FRC, including the early (1950s-1970s) Astron device, the C2/C2U/C2W series being developed by Tri Alpha company in recent years (2000-), and the collisional fusion compression FRC scheme being developed by Helion company. The Astron device proposed by Nick Christofilos aimed to use high-energy electrons rotating at high speed to generate the field-reversed configuration and achieve confinement of the plasma in the central region of the FRC. However, the experiment did not completely achieve the field reversal in the end, and it was not until later that they realized the difficulty of achieving fusion gain with this scheme due to the radiation from high-energy electrons. Therefore, they later considered using high-energy ions, but this posed even greater technological challenges, and there was no time to conduct relevant experiments. The scheme of Tri Alpha company takes into account many advantages of FRC as a fusion reactor and also aims to achieve the concept proposed by Rostoker (1997) of non-thermal hydrogen-boron fusion using high-energy ions rotating in large orbits and colliding with each other. They have made concrete efforts to improve the confinement parameters of FRC as a magnetic confinement route. However, the current actual Lawson triple product of their scheme is still more than four orders of magnitude away from the critical value for deuterium-tritium fusion gain. This is mainly because the energy confinement time has been increased from the 1-100 $\mu$s level in the early FRC to the ms level, but the density has also dropped from $10^{21}-10^{22}{\rm m^{-3}}$ to $10^{19}-10^{20}{\rm m^{-3}}$. The scheme of Helion company uses FRC as the target plasma to achieve magnetically inertial confinement fusion, aiming for transient high density. The difficulties it faces can be referred to in Chapter \ref{chap:mif}.

\subsection{Others}

There are many types of magnetic confinement fusion schemes, but no matter how they vary, they can be divided into two categories: open magnetic field lines and closed magnetic field lines. From an intuitive perspective, many people design various configurations to avoid problems with magnetic confinement fusion. However, in reality, even if a certain problem is avoided, there are usually other more serious problems. This has led to most magnetic confinement devices operating only at low parameters, with temperatures ranging from 1-100 eV and densities ranging from $1e16-1e19{\rm m^{-3}}$. Some devices can even maintain the ionized state of the plasma for a long time, like a fluorescent lamp, but the energy confinement time is very low.In Chapter \ref{chap:lawson}, we analyzed the power density requirements of magnetic confinement fusion reactors and pointed out that the suitable density range is very narrow, only within $10^{19}-10^{22}\mathrm{m^{-3}}$. The Lawson criterion requires a minimum deuterium-tritium fusion value of $n\tau_E>2\times10^{20}\mathrm{m^{-3}\cdot s}$, which means that the energy confinement time $\tau_E$ of the magnetic confinement device cannot be too low. For $n=10^{22}\mathrm{m^{-3}}$, $\tau_E$ should be at least 20 ms. For advanced fuels, a higher confinement time is required. However, in reality, the energy confinement time of most magnetic confinement devices is currently below 10 ms, or even below 1 ms. In particular, open magnetic surface configurations such as mirror and cusp have even lower confinement times compared to closed magnetic surface configurations. This makes it difficult for open magnetic surface configurations to break through in terms of physical parameters, despite their many advantages in engineering. Closed magnetic surface configurations also need to further increase their parameters in order to achieve fusion gain with advanced fuels.

Some common typical magnetic confinement configurations face the aforementioned difficulties. For example, for a mirror configuration, it is believed that a length of 1 km is needed to suppress losses at both ends in order to achieve fusion reactor parameters. For a dipole field, the magnetic field decreases as $1/R^3$ with respect to the major radius, which means that most of the region is a low magnetic field region. It is estimated that a diameter close to 50 meters would be needed to achieve fusion reactor parameters. Similarly, the reversed-field pinch (RFP) currently has low parameters and short discharge times, usually far less than 1 second. The biggest obstacle lies in the energy confinement parameter, which may be related to many factors such as instability and anomalous transport. If this problem cannot be solved, the device can only be made larger, which usually makes it economically unfeasible.

An ideal magnetic confinement scheme would combine the advantages of high beta with reversed field configurations, which are linear, easy to transfer, capable of direct power generation, have high parameters and confinement like a tokamak, and possess the stability of a stellarator without the need for current drive or large disruptions. Currently, a confinement configuration that possesses all these advantages has not been found, and this is the direction that scientists are continuously exploring. To achieve such a configuration, many complex physics issues need to be addressed. These complex physics issues often limit our ability to reach the zeroth-order parameters mentioned in previous chapters, but in some cases, they may exceed people's simple expectations, such as the discovery of the high confinement mode (H mode).

\section{Inertial confinement fusion}
To maintain controllable inertial confinement in a micro-explosion scenario, the allowed fuel mass is limited. The upper limit is generally considered to be in the range of several gigajoules (GJ). Otherwise, even with only a few shots per second, damage would be inflicted on the reactor vessel. The complete combustion of 1 milligram (mg) of deuterium-tritium (DT) releases 341 megajoules (MJ) of fusion energy. Assuming a combustion efficiency of 30\%, the fuel mass can only be a few tens of milligrams. The energy of 1 GJ is approximately equivalent to the energy of a 250-kilogram TNT explosion. Since the destructive force primarily comes from momentum rather than energy, micro-inertial fusion has lower destructive force compared to explosives due to its high velocity and small momentum. For the same kinetic energy $E=\frac{1}{2}mv^2$, with different velocities $v_1$ and $v_2$, the ratio of momentum is given by $p_1/p_2=m_1v_1/m_2v_2=v_2/v_1$. In other words, destructive force is inversely proportional to velocity. The speed of the shock wave produced by conventional explosives is approximately 1-10 ${\rm km/s}$, while the fusion product can reach $10^{3}-10^{4}{\rm km/s}$; therefore, the destructive force for the same energy can be lower by about three orders of magnitude. When the fusion generates a shock wave, the decreased speed leads to an increased destructive force for the same energy.

Considering controllable inertial confinement fusion schemes, the fuel mass $m=\rho V\propto\rho R^\alpha$ should not be too large, where the exponent $\alpha$ satisfies $1<\alpha\leq3$ and depends on the geometric dimension. Also, to achieve fusion gain, $\rho R$ should not be too small, so the density $\rho$ needs to be increased, that is, the compression ratio needs to be increased. This indicates that for a controllable inertial confinement scheme, the driver strength needs to be much higher than that in a hydrogen bomb to achieve a higher compression rate for the target plasma.

\subsection{Laser-driven}

\begin{figure}[htbp]
\begin{center}
\includegraphics[width=14cm]{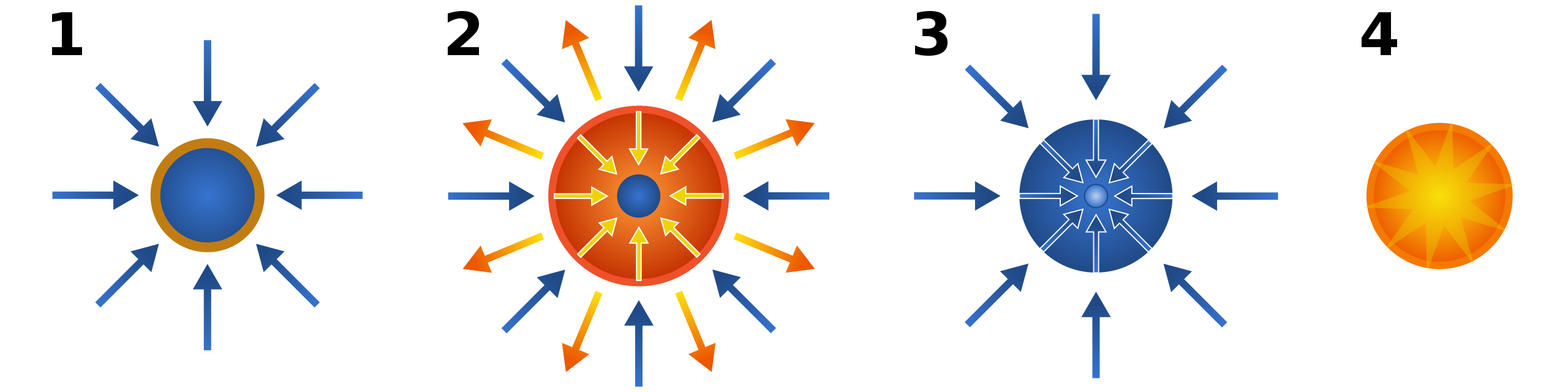}\\
\caption{Laser-driven inertial confinement fusion (wikipedia)}\label{fig:lasericf}
\end{center}
\end{figure}

In general, the schematic diagram of laser-driven inertial confinement fusion is shown in Figure \ref{fig:lasericf}. The blue arrows represent radiation, the orange arrows represent outwardly emitted energy, and the yellow arrows represent inwardly transported heat energy. The laser beam or X-rays generated by the laser rapidly heat the surface of the fusion target, forming a plasma state. Then, the fuel is compressed. After the implosion of the target capsule, the fuel density increases by tens of times, and the temperature reaches the fusion conditions. The thermonuclear fusion undergoes rapid combustion and outward diffusion, resulting in energy output far greater than input.

Laser inertial confinement fusion has many variations in addition to the aforementioned direct drive method, including indirect drive and central hot spot ignition.

Central hot spot ignition refers to increasing the density in the central region of a plasma, such as through a shock wave or rapid compression, so that the density in the central region becomes much higher than the density at the boundaries. Fast ignition is similar, where the central region of a target capsule is rapidly heated and compressed through a guided setup.

Indirect drive is mainly used to achieve more uniform compression. A chamber is placed outside of the target capsule, and when the laser is incident on the chamber, X-rays are reflected and compress the central target capsule more uniformly.

Currently, due to the rapid advancement of laser technology in terms of energy, pulse width, and focusing, the parameters for laser inertial confinement fusion are at the forefront of ICF fusion research. The latest experiments conducted by the National Ignition Facility (NIF) in the United States (in 2014, 2021, and 2022) have already surpassed the critical value for the feasibility of deuterium-tritium fusion. However, considering overall energy efficiency, the gain has not yet exceeded 1. Typical parameters for NIF include a density of $10^{32}{\rm m^{-3}}$, plasma volume of $10^{-7}{\rm cm^{-3}}$, discharge time of $10^{-10}{\rm s}$, laser output energy of 2MJ, and input laser electrical energy of 300-400MJ.

The progress of NIF experiments in fusion energy research has provided valuable insights. After the results of NIF's experiments in 2014 were published, where 1.8MJ of laser energy produced 14kJ of fusion energy (in a 2013 experiment), industry experts and reports from the U.S. Department of Energy concluded that achieving fusion energy gain with the existing NIF device was not possible. However, in an unexpected experiment in August 2021, 1.8MJ of laser energy produced 1.3MJ of fusion energy, exceeding the previous fusion energy output by an order of magnitude. In December 2022, it was further achieved that 2.05MJ of laser energy produced 3.15MJ of fusion energy, marking the first clear achievement of controlled fusion energy gain. The pessimistic outlook for NIF from industry experts was primarily based on calculations of first order and secondary quantities, such as instability. However, the experiments showed that these obstacles could be overcome. The NIF experiments, within the scope of calculations for fundamental quantities in this book, also indicate that for fusion research, it is a priority to focus on fundamental quantities, establish clear constraints, and determine research directions, while difficulties related to first order and secondary quantities can usually be overcome.

\subsection{Particle Beam Drive}

In addition to lasers, particle beams are also used as drivers for inertial confinement fusion. The coupling efficiency of particle beams to plasmas is usually higher than that of lasers and can exceed 20\%. Commonly used particle beams include high-energy electron beams, light ion beams, and heavy ion beams. Wang Gan-chang (2005) discussed the advantages and disadvantages of several drive methods, and p.47 also discussed the scheme of high-speed collision.These schemes each have their drawbacks. Furthermore, when extrapolated to the parameters of a fusion reactor, the cost is not low either. High-energy electron beams have relatively good cost and focusing, but high-energy electrons have strong radiation, which results in energy loss that is detrimental to fusion gain.

\subsection{Other Drivers}

Z-pinch, which uses a series of strong current filaments, heats up rapidly to a plasma state due to the strong current, while also generating a strong magnetic field that attracts each other, resulting in focusing compression and achieving high temperature and high density states to achieve fusion gain. Its driver is generally a high-current pulse capacitor.

The Dense Plasma Focus (DPF) approach [Gallardo (2022)] also uses the pinch effect to increase plasma density and temperature.

\section{Magneto-inertial Confinement}

There are also various schemes for magneto-inertial confinement, which can be classified into magnetic confinement, wall confinement, and inertial confinement.

\subsection{FRC Magnetized Target}
FRC-MTF, also known as a magnetized target fusion scheme based on field-reversed configuration, as mentioned earlier, can barely achieve the D-T critical condition, with great difficulty. The main limitations are the $\beta$ ratio and the compression time $\tau_{dw}$. Currently, no effective breakthrough has been found, and the principle has not been fully explored. A schematic diagram is shown in Figure 1. In practice, conducting a compression experiment may require long time for cleaning, from days to months. There is still a considerable distance from actual fusion energy research. Helion Energy (https://www.helionenergy.com/) is based on this scheme.\begin{figure}[htbp]
\begin{center}
\includegraphics[width=14cm]{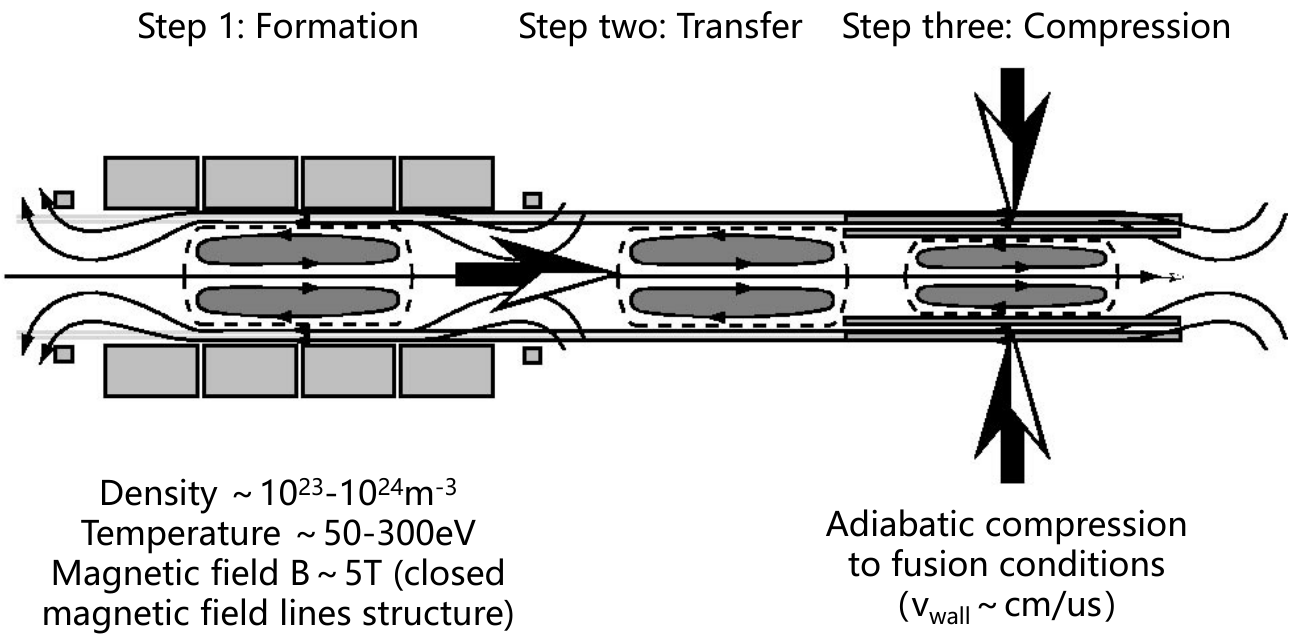}\\
\caption{Schematic diagram of the MTF scheme for FRC [Taccetti (2003)].}\label{fig:frcmtf}
\end{center}
\end{figure}

\subsection{Z-pinch magnetized target}
\begin{figure}[htbp]
\begin{center}
\includegraphics[width=12cm]{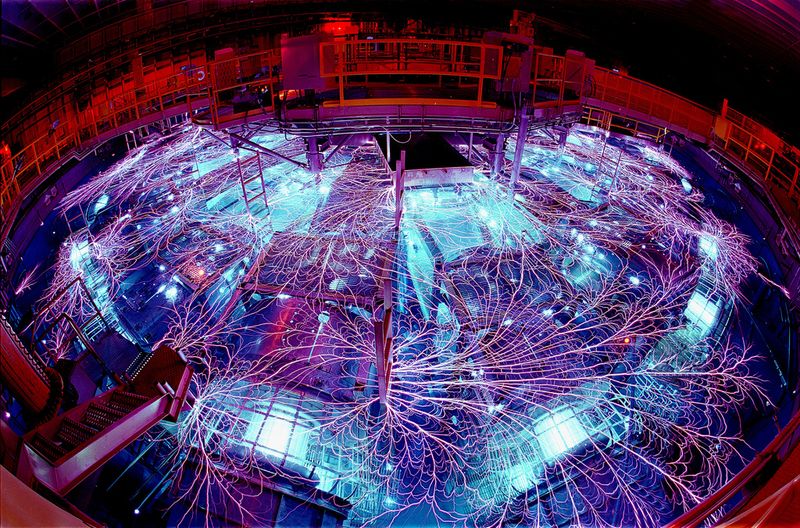}\\
\caption{Z-pinch device (Sandia) used in MagLIF experiments.}\label{fig:sandiaz}
\end{center}
\end{figure}

The MagLIF scheme developed by Sandia National Laboratories in the United States, as shown in Figure \ref{fig:sandiaz}, uses Z-pinch as the plasma target and combines it with laser technology to form a magnetically inertial confinement scheme. Essentially, it is still an inertial confinement scheme that follows the requirements of inertial confinement. For deuterium-tritium fusion, the gain also needs to exceed 100. Currently, its parameters are approaching the critical line for deuterium-tritium fusion gain. Yager-Elorriaga (2022) reviewed its progress. Typical parameters include a density of $10^{29}{\rm m^{-3}}$, plasma volume of $10^{-4}{\rm cm^{-3}}$, discharge time of $10^{-9}{\rm s}$, and a magnetic field of 100MG.

\subsection{Wall-Confined Magnetized Target}
The Russian MAGO device uses the method of explosive compression to achieve high parameters, but it is difficult to achieve $Q>1$. The power generation method is also a challenge. This experiment had made significant progress in the 1990s, but there has been less subsequent research. Garanin (2015) has a detailed summary of this scheme. The MAGO scheme belongs to the wall-confinement magnetized target method, with certain wall constraints, but it cannot be completely classified as wall-confinement. Chirkov (2019) mentioned the MAGO parameters, $\beta\simeq1$.

\section{Other Schemes}
This section discusses some other schemes that have not been discussed previously. Some of them can be assessed for feasibility using conventional Lawson criteria, while others require alternative approaches for assessment.\subsection{Wall Confinement}

In previous discussions, wall confinement, as well as inertial confinement, were mentioned in relation to magnetic inertial confinement. In order to surpass the $\beta$ pressure limit, it is believed that either wall confinement or inertial confinement must be adopted. However, through simple analysis, we can conclude that wall confinement is not feasible. This is mainly due to the fact that the collision cross section and ionization cross section of high-energy particles with wall materials are much larger than the Coulomb cross section, and even larger than the fusion reaction cross section. From the perspective of mean free path or mean collision time, after one fusion reaction has occurred, there have already been more than $10^4$ Coulomb scatterings and more than $10^5$ collisions with wall materials, each collision resulting in the loss of fusion fuel energy, leading to rapid cooling of the fusion fuel. As a result, this process cannot produce fusion energy, but instead heats the wall materials with fusion feedstock.

Therefore, in order to achieve fusion energy, direct contact between the fuel and the wall must be avoided as much as possible. However, the wall is not entirely negative. If the wall has excellent reflectivity to radiation, it can be used to create radiation pressure to confine the plasma from another perspective. This is the method used in thermonuclear bombs and inertial confinement fusion.

\subsection{Several Beam Confinement Schemes}

Regardless of whether it is accelerator-target impact or beam-beam collision, achieving fusion gain is either impossible due to the Coulomb scattering cross section being much larger than the fusion cross section, or by converting kinetic energy into internal energy to approach thermonuclear fusion. For a typical fusion temperature of 10 keV, the corresponding speed of deuterium nuclei is $v=6.8 \times 10^5$ m/s, which requires a collision speed of at least 700 km/s. For non-deuteron-tritium fusion, an even higher temperature is needed, and the corresponding speed is usually higher after accounting for the difference in mass numbers. If such high speeds are achieved through capacitor acceleration, for example, 0.1 mg of deuterium corresponds to approximately $3 \times 10^{22}$ particles, with an energy of 46 MJ. This is the limit of many current fast (less than microseconds) drivers, unless the cost is further increased, such as exceeding one billion US dollars per unit. Therefore, like other inertial confinement fusion schemes, the ultimate question becomes whether it is economically viable. Technological developments may change the assessment of its difficulty.

Below, we quantitatively evaluate several typical beam confinement schemes. These conclusions were already understood in early literature, such as Glasstone (1960). Different people have reintroduced the beam scheme from different perspectives in the hopes of achieving high fusion gain. However, for most schemes, their essence does not surpass this early understanding, meaning their practical feasibility is not high.\subsubsection{Beam-Target Fusion}

When a heated high-energy proton with an energy above 200 keV collides with a thick and wide boron target, according to the Lawson criterion, the temperature and density are already sufficient. The only requirement is for the penetration time to be long enough to exceed the value required by the Lawson criterion. However, this idea is not very feasible mainly due to energy loss during the penetration process. The energy of the high-temperature protons decreases to below 10 keV within a short distance, making it ineffective for fusion reactions. In other words, the proton energy is mainly used for ionization and target heating, rather than effective fusion. Atzeni (2004) has discussed this in detail.

As discussed in the preceding section on mean free path, it is inversely proportional to the reaction cross section. The fusion cross section is only about $1{\rm b}$, while the scattering ionization cross section of high-energy ions bombarding a low-temperature solid target is about $10^{7}{\rm b}$. Therefore, before a fusion reaction occurs, there have already been approximately $10^{7}$ collision scatterings, each scattering losing an energy of about $m_e/m_i$. For instance, in the case of deuterium-deuterium fusion, with an initial energy of 100 keV, an average of $13.4$ MeV is released in one fusion reaction, while for deuterium-tritium fusion, one fusion reaction releases $17.6$ MeV. However, the energy loss is $10^{7}\times0.1/3672=272$ MeV, which is far greater than the energy released in fusion. Furthermore, this process does not yet take into account the decrease in fusion reaction cross section and the increase in scattering cross section during the energy loss process, nor does it account for multiple small-angle scatterings. When these additional factors are taken into account, the energy loss will be even greater than the fusion energy release. Therefore, this method of bombarding a solid target with a beam cannot become a fusion energy source.

\subsubsection{Beam-Plasma Fusion}

For solid targets, the high ionization cross section leads to high energy loss. What if we replace it with an already ionized plasma target? We expect that the situation would improve. Let's evaluate the fusion gain brought about by injecting neutral beams or high-energy ion beams into a low-temperature plasma to see if this method is feasible.

In Chapter \ref{chap:lawson}, we evaluated this approach. When the background electron temperature is high and radiative and transport losses are not considered, rough critical gains of about $F=E_{fus}/E_{beam}\simeq4$ can be achieved for deuterium-tritium fusion, and gains less than 0.5 can be achieved for hydrogen-boron fusion. This is similar to the results calculated by Dawson (1971) and Moreau (1977). Therefore, the feasibility of this method cannot be completely ruled out at the moment, but implementing it is difficult from a zero order perspective. This approach is mainly due to the high initial kinetic energy of the beam $E_{beam}$, which, limited by the slowing-down time, leads to limited fusion energy production $E_{fus}$ and a small gain $F$. Dolan (1981) and Morse (2018) can also be referenced.Although the fusion scheme of injecting low-density high-energy ion beams into a background plasma at low temperature is difficult to achieve fusion energy, when the number of beam particles is sufficiently large, heating the background plasma to the thermalized state and achieving higher fusion gain is possible under sufficient constraint, with the high-energy ion beam mainly serving as a heating method.

\subsubsection{Ion Beam Heating of Solid Targets}
Let us calculate the energy loss caused by the ion beam bombarding a solid target, to determine whether it is possible to heat the solid target to fusion temperature, such as 10 keV, and thus achieve thermonuclear fusion.

Assume a 1 MeV deuterium beam with a current density of $100{\rm A/cm^2}$, which is already a strong beam. Therefore, the ion flux is $100/(1.6\times10^{-19})\simeq6\times10^{20}{\rm cm^{-2}\cdot{}s^{-1}}$, resulting in an energy flux of $6\times10^{26}{\rm eV\cdot cm^{-2}\cdot{}s^{-1}}$. Let us assume that the deuterium target bombarded by the 1 MeV deuterium consists of approximately 1.5${\rm mg/cm^{2}}$. The number of particles (ions + electrons) in 1 g of deuterium is approximately $2\times\frac{1}{2}\times6\times10^{23}$ (Avogadro's constant). Therefore, the number of particles in a volume of 1 $\rm cm^2$ of the target is $6\times10^{23}\times1.5\times10^{-3}=9\times10^{20}$. Let us assume that the energy accelerated by the deuterium is deposited in the deuterium target within an optimistic time of $10^{-6}$ s, which is definitely larger than the actual value, but let us first investigate what can be achieved in an optimistic scenario. In this time period, the total deposited energy is $6\times10^{26}\times10^{-6}=6\times10^{20}$ eV, resulting in an average energy per target particle of $6\times10^{20}/(9\times10^{20})\simeq0.7$ eV. Obviously, this energy is much lower than the several tens of keV required for fusion. In other words, in this beam-target heating method, most of the energy only slightly heats the target plasma and cannot effectively achieve fusion to realize energy gain [Glasstone (1960) p66].

\subsubsection{Beam-Beam Collision}

The scheme of beam-beam collision can increase the center-of-mass energy compared to the beam-target method, thereby achieving a higher fusion reaction rate. At the same time, it is possible to reduce energy losses, such as those lost to the wall, by confining all ions that produce fusion reactions through the beam. Quantitative calculations show that this method is not feasible. First, there is still the problem of collision scattering, where the Coulomb collision cross section is more than $10^3$ times larger than the fusion cross section, causing the particles to scatter instead of undergoing fusion. Fusion only accounts for a small proportion.

The second reason is that the beam current is usually very low, resulting in low fusion power and thus no economic viability. Assuming a collision of 50keV deuterium ions, the center-of-mass energy is 100keV. Assuming a beam current density of $100{\rm A/cm^2}$, the particle flux is $6\times10^{20}{\rm cm^{-2}\cdot{}s^{-1}}$. The velocity of the 50keV deuterium nucleus is about $2\times10^{6}{\rm m/s}$, resulting in a particle number density of deuterium of $3\times10^{18}{\rm m^{-3}}$. According to our previous calculation of power density, the fusion power at this density is only about $10^{2}{\rm W/m^3}$, which is not economically viable. For deuterium-tritium fusion, it is also only about $10^{4}{\rm W/m^3}$. Optimizing the collision energy would also be difficult to fundamentally improve.

This is mainly due to the limitation of the beam current of the charged beam produced by the accelerator, which is essentially caused by the repulsive interaction between particles of the same charge, making it difficult for the beam current to be large. In a thermalized plasma, due to the quasi-neutrality condition or when using a neutral beam injection instead of an ion beam injection, higher beam currents can be achieved, but ultimately it still leads to thermal nuclear fusion, and the feasibility is determined by the criterion for thermal nuclear fusion.

\subsubsection{Migma}

The Migma scheme is a well-known fusion scheme that uses accelerator beam collision, and several devices were constructed between the 1970s and 1980s, achieving significant progress. This scheme can easily accelerate ions to several tens of keV for deuterium-tritium fusion, and even 300keV for hydrogen-boron fusion. The beam density in this scheme is relatively low, but it can be circulated in a storage ring to achieve better confinement, thereby allowing for more collisions and increasing fusion probability. However, the beam is usually unstable, and elastic scattering processes result in significant synchrotron radiation energy loss.

It is claimed (Wikipedia) that Migma IV achieved a record confinement time of about 25 seconds in 1982, and the product of fusion temperature, density, and confinement time reached a record of $4\times10^{14}  {\rm keV\cdot sec\cdot cm^{-3}}$, which was only matched by the JET tokamak in 1987 when it reached $3\times10^{14} {\rm keV\cdot sec\cdot cm^{-3}}$.By ingeniously designed particle storage rings, ions can be captured to achieve good confinement, so that the ion beam can undergo repeated collisions, thereby increasing fusion gain. However, this kind of scheme encounters significant difficulties when extrapolated to fusion reactors. In the 1990s, Norman Rostoker proposed using FRC to confine the colliding ion beam, which is the origin of the TAE company. In principle, these schemes can achieve high energy and even high confinement time. However, the confinement time is usually incompatible with high density, making it difficult to meet the Lawson criterion.

\subsection{Electric field confinement}

In electrostatic confinement, ions and electrons have opposite charges, so it is difficult to confine one while confining the other. In other words, steady-state confinement cannot be achieved. From another perspective, in order to achieve steady-state confinement, the electric field pressure needs to balance with the thermal pressure, $E^2$, which typically has a value of 3e9V/m. Some discussions on the difficulty of electrostatic confinement can also be found in [Glasstone (1960) chap3] and [Roth (1993) p267].

Electrostatic confinement fusion is more popular among science enthusiasts and hobbyists. The most typical device is the fusor. With only a high-voltage power supply, some metal grids, and deuterium gas, fusion reactions can be achieved at low cost. The principle is that the outer metal grid is grounded, while the inner metal grid is negatively charged with a potential difference of tens of kV or hundreds of kV. This causes the gas to ionize on one hand, and accelerates the positively charged ions towards the center on the other hand. As more positive ions gather around the central metal grid, the potential becomes positive and the ions start to accelerate in reverse, resulting in oscillatory motion between the two metal grid spheres. During this process, the ions collide and undergo fusion reactions. However, the energy gain in this scheme is much less than 1, making it unable to achieve the economic output of fusion energy.\subsection{Hybrid Reactor}

The main idea behind the hybrid reactor is to combine the advantages of fusion reactors, which do not favor neutrons, with fission reactors, which require neutrons, in order to achieve better energy output. The success of the hybrid reactor primarily relies on the progress of fusion reactors. If the essential nature of the reactor is more similar to that of a fusion reactor, it still needs to meet the corresponding fusion criteria in terms of scientific feasibility. The main difference lies in the engineering aspect, specifically the use of neutrons produced from deuterium-tritium reactions as the startup neutrons for the fission reactor. If the essential nature of the reactor is closer to that of a fission reactor, then the initial neutrons produced by other methods currently available are already sufficient and low-cost for the fission reactor, eliminating the need for high-cost fusion reactors to generate neutrons.

Therefore, this approach is theoretically promising. However, in practical implementation, many details need to be carefully considered and studied. There are various designs for hybrid reactors both domestically and internationally, with some fusion components based on magnetic confinement such as tokamaks or spherical tokamaks, and others based on inertial confinement like Z-pinch.

\subsection{Cold Fusion}

Since the "cold fusion" event in 1989, despite its notorious reputation, research in this field has never ceased. The main characteristic of cold fusion is its lack of reproducibility. From the perspective of claimed energy output, it is not higher than that of chemical batteries. This fusion method has later been collectively referred to as "low-energy nuclear reactions (LENR)". Similarly, bubble fusion claims the observation of small point-like bubbles collapsing at extremely high temperatures after injecting MHz sound waves into a liquid, resulting in neutron production in deuterium-containing liquids. However, these claims have yet to be verified through repeatable experiments. Therefore, it is more reminiscent of a farce.

\subsection{Others}

Recently, lattice-confinement fusion (LCF) has received some attention, which refers to fusion reactions occurring within the metal lattice of a fuel. Due to the high electron density of the conductive metal, the possibility of repulsion between two light atomic nuclei is reduced, and lattice constraints can induce fusion reactions involving positively charged atoms. By bombarding deuterium-saturated erbium or titanium samples with gamma rays, the gamma rays sometimes cause the deuteron nuclei in the metal lattice to decompose into protons and neutrons. The split neutrons collide with the deuterium nuclei in the lattice, transferring some momentum to them. Normally, deuterium nuclei repel each other, but the electron-shielded deuterium nuclei have enough energy to overcome the Coulomb barrier. Most of the heating in LCF reactions occurs in an area with a diameter of only a few tens of micrometers. From the perspective of fusion energy, the actual parameters of this scheme still have a long way to go.

In a previous section, we mentioned muon-catalyzed fusion and pointed out its economic challenges. However, some literature has optimistic estimates. For example, Gross (1984), on page 16, discussed some muon-catalyzed fusion and mentioned that some literature believes that gain can be achieved.

Lightning is a natural form of high-voltage, intense ionization, which to some extent can provide fusion conditions. If lightning is to be used as a fusion driver, this method requires more detailed analysis based on the criteria of inertial confinement. A typical lightning strike has a voltage of 1-100 MV, a current of 0.1 MA, and a duration of 10-100 μs. Based on these calculations, the energy is about the 10 MJ level. This means that although lightning is powerful, the stored energy is not very high, only equivalent to a few kilowatt-hours.

\section{Brief Evaluation of Some Private Fusion Company Schemes}

In the past two decades, there have been an increasing number of private companies targeting fusion energy research and development. They have obtained funding from the private sector and carried out research and development for certain fusion schemes, while setting relatively aggressive roadmaps. Especially in recent years, private investments have accelerated. In the years 2020 to 2022 alone, over \$4 billion in venture capital has been received.We do not intend to provide a detailed evaluation of various enterprise plans here. This is because most of these plans aim to abandon mainstream schemes that have been extensively studied with higher parameters, hoping to achieve breakthroughs through some special technology. From our discussions, it is clear that the successful development of fusion energy does not rely solely on one particular technology. Even if a certain technology significantly improves a specific parameter, the overall performance may still have a significant gap compared to the requirements of fusion energy.

In order to be representative, we can discuss the difficulties of some typical private fusion enterprise plans from the perspectives of funding amount and research and development team size.

Tri Alpha is one of the most representative private fusion enterprises in the past twenty years. It was founded around 1998 in California, United States. The current team has over 100 members and has received a total funding of over 800 million USD. Their target is hydrogen-boron fusion energy based on the field-reversed configuration (FRC) device. As mentioned earlier, their main progress lies in achieving a significant increase in the energy confinement time of FRC, even reaching a level of 10 milliseconds, which greatly advances people's understanding of FRC. However, from the perspective of fusion triple product, their operating device C2W has a length exceeding 25 meters, which is still about four orders of magnitude away from the critical conditions for deuterium-tritium fusion gain, not to mention hydrogen-boron fusion gain. This is mainly because during the process of increasing the energy confinement time, the density has decreased and there has been no significant breakthrough in temperature. By a simple extrapolation, it is difficult to expect to achieve the conditions for deuterium-tritium fusion gain in a few years with a tenfold increase in funding and several times the size.

General Fusion, located in Canada, is another private fusion enterprise that has been researching for almost twenty years. With a funding of hundreds of millions of USD and a team of dozens of people, they adopt a compressed magnetized target fusion scheme, mainly using mechanical means to compress the spherical target plasma. Currently, the main focus of this scheme is to address instability issues to improve the parameters. However, their Lawson triple product parameters are still orders of magnitude away from the critical value for deuterium-tritium fusion. As discussed in the chapter on magnetically inertial confinement, we have pointed out that this scheme can barely reach the conditions for deuterium-tritium fusion. Even if the parameter conditions are met in physics, challenges still exist in tritium breeding, high-energy neutron shielding, and economic viability for power generation.

Similarly, Helion Energy in the United States adopts the collision fusion and magnetically compressed inertial confinement scheme based on FRC. In 2021, they received nearly 500 million USD in investment. The difficulties they face are inherent in magnetically compressed inertial confinement schemes.First Light, established in the UK in 2009, adopts the inertial confinement fusion scheme employing shock wave compression. This scheme accelerates the plasma by capacitors to achieve velocities of several hundred km/s, which corresponds to enough energy to generate fusion reactions during collision or compression processes. Currently, the company has achieved plasma velocities of approximately 25 km/s (data from 2019). It should be noted that for deuterium at 10 keV, the corresponding velocity is 670 km/s, meaning that in order to achieve deuterium-tritium fusion, the velocity needs to be increased at least 20 times. Since energy is proportional to the square of velocity, the energy of the capacitors needs to be increased by 400 times to accelerate the same mass to fusion conditions. This would result in a significant increase in cost. Alternatively, the mass of the accelerated plasma could be reduced, but this would also lower the energy released during each fusion event. Not to mention the difficulty of tritium breeding and the cost-effectiveness of power generation. In comparison, power generation using inertial confinement and magnetized inertial confinement fusion is more challenging than steady-state magnetic confinement.

Tokamak Energy in the UK is the first private fusion company to approach mainstream fusion routes. Established in 2009, they aim to use high-temperature superconducting magnets in the spherical tokamak configuration, taking advantage of strong magnetic fields and the compactness of low aspect ratio designs. Additionally, their research suggests that the empirical scaling law for spherical tokamaks extrapolates better than ITER's scaling law. Currently, they have achieved plasma temperatures of approximately 100 million degrees Celsius through neutral beam injection and magnetic reconnection. The company's progress demonstrates the feasibility of achieving high plasma parameters in compact small devices. With sufficient funding, it is expected that their parameters can be further improved, possibly reaching the conditions for deuterium-tritium fusion. The key challenge for fusion energy lies in solving the economic power generation problem of deuterium-tritium fusion.

The Commonwealth Fusion Systems (CFS) company, spun off from MIT in the United States, was established in 2018. It secured an initial \$200 million in funding and is developing high-temperature superconducting magnets. Their first device, SPARC, is a high-field tokamak designed to achieve net energy gain from deuterium-tritium fusion. The central magnetic field is designed to be 12.2 Tesla, and the auxiliary heating relies solely on ion cyclotron resonance heating. The volume of the vacuum chamber is only about 1/25th of ITER's. This approach fully utilizes MIT's decades of experience in high-field tokamak development and ion cyclotron resonance heating. In 2021, they successfully developed the prototype of a toroidal coil with a surface magnetic field of 20 Tesla. Shortly after that, they secured \$1.8 billion in new investments, making them the fusion company with the largest funding. It is widely believed in the industry that high magnetic fields can allow tokamaks to achieve high parameters in compact devices, and there is no doubt that CFS's approach can surpass the conditions for deuterium-tritium fusion gain. As for fusion energy, the key challenge remains how to solve the economic power generation problem of deuterium-tritium fusion, and currently, the company has not provided a clear solution.In recent years, there have been several private fusion enterprises emerging in China. In 2018, the ENN Group (Langfang, Hebei Province), an energy company, officially launched a fusion energy research project, but without a fixed route, instead exploring, researching, and constructing multiple devices including reversed-field pinch configurations. Its largest current device is a spherical tokamak (spherical torus). The annual research and development budget is about 200-300 million RMB, with a team of about 100 people. The ultimate goal is to achieve clean and pollution-free commercial fusion energy, thus targeting hydrogen-boron fusion. In 2022, ENN confirmed the roadmap for spherical torus hydrogen-boron fusion. 

In 2021, Energy Singularity (Shanghai) and Startorus Fusion (Xi'an, Shaanxi) were established domestically, securing financing of 400 million RMB and several hundred million RMB respectively in 2022. Energy Singularity's route is similar to the high-temperature superconducting, strong magnetic field tokamak of the US CFS. Startorus Fusion's route is similar to Tokamak Energy in the UK, but it prioritizes the magnetic reconnection compression technology, with its original team coming from the SUNIST laboratory at Tsinghua University. Due to China's rapid advancement in the entire industry chain capabilities over the past 10 to 20 years, as well as its economic development and pursuit of technological innovation, Chinese private fusion enterprises may surpass their international counterparts. It is still difficult to predict how to achieve the target parameters and how to generate electricity economically in the future.

According to statistics, there are currently more than 30 fusion private enterprises internationally, which cannot all be discussed here. However, most of the proposed solutions have low feasibility and are located in the lower left corner of the Lawson diagram, not attracting mainstream attention. For example, the magnet mirror-like concept proposed by Lockheed Martin in 2014 has been recognized by few mainstream fusion researchers and currently has extremely low parameters. The inertial confinement fusion concept based on hydrogen-boron avalanche reaction proposed by Australian company HB11 has yet to scientifically confirm the authenticity of the avalanche mechanism.

\section{Promising New Technologies that Could Change the Landscape}

In response to the key challenges of fusion energy, breakthroughs in some new technologies may have the potential to change the research and development landscape or progress of fusion energy. From the perspective of magnetic confinement, apart from the cyclotron radiation related to the magnetic field, steady-state strong magnetic fields have overall advantages and many positive effects on confinement and parameter enhancement. Even when the thermal power is fixed, a simple scaling law indicates that the fusion power $P_{fus}\propto B^4$, while the relationship between the volume and the radius of the device is usually $V\propto R^3$. In other words, increasing the magnetic field can significantly reduce the size of the device. In existing magnetic confinement devices such as tokamaks, mature technologies such as cooled copper conductors or low-temperature superconducting magnets are commonly used. The former, due to the heat generated by resistance, can only operate in short pulses, such as discharge times below 10 seconds; the latter is limited by current density, making it difficult to achieve high magnetic fields, and also requires a large and complex low-temperature system. The latest high-temperature superconducting magnet technology may change this situation, as it allows for large current density, strong steady-state magnetic fields, and relatively less demanding low-temperature systems, such as liquid nitrogen cooling. The development of high-temperature superconducting magnet technology may make long-pulse tokamaks with center magnetic fields exceeding 10T possible in the near future. Under strong magnetic fields, the physical properties of the plasma, including collision properties, may also undergo some changes, which may make it easier to achieve fusion parameters.

It is difficult to achieve precise implementation of the three-dimensional magnetic field in a stellarator, and the development of 3D printing technology may improve this situation.

For inertial confinement or magnetically inertial confinement fusion, the development of driver technology is crucial, including intensity, efficiency, and focusing, among other factors.

The process of extracting fusion energy is closely related to material technology, whether it is neutron-resistant materials required for neutron-containing fusion or heat-resistant materials required for non-neutron fusion.

In addition, if fusion energy gain is not pursued, the integrated technologies generated during fusion research and development can be used for neutron sources, space propulsion, and so on.

\vspace{30pt}
Key points of this chapter:
\begin{itemize}
\item Various fusion methods currently available are still a considerable distance away from commercialization and need to overcome a series of obstacles;
\item New progress and the development of new technologies have given us much hope;
\item Stellar gravitational confinement fusion has an extremely long confinement time and can confine radiation, but this method cannot be realized on the ground;
\item In principle, fusion energy can be achieved using the method of a hydrogen bomb, and even hydrogen-boron fusion can be achieved under existing technological conditions, but the difficulty lies in the enormous uncontrolled amount of fusion energy released by the fusion pile.
\end{itemize} 

\chapter{Summary and Outlook}\label{chap:summary}

Fusion energy presents immense challenges, but it is still worth pursuing until it is achieved. From the theoretical analysis of fusion, it is interesting that almost all key requirements are close to critical values, neither too low to be easily achievable nor too high to be impossible to attain [Dawson (1983)].

\section{Summary}

This book takes stock of the fundamental physics processes and parameter dependencies involved in fusion energy research, focusing on the relationships among various energies. It summarizes some of the core zero order conditions required for different fusion approaches, such as pointing out that temperature, density, and confinement time can only take certain values within a certain range, and it discusses the requirements of the Lawson criterion for different perspectives.

The most crucial factor influencing fusion is the value of the fusion reaction cross section, which limits the selection of fusion fuels to only a few types. Then there are the fusion materials, such as the scarcity of raw materials, the energy and proportion of product neutrons, which further narrows down the selection of materials. The limitation of fusion power density also determines that the density cannot be too high or too low. The fact that the Coulomb scattering cross section is much larger than the fusion cross section means that fusion mainly relies on thermonuclear fusion. The ratio of Bremsstrahlung radiation to reaction rate determines that the fusion temperature can only be selected within a certain range. The energy gain requires the product of density and energy confinement time to have a minimum value at a given temperature. When fusion power density limits the plasma density, the energy confinement time has a minimum value. The energy confinement time cannot be too long either, as it would cause product accumulation and hinder further effective fusion reactions.

We have also established more typical models for magnetic confinement fusion, inertial confinement fusion, and magneto-inertial confinement fusion, and analyzed their specific parameter requirements. We have pointed out that magnetically confined deuterium-tritium fusion is already physically feasible, with the main difficulties lying in tritium breeding, high-energy neutron shielding, and the economic viability of power generation. In the case of inertial confinement, the main challenges lie in compression ratio and driver efficiency. For deuterium-tritium fusion, the current condition of physical viability has been achieved, but there is still a large gap to achieve engineering and commercial feasibility, and the economic viability of power generation remains a challenge. For magneto-inertial confinement fusion, if it is based on magnetically confined target plasma, it will exceed the beta ratio limit and have difficulty meeting the requirements of energy confinement time. However, if it is based on inertially confined target plasma, it can be analyzed based on the criteria of inertial confinement fusion.At the same time, we summarized various fusion schemes and pointed out that the feasibility of solar fusion is due to the extremely high density achieved by the strong gravitational field, which can effectively confine radiation for over a million years, conditions that cannot be achieved on the ground. As for the feasibility of hydrogen bombs, the key lies in the high compression ratio provided by atomic bombs and the ability to confine radiation to a certain extent. The most crucial aspect is the amount of fuel, which reduces the requirement for compression ratio compared to controlled inertial confinement fusion. The characteristic of hydrogen bombs is that they are easy to make large but difficult to make small, resulting in uncontrollable energy output.

Among the various magnetic confinement schemes, tokamak parameters are currently leading and can achieve scientific feasibility conditions for deuterium-tritium fusion, but they still face the difficulty of major plasma current disruptions. Stellarator parameters rank second among magnetic confinement schemes and do not have the risk of current disruptions, but the difficulty lies in the extremely complex engineering. Other magnetic confinement schemes primarily have relatively low confinement parameters. As for inertial confinement fusion, laser-driven approaches have already achieved conditions for deuterium-tritium fusion gain, but the overall driver efficiency needs to be improved, as well as solving power generation issues. For magnetized inertial confinement fusion schemes, parameters for deuterium-tritium fusion are expected to be achieved, such as MagLIF and MAGO schemes, but they are difficult to advance with advanced fuels and pulsed power generation has the same difficulty as inertial confinement fusion.

For other schemes, each has its own challenges. Beam fusion has limitations in low gain or low energy density. Electric field confinement requires extremely strong electric fields or approaches inertial confinement, making it difficult to achieve economic viability. Cold fusion, lattice fusion, and other methods still lack effective, repeatable experimental evidence or have too low energy output. Muon catalyzed fusion is not currently economically viable.

In summary, for any fusion energy scheme based on existing optimization or newly proposed, it is necessary to first quantitatively test the zero order quantities mentioned in this book. Otherwise, it is easy to emphasize the advantages of a certain aspect and ignore the disadvantages of other aspects, resulting in lower overall performance. The analysis of zero order quantities can also indicate the key breakthrough points for specific schemes, such as high density or high confinement capability, increasing beta or improving magnetic fields, etc.

As an example, we list a list of questions for reference: 

\begin{enumerate}

\item Nuclear reaction:\begin{itemize}
\item What fusion fuel is being used: deuterium-tritium, deuterium-deuterium, deuterium-helium, hydrogen-boron, or others?
\item What are the advantages and disadvantages of each corresponding fuel?
\item Is it a thermonuclear fusion or close to thermonuclear fusion? If it is non-thermonuclear fusion, how is it achieved and maintained?
\item Is there a need to increase the reaction rate, and if so, by how much and how is it achieved?
\item Are there any high-energy neutron issues and how are they shielded?
\item Are there fuel breeding issues (tritium, helium-3) and how are they solved?
\item What are the temperature, density, and energy confinement for the Lawson criterion?
\item What is the target value for the gain factor Q, which is the ratio of fusion output energy to input energy?
\item What is the proportion of fusion fuel and how is it determined?
\end{itemize}

\item Confinements:

\begin{itemize}
\item What kind of confinement is being used: magnetic confinement, inertial confinement, magnetoinertial confinement, or others?
\item Is bremsstrahlung radiation transparent? If not, what is the opacity? If it is transparent, can bremsstrahlung radiation be utilized?
\item Are there strong synchrotron radiation losses and can they be reflected, absorbed, or utilized in some ways, and what is the utilization efficiency?
\item What are the other necessary parameter conditions, such as the need for a hot ion mode, the required magnetic field strength, and the desired beta, and how are they achieved?
\item How is the desired ion temperature achieved?
\item How long can the required density condition be maintained?
\item Can the confinement time requirements be met?
\item What is the burning rate of the fusion fuel and are there impurity issues?
\item What is the spatial distribution of temperature and density: uniform ignition or localized ignition?
\item Are there any other technical requirements, such as current drive, strong driver, etc., and how are they achieved?
\end{itemize}

\item Power generation:

\begin{itemize}
\item Is it mainly thermal conversion power generation or direct power generation, and what is the efficiency?
\item Can energy other than power generation be utilized, and how is it utilized?
\item What is the power density and total power?
\item What is the size of the device and total output power?
\item How is the overall economic viability?
\end{itemize}

\end{enumerate}

Asking the right questions will help us gain a deeper understanding of the difficulties we face and find the correct direction for research and development.\section{Prospects}

The challenge of controlled fusion energy is enormous, with demanding parameter space. Currently, numerous experiments, such as magnetic confinement fusion and inertial confinement fusion, have demonstrated the scientific feasibility of deuterium-tritium fusion gain. However, there are still challenges in terms of fuel scarcity, high-energy neutron shielding, power generation, and engineering and economic difficulties. Advanced fuels, such as deuterium-deuterium, deuterium-helium, and hydrogen-boron, have even more extreme parameter requirements and greater difficulty.

However, through the analysis presented in this book, there is no evidence to suggest that controlled fusion energy is infeasible in principle. Even hydrogen-boron fusion has achievable parameter space. As mentioned at the beginning of this book, the historical experience of science tells us that "as long as it does not violate the laws of physics, even the most difficult tasks will eventually be accomplished." Therefore, controlled fusion energy is still worthy of investment until realization. On the other hand, due to the pursuit of extreme parameters in fusion research and development, derived technologies may find significant applications in other fields.

In every choice made in the history of fusion energy research, the selected route was not because it was easy, but because other routes were even more difficult. Breakthroughs in new technologies may reevaluate the difficulty of routes and lead to new choices. The future of fusion energy depends on collective human wisdom. This is not a challenge only for fusion scientists to solve, but also requires breakthroughs in other scientific and technological fields, as well as support from political and economic environments.

Regarding fusion energy research, perhaps we can borrow a phrase from mathematician Hilbert, "We must achieve, we will achieve"\footnote{Original quote by Hilbert: "We must know, we will know."}.

\vspace{30pt}
Key points of this chapter:
\begin{itemize}
\item Summarized the entire book, with no evidence suggesting that controlled fusion energy is infeasible in principle.
\item The future of fusion energy is promising, but it still requires collective human wisdom.
\end{itemize}

\appendix

\chapter{Appendix 1: Some Basic Information}\label{chap:appendix}

\section{Basic Constants}

Vacuum permeability: $\mu_0 = 4\pi\times10^{-7}{\rm H/m}$

Boltzmann constant: $k_B = 1.3807\times10^{-23}{\rm J/K}$

Avogadro's constant: $N_A = 6.0221\times10^{23}{\rm mol^{-1}}$

Gravitational constant: $G = 6.6726\times10^{-11}{\rm m^3\cdot{}s^{-2}\cdot{}kg^{-1}}$

Elementary charge: $e = 1.6022\times10^{-19}{\rm C}$

Proton mass: $m_p = 1.6726\times10^{-27}{\rm kg} = 938.27{\rm MeV}$

Neutron mass: $m_n = 1.6749\times10^{-27}{\rm kg} = 939.57{\rm MeV}$

Electron mass: $m_e = 9.1094\times10^{-31}{\rm kg} = 0.511{\rm MeV}$

Vacuum permittivity: $\epsilon_0 = 8.8542\times10^{-12}{\rm F/m}$

Speed of light in vacuum: $c = 2.99792458\times10^{8}{\rm m/s}$

Planck's constant: $h = 6.6261\times10^{-34}{\rm J\cdot{}s}$

Classical electron radius: $r_e = e^2/4\pi\epsilon_0m_ec^2 = 2.8179\times10^{-15}{\rm m}$

Fine-structure constant: $\alpha = e^2/2\epsilon_0hc = 1/137.038$

Proton-electron mass ratio: $m_p/m_e = 1836.1$

Stefan-Boltzmann constant\footnote{To distinguish it from the fusion reaction cross-section $\sigma$, in the main text, we use $\alpha$ instead of the Stefan-Boltzmann constant.}: $\sigma = 5.6705\times10^{-8}{\rm W\cdot m^{-2}\cdot K^{-4}}$

Gravity acceleration $g=9.8067{\rm m\cdot s^{-2}}$

Unit conversion ${\rm 1eV=1.6022\times10^{-19} J=1.1604\times10^{4} K}$

\section{Basic Data}

1 kWh = 1 kW$\cdot$h = $3.6\times10^6$ J = 3.6 MJ (approximately equivalent to lifting 3.6 tons of water to a height of 100 meters)

1 cal = energy required to raise the temperature of 1 g of water by 1$^{\circ}$C = 4.1868 J

TNT (trinitrotoluene) = 4.184 MJ/kg

The energy of oil and natural gas is about 40-55 MJ/kg, with an average of about 45 MJ/kg.

Coal has a large variation, about 20 MJ/kg.

Dry wood: 16 MJ/kg

Number density of atmospheric particles $2.69\times10^{25}{\rm m^{-3}}$

1 atmosphere $\simeq 10^5{\rm Pa}$

1 atmosphere $\simeq$ 10 meters of water pressure

Depth of the Mariana Trench $\simeq 10^4$ meters $\simeq 10^3$ atmospheres

Material pressure limit: about $10^3$ MPa

Natural uranium (0.7\% U-235 and 99.3\% U-238), light-water reactor (conventional reactor), 500 GJ/kg

Natural uranium, U and Pu cycles in light-water reactors, 650 GJ/kg

Natural uranium, fast-neutron reactor, 28,000 GJ/kg

\section{Some Characteristic Scales}The gyroradius will restrict the steady-state magnetic confinement device and cannot be too small, generally requiring $r>\rho_{ci}$.

Debye length
\begin{equation}
\lambda_D=\Big(\frac{\epsilon_0k_BT_e}{n_ee^2}\Big)^{1/2}=2.35\times10^5(T_e/n_e)^{1/2} ~{\rm m}, ~T_e~{\rm in~ keV}.
\end{equation}

Larmor radius
\begin{equation}
\rho_{cs}=\frac{v_\perp}{\omega_{cs}}=\frac{(2m_sT_s)^{1/2}}{q_sB}.
\end{equation}
Here, we take $v_\perp^2=2v_T^2$, thermal velocity $v_T=\sqrt{k_BT/m}$, and cyclotron frequency $\omega_{cs}=qB/m_s$. For electrons,
\begin{equation}
\rho_{ce}=1.07\times10^{-4}\frac{T_e^{1/2}}{B}~{\rm m}, ~T_e~{\rm in ~keV}.
\end{equation}
For ions,
\begin{equation}
\rho_{ci}=4.57\times10^{-3}\Big(\frac{m_i}{m_p}\Big)^{1/2}\frac{T_i^{1/2}}{Z_iB}~{\rm m}, ~T_i~{\rm in~ keV}.
\end{equation}

Cyclotron frequency
\begin{equation}
\omega_{cs}=\frac{q_sB}{m_s}, ~f_{cs}=\frac{\omega_{cs}}{2\pi}.
\end{equation}
For electrons,
\begin{equation}
\omega_{ce}=\frac{eB}{m_e}=0.176\times10^{12}B~{\rm~ s^{-1}}, ~f_{ce}=\frac{\omega_{ce}}{2\pi}=28.0\times10^9 B~{\rm ~Hz}.
\end{equation}
For ions,
\begin{equation}
\omega_{ci}=\frac{ZeB}{m_i}=95.5\times10^{6}\frac{Z}{A}B~{\rm s^{-1}}, ~f_{ci}=\frac{\omega_{ci}}{2\pi}=15.2\times10^6\frac{Z}{A} B~{\rm Hz}.
\end{equation}

Thermal velocity (note: some definitions may differ by $\sqrt{2}$)
\begin{equation}
v_{Ts}=\sqrt{k_BT_s/m_s}.
\end{equation}Electrons
\begin{equation}
v_{Te}=1.33\times10^7\sqrt{T_e} \text{ m/s},~T_e~\text{in~keV}.
\end{equation}
Ions
\begin{equation}
v_{Ti}=3.09\times10^5\sqrt{T_i/A} \text{ m/s},~T_i~\text{in~keV}.
\end{equation}

The mean free path (collision time multiplied by $v_T$), for ions with $Z=1$ , electrons
\begin{equation}
\lambda_e=v_{Te}\tau_e=\Big(\frac{k_BT_e}{m_e}\Big)^{1/2}\tau_e=1.44\times10^{23}\frac{T_e^2}{n\ln\Lambda}~\text{m}=8.5\times10^{21}\frac{T_e^2}{n}~\text{m}, ~T_e~\text{in~keV}.
\end{equation}
Where $\ln\Lambda\simeq17$. For ions
\begin{equation}
\lambda_i=v_{Ti}\tau_i\simeq v_{Te}\tau_e=\lambda_e.
\end{equation}

\section{Fusion Reaction Cross Section and Reactivity}

When two nuclei ($X_1$ and $X_2$) merge, a fusion reaction occurs, forming heavy nucleus $X_3$ and light nucleus $X_4$, usually represented as
\begin{equation}
X_1+X_2 \to X_3+X_4,
\end{equation}
or
\begin{equation}
X_1(x_2,x_4)X_3.
\end{equation}
Due to conservation of momentum and energy, the energy released is inversely proportional to the mass of the fusion products $X_3$ and $X_4$.

\subsection{Reaction Cross Section}
The rate of nuclear reactions is represented by the reaction cross section $\sigma$, which represents the probability of the reaction by the equivalent target area of the incident particles colliding with the target. Its value is related to the energy. The fusion reaction cross section is usually expressed in the following form:
\begin{equation}
\sigma(E)=\frac{S(E)}{E}e^{-\sqrt{E_G/E}},
\end{equation}
where $E$ is the center-of-mass energy,
\begin{equation}
E=\frac{1}{2}m_rv^2,~~v=|{\bm v}|=|{\bm v}_1-{\bm v}_2|,~~m_r=\frac{m_1m_2}{m_1+m_2},
\end{equation}
where ${\bm v}_1$ and ${\bm v}_2$ are the velocities of the two reacting nuclei in the laboratory coordinates.
The atomic structure factor $S(E)$ (astrophysical factor) is usually slow-varying with energy. The Gamow energy representing the strength of the Coulomb barrier is given as
\begin{equation}
E_G=(Z_1Z_2\pi e^2)^2{\frac{2m}{{\hbar}^2}}=(31.39Z_1Z_2)^2A_r[{\rm keV}],~~A_r=\frac{m_r}{m_p}.
\end{equation}

\begin{figure}[htbp]
\begin{center}
\includegraphics[width=15cm]{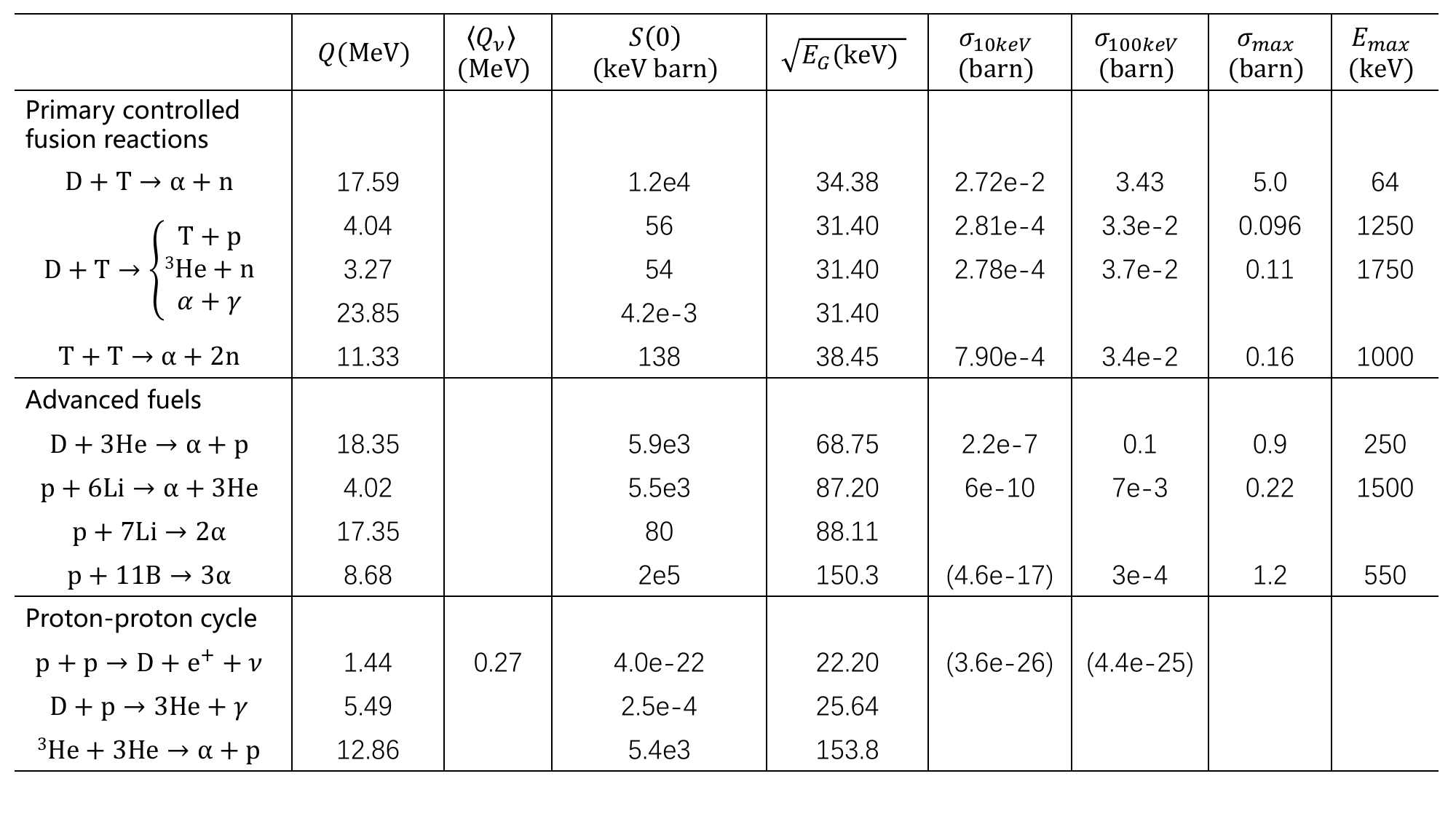}\\
\caption{Fusion cross section data [Atzeni (2004)].}\label{fig:sigmadat1}
\end{center}
\end{figure}

Different data sources have slight differences in the reaction cross section. Cox (1990) lists a relatively complete set of nuclear reaction channels related to thermonuclear fusion and gives cross section data for four advanced fuels. Here, we are only interested in a few representative data with high reaction cross section, which are shown in Figure \ref{fig:sigmadat1}.

\begin{figure}[htbp]
\begin{center}
\includegraphics[width=15cm]{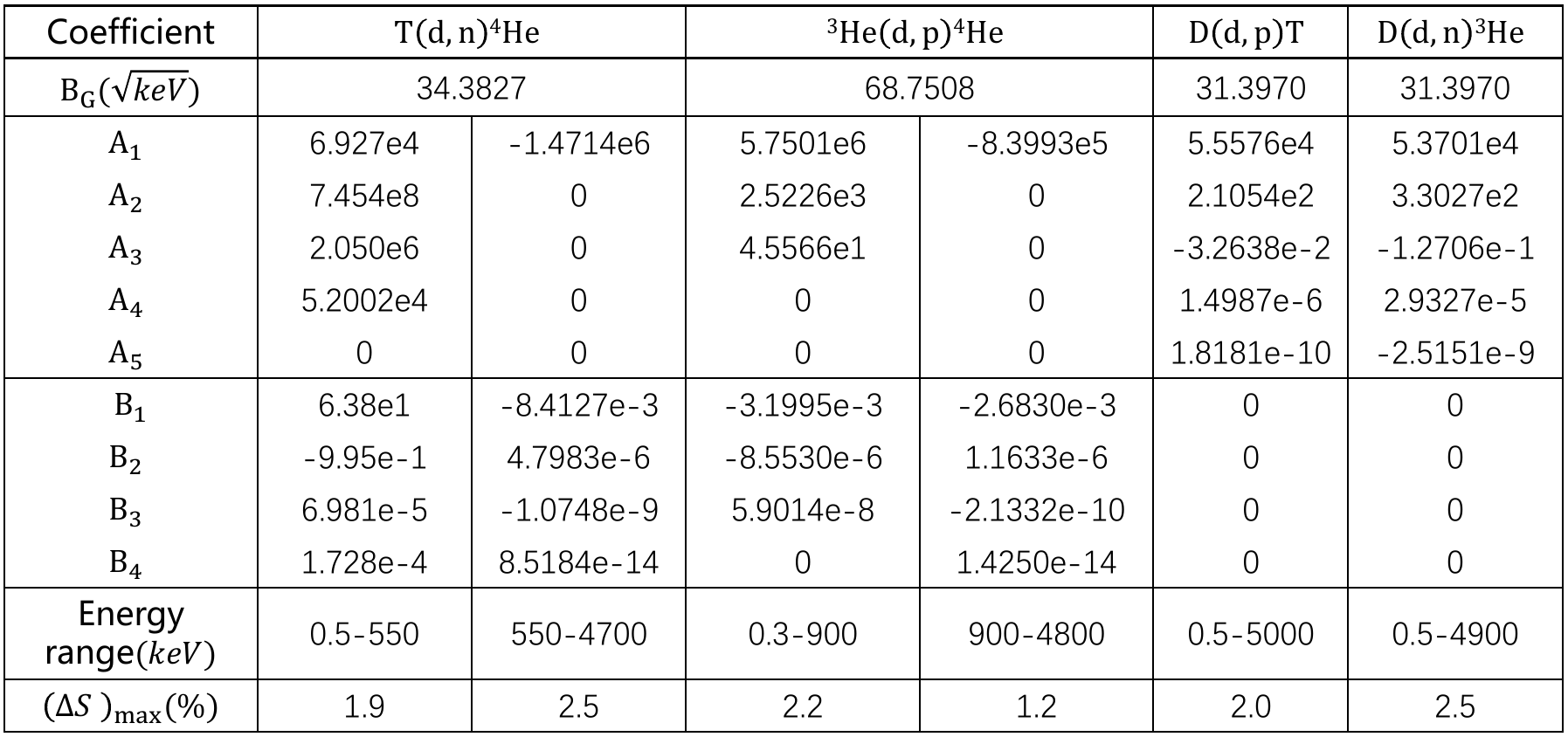}\\
\caption{Fitting coefficients for dt, ddn, ddp, and dhe3 fusion cross section data from Bosch (1992), $B_G=\sqrt{E_G}$.}\label{fig:Bosch92dat}
\end{center}
\end{figure}

We are mainly interested in five reactions, namely dt, ddn, ddp, dhe3, pb11, and pp, and we prefer to use fitting formulas. The NRL Handbook [Richardson (2019)] provides some reaction data and fitting formulas for light elements, but according to Bosch (1992), the fitting accuracy is not sufficient. Therefore, for dt, ddn, ddp, dhe3, we use the data from Bosch (1992)
\begin{equation}
S(E)=\frac{A_1+A_2E+A_3E^2+A_4E^3+A_5E^4}{1+B_1E+B_2E^2+B_3E^3+B_4E^4}.
\end{equation}
The energy E is measured in keV, and the cross section is measured in mb ($\rm{mb=10^{-31}m^2}$).

For the reaction pb11, we use the fitting data from Nevins (2000), which is described in segments due to the resonance peak. Additionally, we also use the data from Sikora (2016) for comparison.

For the reaction pp, we use the data from Angulo (1999),
\begin{equation}
S(E)=3.94\times10^{-22}(1+11.7\times10^{-3}E+75\times10^{-6}E^2),~~B_G=22.20.
\end{equation}
Some of these fitting formulas are listed in Atzeni (2004). For some reaction data not listed here, they are mostly provided in the comprehensive review article by Angulo (1999), including the sub-reactions of the aforementioned reactions and other possible advanced fuels. It should be noted that the reactions ${\rm T+T}$, ${\rm T+^3He}$, ${\rm ^3He+^3He}$, and ${\rm p+^6Li}$ also have significant reaction cross sections. When calculating the reaction rates in a complete fusion reactor, these sub-reactions should also be taken into account. This book mainly discusses the zero-order quantity of fusion and does not focus on these sub-reactions.

Figure \ref{fig:fusion_cross_section} shows several major fusion reaction cross sections.

\subsection{Reactivity}

The fusion reaction rate (reactivity) is given by,
\begin{equation}
\langle\sigma v\rangle=\int\int d{\bm v}_1d{\bm v}_2\sigma(|{\bm v}_1-{\bm v}_2|)|{\bm v}_1-{\bm v}_2|f_1({\bm v}_1)f_2({\bm v}_2),
\end{equation}
where $f_1$ and $f_2$ are the normalized distribution functions of the two ions, i.e., $\int f_{1,2}d{\bm v}=1$.

The number of nuclear reactions occurring per unit volume per unit time is given by,
\begin{equation}
R_{12}=\frac{n_1n_2}{1+\delta_{12}}\langle\sigma v\rangle,
\end{equation}
where $n_1$ and $n_2$ are the number densities of the two nuclei. If the nuclei are different, $\delta_{12}=0$; if they are the same, $\delta_{12}=1$ because the reaction is calculated twice when the nuclei are identical.

For both ions following the Maxwellian distribution
\begin{equation}
f_j(v)=\Big(\frac{m_j}{2\pi k_BT_j}\Big)^{3/2}\exp\Big(-\frac{m_jv^2}{2k_BT_j}\Big),
\end{equation}
where $m_j$ is the mass of particle $j$, $k_B$ is the Boltzmann constant, and $T_j$ is the temperature of particle $j$ [Nevins00], we have
\begin{equation}
\langle\sigma v\rangle_M=\sqrt{\frac{8}{\pi m_r}}\frac{1}{(k_BT_{r})^{3/2}}\int_0^{\infty}\sigma(E)E\exp\Big(-\frac{E}{k_BT_{r}}\Big)dE,
\end{equation}
where the effective temperature $T_r$ is defined as 
\begin{equation}
T_{r}=\frac{m_1T_2+m_2T_1}{m_1+m_2}.
\end{equation}
Since the fitting range for $\langle\sigma v\rangle_M$ in literature such as Bosch (1992) is primarily below 100 keV, this book mainly utilizes a direct numerical integration method for calculating the reaction cross-section rather than using fitted formulas, in order to ensure accuracy within the range of 1-1000 keV.

For the pp reaction (since the reaction rate is very low and difficult to measure experimentally, it is determined through theoretical models), the approximate expression is given by
\begin{eqnarray}\nonumber
\langle\sigma v\rangle_{pp}&\simeq&1.32\times10^{-43}T^{-2/3}\exp\Big(-\frac{14.93}{T^{1/3}}\Big)\\
&&\times(1+0.044T+2.03\times10^{-4}T^2+2.25\times10^{-7}T^3-2.0672\times10^{-10}T^4)~{\rm m^3/s}.
\end{eqnarray}
In the range of 1-10 keV, it is lower by about 24-25 orders of magnitude compared to DT fusion. At the center of the Sun, the energy released by fusion reactions is approximately 0.018 W/kg, roughly 1/50 of the energy generated by human metabolism.

For the dt reaction, within the range of 8-25 keV, a simplified fitting formula is often used, with an error within 15\%
\begin{equation}
\langle\sigma v\rangle_{dt}\simeq1.1\times10^{-24}T^2~{\rm m^3/s}.
\end{equation}

It is also important to note that the reaction rate may vary under conditions such as high density, polarization, and charged outer shell of the nucleus. Atzeni (2004) [Atzeni2004] provides some discussion on this.

\section{Magnetic pressure and plasma beta ($\beta$)}
The thermal pressure of the plasma is given by
\begin{equation}
P=nk_BT=\sum_sn_sk_BT_s=1.6\times10^{4}n_{20}T_{keV}{\rm [Pa]}, 
\end{equation}
where $n$ is the number density, $k_B$ is the Boltzmann constant, and $T$ is the plasma temperature.When there are multiple components, the pressures of each component should be added together. 

Magnetic pressure is given by
\begin{equation}
P_B=\frac{B^2}{2\mu_0}=3.98\times10^{5}B_{T}^2{\rm [Pa]},
\end{equation}

The total pressure ratio beta and the pressure ratio of each component can be calculated using
\begin{equation}
\beta=\frac{P}{P_B}=\frac{2\mu_0P}{B^2},~~\beta_s=\frac{P_s}{P_B}=\frac{2\mu_0P_s}{B^2}.
\end{equation}
It should be noted that for temperature, the unit conversion is given by $T_{eV}=k_BT/e$, where $T_{keV}$ is the temperature in keV, $n_{20}$ is the density in units of $10^{20}{\rm m^{-3}}$, and $B_T$ is the magnetic field in Tesla. The standard atmospheric pressure is approximately $\rm 1atm=1.013\times10^5Pa\simeq10^5Pa$. Therefore, 1 atmosphere is equivalent to a magnetic field of approximately 0.5T.

\begin{table}[htp]
\caption{Magnetic pressure and thermal pressure under typical fusion reactor parameters, note that the following density refers to the sum of the electron and ion densities. }
\begin{center}
\begin{tabular}{c|c|c|c|c|c|c}
\hline\hline
Typical device& $T$ (keV) & $n$ ($\rm m^{-3}$)  & $B$ (T) &  $P_B$ (Pa)  &  $P$ (Pa) & $\beta$ \\\hline
Physics experiment & 2 & $5\times10^{19}$ & 1 & $4.0\times10^5$ & $1.6\times10^4$  & 0.04  \\
D-T reactor & 10 & $10^{20}$ & 5 & $1.0\times10^7$ & $1.6\times10^5$ & 0.016  \\
D-D/D-He reactor & 50 & $1\times10^{21}$ & 10 & $4.0\times10^7$ & $8.0\times10^6$ & 0.2  \\
p-B reactor & 200 & $2\times10^{21}$ & 20 & $1.6\times10^8$ & $6.4\times10^7$ & 0.4  \\
\hline\hline
\end{tabular}
\end{center}
\label{tab:Pandbeta}
\end{table}%

From Table \ref{tab:Pandbeta}, it can be seen that, considering the constraint that $\beta<1$ typically for magnetic confinement devices, the current engineering of magnetic technology is sufficient for a deuterium-tritium reactor. However, for deuterium-deuterium or deuterium-helium reactors, increased confinement, decreased density, or development of stronger magnetic field technology is required. As for a hydrogen-boron reactor, the conditions are very demanding, and both low density (high confinement) and strong magnetic field technology need to approach the technological limits. Even if these conditions are achieved, there will still be other issues, which are discussed more systematically in the main text.

\section{The $\beta$ Limitation of Magnetic Confinement}

The $\beta$ limitation of magnetic confinement arises from two aspects: equilibrium and instability. We can prove that the average $\beta$ is less than 1 based on one-dimensional equilibrium.

\begin{figure}[htbp]
\begin{center}
\includegraphics[width=15cm]{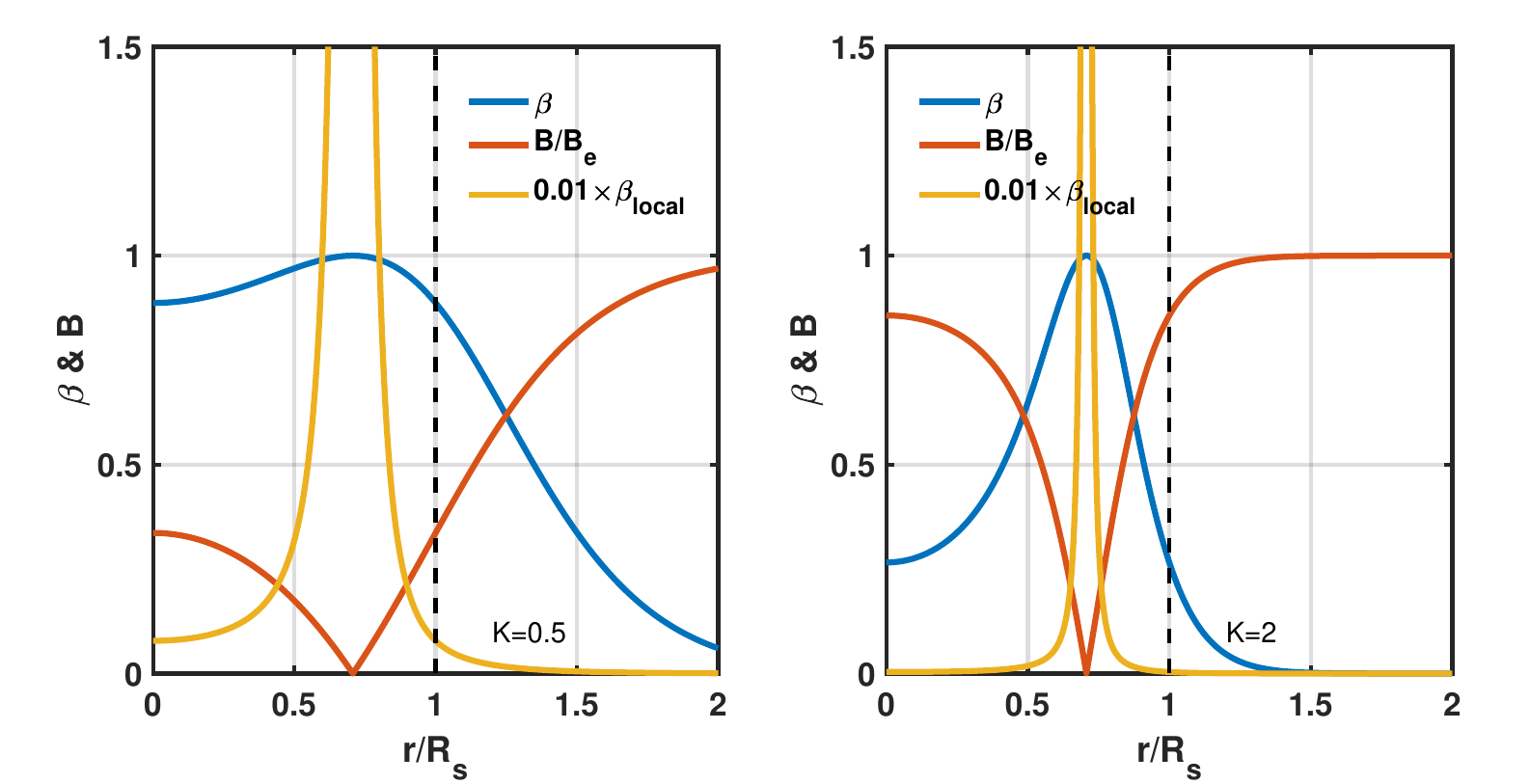}\\
\caption{In the equilibrium model of a rigid rotor, the local $\beta_{local}$ can be much greater than 1, allowing for the breakthrough of $\beta$ limitation and the reduction of synchrotron radiation caused by magnetic field.}\label{fig:betaRR}
\end{center}
\end{figure}

The magnetohydrodynamic equilibrium equation is given by
\begin{eqnarray}
J \times B=\nabla P.
\end{eqnarray}
In the cylindrical coordinates $(r, \theta,z)$, the one-dimensional equilibrium can be written as
\begin{equation}
J_{\theta} B_{z}-J_{z} B_{\theta}=\frac{\partial P}{\partial r}. \end{equation}
For an ideal FRC (Field-Reversed Configuration), $B_\theta=0$. Using Ampere's law, $ \mu_{0} J=\nabla \times \boldsymbol{B}$, and integrating, we obtain
\begin{equation}
P+\frac{B_{z}^{2}}{2 \mu_{0}}=\frac{B_{e}^{2}}{2 \mu_{0}}+\int \frac{B_{z}}{\mu_{0}}\left(\frac{\partial B_{r}}{\partial z}\right) d r,
\end{equation}
where $B_e$ is the external magnetic field magnitude. The last term represents the magnetic field curvature effect, which can be neglected in the midplane quasi-one-dimensional equilibrium. Thus, we have
\begin{equation}
P+\frac{B_{z}^{2}}{2 \mu_{0}}=\frac{B_{e}^{2}}{2 \mu_{0}}.
\end{equation}
Furthermore, by using 
\begin{equation}
\beta_{local} = \frac{2\mu_0P}{B^2},
\end{equation}
it can be seen that near the magnetic axis, $B \simeq 0$, hence the local beta pressure can be significantly larger than 1, $\beta_{local} \gg 1$. However, for a plasma or fusion device, the beta pressure commonly used is defined as
\begin{equation}
\beta = \frac{2\mu_0P}{B_e^2}.
\end{equation}
Since the thermal pressure $P \geq 0$ and the magnetic pressure $B^2/2\mu_0 \geq 0$, it follows that:
\begin{equation}
0 \leq \beta \leq 1.
\end{equation}
In terms of mechanical equilibrium, the upper limit for beta pressure is $\beta_{max}=1$. In practice, FRC can achieve $\beta \simeq 1$, spherical tokamaks can achieve $\beta \simeq 0.4$, and conventional tokamaks, due to stability limitations, typically have $\beta \leq 0.1$.

In the book, a zero-dimensional model with $\beta_{max}=1$ is employed to discuss the parameters of magnetic confinement and magnetic inertia constraints. It is also mentioned that the magnetic field should not be too high when discussing the issue of synchrotron radiation. If an FRC configuration can be created in which the central region has high thermal pressure and low magnetic field, and the boundary region has low thermal pressure and high magnetic field, it would satisfy the requirements of magnetic confinement equilibrium and stability, and significantly reduce the limitations posed by beta and synchrotron radiation.

This is one of the reasons why some researchers believe that FRC is a promising configuration for achieving economical fusion energy. However, in reality, the energy confinement time for current FRC experiments is still relatively short, less than 10ms. The radial balance model for a rigid rotor FRC is given by
\begin{equation}
P = \frac{B_e^2}{2\mu_0} \cdot \text{sech}^2(K \cdot u), ~~u = 2r^2/R_s^2 - 1,
\end{equation}
where $u$ is the radial minor radius variable, $R_s$ is the separatrix radius, $B_e$ is the magnitude of the magnetic field outside the vacuum region, and $K$ is a free parameter. Figure \ref{fig:betaRR} shows typical profiles for beta pressure and magnetic field.

\section{Electric Field in Non-neutral Plasmas}When considering a quasi-steady-state plasma, the Poisson's equation can be used to calculate the relationship between the electric field and charge density:
\begin{equation}
\nabla\cdot{\bm E}=\frac{\rho}{\epsilon_0},~~\rho=\sum_jq_jn_j,
\end{equation}
where ${\bm E}=-\nabla\phi$ is the electric field, $\phi$ is the electric potential, $\rho$ is the charge density, $q_j$ and $n_j$ are the charge and number density of particles of species $j$, and $\epsilon_0$ is the vacuum permittivity. $e$ represents the charge of an electron.

Under cylindrical symmetry, we have:
\begin{equation}
\nabla\cdot{\bm E}=\frac{1}{r}\frac{\partial}{\partial r}(rE_r)=\frac{\sum_jq_jn_j}{\epsilon_0}.
\end{equation}

Assuming only electrons and ions with charge $Z$ are present:
\begin{equation}
\frac{1}{r}\frac{\partial}{\partial r}(rE_r)=\frac{e(Zn_i-n_e)}{\epsilon_0}=\frac{e\delta n(r)}{\epsilon_0},
\end{equation}
where $\delta n=Zn_i-n_e$ represents the deviation of the particle number density from quasi-neutrality. Solving the above equation, we obtain:

\begin{equation}
E_r(r)=\frac{1}{r}\frac{e}{\epsilon_0}\cdot\int r\delta n(r) dr.
\end{equation}

To have an order of magnitude understanding, let's assume $\delta n(r)$ is constant for $r\in[0,a)$. The radial electric field becomes:
\begin{equation}
E_r=\frac{er\delta n}{2\epsilon_0},
\end{equation}
and the radial potential difference is:
\begin{equation}
\phi(r)=\int E(r) dr=\frac{er^2\delta n}{4\epsilon_0}.\end{equation}
That is,
\begin{equation}
\delta n(r)_{[m^{-3}]}=\frac{4\epsilon_0}{ea^2}\phi(a)=2.21\times10^8\frac{\phi_{[V]}}{a^2_{[m]}}.
\end{equation}
For a plasma with $a=1{\rm m}$, if the potential difference is $100{\rm kV}$, which means electrons can be accelerated to $100{\rm keV}$, the corresponding density difference is
\begin{equation}
\delta n(r)_{[m^{-3}]}=2.21\times10^8\frac{10^5}{1^2}=2.21\times10^{13}~{\rm m^{-3}}.
\end{equation}

Conclusion and Discussion:
(1) For a plasma with a density of $10^{17}{\rm m^{-3}}$, even if an electric field of up to $100{\rm kV}$ can be formed internally, the deviation from quasi-neutrality is only about one ten-thousandth. This is why in most cases, it can be assumed that the fusion plasma is quasi-neutral, that is, $Zn_i\simeq n_e$.
(2) Non-neutral plasmas generally have low densities, $n<10^{14}{\rm m^{-3}}$.
(3) If it is possible to achieve high-density non-neutral plasmas, it can indeed bring new possibilities to fusion. The problem is how to achieve high-density non-neutral plasmas and what difficulties there may be.

\section{Plasma Current and Current Density}

We discuss the nature of current and look at the quantitative relationship. The current density $J$ comes from the directed drift motion of charge $v_j$, that is,
\begin{equation}
J=\sum_jq_jn_jv_j.
\end{equation}
The total current $I=J\cdot S$, where $S$ is the current-carrying area. If only electrons are considered and it is assumed that all electrons have an average drift velocity $v_d$, then
\begin{equation}
J = en_ev_d = 1.6022 * n_e [10^18 m^{-3}] * v_d [10^7 m/s] [MA/m^2].
\end{equation}
Notice that the relativistic energy of an electron is $0.511{\rm MeV}$, thus the energy corresponding to velocity $v_d$ is 
\begin{equation}
E = 0.511 * v_d^2 / c^2 MeV = 0.568 v_d [10^7 m/s]^2 keV.
\end{equation}
In other words, electrons with a density of $10^{18}{\rm m^{-3}}$ and a drift energy of $568{\rm eV}$ (i.e., drift velocity of $10^7{\rm m/s}$) can generate a current of $1.6022{\rm MA/m^2}$.
For protons, the relativistic energy is $938{\rm MeV}$, namely
\begin{equation}
E = 938 * v_d^2 / c^2 MeV = 10.4 v_d [10^6 m/s]^2 keV.
\end{equation}

Table 1: Current density at typical densities and drift velocities.
\begin{center}
\begin{tabular}{c|c|c|c|c}
\hline\hline
& Density (${\rm m^{-3}}$) & \thead {\normalsize Drift \\ \normalsize Energy (${\rm keV}$)} & \thead {\normalsize Drift \\ \normalsize Velocity (${\rm m/s}$)} & \thead {\normalsize Current\\ \normalsize Density (${\rm MA/m^2}$)} \\
\hline
Electron & $10^{18}$ & $0.568$ & $10^{7}$ &1.6 \\
Electron & $10^{19}$ & $56.8$ & $10^{8}$ & 160 \\
Proton & $10^{19}$ & $10.4$ & $10^{6}$ & 16 \\
Proton & $10^{19}$ & $200$ & $4.4\times10^{6}$ & 70   \\
\hline\hline
\end{tabular}
\end{center}
\label{tab:Jv}

Table \ref{tab:Jv} lists the current density under typical density and drift velocity conditions. It can be seen that the current carried by electrons with the same energy is much larger than that of ions. To increase the current density, either the drift velocity or the density needs to be increased. The upper limit of velocity is the speed of light. To further increase the current with a certain energy, it is necessary to increase the current density or the area.

It should also be noted that drift current generates a corresponding magnetic field in the perpendicular direction. The magnetic field generated by a typical straight current is given by
\begin{equation}
B=\frac{\mu_0 I}{2\pi r}=0.2\frac{I_{\rm [MA]}}{r_{\rm [m]}}~{\rm [T]}.
\end{equation}
And the magnetic field generated by a circular current loop at the geometric center is given by
\begin{equation}
B=\frac{\mu_0 I}{2 r}=0.628\frac{I_{\rm [MA]}}{r_{\rm [m]}}~{\rm [T]}.
\end{equation}
From the above values, it can be seen that for a typical pulse fusion device with a size of 0.1 m, if the magnetic field is around 1 T, the corresponding current is approximately in the MA range. To further improve the parameters, either the size needs to be smaller or a larger driver is required.

\section{Fusion Heating}

The three key parameters for fusion are density, energy confinement time, and ion temperature. Achieving high density is not difficult itself, and the confinement time is determined by the confinement method. However, the ion temperature is a parameter that needs to be directly overcome, and it is necessary to explore which technological methods can help achieve the required temperature conditions for fusion. For deuterium-tritium fusion, an ion temperature of approximately 5-30 keV is needed; for deuterium-deuterium and deuterium-helium-three fusion, an ion temperature of approximately 30-100 keV is required; for hydrogen-boron fusion, approximately 150-400 keV is needed. How can such high temperatures be achieved?

For inertial confinement fusion, high-temperature plasma is usually achieved through the driver itself and rapid compression methods, such as strong laser energy, Z-pinch current, and X-rays from a hydrogen bomb.

For magnetic confinement, the most effective method is Ohmic heating. Due to the resistance of the plasma, voltage is applied to the plasma through the principle of a transformer, generating current, and then the energy is transferred to the plasma for heating through resistive dissipation. Both Ohmic heating and magnetic compression heating utilize the rapid rise of external coil currents for driving and can be classified as electromagnetic heating. However, since the plasma resistance decreases rapidly with the increase in temperature, Ohmic drive can usually only heat the plasma to the keV level.

Therefore, auxiliary heating is particularly important for magnetic confinement fusion. Common methods include wave heating and high-energy particle heating. The former includes plasma waves in various frequency ranges, such as electron cyclotron resonance heating (ECRH, usually above 50 GHz), lower hybrid waves (LHW), ion cyclotron resonance heating (ICRF), and Alfven waves (usually below 10 MHz). The latter includes fusion charged products and neutral beam injection (NBI).

In the case of magnetic inertial confinement, such as Field-Reversed Configuration (FRC), coil electromagnetic induction, heating, and compression are mainly achieved through the pulsed circuit of the capacitor.

Existing heating methods have made it relatively easy to heat to the 10 keV level required for deuterium and tritium, mainly because of the cost. However, it is currently highly challenging to heat ions to the 50 keV or even 200 keV required for advanced fuels. For inertial confinement and magneto-inertial confinement, the main challenge lies in the strength of the drivers. For magnetic confinement, the possible methods currently include neutral beam injection and ion cyclotron resonance heating. Neutral beam injection generates high-energy beam currents through an accelerator, which are neutralized and injected into the plasma, transferring energy to the main plasma through processes such as ionization and collision. Accelerators can easily achieve beam energies of several tens or even hundreds of keV. JT-60U has demonstrated the achievement of plasma temperatures close to 5 billion degrees using neutral beams. Ion cyclotron wave heating is the only plasma wave heating method that can effectively directly heat ions, while other wave heating methods mainly heat electrons and then indirectly heat ions through electron-ion collisions. However, there is currently no experimental demonstration that ion cyclotron resonance heating can heat ions to above 30 keV. The main issue with neutral beams is their high cost, and under fusion reactor conditions, high-energy beams are required due to penetration depth issues, which rely on negative ion source technology to increase the neutralization rate.

The ultimate goal is to achieve the direct heating of charged fusion products and achieve self-sustainment.

\section{Thermodynamics Laws and Fusion Energy}

The first law of thermodynamics is the law of energy conservation. This law indicates that a "first kind of perpetual motion machine" is not possible. In this book, the law of energy conservation is almost everywhere, including the distribution of fusion product energy, the derivation of the Lawson criterion's energy balance, and so on.

The second law of thermodynamics has several expressions: Clausius' formulation states that heat can spontaneously transfer from a higher temperature object to a lower temperature object, but it is not possible for heat to spontaneously transfer from a lower temperature object to a higher temperature object; Kelvin-Planck's formulation states that it is impossible to take heat from a single heat source and completely convert it into work without producing other effects. And the entropy increase formulation states that the entropy of an isolated system never decreases. This law is also evident in fusion, for example, it explains why fusion energy research is essentially approaching thermal equilibrium, why the hot ion mode is difficult to sustain, and why the temperature and velocity differences between different components tend to converge without an external source. This law also points out that a "second kind of perpetual motion machine" is not possible, and that the thermal efficiency of a heat engine cannot reach 100\%.

We usually do not need to pay extra attention to the other two thermodynamic laws in fusion research. The third law of thermodynamics states that the entropy value of a perfect crystal of all pure substances is zero at absolute zero temperature, or absolute zero temperature (T=0K) is unattainable. The zeroth law of thermodynamics states that if two thermodynamic systems are in thermal equilibrium with a third thermodynamic system, then they must also be in thermal equilibrium with each other, meaning that thermal equilibrium is transitive.

In general, some potential breakthrough directions in fusion energy research are mainly related to the second law of thermodynamics, such as improving power generation efficiency, increasing fusion reaction rate through non-thermalized plasma, and reducing fusion conditions through the hot ion mode, etc. The extent to which these potential solutions can achieve breakthroughs is also limited by the second law, and its quantitative value is mainly determined by solving the Fokker-Planck equation for the evolution of particle distribution function in collisions. Although this book mainly uses its simplest reduced model for discussion, such as most calculations can be replaced by characteristic times such as collisions, heat exchange, and slowing down. Solutions that violate this law are not feasible. Rider (1995) excluded the feasibility of some non-thermalized fusion methods through the calculation of the Fokker-Planck equation, pointing out that the energy consumption required to maintain the corresponding non-thermalized state is greater than the energy output of fusion.

\section{Related Major References}

Bishop (1958) is the first decrypted document systematically explaining the US Sherwood fusion energy research program. We can see that most of the core issues of fusion discussed today, such as fusion schemes, temperature and density requirements, heating, diagnostics, radiation, etc., have been addressed. Glasstone (1960) is an early and more comprehensive book introducing the basic principles and methods of controlled fusion.
Freidberg (2007) is a good textbook and has excellent descriptions of collisions, transport, fusion physics and engineering, and some fusion schemes.
Roth (1986) systematically elaborated on fusion energy schemes based on engineering constraints.
Teller (1981) discusses various aspects of magnetic confinement in detail, and the last chapter of the book, written by Dawson, is an important literature on non-deuterium-tritium fusion research. Wesson (2011) elaborates on various aspects of tokamaks, mainly focusing on deuterium-tritium fusion.
Atzeni (2004) focuses on inertial confinement fusion, and its first chapter provides a detailed and comprehensive explanation of fusion nuclear reactions, making it a comprehensive reference. Parisi (2018) describes the status of fusion energy research and the basic problems and challenges involved from a popular science perspective.
Long (2018) provides a systematic evaluation of the prospects of magnetically inertially confined fusion.Ryzhkov (2019) studied magneto-inertial confined fusion. Reinders (2021) provided a systematic review of fusion research, but the conclusions were extremely pessimistic and disappointing about fusion energy. Chen (2011) provided a popular science introduction to various aspects of fusion energy research.
Hartwig (2017) discussed the underlying logic of fusion and concluded that the study of tokamak with strong magnetic field is the most worthwhile direction.
Kikuchi (2012) provided a detailed introduction to the physics of fusion. Dolan (1981) extensively discussed the principles, experiments, and engineering issues of fusion, while Dolan (2013) focused on the engineering aspects of magnetically confined fusion. Gross (1984) summarized a large amount of fusion and plasma related formulas and data. Choi (1977) published conference proceedings that provided a detailed summary of research on advanced fuel fusion at the time. Rose (1961), Artsimovich (1965), Kammash (1975), Raeder (1986) were early and comprehensive summaries of macroscopic fusion research. Lu Hefu (1960) published the first detailed summary of fusion research in China. There have also been publications in China that range from popular science to professional introductions to the principles of fusion, such as Hu Xiwei (1981), Zhu Shiyao (1992), Li Yin'an (1992), Wang Naiyan (2001), and Wang Ganchang (2005). 
\chapter{Appendix II: Tokamak System Code}\label{chap:tokamak}

The system code refers to a complete mathematical model that combines the basic physical, engineering, and economic parameters and constraints to design the device's fundamental parameters. It is usually a zero-dimensional model that involves simple function equations and does not involve complex differential equations. In literature, system codes for Tokamak parameter design are frequently used. Due to the use of some complex but unreliable scaling laws, they often have parameter sensitivity, which means that a slight change in one parameter may lead to significant differences in the results.

A complete system code generally includes three modules: physics, engineering, and economics. Here, we establish a physical design model for Tokamak-based devices, without considering detailed engineering and economic modules. We examine the parameter dependency relationships and use definitions, nuclear reaction cross-sections, radiation, engineering limits, among others, as the main model basis. We treat some core parameters as input variables of the model, and for less reliable formulas (such as ITER98 confinement time scaling rate), we adopt a posteriori method to avoid parameter sensitivity. In this model, whether it is the L-mode, H-mode, or other confinement modes, they are all dealt within the same model. Their differences are only reflected in a few input parameters, such as profile factors and confinement time factors, or a few model equations. The results are eventually presented in the form of Plasma Operating Contours (POPCON), so as to form an optimized design for the parameters.

Compared to the zero-dimensional model mentioned earlier, this model mainly adds the effects of profile non-uniformity and introduces some physics related to actual experiments, making it more consistent with the actual device.

\section{Model}

The model is divided into two parts: geometric configuration and physics, as well as some main posterior physical quantities.

\subsection{Configuration and Geometric Relationships}
\begin{figure}[htbp]
\begin{center}
\includegraphics[width=10cm]{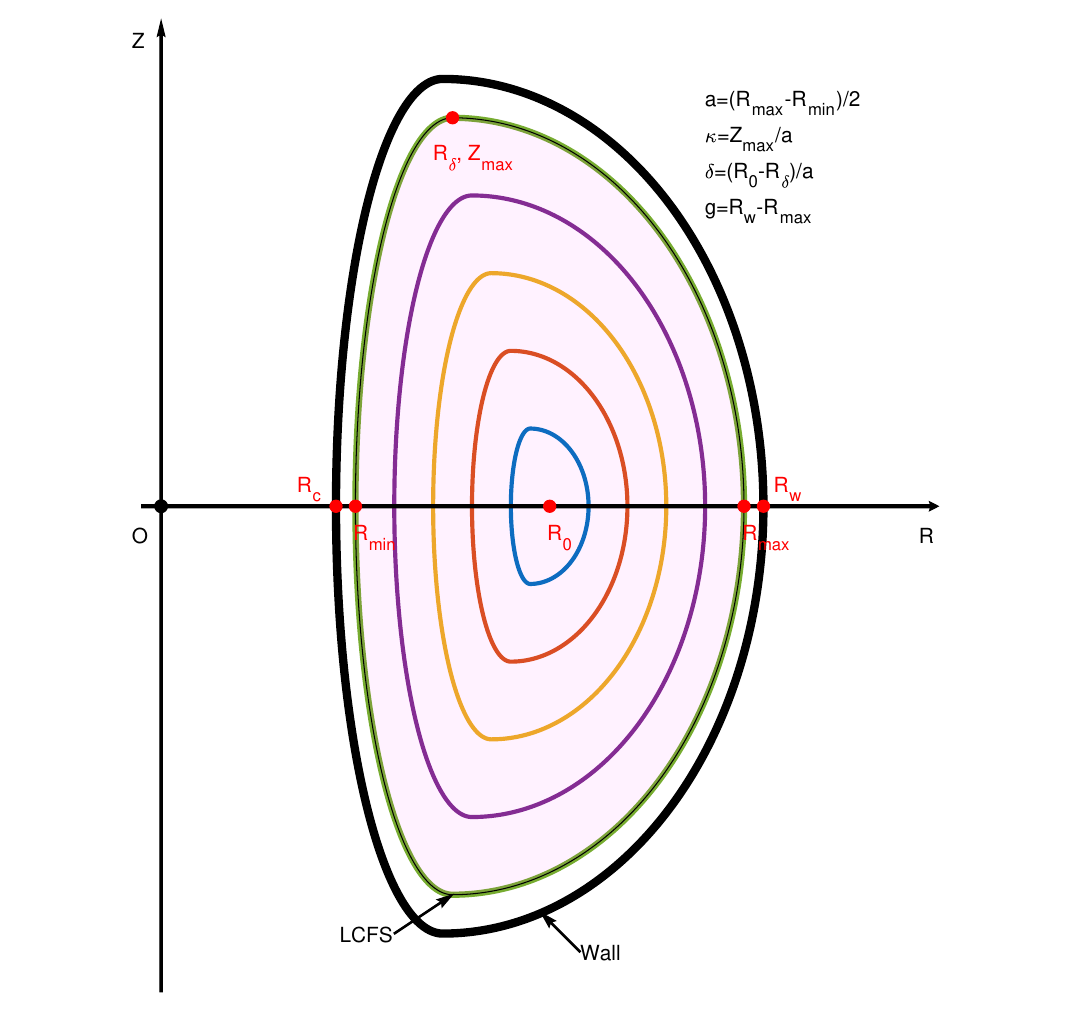}\\
\caption{A poloidal cross-sectional model of a tokamak configuration.}\label{fig:tokamakshape}
\end{center}
\end{figure}

In this model, as shown in Figure \ref{fig:tokamakshape}, we assume that the inner wall of the device and the outer closed magnetic surface are both composed of two concatenated ellipses on the high-field side and the low-field side. The geometric relationship of the device satisfies the references [Stambaugh (1998, 2011), Petty (2003)]. The equation satisfied by the outer closed magnetic surface curve on the left side of $(R_\delta, Z_{max})$ is given by
\begin{eqnarray}
    (R-R_0+\delta a)^2+(\frac{1-\delta}{\kappa})^2Z^2=a^2(1-\delta)^2,
\end{eqnarray}
The design shape of the left wall of the device is given by
\begin{eqnarray}
    [R-R_0+\delta (a+g)]^2+(\frac{1-\delta}{\kappa})^2Z^2=(a+g)^2(1-\delta)^2.
\end{eqnarray}
The equation satisfied by the outer closed magnetic surface curve on the right side of $(R_\delta, Z_{max})$ is given by
\begin{eqnarray}
    (R-R_0+\delta a)^2+(\frac{1+\delta}{\kappa})^2Z^2=a^2(1+\delta)^2,
\end{eqnarray}
The design shape of the right wall of the device is given by
\begin{eqnarray}
    [R-R_0+\delta (a+g)]^2+(\frac{1+\delta}{\kappa})^2Z^2=(a+g)^2(1+\delta)^2.
\end{eqnarray}
In the above model, the semi-widths of the ellipses on the left and right sides of the plasma are $a(1-\delta)$ and $a(1+\delta)$, respectively, and the height is $Z_{max}=\kappa a$. The center in the major radius direction is $R_0$. For the device wall, similar to the plasma, the center in the major radius direction is also $R_0$, and the semi-width and height of the left and right ellipses are replaced by $a+g$ in the expressions.

As the geometric parameters for the input are $(R_0, A, \delta, \kappa, g)$, where $R_0$ is the major radius of the plasma, $a$ is the minor radius, $\delta$ is the triangularity, $\kappa$ is the elongation, $A=R_0/a$ is the aspect ratio, and $g$ is the distance from the plasma boundary to the device wall. Thus, the ratio of the radius from the vacuum chamber inner wall to the center and the device aspect ratio are respectively given by
\begin{eqnarray}
    R_c=R_0-a-g,~~~ A_d=\frac{R_0}{g+a},
\end{eqnarray}
The expressions for the plasma volume and the device volume are respectively given by
\begin{eqnarray}
    V_p=[2\pi^2\kappa(A-\delta)+\frac{16}{3}\delta\pi\kappa] a^3,\\
    V_d=[2\pi^2\kappa(A_d-\delta)+\frac{16}{3}\delta\pi\kappa] (a+g)^3,
\end{eqnarray}
The approximate expressions for the plasma and device surface areas are respectively given by (where Petty (2003) used $A_{wall}=(2\pi R)(2\pi a)\sqrt{0.5(1+\kappa^2)}$, the surface area is only used when calculating wall loads, and does not affect the calculation of other quantities)
\begin{eqnarray}
    S_p=(4\pi^2A\kappa^{0.65}-4\kappa\delta) a^2,\\
    S_w=(4\pi^2A_d\kappa^{0.65}-4\kappa\delta) (a+g)^2.
\end{eqnarray}
If $g=0$, it means that the plasma is in tight contact with the inner wall of the device, $A_d=A$, $V_p=V_d$, $S_p=S_w$. It should be noted that the major radius $R_0$ defined here is the radius of the geometric center of the plasma, while the actual equilibrium configuration usually has a certain difference in the radius of the magnetic axis, which comes from the Shafranov shift. This difference has a slight impact on the design parameters, but does not bring about any essential effect on the zero-dimensional design parameters. Here for simplicity, we assume that $R_0$ is the magnetic axis itself.

\begin{figure}[htbp]
\begin{center}
\includegraphics[width=10cm]{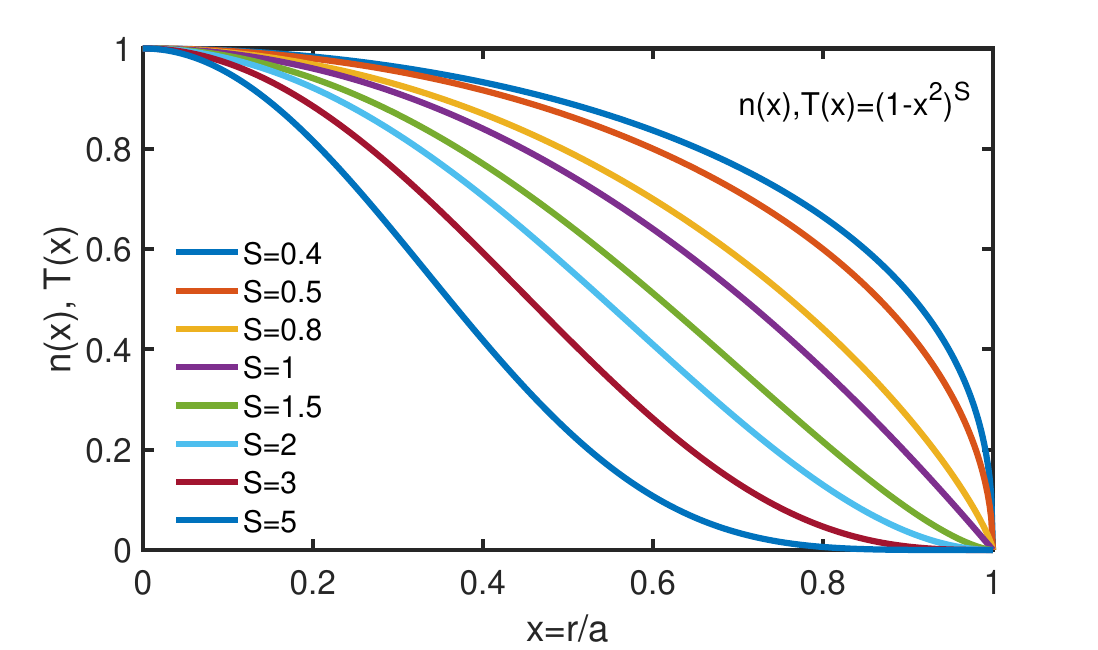}\\
\caption{Radial profiles of temperature and density.}\label{fig:nTprofile}
\end{center}
\end{figure}In this model, we assume the form of temperature and density profiles shown in Figure \ref{fig:nTprofile}. The model equations are given by
\begin{eqnarray}
    n(x)=n_0(1-x^2)^{S_n},\\
    T(x)=T_0(1-x^2)^{S_T},
\end{eqnarray}
which approximate the shape of the plasma profile with pedestal under the H-mode. Here, $x=r/a$, where $r$ is the scale of the minor radius direction away from the magnetic axis $R_0$, and the subscript '0' represents the value on the magnetic axis. The pedestal effect is enhanced as $S_n$ and $S_T$ decrease. By taking averages over the profile, we obtain volume-averaged and line-averaged values
\begin{eqnarray}
    \langle n\rangle=\int_0^1n_0(1-x^2)^{S_n}2xdx=\frac{n_0}{1+S_n},\\
    \langle T\rangle=\int_0^1T_0(1-x^2)^{S_T}2xdx=\frac{T_0}{1+S_T},\\
    \langle n\rangle_l=\int_0^1n_0(1-x^2)^{S_n}dx=\frac{\sqrt{\pi}}{2}\frac{\Gamma(S_n+1)}{\Gamma(S_n+1.5)}n_0,\\
    \langle T\rangle_l=\int_0^1T_0(1-x^2)^{S_T}dx=\frac{\sqrt{\pi}}{2}\frac{\Gamma(S_T+1)}{\Gamma(S_T+1.5)}T_0.
\end{eqnarray}
We assume that the density profile factor for all components is the same $S_n$, and the temperature profile factor is also the same $S_T$.

\subsection{Physical Relationships}

The plasma ion composition consists of two types of ions participating in nuclear reactions, helium ash, and other impurities, with corresponding densities of $n_1, n_2, n_{He}, n_{imp}$ and electric charges of $Z_1, Z_2, Z_{He}, Z_{imp}$, where $Z_{He}=2$. According to the quasi-neutrality condition, the ion and electron densities are given by
\begin{eqnarray}
n_i=(n_1+n_2)/(1+\delta_{12})+n_{He}+n_{imp},\\
    n_e=(n_1Z_1+n_2Z_2)/(1+\delta_{12})+n_{He}Z_{He}+n_{imp}Z_{imp},
\end{eqnarray}
where $\delta_{12}=1$ for the same type of ion and $\delta_{12}=0$ for different types of ions. The average charge number and effective charge number are defined as
\begin{eqnarray}
    Z_{i}=n_e/n_i,\\
    Z_{eff}=[(n_1Z_1^2+n_2Z_2^2)/(1+\delta_{12})+n_{He}Z_{He}^2+n_{imp}Z_{imp}^2]/n_e.
\end{eqnarray}
We assume that the temperature of all ions is $T_i$ and the electron temperature is $T_e=f_TT_i$. The densities of the two main ions participating in the reaction are given by $n_{12}=(n_1+n_2)/(1+\delta_{12})$, with the fractions $f_{12}=n_{12}/n_i$ for the mixed ions, $f_{He}=n_{He}/n_i$ for the helium ions, and $f_{imp}=n_{imp}/n_i$ for the impurities. Furthermore, we set the proportion of the first type of ion as $x_1=n_1/n_{12}$ and the proportion of the second type of ion as $x_2=n_2/n_{12}$. If the two ions are the same, $x_1=x_2=1$; if they are different, $x_2=1-x_1$. Therefore, we have
\begin{eqnarray}
f_{12}+f_{He}+f_{imp}=1,\\
f_{12}(x_1Z_1+x_2Z_2)/(1+\delta_{12})+f_{He}Z_{He}+f_{imp}Z_{imp}=Z_i.
\end{eqnarray}

Total fusion power is given by 
\begin{eqnarray}
P_{fus}=\frac{Y}{1+\delta_{12}}\int n_1n_2\langle\sigma v\rangle dV=\frac{Y}{1+\delta_{12}}n_{10}n_{20}\Phi \cdot V_p,\\
\Phi=2\int_0^1(1-x^2)^{2S_n}\langle\sigma v\rangle xdx,
\end{eqnarray}
where $P_{fus}$ is the fusion power, $Y$ is the energy per single nuclear reaction, $\delta_{12}$ is the dilution factor, $n_1$ and $n_2$ are the number densities of the reacting species, $\langle\sigma v\rangle$ is the reaction rate, $V$ is the plasma volume, and $\Phi$ is a factor related to the reaction rate. The reaction rate $\langle\sigma v\rangle$ is a function of the ion effective temperature $T_i(x)$. Taking into account the potential reaction rate enhancement, we assume that these factors can be combined into an amplification factor $f_\sigma$, relative to the Maxwellian distribution, such that
\begin{eqnarray}
\langle\sigma v\rangle=f_\sigma \langle\sigma v\rangle_M.
\end{eqnarray}

Bremsstrahlung power [Nevins (1998)]
\begin{eqnarray}\nonumber
P_{brem}=C_Bn_{e0}^2\sqrt{k_BT_{e0}}V_p\Big\{Z_{eff}\Big[\frac{1}{1+2S_n+0.5S_T}+\frac{0.7936}{1+2S_n+1.5S_T}\frac{k_BT_{e0}}{m_ec^2}\\
+\frac{1.874}{1+2S_n+2.5S_T}\Big(\frac{k_BT_{e0}}{m_ec^2}\Big)^2\Big]+\frac{3}{\sqrt{2}(1+2S_n+1.5S_T)}\frac{k_BT_{e0}}{m_ec^2}\Big\} ~{\rm [MW]}.
\end{eqnarray}
where $C_B=5.34\times10^{-43}$, $n_{e0}$ is the electron density, $T_{e0}$ is the electron temperature, $Z_{eff}$ is the effective charge, $S_n$ is a parameter related to the shape of the electron temperature profile, and $S_T$ is a parameter related to the shape of the electron density profile. The temperature $k_BT_e$ and the energy $m_ec^2$ are in keV, and the density $n_e$ is in ${\rm m^{-3}}$.

Synchrotron radiation power is taken as in
[Costley (2015)] [from Kukushkin (2009)]
\begin{eqnarray}
P_{cycl}=4.14\times10^{-7}n_{eff}^{0.5}T_{eff}^{2.5}B_{T0}^{2.5}(1-R_w)^{0.5}\Big(1+2.5\frac{T_{eff}}{511}\Big)\cdot\frac{1}{a_{eff}^{0.5}}V_p~{\rm [MW]},\\
n_{eff}=\langle n_e\rangle=n_{e0}/(1+S_n),~~
a_{eff}=a\kappa^{0.5},\\
T_{eff}=T_{e0}\int_0^1(1-x^2)^{S_T}dx\neq \langle T_e\rangle=2T_{e0}\int_0^1(1-x^2)^{S_T}xdx,
\end{eqnarray}
where $n_{eff}$ is the effective electron density, $T_{eff}$ is the effective electron temperature, $B_{T0}$ is the magnetic field, $R_w$ is the wall reflection coefficient, $a_{eff}$ is the effective minor radius, and $\kappa$ is a parameter related to the elongation of the plasma. The temperature $T_e$ is in keV, the density $n_{eff}$ is in $10^{20}{\rm m^{-3}}$, and the magnetic field $B_{T0}$ is in T.

It is difficult to accurately estimate the magnitude of the synchrotron radiation, so for the purposes of the conclusions drawn, we can simply attribute it to the requirement for a higher wall reflection coefficient $R_w$, i.e., if the actual synchrotron radiation is greater than the value given by the above equation, it means that a higher wall reflection coefficient $R_w$ is required.

Internal energy in plasma
\begin{eqnarray}
    E_{th}=\frac{3}{2}k_B\int(n_eT_e+n_iT_i)dV=\frac{3}{2}k_B\frac{n_{i0}T_{i0}+n_{e0}T_{e0}}{1+S_n+S_T}V_p.
\end{eqnarray}
Energy balance at steady state
\begin{eqnarray}
\frac{dE_{th}}{dt}=-\frac{E_{th}}{\tau_E}+f_{ion}P_{fus}+P_{heat}-P_{rad}=0,
\end{eqnarray}
We adopt the approach from the conventional system modeling code [Costley (2015)], where the energy of the charged products in fusion power $P_{fus}$ is fully considered as energy input. For example, in the case of deuterium-tritium fusion, the energy of the charged products accounts for only 1/5 of the total fusion energy, $f_{ion}=0.2$; for hydrogen-boron fusion, the energy of the fusion products is carried entirely by the charged alpha particles, $f_{ion}=1$. The radiation power only considers bremsstrahlung and cyclotron radiation, $P_{rad}=P_{brem}+P_{cycl}$. $P_{heat}$ represents the external heating power, thus we have
\begin{eqnarray}
P_{heat}=P_{rad}+\frac{E_{th}}{\tau_E}-f_{ion}P_{fus}.
\end{eqnarray}
Fusion gain factor is defined as
\begin{eqnarray}
Q_{fus}\equiv \frac{P_{fus}}{P_{heat}}.
\end{eqnarray}

\subsection{Posterior parameters}

Plasma beta
\begin{eqnarray}
    \beta_T=\frac{2\mu_0k_B\int(n_iT_i+n_eT_e)dV}{B_{T0}^2}=\frac{2\mu_0k_B(n_iT_i+n_eT_e)}{(1+S_n+S_T)B_{T0}^2}.
\end{eqnarray}
Greenwald density limit
\begin{eqnarray}
    n_{Gw}=10^{20}\times\frac{I_p}{\pi a^2},
\end{eqnarray}
where the minor radius $a$ is in meters, plasma current $I_p$ is in mega-amperes (MA), and $n_{Gw}$ is in ${\rm m^{-3}}$. Line-averaged density
\begin{eqnarray}
    {\bar{n}_0}=n_{e0}\frac{\sqrt{\pi}}{2}\frac{\Gamma(S_n+1)}{\Gamma(S_n+1.5)}\simeq\frac{2n_{e0}}{2+S_n}.
\end{eqnarray}
The approximate expression commonly used in the later reference is exact for $S_n=0$ and $1$, and has certain accuracy for $0<S_n<2$, where $\Gamma$ is the Euler gamma function. For a tokamak, usually ${\bar{n}_0}/{n_{Gw}}<1$, which requires that the current $I_p$ cannot be too small.
We can use a simple formula for the safety factor [Costley (2015)]
\begin{eqnarray}
    q=\frac{5B_{T0}a^2\kappa}{R_0I_p},
\end{eqnarray}
or a more complex fitting formula with a wide range [Petty (2003)]
\begin{eqnarray}
    q_{95}=\frac{5B_{T0}a^2G}{R_0I_p},~G=0.5[1+\kappa^2(1+2\delta^2-1.2\delta^3)]\frac{(1-0.26255/A+1.3333/A^2)}{(1-1/A^2)^{1.462378}},
\end{eqnarray}
Since $q_{95}>2$ is usually required to avoid plasma disruption, this implies that the current $I_p$ cannot be too large.
Normalized beta pressure
\begin{eqnarray}
    \beta_N=100\beta_T\frac{aB_{T0}}{I_p},
\end{eqnarray}
where the minor radius $a$ is in meters, plasma current $I_p$ is in MA, and magnetic field $B_{T0}$ is in Tesla (T). Instability typically requires $\beta_N<12/A$, which also implies that the current $I_p$ cannot be too large.
Poloidal beta pressure
\begin{eqnarray}
    \beta_p=\frac{25}{\beta_T}\frac{1+\kappa^2}{2}(\beta_N/100)^2.
\end{eqnarray}
Thermal load on the unit area of the wall
\begin{eqnarray}
    P_{wall}=\frac{P_{fus}+P_{heat}}{S_w},
\end{eqnarray}
usually has an upper limit.

The above formulas and models are based on standard geometric relationships or definitions, or are calculated based on more fundamental physical processes. They have strong generality and are not highly sensitive to parameters or detailed operating modes of the device. There are also some physical relationships for which there are currently no reliable models. Therefore, we use these formulas as post-formulas, meaning that we compare the results obtained from the models with these formulas. This includes energy-constrained time scaling laws, current drive efficiency, L-H transition power threshold, instability conditions, etc.ITER98 Energy Confinement Scaling Law and H98 Factor
\begin{eqnarray}
    \tau_E^{IPB98}=0.145I_p^{0.93}R_0^{1.39}a^{0.58}\kappa^{0.78}{\bar{n}}_{20}^{0.41}B_{T0}^{0.15}M^{0.19}P_L^{-0.69},\\
    H_{98}=\tau_E/\tau_E^{IPB98},
\end{eqnarray}
where the total power
\begin{eqnarray}
    P_L=P_{heat}+f_{ion}P_{fus},
\end{eqnarray}
is measured in MW, and the average mass number $M=(x_1A_1+x_2A_2)/(1+\delta_{12})$. The plasma current $I_p$ is measured in MA, the major and minor radii $R_0$ and $a$ are measured in meters, the magnetic field $B_{T0}$ is measured in T, and the average density is measured in units of ${\rm 10^{20}m^{-3}}$. For the design of conventional tokamaks, the energy confinement scaling law ITER98 can be used for post-dictions, while for some special cases, other energy confinement scaling laws, such as for low aspect ratio spherical tori, can be used[Kurskiev (2022)].

\vspace{10pt}
In conclusion, the model input parameters are: $T_{i0}, n_{e0}, f_{T}, x_1, f_{He}, f_{imp}, Z_{imp}, B_{T0},\\ \tau_E, R_w, f_\sigma$, as well as geometrical and profile-related parameters $S_n, S_T, R_0, A, g, \kappa, \delta$, and the plasma current $I_p$. For $I_p$, it can also be replaced with $\beta_N$ as an input parameter; for $B_{T0}$, it can also be replaced with $\beta_T$ as an input parameter. Here, we set the energy confinement time $\tau_E$ as an input parameter to avoid the sensitivity brought by complex energy confinement scaling laws. Finally, the energy confinement scaling law is used as a post-diction parameter to determine whether $H_{98}$ is within a reasonable range.

The main output parameters are: $Q_{fus}, P_{fus}, P_{brem}, P_{cycl}, E_{th}, P_{heat}, \beta_N, \beta_T, {\bar n}, q_{95},\\ V_p, S_p$, etc.

\section{Results}

We calculate the typical results for deuterium-tritium fusion and hydrogen-boron fusion.

\begin{figure}[htbp]
\begin{center}
\includegraphics[width=15cm]{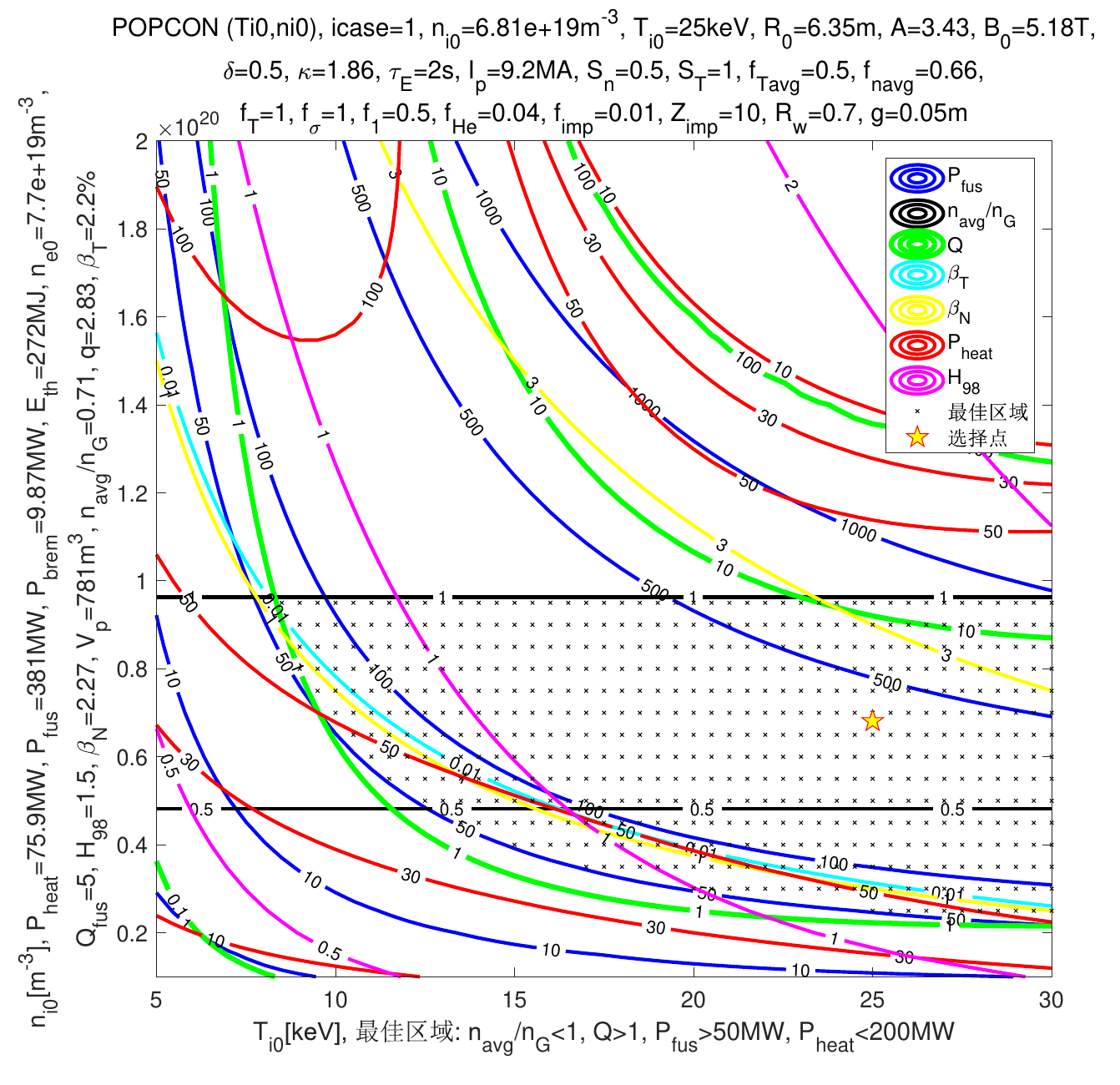}\\
\caption{System model parameter calculation for deuterium-tritium parameters in ITER tokamak.}\label{fig:tokamakdt}
\end{center}
\end{figure}

\begin{figure}[htbp]
\begin{center}
\includegraphics[width=15cm]{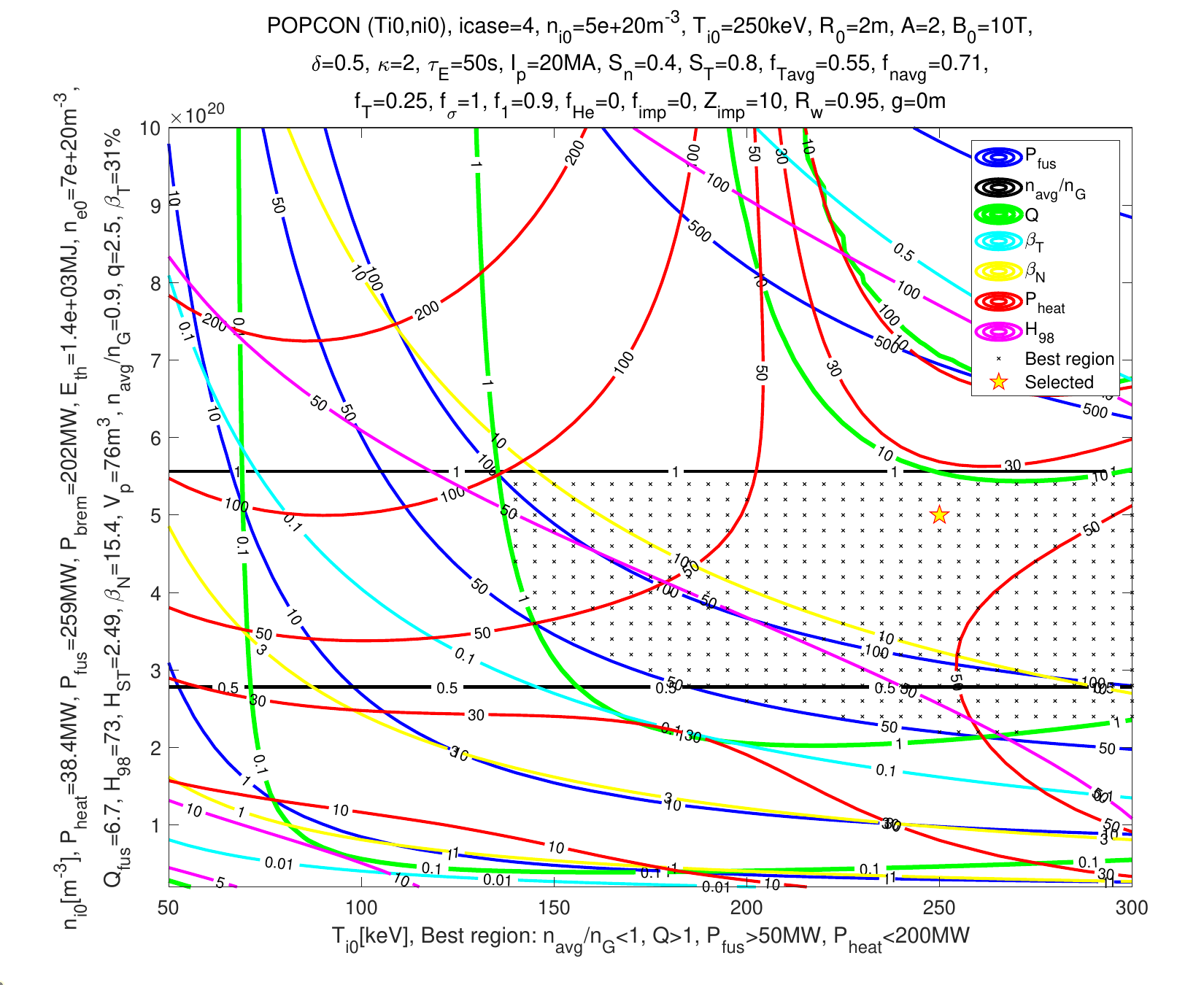}\\
\caption{System model calculation of typical hydrogen-boron tokamak configurations.}\label{fig:tokamakpb}
\end{center}
\end{figure}

For deuterium-tritium fusion parameters, we use the parameters from ITER, and the results are shown in Figure \ref{fig:tokamakdt}, which is similar to the results in the appendix of Costley (2015). For hydrogen-boron fusion, as shown in Figure \ref{fig:tokamakpb}, it can be seen that there are high requirements for energy confinement and other constraints. If the ITER98 scaling law is adopted, the confinement factor $H_{98}=73\gg1$, which is challenging with the current technological capabilities. However, if the spherical tori scaling law is applicable at high parameters, the required confinement factor $H_{ST}\simeq2.49$, indicating that it is not completely infeasible.

\printindex

\end{CJK*}
\end{document}